\documentclass[pra,longbibliography,twocolumn,showpacs,floatfix,amsmath,amssymb,superscriptaddress]{revtex4-1}
\usepackage{graphicx}
\usepackage{tabularx}
\usepackage{times}
\usepackage{color}
\usepackage{latexsym,amsmath,amssymb,bm,dsfont,braket}
\usepackage[matrix,arrow,curve]{xy}
\usepackage{hyperref}

\def\usearXivsubfig{\relax}
\ifdefined\usearXivsubfig
	\usepackage{subfigure}
	\newcommand{\subfigref}[1]{\ref{#1}}
\else
	\usepackage[caption=false,labelformat=parens,subrefformat=parens]{subfig}
	\newcommand{\subfigure}[2][{}]{\subfloat[{#1}]{#2}}
	\newcommand{\subfigref}[1]{\subref*{#1}}
\fi

\newcommand{\bbone}{ {\mathds{1}} }
\newcommand{\I}{ {\mathds{1}} }
\newcommand{\SUtwo}{ {\mathrm{SU(2)}} }

\begin{document}


\title{Universal topological quantum computation from a superconductor/Abelian quantum Hall heterostructure}

\author{Roger S. K. Mong}
\affiliation{Department of Physics, California Institute of Technology, Pasadena, CA 91125, USA}

\author{David J. Clarke}
\affiliation{Department of Physics, California Institute of Technology, Pasadena, CA 91125, USA}

\author{Jason Alicea}
\affiliation{Department of Physics, California Institute of Technology, Pasadena, CA 91125, USA}

\author{Netanel H. Lindner}
\affiliation{Department of Physics, California Institute of Technology, Pasadena, CA 91125, USA}

\author{Paul Fendley}
\affiliation{Department of Physics, University of Virginia, Charlottesville, VA 22904, USA}

\author{Chetan Nayak}
\affiliation{Microsoft Research, Station Q, University of California, Santa Barbara, CA 93106, USA}
\affiliation{Department of Physics, University of California, Santa Barbara, CA 93106, USA}

\author{Yuval Oreg}
\affiliation{Department of Condensed Matter Physics, Weizmann Institute of Science, Rehovot, 76100, Israel}

\author{Ady Stern}
\affiliation{Department of Condensed Matter Physics, Weizmann Institute of Science, Rehovot, 76100, Israel}

\author{Erez Berg}
\affiliation{Department of Condensed Matter Physics, Weizmann Institute of Science, Rehovot, 76100, Israel}

\author{Kirill Shtengel}
\affiliation{Department of Physics and Astronomy, University of California, Riverside, California 92521, USA}
\affiliation{Institute for Quantum Information, California Institute of Technology, Pasadena, CA 91125, USA}

\author{Matthew P. A. Fisher}
\affiliation{Department of Physics, University of California, Santa Barbara, CA 93106, USA}

\begin{abstract}
{Non-Abelian anyons promise to reveal spectacular features of quantum mechanics that could ultimately provide the foundation for a decoherence-free quantum computer.  A key breakthrough in the pursuit of these exotic particles originated from Read and Green's observation that the Moore-Read quantum Hall state and a (relatively simple) two-dimensional $p+ip$ superconductor both support so-called Ising non-Abelian anyons.  Here we establish a similar correspondence between the $\mathbb{Z}_3$ Read-Rezayi quantum Hall state and a novel two-dimensional superconductor in which charge-$2e$ Cooper pairs are built from fractionalized quasiparticles.  In particular, both phases harbor Fibonacci anyons that---unlike Ising anyons---allow for universal topological quantum computation solely through braiding.  Using a variant of Teo and Kane's construction of non-Abelian phases from weakly coupled chains, we provide a blueprint for such a superconductor using Abelian quantum Hall states interlaced with an array of superconducting islands.  Fibonacci anyons appear as neutral deconfined particles that lead to a two-fold ground-state degeneracy on a torus.  In contrast to a $p+ip$ superconductor, vortices do not yield additional particle types yet depending on non-universal energetics can serve as a trap for Fibonacci anyons.  These results imply that one can, in principle, combine well-understood and widely available phases of matter to realize non-Abelian anyons with universal braid statistics.  Numerous future directions are discussed, including speculations on alternative realizations with fewer experimental requirements.
} 
\end{abstract}

\maketitle

\section{Introduction}

The emergence of anyons that exhibit richer exchange statistics than the constituent electrons and ions in a material is among the most remarkable illustrations of `more is different'.  Such particles fall into two broad categories: Abelian and non-Abelian.  Interchanging Abelian anyons alters the system's wavefunction by a phase $e^{i\theta}$ intermediate between that acquired for bosons and fermions~\cite{Leinaas,WilczekAnyons}.  Richer still are non-Abelian anyons, whose exchange rotates the system's quantum state amongst a degenerate set of locally indistinguishable ground states produced by the anyons~\cite{Bais,Goldin81,Goldin84,MooreSeiberg1,MooreSeiberg2,Witten,Fredenhagen,Froehlich,Imbo89,Alford1,Alford2}.  The latter variety realize the most exotic form of exchange statistics that nature in principle permits, which by itself strongly motivates their pursuit.  Non-Abelian anyons are further coveted, however, because they provide a route to fault-tolerant topological quantum computation~\cite{kitaev,Freedman98,Freedman03,MeasurementOnlyTQC,TQCreview}. Here, qubits are embedded in the system's ground states and, by virtue of non-Abelian statistics, manipulated through anyon exchanges.  The non-locality with which the information is stored and processed elegantly produces immunity against decoherence stemming from local environmental perturbations.  One thereby sidesteps the principal bottleneck facing most quantum computing approaches, but at the expense of introducing a rather different challenge: identifying suitable platforms for non-Abelian excitations.  

The quantum Hall effect catalyzed numerous breakthroughs in the search for anyons in physical systems \cite{TQCreview,SternReview}.  Quantum Hall states supporting fractionally charged Abelian anyons are by now widely believed to surface in a myriad of settings including $\mathrm{GaAs}$ \cite{GaAsFQHE}, graphene \cite{GrapheneFQHE1,GrapheneFQHE2}, oxide interfaces \cite{OxideFQHE1,OxideFQHE2}, and $\mathrm{CdTe}$ \cite{CdTeFQHE} among others.
Moreover, Moore and Read suggested in 1991 that the quantum Hall regime could support non-Abelian anyons and constructed a candidate state---a quantum Hall fluid in which composite fermions undergo $p+ip$ pairing \cite{MooreReadNonabelion91}.
This phase supports chiral edge states consisting of a neutral Majorana sector coupled to a bosonic charge mode \cite{Milovanovic96}, along with Ising non-Abelian anyons \footnote{The term `Ising anyon' refers to a non-Abelian particle whose nontrivial braiding statistics derives from bound Majorana zero-modes.  Strictly speaking Ising anyons have a particular overall $\mathrm{U(1)}$ phase associated with their braiding, though we will use this terminology even when this overall phase is ill-defined.} carrying charge $e/4$ in the bulk \cite{Nayak96c,Gurarie1997,Tserkovnyak03,Seidel08,Read2008,Baraban09,Prodan09,Bonderson11b}.
A variety of experiments support the onset of the Moore-Read state (or its particle-hole conjugate~\cite{antiPfaffian1,antiPfaffian2}) at filling factor $\nu = 5/2$ in $\mathrm{GaAs}$ quantum wells~\cite{FiveHalvesDiscovery,Willett1,Willett2,FiveHalvesTunneling,FiveHalvesCharge,FiveHalvesNeutralModes,FiveHalvesKang,FiveHalvesSpinPolarization,Willett3}.  It is important to remark, however, that braiding Ising anyons does not produce a gate set sufficient for universal topological quantum computation.  Thus more exotic non-Abelian phases that do not suffer from this shortcoming are highly desirable.

Quantum Hall systems can, in principle, host non-Abelian anyons with universal braid statistics (i.e., that allow one to approximate an arbitrary unitary gate with braiding alone).  In this context the $\mathbb{Z}_3$ Read-Rezayi state~\cite{ReadRezayi}, which generalizes the pairing inherent in the Moore-Read phase to clustering of triplets of electrons~\cite{FendleyFisherNayak2}, constitutes the `holy grail'.  Chiral edge states with a very interesting structure appear here: a charged boson sector that transports electrical current (as in all quantum Hall states) in this case coexists with a neutral sector that carries only energy and is described by the chiral part of $\mathbb{Z}_3$ parafermion conformal field theory.  As a byproduct of this neutral sector the bulk admits vaunted `Fibonacci' anyons---denoted $\varepsilon$---that obey the fusion rule $\varepsilon \times \varepsilon \sim \I + \varepsilon$.  This fusion rule implies that the low-energy Hilbert space for $n$ $\varepsilon$ particles with trivial total topological charge has a dimension given by the $(n-1)$\textsuperscript{th} Fibonacci number.  Consequently, the asymptotic dimension per particle,
		  usually called the quantum dimension, is the golden ratio $\varphi \equiv (1+\sqrt{5})/2$.  Perhaps the most remarkable feature of Fibonacci anyons is that they 
		  allow for universal topological quantum computation in which a single gate---a counterclockwise exchange of two Fibonacci anyons---is sufficient to approximate any unitary transformation to within desired accuracy (up to an inconsequential overall phase).  Such particles remain elusive, though the $\mathbb{Z}_3$ Read-Rezayi state and its particle-hole conjugate~\cite{AntiReadRezayi} do provide plausible candidate ground states for fillings $\nu = 13/5$ and $12/5$.  Intriguingly, a plateau at the latter fraction has indeed been measured in $\mathrm{GaAs}$, though little is presently known about the underlying phase; at $\nu = 13/5$ a well-formed plateau has so far eluded observation~\cite{Pan,Xia,SecondLL}.  

		  Read and Green~\cite{ReadGreen} laid the groundwork for the pursuit of non-Abelian anyons outside of the quantum Hall effect by demonstrating a profound correspondence between the Moore-Read state and a spinless 2D $p+ip$ superconductor
		  \footnote{Throughout, when referring to spinless $p$-wave superconductivity we implicitly mean the topologically nontrivial weak pairing phase.}.
		  Many properties that stem from composite-fermion pairing indeed survive in the vastly different case where physical electrons form Cooper pairs.  In particular, both systems exhibit a chiral Majorana edge mode at their boundary and support Ising non-Abelian anyons in the bulk.  Several important distinctions between these phases do, nevertheless, persist:  $(i)$ Their edge structures are not identical---a $p+ip$ superconductor lacks the chiral bosonic charge mode found in Moore-Read.  $(ii)$ Different classes of topological phenomena arise in each case.  On one hand a $p+ip$ superconductor realizes a topological superconducting phase with short-range entanglement; the Moore-Read state, on the other, exhibits true topological order, long-range entanglement, and hence nontrivial ground-state degeneracy on a torus.  This important point closely relates to the next two distinctions.  $(iii)$ In contrast to the paired state of composite fermions, an electronic $p+ip$ superconductor is characterized by a local order parameter.  Defects in that order parameter---i.e., neutral $h/2e$ vortices---bind Majorana zero-modes and, accordingly, constitute the Ising anyons akin to charge-$e/4$ quasiparticles in the Moore-Read state~\cite{ReadGreen,Ivanov}.  $(iv)$ Because of the energy cost associated with local order parameter variations, superconducting vortices are strictly speaking confined (unlike $e/4$ quasiparticles).  This does not imply inaccessibility of non-Abelian anyons in this setting, since the `user' can always supply the energy necessary to separate vortices by arbitrary distances.  Non-Abelian braiding statistics is, however, realized only projectively~\cite{ProjectiveStatistics,BarkeshliParafendleyons1} as a result---i.e., up to an overall phase that for most purposes is fortunately inessential.  The existence of an order parameter may actually prove advantageous, as experimental techniques for coupling to order parameters can provide practical means of manipulating non-Abelian anyons in the laboratory.  

		  Shortly after Read and Green's work, Kitaev showed that a 1D spinless $p$-wave superconductor forms a closely related topological superconducting phase~\cite{1DwiresKitaev} (which one can view as a 2D $p+ip$ superconductor squashed along one dimension).  Here domain walls in the superconductor bind Majorana zero-modes and realize confined Ising anyons whose exotic statistics can be meaningfully harvested in wire networks~\cite{AliceaBraiding,HalperinBraiding,ClarkeBraiding,BondersonBraiding}.  Although such nontrivial one-dimensional (1D) and two-dimensional (2D) superconductors are unlikely to emerge from a material's intrinsic dynamics, numerous blueprints now exist for engineering these phases in heterostructures fashioned from ingredients such as topological insulators, semiconductors, and $s$-wave superconductors~\cite{FuKane,MajoranaQSHedge,Sau,Alicea,1DwiresLutchyn,1DwiresOreg,CookFranz} (see Refs.~\onlinecite{BeenakkerReview} and \onlinecite{AliceaReview} for recent reviews).  These proposals highlight the vast potential that `ordinary' systems possess for designing novel phases of matter and have already inspired a flurry of experiments.  Studies of semiconducting wires interfaced with $s$-wave superconductors have proven particularly fruitful, delivering numerous possible Majorana signatures~\cite{mourik12,das12,Rokhinson,deng12,finck12,Churchill}.  

		  These preliminary successes motivate the question of whether one can---even in principle---design blueprints for non-Abelian anyons with richer braid statistics compared to the Ising case.  Several recent works demonstrated that this is indeed possible using, somewhat counterintuitively, \emph{Abelian} quantum Hall states as a canvas for more exotic non-Abelian anyons~\cite{ClarkeParafendleyons,LindnerParafendleyons,ChengParafendleyons,VaeziParafendleyons,BarkeshliParafendleyons1,BarkeshliParafendleyons2} (see also Refs.~\onlinecite{ChernInsulatorParafendleyons,QuantumWiresParafendleyons}).  Most schemes involve forming a fractionalized `wire' out of counterpropagating Abelian quantum Hall edge states.  This `wire' can acquire a gap via competing mechanisms, e.g., proximity-induced superconductivity or electronic backscattering.
		  Domain walls separating physically distinct gapped regions bind $\mathbb{Z}_n$ generalizations of Majorana zero-modes
		  \cite{Fendley}\footnote{Some references refer to these generalizations as parafermion zero-modes.  We intentionally avoid this nomenclature here to avoid confusion with the rather different (though related) parafermions that appear in conformal field theory, particularly since both contexts frequently arise in this paper.}
		  and consequently realize non-Abelian anyons of a more interesting variety than those in a 1D $p$-wave superconductor.
		  Unfortunately, however, they too admit non-universal braid statistics, though achieving universal quantum computation requires fewer unprotected operations \cite{ClarkeParafendleyons,Hastings}.

\begin{figure}[b]
	\centering
	\includegraphics[width = \columnwidth]{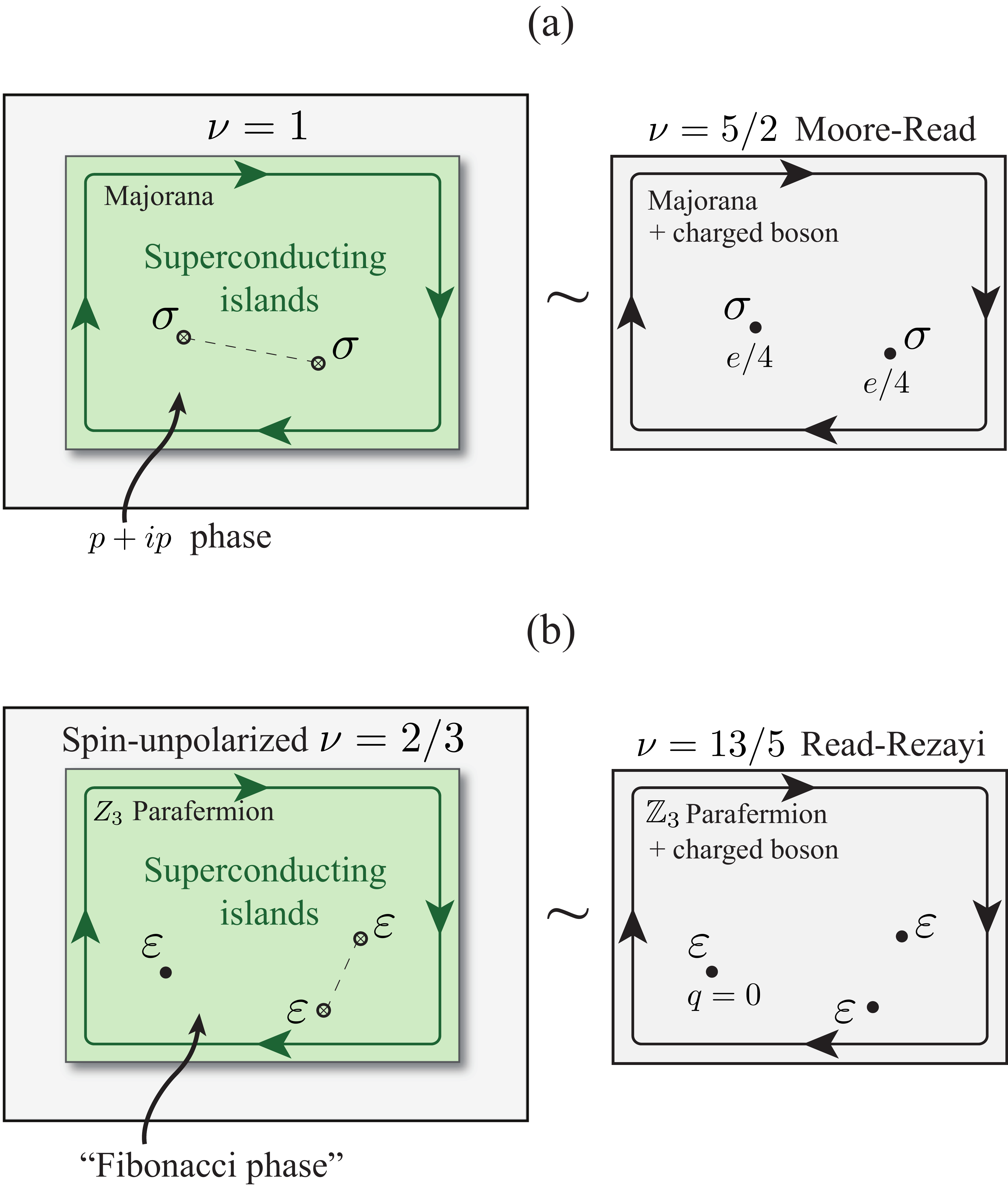}
	\caption{%
		Schematic illustration of main results.  Abelian quantum Hall states interlaced with an array of superconducting islands (left column) realize analogues of exotic non-Abelian quantum Hall states (right column).  The interface between the superconducting regions and surrounding Abelian quantum Hall fluids supports chiral modes similar to those on the right, but without the bosonic charge sector.  (We suppress the edge states at the outer boundary of the Abelian quantum Hall states for simplicity.)  Solid circles denote deconfined non-Abelian excitations, while open circles connected by dashed lines represent confined $h/2e$ superconducting vortices.  Quasiparticle charges are also listed for the non-Abelian quantum Hall states.
		In (a) $\sigma$ particles represent Ising anyons, which in the $p+ip$ phase on the left correspond to confined vortex excitations.  In (b) $\varepsilon$ is a Fibonacci anyon that exhibits universal braid statistics.  The superconducting Fibonacci phase is topologically ordered and supports \emph{deconfined} $\varepsilon$ particles---similar to the Read-Rezayi state.  Vortices in this nontrivial superconductor do not lead to new quasiparticle types, but can in principle trap Fibonacci anyons.
	}
	\label{Summary}
\end{figure}

In this paper we advance this program one step further and pursue a similar strategy towards non-Abelian anyons with \emph{universal} braid statistics.  More precisely, our goal is to construct a new 2D superconductor that bears the same relation to the $\mathbb{Z}_3$ Read-Rezayi state as a spinless $p+ip$ superconductor bears to Moore-Read.  With this analogy in mind it seems reasonable to demand that such a phase satisfy the following basic properties.  First, the boundary should host a chiral $\mathbb{Z}_3$ parafermion edge mode, but lack the Read-Rezayi state's bosonic charge sector.  And second, the bulk should exhibit essentially the same non-Abelian content as the Read-Rezayi phase---particularly Fibonacci anyons.  

We show that one can nucleate a phase with precisely these properties, not in free space but rather in the interior of a fractionalized medium.   Our approach resembles that of Refs.~\onlinecite{InteractingAnyons1} and \onlinecite{InteractingAnyons2} which demonstrated that hybridizing a finite density of non-Abelian anyons produces new descendant phases in the bulk of a parent non-Abelian liquid.  In the most experimentally relevant cases of the Moore-Read state and a 2D spinless $p+ip$ superconductor these descendants were found to be Abelian.  
We describe what amounts, in a sense, to an inverse of this result. 
The specific construction we follow relies on embedding an array of superconducting islands in an \emph{Abelian} quantum Hall system to proximity-induce Cooper pairing in the fluid.  When the islands remain well-separated, each one binds localized zero-modes that collectively encode a macroscopic ground state degeneracy spanned by different charge states on the superconductors.  Hybridizing these zero-modes can then lift this degeneracy in favor of novel \emph{non-Abelian} 2D superconducting phases---including the Read-Rezayi analogue that we seek.

As an illustrative warm-up, Sec.~\ref{MajoranaCase} explores the simplest trial application corresponding to an integer quantum Hall system at filling $\nu = 1$.  Here the superconducting islands trap Majorana modes that, owing to broken time-reversal symmetry, rather naturally couple to form a 2D spinless $p+ip$ superconducting phase within the fluid.  In other words, imposing Cooper pairing provides a constructive means of generating the non-Abelian physics of the Moore-Read state starting from the comparatively trivial \emph{integer} quantum Hall effect.  This result is fully consistent with earlier studies of Refs.~\onlinecite{QAHproposal1} and \onlinecite{QAHproposal2} that explored similar physics from a complementary perspective.

One can intuitively anticipate richer behavior for a superconducting array embedded in an Abelian \emph{fractional} quantum Hall state.  In particular, since here charge-$2e$ Cooper pairs derive from conglomerates of multiple fractionally charged quasiparticles, such a setup appears natural for building in the clustering properties of Read-Rezayi states.  This more interesting case is addressed in the remainder of the paper.  We focus specifically on the experimentally observed spin-unpolarized $\nu = 2/3$ state~\cite{SpinUnpolarizedStates}---also known as the (112) state---for which superconducting islands bind $\mathbb{Z}_3$ generalizations of Majorana modes.  [Note that various other quantum Hall phases, e.g., the bosonic (221) state, yield the same physics.]  Hybridization of these modes is substantially more difficult to analyze since the problem can not, in contrast to the integer case, be mapped to free fermions.  Burrello \emph{et al}.\ recently addressed a related setup consisting of generalized Majorana modes coupled on a 2D lattice, capturing Abelian phases including a generalization of the toric code~\cite{ParafendleyonLattice}.  We follow a different approach inspired by Teo and Kane's method of obtaining non-Abelian quantum Hall phases from stacks of weakly coupled Luttinger liquids~\cite{TeoKaneChains}.  Though their specific coset construction is not applicable to our setup, a variant of their scheme allows us to leverage theoretical technology for 1D systems---i.e., bosonization and conformal field theory---to controllably access the 2D phase diagram.  

With the goal of bootstrapping off of 1D physics, Secs.~\ref{Z3CFT} and \ref{QH_Z3_criticality} develop the theory for a single chain of superconducting islands in a $\nu = 2/3$ state.  There we show, by relating the setup to a three-state quantum clock model, that this chain can be tuned to a critical point described by a non-chiral $\mathbb{Z}_3$ parafermion conformal field theory.  Section~\ref{RRsection} then attacks the 2D limit coming from stacks of critical chains.  (A related approach in which the islands are `smeared out' is discussed in Sec.~\ref{UniformTrenches}.)  Most importantly, we construct an interchain coupling that generates a gap in the bulk but leaves behind a gapless \emph{chiral} $\mathbb{Z}_3$ parafermion sector at the boundary, thereby driving the system into a superconducting cousin of the $\mathbb{Z}_3$ Read-Rezayi state that we dub the `Fibonacci phase'.  

The type of topological phenomena present here raises an intriguing question.  That is, should one view this state as analogous to a spinless $p+ip$ superconductor (which realizes short-ranged entanglement) or rather an intrinsic non-Abelian quantum Hall system (which exhibits true topological order)?  Interestingly, although superconductivity plays a key role microscopically for our construction, we argue that the Fibonacci phase is actually topologically ordered with somewhat `incidental' order parameter physics.  Indeed we show that Fibonacci anyons appear as \emph{deconfined} quantum particles, just like in the $\mathbb{Z}_3$ Read-Rezayi state, leading to a two-fold ground-state degeneracy on a torus that is the hallmark of true topological order.  Moreover, superconducting vortices do not actually lead to new quasiparticle types in sharp contrast to a $p+ip$ superconductor where vortices provide the source of Ising anyons.  In this sense the fact that the Fibonacci phase exhibits an order parameter is unimportant for universal topological physics.  Vortices can, however, serve as one mechanism for trapping Fibonacci anyons---depending on non-universal energetics---and thus might provide a route to manipulating the anyons in practice.  Section~\ref{TQFTsection} provides a topological quantum field theory interpretation of the Fibonacci phase that sheds light on the topological order present, and establishes a connection between our construction and that of Refs.~\onlinecite{InteractingAnyons1,InteractingAnyons2}.


Figure~\ref{Summary} summarizes our main results for the $\nu = 1$ and $\nu = 2/3$ architectures as well as their relation to `intrinsic' non-Abelian quantum Hall states.  (For a more complete technical summary see the beginning of Sec.~\ref{Discussion}.)  
On a conceptual level, it is quite remarkable that a phase with Fibonacci anyons can emerge in simple Abelian quantum Hall states upon breaking charge conservation by judiciously coupling to ordinary superconductors.  Of course experimentally realizing the setup considered here will be very challenging---certainly more so than stabilizing Ising anyons.  It is worth, however, providing an example that puts this challenge into proper perspective.  As shown in Ref.~\onlinecite{Baraban10} a 128-bit number can be factored in a fully fault-tolerant manner using Shor's algorithm with $\approx 10^3$ Fibonacci anyons. In contrast, performing the same computation with Ising anyons would entail much greater overhead since the algorithm requires $\pi/8$ phase gates that would need to be performed non-topologically, and then distilled, e.g., according to Bravyi's protocol~\cite{Bravyi06}.
For a $\pi/8$ phase gate with $99\%$ fidelity, factoring a 128-bit number would consequently require $\approx 10^9$ Ising anyons in the scheme analyzed in Ref.~\onlinecite{Baraban10}
	\footnote{This is dependent on the specific protocol, and the precise numbers will vary \cite{Svore}.}.
Thus overcoming the nontrivial fabrication challenges involved could prove enormously beneficial for quantum information applications.  In this regard, inspired by recent progress in Majorana-based systems we are optimistic that it should similarly be possible to distill the architecture we propose to alleviate many of the practical difficulties towards realizing Fibonacci anyons.  Section~\ref{Discussion} proposes several possible simplifications---including alternate setups that do not require superconductivity---along with numerous other future directions that would be interesting to explore.  The abundance of systems known to host Abelian fractional quantum Hall phases and the large potential payoff together provide strong motivation for further pursuit of this avenue towards universal topological quantum computation.

\section{\texorpdfstring{Trial application: $p+ip$ superconductivity from the integer quantum Hall effect}{Trial application: p+ip from IQHE}}
\label{MajoranaCase}

The first proposal for germinating Ising anyons in an integer quantum Hall system was introduced by Qi, Hughes, and Zhang~\cite{QAHproposal1}; these authors showed that in the vicinity of a plateau transition, proximity-induced Cooper pairing effectively generates spinless $p+ip$ superconductivity in the fluid.  In this section we will establish a similar link between these very different phases from a viewpoint that illustrates, in a simplified setting, the basic philosophy espoused later in our pursuit of a Read-Rezayi-like superconductor that supports Fibonacci anyons.  Specifically, 
here we investigate weakly coupled \emph{critical} 1D superconducting regions embedded in a $\nu = 1$ quantum Hall system, following the spirit of Ref.~\onlinecite{TeoKaneChains} (see also Ref.~\onlinecite{CoupledChainsNagaosa}).  This quasi-1D approach gives one a convenient window from which to access various states present in the phase diagram---including a spinless 2D $p+ip$ superconductor analogous to the Moore-Read state~\cite{ReadGreen}.  There are, of course, experimentally simpler ways of designing superconductors supporting Ising anyons, but we hope that this discussion is instructive and interesting nonetheless.  Two complementary approaches will be pursued as preliminaries for our later treatment of the fractional quantum Hall case.  

\begin{figure}[t]
	\centering
	\subfigure[\label{fig:Maj1a}]{ \includegraphics[width = 0.99\columnwidth]{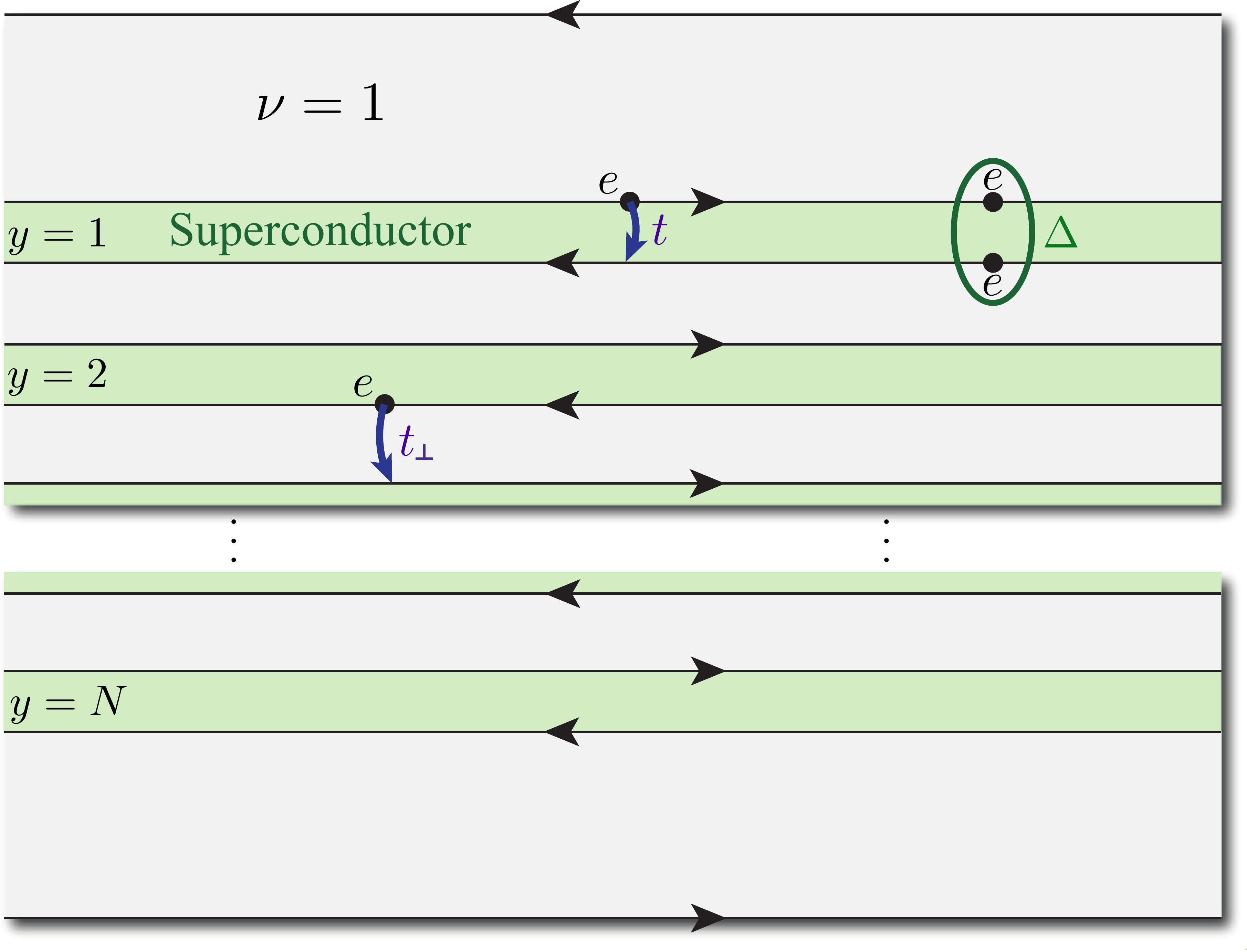} }
	\\
	\subfigure[\label{fig:Maj1b}]{ \qquad \includegraphics[width = 0.7\columnwidth]{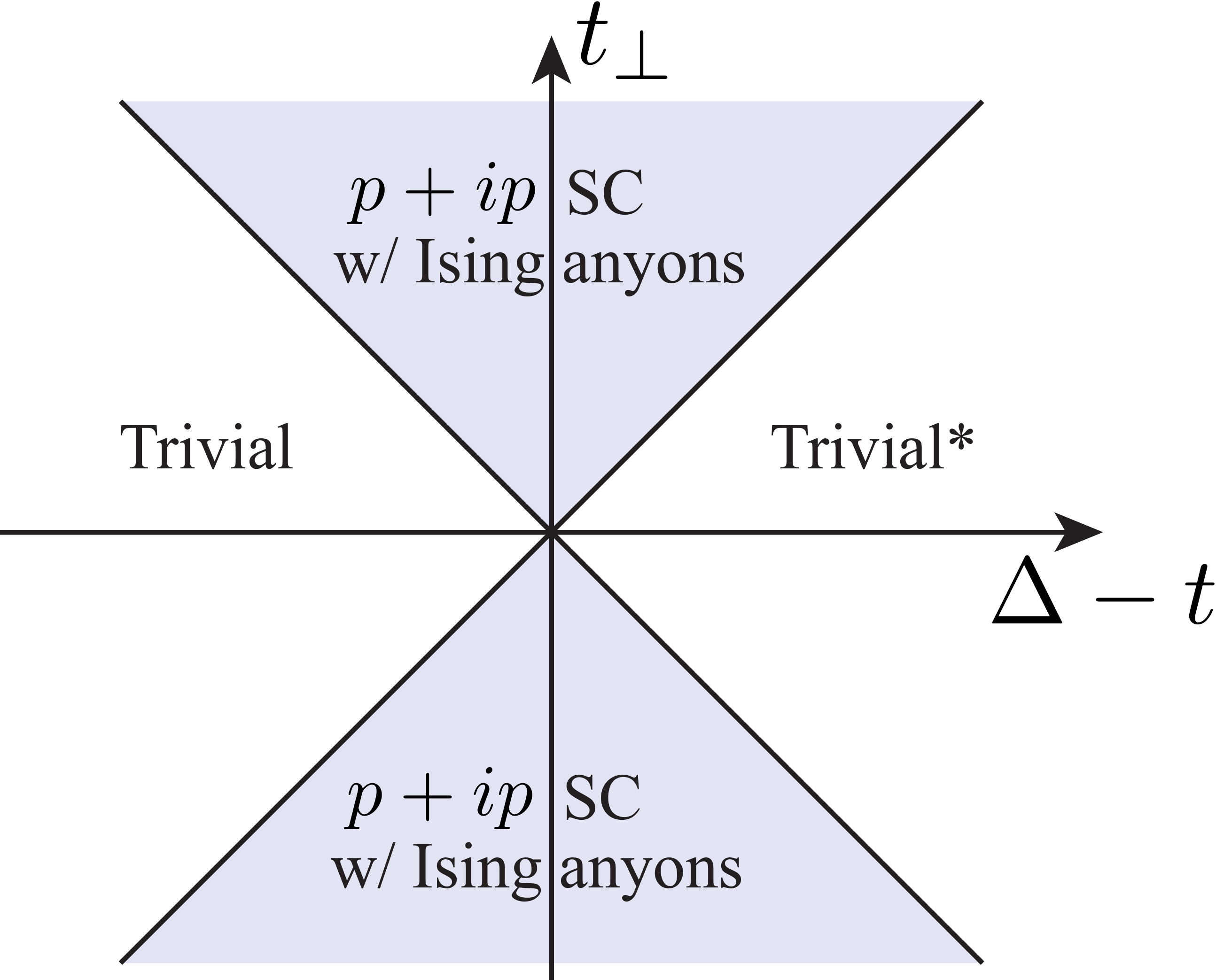} }
	\caption{(a) Setup used to nucleate a $p+ip$ superconducting state with Ising anyons inside of a $\nu = 1$ quantum Hall fluid.  The arrows indicate integer quantum Hall edge states.  Uniform superconductors fill each of the $N$ trenches shown.  The edge states opposite a given trench can hybridize either through electron backscattering $t$ or Cooper pairing $\Delta$ mediated by the intervening superconductor; both processes favor gapping the edge modes, but in competing ways.  Adjacent trenches are assumed to couple weakly via electron tunneling $t_\perp$.  With $t_\perp = 0$ and $t = \Delta$ each trench resides at a critical point at which the adjacent quantum Hall edge states evolve into counterpropagating Majorana modes.  Turning on $t_\perp$ then mixes these modes in such a way that `unpaired' chiral Majorana edge states survive at the boundary, thus triggering a $p+ip$ phase.  (b) Phase diagram for the weakly coupled trenches near criticality.  States that smoothly connect to the limit of decoupled chains are labeled `trivial'; see text for a more detailed description of their properties.}
	\label{Majorana1}
\end{figure}

\subsection{Uniform trench construction}
\label{UniformCase}

Consider first the setup in Fig.~\subfigref{fig:Maj1a}, wherein a $\nu = 1$ quantum Hall system contains a series of trenches (labeled by $y = 1,\ldots,N$) filled with some long-range-ordered superconducting material.  As the figure indicates the boundary of each trench supports spatially separated right/left-moving integer quantum Hall edge states described by operators $f_{R/L}(y)$.
We assume that adjacent counterpropagating edge modes hybridize and are therefore generically unstable, due either to ordinary electron backscattering or Cooper pairing mediated by the superconductors
	\footnote{If the quantum Hall edge states are completely spin-polarized then the superconductor should have a triplet component in order to achieve the desired proximity effect.}.
Let the Hamiltonian governing these edge modes be $H = H_\textrm{KE} + \delta H + H_\perp$.  Here
\begin{align}
  H_\textrm{KE} = \sum_{y = 1}^N \int_x\!\big[ {-i}v f_R^\dag(y) \partial_x f_R(y) + i v f_L^\dag(y) \partial_x f_L(y) \big]
  \label{H_KE}
\end{align}
captures the kinetic energy for right- and left-movers, with $x$ a coordinate along the trenches (which we usually leave implicit in operators throughout this section).  The second term, $\delta H$, includes electron tunneling and Cooper pairing perturbations acting separately within each trench:
\begin{align}
  \delta H = \sum_{y = 1}^N \int_x\! \big[ {-t} f_R^\dag(y) f_L(y) + \Delta f_R(y) f_L(y) + H.c. \big]
  \label{deltaH_Majorana}
\end{align}
where $t>0$ and $\Delta>0$ denote the tunneling and pairing strengths.  Finally, $H_\perp$ incorporates electron tunneling between neighboring trenches with amplitude $t_\perp$,
\begin{align}
  H_\perp = -t_\perp \sum_{y = 1}^{N-1}\int_x\! \big[f_L^\dag(y) f_R(y+1) + H.c.\big].
  \label{H_perp_Majorana}
\end{align}
Figure~\subfigref{fig:Maj1a} illustrates all of the above processes.  

Hereafter we assume $|t_\perp |\ll t,\Delta$ corresponding to the limit of weakly coupled trenches.  It is then legitimate to first treat $H_\textrm{KE} + \delta H$, which is equivalent to the Hamiltonian for $N$ independent copies of quantum spin Hall edge states with backscattering generated by a magnetic field and proximity-induced pairing~\cite{MajoranaQSHedge}.  As in the quantum spin Hall problem, the $t$ and $\Delta$ perturbations favor physically distinct gapped phases that cannot be smoothly connected without crossing a phase transition.  For $\Delta >t$ each trench realizes a 1D topological superconductor with Majorana zero-modes bound to its endpoints, while for $\Delta <t$ trivial superconductivity appears.  Deep in either gapped phase small hopping $t_\perp$ between trenches clearly yields only minor quantitative effects on the bulk.  

We therefore focus on the critical point $t = \Delta$ at which these opposing processes balance.  Here arbitrarily weak $t_\perp$ can play an important role as each trench remains gapless.  In this limit one can factorize $\delta H$ in a revealing way:
\begin{align}
	\delta H_{t = \Delta} = -t\sum_{y = 1}^N \int_x\! \big[ f_R^\dag(y) -f_R(y) \big] \big[ f_L(y) + f_L^\dag(y) \big].
\end{align}
At the transition the `real part' of $f_R(y)$ and the `imaginary part' of $f_L(y)$ are thus unaffected by the perturbations in $\delta H$, while the other components hybridize and gap out.  Hence the important low-energy operators at the critical point correspond to right- and left-moving gapless Majorana fields $\gamma_{R/L}(y)$, defined as
\begin{align}\begin{split}
	\gamma_{R}(y)&= \frac{1}{2}\left[f_{R}^\dagger(y)+f_{R}(y)\right] ,
\\ \qquad\gamma_{L}(y)&= \frac{i}{2}\left[f_{L}^{\dagger}(y)-f_{L}(y)\right] .
	\label{fRL}
\end{split}\end{align}
Notice that, like the original quantum Hall edge states, the chiral Majorana modes emerging at criticality are spatially separated across each trench.  
Using Eq.~\eqref{fRL} one can straightforwardly derive an effective low-energy Hamiltonian that incorporates small deviations away from criticality as well as weak inter-trench coupling $t_\perp$; this reads
\begin{align}
	H_\textrm{eff} &= \sum_{y = 1}^N \int_x
		\begin{bmatrix} -iv \gamma_R(y) \partial_x \gamma_R(y) + i v \gamma_L(y) \partial_x \gamma_L(y)
	  	\\+ i m\gamma_R(y)\gamma_L(y) \end{bmatrix}
	\nonumber\\
		&\quad + 2i t_\perp \sum_{y = 1}^{N-1}\int_x\gamma_L(y)\gamma_R(y+1) ,
	\label{Heff}
\end{align}
where $m = 2(\Delta-t)$.  [To obtain this result one can simply replace $f_R(y) \rightarrow \gamma_R(y)$ and $f_L(y) \rightarrow i \gamma_L(y)$ in $H$ since the imaginary part of the former and the real part of the latter are gapped; note the consistency with Eq.~\eqref{fRL}.]

The structure of the phase diagram for $H_\textrm{eff}$, which appears in Fig.~\subfigref{fig:Maj1b}, can be deduced by examining limiting cases.  First, in the limit $|m| \gg t_\perp$ perturbations within each trench dominate and drive gapped phases determined by the sign of $m$.  With $m<0$ tunneling $t$ yields a trivially gapped superconducting state within the quantum Hall system.  Conversely, for $m>0$ Cooper pairing $\Delta$ produces a chain of Majorana modes at the left and right ends of the trenches that form a dispersing band due to small $t_\perp$.  We also refer to the resulting 2D superconductor as trivial since it smoothly connects to the decoupled-chain limit.
(This phase nevertheless retains some novel features and is characterized by nontrivial `weak topological indices' \cite{CoupledChainsNagaosa}.  For instance, lattice defects can bind Majorana zero-modes~\cite{CoupledChainsNagaosa}, and the dispersing 1D band of hybridized Majorana modes can be stable if certain symmetries are present on average~\cite{RingelKrausStern:WTI:12,MongBardarsonMooreWTI12,FuKaneSymplectic,FulgaAkhmerovEdge}.
Hence we denote this trivial state with a star in the phase diagram~\footnote{This state is also sometimes referred to as a `weak 2D topological superconductor', not to be confused with the weak pairing phase of a spinless 2D $p+ip$ superconductor.}.)
More interesting for our purposes is the opposite limit where $t_\perp$ dominates so that genuinely 2D phases can arise.  Upon inspecting the last term in Eq.~\eqref{Heff} one sees that when $m = 0$ inter-trench hopping gaps out all Majorana fields in the bulk, but leaves behind gapless chiral Majorana edge states described by $\gamma_R(y = 1)$ on the top edge and $\gamma_L(y = N)$ on the bottom.  This edge structure signifies the onset of spinless $p+ip$ superconductivity with vortices that realize Ising anyons.  By passing to momentum space and identifying where the bulk gap closes, one can show that the transitions separating the states above occur at $|\Delta-t| = |t_\perp|$, yielding the phase boundaries of Fig.~\subfigref{fig:Maj1b}.  

We have thereby established the correspondence illustrated in Fig.~\ref{Summary}(a) between an integer quantum Hall system with (long) superconducting islands and the Moore-Read state.  Towards the end of this paper, Sec.~\ref{UniformTrenches} will discuss a similar uniform-trench setup in the fractional quantum Hall case.  For technical reasons, however, it will prove simpler to analyze a fractional quantum Hall system with superconductivity introduced \emph{non-uniformly} within each trench.  In fact most of our treatment will be devoted to such an architecture.  As a preliminary, the next subsection analyzes spatially modulated trenches in an integer quantum Hall system, once again recovering spinless $p+ip$ superconductivity from weakly coupled chains.  


\subsection{Spatially modulated trenches}
\label{ModulatedCase}

\begin{figure}[t]
	\centering
	\subfigure[\label{fig:Maj2a}]{ \includegraphics[width = 0.99\columnwidth]{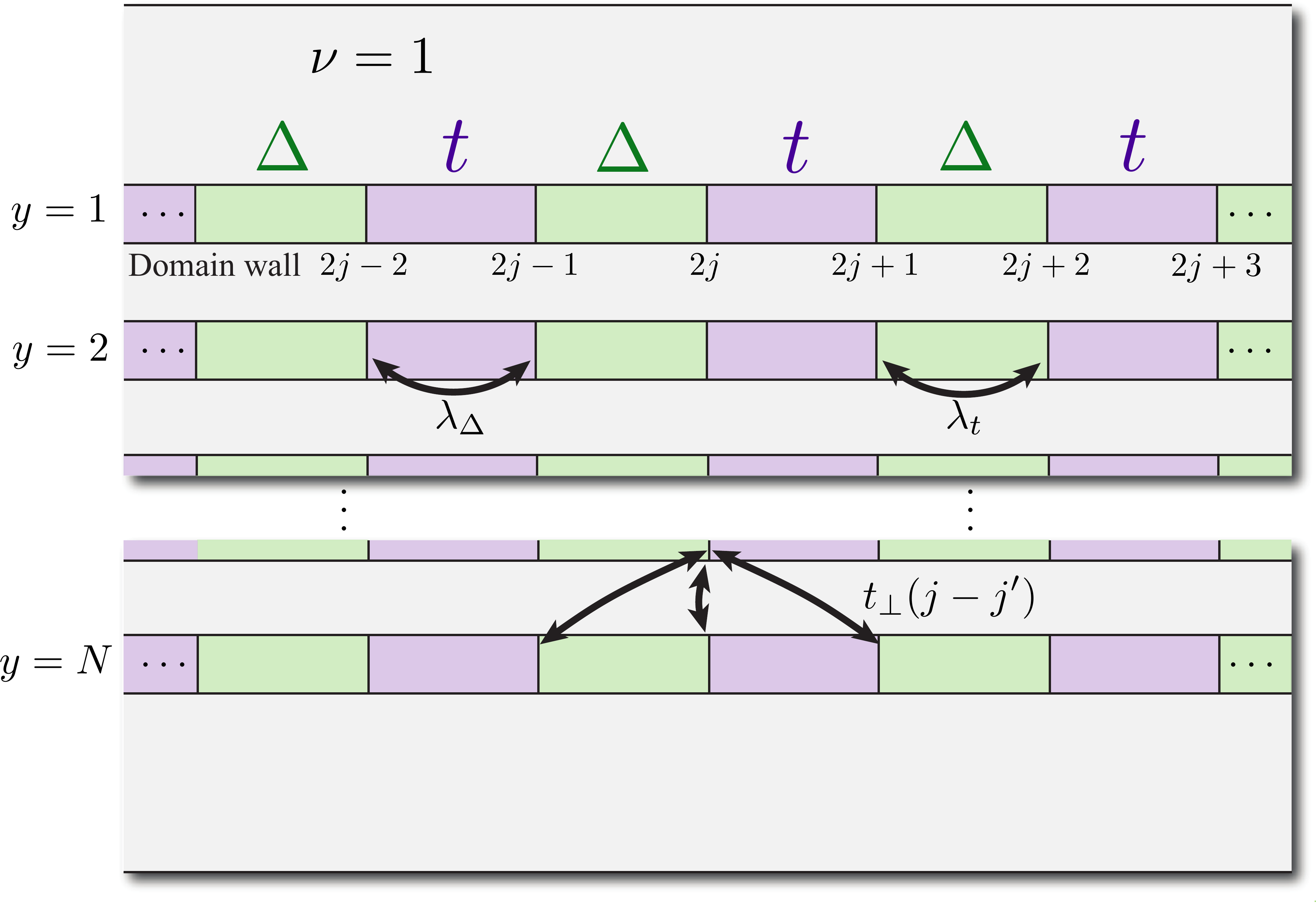} }
	\\
	\subfigure[\label{fig:Maj2b}]{ \qquad \includegraphics[width = 0.7\columnwidth]{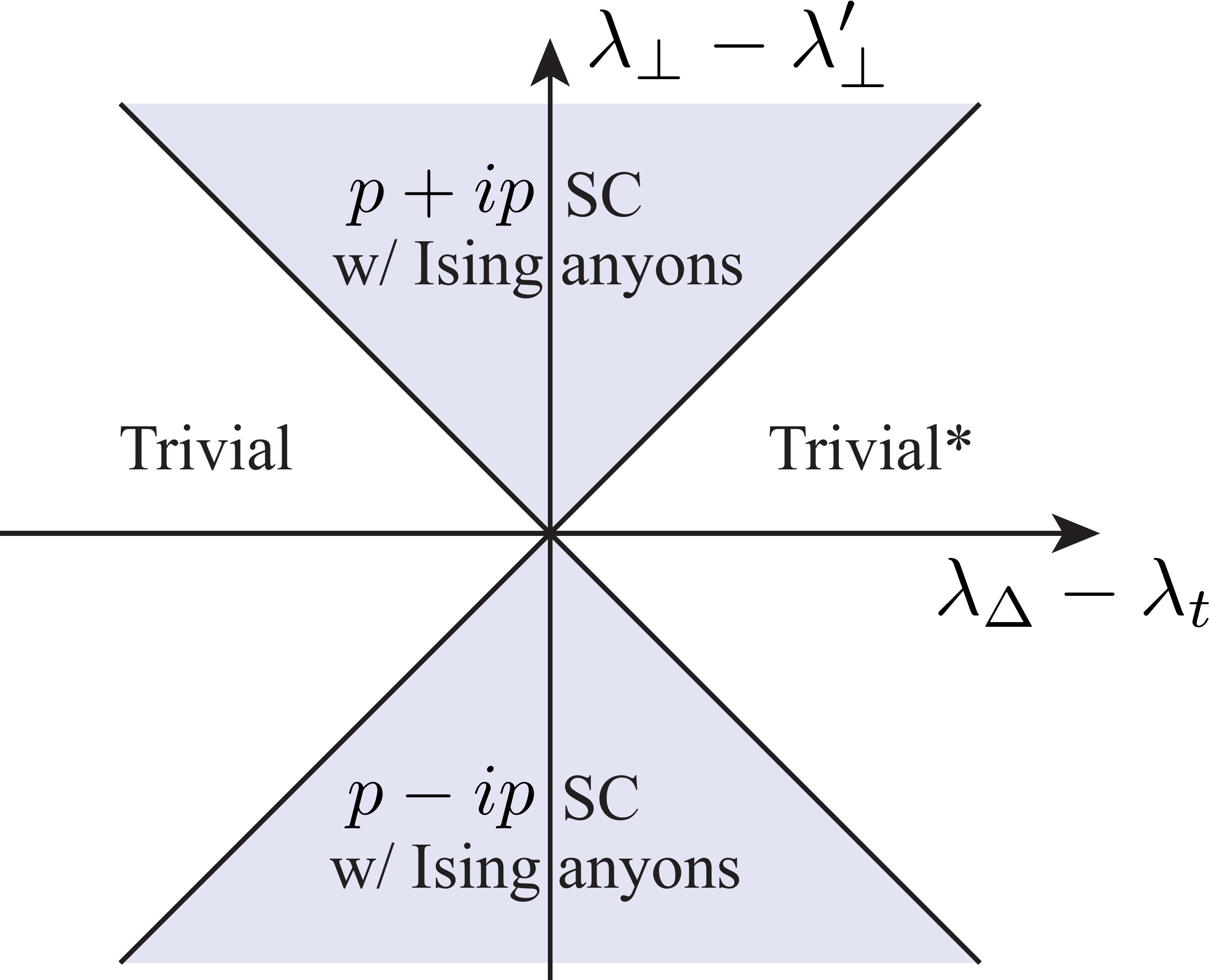} }
	\caption{(a) Variation on the setup of Fig.~\subfigref{fig:Maj1a} that also supports a $p+ip$ superconducting state with Ising anyons.  Here a $\nu = 1$ quantum Hall system hosts spatially modulated trenches whose edge states are gapped in an alternating fashion by backscattering $t$ and Cooper pairing $\Delta$.  When the trenches decouple and the gapped regions are `large', each domain wall binds a Majorana zero-mode.  Electron hopping across the domains hybridizes the chain of Majorana modes in each trench through couplings $\lambda_\Delta$ and $\lambda_t$ shown above.  These couplings favor competing gapped phases, and when $\lambda_\Delta = \lambda_t$ each chain realizes a critical point with counterpropagating gapless Majorana modes in the bulk---similar to the uniform trench setup of Fig.~\subfigref{fig:Maj1a}.  Turning on weak coupling $t_\perp(j-j')$ between domain walls $j$ and $j'$ in adjacent trenches then generically drives the system into a $p+ip$ phase (or a $p-ip$ state with opposite chirality).  (b) Phase diagram for the 2D array of coupled Majorana modes near criticality.  Here $\lambda_\perp$ and $\lambda_\perp'$ represent interchain couplings between gapless Majorana fermions at the critical point, which follow from $t_\perp(j-j')$ according to Eq.~\eqref{lambda_perp}.}
	\label{Majorana2}
\end{figure}

We now explore the modified setup of Fig.~\subfigref{fig:Maj2a} in which the $\nu = 1$ edge states within each trench are sequentially gapped by pairing $\Delta$ and electron tunneling $t$, creating an infinite, periodic array of domain walls labeled according to the figure.  This setup can again be described by a Hamiltonian $H = H_\textrm{KE} + \delta H + H_\perp$ as defined in Eqs.~\eqref{H_KE} through \eqref{H_perp_Majorana}, but now with $t$ and $\Delta$ varying in space.  For simplicity we will assume $t = 0$ in the pairing-gapped regions and $\Delta = 0$ in the tunneling-gapped regions (one can easily relax this assumption if desired).  

Suppose for the moment that each domain is long compared to the respective coherence length, and that the trenches are sufficiently far apart that they decouple.  
In this case the Cooper-paired regions constitute 1D topological superconductors that produce a Majorana zero-mode exponentially bound to each domain wall~\cite{MajoranaQSHedge}.  An explicit calculation reveals that the Majorana operator for domain wall $j$ at position $x_j$ in trench $y$ takes the form (up to normalization)
\begin{align}
  \gamma_j(y) &\propto \int_x e^{-\frac{|x-x_j|}{\xi(x-x_j)}}\left[f_R(y)-i(-1)^j f_L(y)+H.c.\right].
  \label{MajoranaOps}
\end{align}
Here $\xi(x-x_j)$ denotes the decay length for the Majorana mode and is given either by $v/t$ or $v/\Delta$ depending on the sign of $x-x_j$.  The 2D array of zero-modes present in this limit underlies a \emph{macroscopic} ground-state degeneracy, since one can combine each pair of Majoranas into an ordinary fermion that can be vacated or filled at no energy cost.  Next, imagine shrinking the width of the tunneling- and pairing-gapped regions, as well as the spacing between trenches, such that domain walls couple appreciably.  Our objective here is to investigate how the resulting hybridization amongst nearby Majorana modes resolves the massive degeneracy present in our starting configuration.

Focusing again on the weakly coupled chain limit, we first incorporate hybridization within each trench.  
The simplest intra-chain perturbation consistent with the symmetries of the problem tunnels right- and left-moving electrons between neighboring domain walls and reads \footnote{We specifically enforced an anti-unitary symmetry that sends $f_{R/L} \rightarrow i f_{L/R}$, which preserves Eqs.~\eqref{H_KE} and \eqref{deltaH_Majorana}.  Consequently, the prefactors in front of the $f_R$ and $f_L$ hopping terms in Eq.~\eqref{DiscreteHoppings} must be complex conjugates, as written.}  
\begin{align}
  H_\textrm{intra} &=  \frac{1}{4}\sum_{y = 1}^N \sum_j \lambda_j
		\begin{bmatrix} -i f_R^\dagger(x_j,y)f_R(x_{j+1},y) \\ + i f_L^\dagger(x_j,y)f_L(x_{j+1},y) \\ + H.c. \end{bmatrix} .
  \label{DiscreteHoppings}
\end{align}
[This is just a discrete version of the kinetic energy in Eq.~\eqref{H_KE}.]  The $x$ coordinate in the argument of $f_{R/L}$, usually left implicit, has been explicitly displayed since it is now crucial.  We define the real couplings appearing above as $\lambda_j \equiv \lambda_\Delta$ for $j$ even and $\lambda_j \equiv \lambda_t$ for $j$ odd.  Physically, $\lambda_\Delta$ and $\lambda_t$ respectively arise from coupling adjacent pairing- and tunneling-gapped regions [see Fig.~\subfigref{fig:Maj2a}], and thus clearly need not be identical.  We assume however that $\lambda_\Delta,\lambda_t \geq 0$.  

According to Eq.~\eqref{MajoranaOps}, projection of $H_\textrm{intra}$ into the low-energy manifold spanned by the Majorana operators is achieved (up to an unimportant overall constant that we will neglect) by replacing 
\begin{align}
  f_R(x_j,y) \rightarrow \gamma_j(y), \quad f_L(x_j,y) \rightarrow i(-1)^j \gamma_j(y).  
  \label{fRLprojection}
\end{align}
This projection yields the following effective Hamiltonian for the decoupled trenches,
\begin{align}
  H_\textrm{intra} &\rightarrow -i\sum_{y = 1}^N \sum_j
	\begin{bmatrix} \lambda_t \gamma_{2j-1}(y)\gamma_{2j}(y) \phantom{m}\\\phantom{|} + \lambda_\Delta \gamma_{2j}(y)\gamma_{2j+1}(y) \end{bmatrix} .
  \label{HintraMajorana}
\end{align}
which is equivalent to $N$ independent Kitaev chains~\cite{1DwiresKitaev}.  As written above $\lambda_\Delta$ and $\lambda_t$ favor distinct dimerization patterns for the Majorana operators that can not be smoothly connected without closing the bulk gap.  Alternatively, one can view the problem in more physical terms by implementing a basis change to ordinary fermions $c_j(y) = [\gamma_{2j-1}(y) +i \gamma_{2j}(y)]/2$.  Equation~\eqref{HintraMajorana} then describes decoupled 1D $p$-wave-paired wires.  If $\lambda_\Delta$ dominates the superconducting wires reside in a gapped topological phase with protected Majorana end-states, whereas if $\lambda_t$ dominates a trivially gapped state emerges.  

The transition separating these 1D phases arises when $\lambda_\Delta = \lambda_t$.  Viewed in terms of superconductors this limit corresponds to the situation where the chemical potential for the $c_j$ fermions is fine-tuned to the bottom of the band, so that gapless bulk excitations remain at zero momentum despite the $p$-wave pairing.  As in the preceding subsection we will concentrate on this transition point since here even weak inter-trench coupling (to which we turn shortly) can qualitatively affect the physics.  When $\lambda_\Delta = \lambda_t$ one can solve either Eq.~\eqref{HintraMajorana} directly, or the equivalent superconducting problem, by diagonalizing the Hamiltonian in momentum space.  This exercise shows that at criticality right- and left-moving Majorana fields $\gamma_{R/L}(y)$ form the relevant low-energy degrees of freedom---precisely as in the uniform-trench construction examined earlier.  Moreover, these continuum fields relate to the lattice Majorana operators via
\begin{align}
  \gamma_j(y) \sim \gamma_R(y) + (-1)^j \gamma_L(y).
  \label{FieldExpansion}
\end{align}
Using Eq.~\eqref{FieldExpansion} to rewrite Eq.~\eqref{HintraMajorana} and taking the continuum limit yields
\begin{align}
	H_\textrm{intra} &\sim \sum_{y = 1}^N \int_x
		\begin{bmatrix} -i\tilde v \gamma_R(y) \partial_x \gamma_R(y) + i \tilde v \gamma_L(y) \partial_x \gamma_L(y)
  		\\ + i \tilde m\gamma_R(y)\gamma_L(y) \end{bmatrix} ,
  \label{HintraMajorana2}
\end{align}
where the velocity $\tilde v$ follows from the tunnelings in Eq.~\eqref{HintraMajorana} and $\tilde m \propto \lambda_\Delta-\lambda_t$ reflects small deviations away from criticality.  Note that Eq.~\eqref{HintraMajorana2} exhibits an identical structure to the intra-chain terms in Eq.~\eqref{Heff}, which were derived for spatially uniform trenches.  The appearance of common physics near criticality in the two setups is quite natural; indeed, in a coarse-grained picture appropriate for the critical point the spatial modulations in the trenches are effectively blurred away.  

One can now readily restore weak coupling between neighboring trenches.  Consider the following inter-trench Hamiltonian,
\begin{align}
  H_\perp &= -\sum_{y = 1}^{N-1} \sum_{j,j'} t_\perp(j-j')
		\begin{bmatrix} f_{L}^\dagger(x_j,y) f_R(x_{j'},y+1) \\ + H.c. \end{bmatrix} ,
  \label{HperpMajorana}
\end{align}
which encodes generic electron hoppings from the bottom of domain wall $j$ in one trench, to the top of domain wall $j'$ in the trench just below.  We have assumed that the tunneling strengths $t_\perp(j-j')$ above are real and depend only on the spacing $j-j'$ between domain walls.  These hoppings should be reasonably short-ranged as well; see Figure~\subfigref{fig:Maj2a} for examples of significant processes.  Since we are interested in weak interchain coupling near criticality it is useful to filter out high-energy physics, employing Eqs.~\eqref{fRLprojection} and \eqref{FieldExpansion} to project $H_\perp$ onto the low-energy manifold:
\begin{align}
	H_\perp &\sim i \sum_{y = 1}^{N-1} \int_x \big[ \lambda_\perp\gamma_L(y)\gamma_R(y+1) + \lambda_\perp' \gamma_R(y) \gamma_L(y+1) \big].
  \label{HperpMajorana2}
\end{align}
The coupling constants here are defined as
\begin{equation}
	\lambda_\perp \propto \sum_j t_\perp(j), \quad \lambda_\perp' \propto \sum_j (-1)^j t_\perp(j)
	\label{lambda_perp}
\end{equation}
and, importantly, differ in magnitude unless fine-tuned.

The full low-energy theory describing our weakly coupled, spatially modulated trenches is $H_\textrm{eff} = H_\textrm{intra} + H_\perp$ with the terms on the right side given in Eqs.~\eqref{HintraMajorana2} and \eqref{HperpMajorana2}.  When $\lambda_\perp' = 0$ this effective Hamiltonian is essentially identical to Eq.~\eqref{Heff}
 \footnote{The additional $\lambda_\perp'$ term reflects the reduced translation symmetry in the present setup.  Consequently, the low-energy expansion for $f_{R/L}(y)$ involves both $\gamma_R(y)$ \emph{and} $\gamma_L(y)$ in contrast to the uniform-trench system, so that more terms arise under projection.}.
The phase diagram thus mimics that of the uniform-trench case, and can again be inferred from considering extreme cases.  When the mass term $\tilde m \propto \lambda_\Delta-\lambda_t$ dominates over all other couplings we obtain superconducting states that smoothly connect to the decoupled-chain limit; the cases $\lambda_\Delta<\lambda_t$ and $\lambda_\Delta>\lambda_t$ respectively correspond to the trivial and `trivial*' phases discussed in the previous subsection.  If instead $\lambda_\perp$ dominates then the interchain coupling gaps out all Majorana fields in the bulk, but leaves a gapless right-mover at the top edge and a gapless left-mover at the bottom edge.  This is the spinless $p+ip$ superconducting phase that supports Ising anyons.  Finally, by examining Eq.~\eqref{HperpMajorana2} we see that when $\lambda_\perp'$ provides the leading term we simply obtain a spinless $p-ip$ superconductor with gapless edge states moving in the opposite direction.  All of these phases exhibit a bulk gap; the transition between them occurs when $|\tilde m| = |\lambda_\perp-\lambda_\perp'|$ at which this gap closes.  Figure~\subfigref{fig:Maj2b} illustrates the corresponding phase diagram.  It is worth stressing that when the trenches are each tuned to criticality (so that $\tilde m = 0$), interchain coupling generically drives the system to either the $p+ip$ or $p-ip$ phase since $\lambda_\perp-\lambda_\perp'$ vanishes only with fine-tuning.  

To summarize, in this section we have shown that depositing superconducting islands (either uniformly or non-uniformly) within integer quantum Hall trenches allows one to access nontrivial 2D superconducting states supporting Ising anyons.  This outcome emerges quite naturally from weak interchain perturbations when the individual trenches are tuned to criticality, which can be traced to the fact that time-reversal symmetry is absent and the carriers in the quantum Hall fluid derive from a single fermionic species.  So far the weakly coupled chain approach was convenient but by no means necessary since this section dealt only with free fermions.  One can readily verify, for instance, that the Ising anyon phases we captured survive well away from this regime and persist even in an isotropic system.  The remainder of this paper treats analogous setups where the $\nu = 1$ state is replaced by a strongly correlated fractional quantum Hall fluid.  Throughout numerous parallels will arise with the simpler treatment described here.  We should point out that in the fractional case the weakly coupled chain approach provides the \emph{only} analytically tractable window at our disposal, though we similarly expect isotropic relatives of the physics we capture to exist there as well.

\section{\texorpdfstring{Overview of $\mathbb{Z}_3$ parafermion criticality}{Overview of Z3 parafermion criticality}}
\label{Z3CFT}

One useful way of viewing Sec.~\ref{ModulatedCase} is that we dissected a $\nu = 1$ quantum Hall system to construct a non-local representation of the transverse-field Ising model---i.e., a Majorana chain.  
In preparation for treating the more theoretically challenging $\nu = 2/3$ fractionalized setup, here we review an analogous $\mathbb{Z}_3$-invariant chain corresponding to the three-state quantum clock model. This clock model realizes a critical point  described by a $\mathbb{Z}_3$ parafermion conformal field theory (CFT), which provides the building blocks for the Read-Rezayi wavefunction and plays a central role in describing the edge modes of this state. Studying the chain will enhance our understanding of the symmetries, phase structure, and perturbations of this CFT.  Furthermore, much of the groundwork necessary for our subsequent quantum Hall analysis will be developed here.

The $\mathbb{Z}_3$ quantum clock model is comprised of a chain of three-component `spins'.  Here we assume an infinite number of sites (to avoid subtleties with boundary conditions) and define operators $\sigma_j$ and $\tau_j$ that act nontrivially on the three-dimensional Hilbert space capturing the spin at site $j$.  
These operators
satisfy a generalization of the Pauli matrix algebra,
\begin{eqnarray}
  \sigma_j^3 = \tau_j^3 = 1, \quad &&\sigma_j^\dagger = \sigma_j^2, \qquad\tau_j^\dagger = \tau_j^2,
  \nonumber \\
\sigma_j\tau_j &=& e^{i2\pi/3}\tau_j\sigma_j,
  \label{ClockModelOps}
\end{eqnarray}
while all other commutators aside from the last line above are trivial: $[\sigma_j,\tau_{j'\neq j}]=[\sigma_j,\sigma_{j'}]=[\tau_j,\tau_{j'}]=0$.
It follows that $\sigma_j$ and $\tau_j$ can point in three inequivalent directions separated by an angle of $2\pi/3$, similar to a clock hand that takes on only three symmetric orientations.  Noncommutation of these operators implies that $\tau_j$ `winds' $\sigma_j$ and vice versa.  In other words, each operator can be represented by a matrix with eigenvalues $1,\,e^{i2\pi /3},\,e^{-i2\pi /3}$, but one cannot simultaneously diagonalize $\sigma_j$ and $\tau_j$.
The simplest quantum clock Hamiltonian bears a similar structure to the transverse-field Ising model and reads
\begin{equation}
  H = -J\sum_j(\sigma_j^\dagger \sigma_{j+1}+H.c.)-h\sum_j (\tau_j^\dagger + \tau_j),
  \label{ClockH}
\end{equation}
where we assume couplings $J,h\geq 0$.  This 1D Hamiltonian can be found by taking an anisotropic limit of the 2D classical three-state Potts model, and so the two share essentially identical physical properties.  

The quantum clock model in Eq.~\eqref{ClockH} exhibits the useful property of non-local duality symmetry.  Indeed, upon introducing dual operators
\begin{equation}
  \mu_j = \prod_{k\leq j}\tau_k,\qquad\nu_j = \sigma_j^\dagger \sigma_{j+1}
\end{equation}
that satisfy the same relations as in Eq.~\eqref{ClockModelOps} with $\sigma_j\rightarrow \mu_j$ and $\tau_j\rightarrow \nu_j$, the Hamiltonian takes on an identical form
\begin{equation}
  H_\textrm{dual} = -h\sum_j(\mu_j^\dagger \mu_{j+1}+H.c.)-J\sum_j (\nu_j^\dagger + \nu_j)
\end{equation}
with $h$ and $J$ interchanged.  Equation~\eqref{ClockH} additionally exhibits a number of other symmetries that play an important role in our analysis.  Spatial symmetries include simple lattice translations $T_x$ and parity $P$ (which sends $\sigma_j\rightarrow \sigma_{-j}$ and $\tau_j\rightarrow \tau_{-j}$).  The model also preserves a $\mathbb{Z}_3$ transformation ($\sigma_j\rightarrow e^{i2\pi/3}\sigma_j$) and a corresponding dual operation $\mathbb{Z}_3^\textrm{dual}$ ($\mu_j\rightarrow e^{i2\pi/3}\mu_j$).  Finally, there exists a time-reversal symmetry $\mathcal{T}$ that squares to unity ($\sigma_j\rightarrow\sigma_j, \tau_j\rightarrow \tau_j^\dagger$) and a charge conjugation symmetry $\mathcal{C}$ that flips the sign of the $\mathbb{Z}_3$ charge carried by the clock model operators ($\sigma_j\rightarrow \sigma_j^\dagger, \tau_j\rightarrow\tau_j^\dagger$).  

Like the closely related transverse-field Ising model, the clock Hamiltonian supports two symmetry-distinct phases.  When $J$ dominates, a ferromagnetic phase emerges with $\langle \sigma_j\rangle \neq 0$, thus spontaneously breaking $\mathbb{Z}_3$; increasing $h$ drives a transition to a paramagnetic state that in dual language yields $\langle \mu_j\rangle \neq 0$ and a broken $\mathbb{Z}_3^\textrm{dual}$.  Hence one can view $\sigma_j$ as an order parameter and $\mu_j$ as a `disorder parameter'. Duality implies that the phase transition occurs at the self-dual point $J = h$, and indeed the exact solution shows that this point is critical~\cite{Temperley:1971}.  The scaling limit of the self-dual clock Hamiltonian is described by a $\mathbb{Z}_3$ parafermion (or equivalently three-state Potts) CFT~\cite{ZamolodchikovParafermion}, whose content we discuss further below.

We will describe in the next section a new physical route to this CFT.  In particular, our approach uses $\nu = 2/3$ quantum Hall states to construct a chain of $\mathbb{Z}_3$ generalized Majorana operators that arise from the clock model via a `Fradkin-Kadanoff' transformation~\cite{FradkinKadanoff}.  This transformation---which is analogous to the more familiar Jordan-Wigner mapping in the transverse-field Ising chain---also lends useful intuition for the physical meaning of parafermion fields as we will see.  
The Fradkin-Kadanoff transformation in the clock model allows for two closely related forms of these $\mathbb{Z}_3$ generalized Majorana operators: either
\begin{subequations}\begin{align}
	\alpha_{R,2j-1} = \sigma_j \mu_{j-1}, \qquad \alpha_{R,2j} = e^{i2\pi/3}\sigma_j\mu_j,
	\label{eq:alphaRdef}
\end{align}
or
\begin{align}
	\alpha_{L,2j-1} = \sigma_j \mu^\dagger_{j-1}, \qquad \alpha_{L,2j} = e^{-i2\pi/3}\sigma_j\mu^\dagger_j,
	\label{eq:alphaLdef}
\end{align}\end{subequations}
which differ only in the string of operators encoded in the disorder parameter $\mu_j$.  Note that when applying a Jordan-Wigner transformation to the Ising chain there is no such freedom since there the string is Hermitian.  The above operators satisfy
\begin{equation}
  \alpha_{A,j}^3 = 1, \qquad \alpha_{A,j}^\dagger = \alpha_{A,j}^2
  \label{alphaProperties}
\end{equation}
for $A = R/L$, similar to the clock operators from which they derive.  Because of the strings, however, they exhibit non-local commutation relations,
\begin{align}\begin{split}
  \alpha_{R,j} \alpha_{R,j'} &= e^{i\frac{2\pi}{3} \operatorname{sgn}(j'-j)}\alpha_{R,j'}\alpha_{R,j} \,,
  \\
  \alpha_{L,j} \alpha_{L,j'} &= e^{-i\frac{2\pi}{3} \operatorname{sgn}(j'-j)}\alpha_{L,j'}\alpha_{L,j} \,.
  \label{alphacommutator}
\end{split}\end{align}

Equations~\eqref{alphaProperties} and \eqref{alphacommutator} constitute the defining properties for the $\mathbb{Z}_3$ generalized Majorana operators that will appear frequently in this paper. By using the labels $L$ and $R$ we have anticipated the identification of these operators with left- and right-moving fields in the CFT.  On the lattice, however, $\alpha_{Rj}$ and $\alpha_{Lj}$ are not independent, as one can readily verify that
\begin{align}\begin{split}
	\alpha_{R,2j+1}^\dagger\alpha_{R,2j} &= e^{i2\pi/3}\alpha_{L,2j+1}^\dagger \alpha_{L,2j} \,,
\\	\alpha_{R,2j-1}^\dagger\alpha_{R,2j} &= \alpha_{L,2j}^\dagger \alpha_{L,2j-1} \,.
	\label{LRrelation}
\end{split}\end{align}
Despite this redundancy, it is nevertheless very useful to consider both representations since $\alpha_{Rj}$ and $\alpha_{Lj}$ transform into one another under parity $P$ and time-reversal $\mathcal{T}$.

In terms of $\alpha_{Rj}$, the clock Hamiltonian of Eq.~\eqref{ClockH} reads
\begin{align}\begin{split}
  H &= -J \sum_j (e^{i2\pi/3}\alpha_{R,2j+1}^\dagger \alpha_{R,2j} + H.c.)
  \\&\quad	-h \sum_j (e^{i2\pi/3}\alpha_{R,2j}^\dagger\alpha_{R,2j-1} +H.c.).
  \label{Hparafendleyons1}
\end{split}\end{align}
An equivalent form in terms of $\alpha_{L,j}$ follows from exploiting Eqs.~\eqref{LRrelation}.  The ferromagnetic and paramagnetic phases of the original clock model correspond here to distinct dimer patterns for $\alpha_{R,j}$ (or $\alpha_{L,j}$) favored by the $J$ and $h$ terms above.  On a finite chain, the symmetry-related degeneracy of the ferromagnetic phase is encoded through $\mathbb{Z}_3$ zero-modes bound to the ends of the system \cite{Fendley}, similar to the Majorana end-states in a Kitaev chain \cite{1DwiresKitaev}.  The dimerization appropriate for the paramagnetic phase, by contrast, supports no such edge zero-modes, consistent with the onset of a unique ground state.  In this representation $\mathbb{Z}_3$ parafermion criticality arising at $J = h$ corresponds to the limit where these competing dimerizations balance, leaving the system gapless.  For the remainder of this section we provide an overview of this well-understood critical point.  

The $\mathbb{Z}_3$ parafermion CFT has central charge $c = 4/5$, and is rational.
One of the very useful properties of a rational CFT is that a finite set of operators---dubbed primary fields---characterize the entire Hilbert space.  That is, all states in the Hilbert space can be found by acting with the primary fields and the (possibly extended) conformal symmetry generators on the ground state.  With appropriate boundary conditions, the theory admits independent left- and right-moving conformal symmetries, and so it is useful to consider purely chiral primary fields. These fields exhibit non-local correlations; local operators are found by combining left- and right-movers in a consistent way.

When the conformal symmetry algebra is extended by a spin-3 current into the so-called `$\mathcal{W}_3$-algebra' \cite{ZamolodchikovParafermion,FateevZamo:W3:87}, the $\mathbb{Z}_3$ parafermion CFT possesses six right-moving primary fields.
These consist of the identity field $I_R$, the chiral parts of the spin field $\sigma_R$ and $\sigma_R^\dagger$, parafermion fields $\psi_R$ and $\psi_R^\dagger$, and the chiral part $\epsilon_R$ of the `thermal' operator
The left-moving sector contains an identical set of fields, labeled by replacing $R$ with $L$.  The CFT analysis yields the exact scaling dimensions of these operators---the chiral spin fields each have dimension $1/15$, the parafermions each have dimension $2/3$, while $\epsilon_{R/L}$ has dimension $2/5$.  

Perturbing the critical Hamiltonian by the thermal operator---which changes the ratio of $J/h$ away from criticality---provides a field-theory description of the clock Hamiltonian's gapped ferromagnetic and paramagnetic phases.  Note that in the Ising case, the thermal operator is composed of chiral Majorana fields, which also form the analogue of the parafermions $\psi_{R/L}$.  The fact that here the parafermions and thermal operator constitute independent fields allows for additional relevant perturbations, which in part underlies the interesting behavior we describe in this paper.  More precisely, perturbing the critical Hamiltonian instead by $\psi_L\psi_R + H.c.$ violates $\mathbb{Z}_3$ symmetry, but still results in two degenerate ground states that are \emph{not} symmetry-related \cite{Fateev1990,FateevZamo1991}; see Sec.~\ref{IntegrableLadder} for further discussion. The analogous property in our quantum Hall setup is intimately related to the appearance of Fibonacci anyons.

All of the symmetries introduced earlier in the lattice model are manifested in the CFT.  
Particularly noteworthy are the $\mathbb{Z}_3$ and $\mathbb{Z}_3^\textrm{dual}$ symmetries, whose existence is actually more apparent in the CFT due to independence of the left- and right-moving fields.  The former transformation sends $\psi_A\to e^{i2\pi/3} \psi_A$ and $\sigma_A \to e^{i2\pi/3}\sigma_A$, where $A=L$ or $R$.  (As usual the conjugate fields acquire a phase $e^{-i2\pi/3}$ instead.)  The dual transformation $\mathbb{Z}_3^\textrm{dual}$ similarly takes  $\psi_R\to e^{i2\pi/3} \psi_R$ and $\sigma_R\to e^{i2\pi/3} \sigma_R$, but alters left-movers via $\psi_L\to e^{-i2\pi/3} \psi_L$ and $\sigma_L\to e^{-i2\pi/3} \sigma_L$. Under either symmetry the fields $\epsilon_L$ and $\epsilon_R$ remain invariant; this is required in order for the Hamiltonian to preserve both $\mathbb{Z}_3$ and $\mathbb{Z}_3^\textrm{dual}$ for all couplings $J$ and $h$.

The relation between the lattice operators and primary fields at the critical point provides valuable insight into the physical content of the CFT.  Reference \onlinecite{LatticeCFTrelation} establishes such a correspondence by appropriately matching the spin and symmetry properties carried by a given microscopic operator and the continuum fields.  This prescription yields the following familiar expansions for the lattice order and disorder parameters,
\begin{equation}
  \sigma_j \sim \sigma_R^\dagger \sigma_L^\dagger + \dots \;, \qquad \mu_j \sim \sigma_R^\dagger \sigma_L + \dots \;,
  \label{sigma_mu}
\end{equation}
where the ellipses denote terms with subleading scaling dimension.  One can similarly express the thermal operator as
\begin{equation}
  \sigma_j^\dagger \sigma_{j+1} + H.c. \sim \epsilon_R \epsilon_L + \dots \;.
\end{equation}
Most crucial to us here is the expansion of the $\mathbb{Z}_3$ generalized Majorana operators~\cite{LatticeCFTrelation}, which 
will form the fundamental low-energy degrees of freedom in our quantum Hall construction:
\begin{subequations} \label{eq:alphaCFT} \begin{align}
  \alpha_{R,j} &\sim a\, \psi_{R} + (-1)^j b\, \sigma_{R}\epsilon_L + \dots \;,
  \label{alphaR} \\
  \alpha_{L,j} &\sim a\, \psi_{L} + (-1)^j b\, \sigma_{L}\epsilon_R + \dots \;,
  \label{alphaL}
\end{align}\end{subequations}
with $a,b$ real constants.
[The phases in the definition of $\alpha_{R/L}$ in Eqs.~\eqref{eq:alphaRdef} and \eqref{eq:alphaLdef} are paramount in this lattice operator/CFT field correspondence.]
The above equations endow clear meaning to the parafermion fields---they represent long-wavelength fluctuations in the generalized Majorana operators at the critical point.  Importantly, however, these lattice operators also admit an oscillating component involving products of $\sigma$ and $\epsilon$ fields, which in fact yield a slightly smaller scaling dimension than the parafermion fields.  In Sec.~\ref{RRsection} we will use the link between ultraviolet and infrared degrees of freedom encapsulated in Eqs.~\eqref{alphaR} and \eqref{alphaL} to controllably explore the phase diagram for coupled critical chains.

The physical meaning of the chiral primary fields is further illuminated by their fusion algebra, which describes how the fields behave under operator products.  This property is constrained strongly but not entirely by commutativity, associativity, and consistency with the $\mathbb{Z}_3$ symmetries.  Any fusion with the identity of course is trivial.  As a more enlightening example, two parafermion fields  obey the fusion rule $\psi_R\times \psi_R \sim \psi_R^\dagger$ (and similarly for $\psi_L$). That is, taking the operator product of two parafermion fields contains something in the sector of the conjugate parafermion (i.e., the conjugate parafermion itself or some descendant field obtained by acting with the symmetry generators on the parafermion). This fusion is natural to expect given the properties in Eq.~\eqref{alphaProperties} exhibited by the lattice analogs $\alpha_{R/Lj}$. The complete set of fusion rules involving $\psi_R$ or $\psi_L$ reads
\begin{align}\begin{aligned}
   \psi\times I &\sim \psi, & \psi\times\psi &\sim \psi^\dagger, & \psi\times\psi^\dagger &\sim I, 
	\\
   \psi\times\sigma^\dagger &\sim \epsilon, & \psi\times\sigma &\sim \sigma^\dagger, & \psi\times\epsilon &\sim \sigma\ ;
   \label{psi_rules}
\end{aligned}\end{align}
here and below the fields in such expressions implicitly all belong to either the $L$ or $R$ sectors.  Fusion rules for $\psi_{R/L}^\dagger$ simply follow by conjugation or by fusing again with $\psi_{R/L}$.

The remaining rules for fusion with $\sigma_{R/L}$ are
\begin{align}
		\sigma\times\sigma &\sim \sigma^\dagger + \psi^\dagger,
	&	\sigma\times\epsilon &\sim \sigma + \psi,
	&	\sigma\times\sigma^\dagger \sim I + \epsilon\ ,
  \label{sigma_rules}
\end{align}
with those for $\sigma_{R/L}^\dagger$ given by conjugation. 
A sum on the right-hand side indicates that two particular fields can fuse to more than one type of field, signaling degeneracies.
Finally, the chiral part of the thermal operator exhibits a `Fibonacci' fusion rule,
\begin{equation}
  \epsilon \times \epsilon \sim I + \epsilon\ .
  \label{epsilon_rule}
\end{equation}
Equation~\eqref{epsilon_rule} is especially important: it underlies why the `decorated' fractional quantum Hall setup to which we turn next yields Fibonacci anyons with universal non-Abelian statistics.  (To be precise we reserve $\epsilon$ and $I$ for CFT operators; the related Fibonacci anyon and trivial particle that appear in the forthcoming sections will be respectively denoted $\varepsilon$ and $\I$.)

  
\section{\texorpdfstring{$\mathbb{Z}_3$ parafermion criticality via $\nu = 2/3$ quantum Hall states}{Z3 parafermion criticality via nu=2/3 quantum Hall states}}
\label{QH_Z3_criticality}

Our goal now is to illustrate how one can engineer the non-local representation of the clock model in Eq.~\eqref{Hparafendleyons1}, and with it a critical point described by $\mathbb{Z}_3$ parafermion CFT, using edge states of a spin-unpolarized $\nu = 2/3$ system in the so-called (112) state.  As a primer, Sec.~\ref{EdgeTheorySection} begins with an overview of the edge theory for this quantum Hall phase (see Ref.~\onlinecite{KaneFisherPolchinski} for an early analysis).  Section~\ref{ParafendleyonSection} then constructs $\mathbb{Z}_3$ generalized Majorana zero-modes from counterpropagating sets of $\nu = 2/3$ edge states, while Sec.~\ref{Z3criticality} hybridizes these modes along a 1D chain to generate $\mathbb{Z}_3$ parafermion criticality.  Results obtained here form the backbone of our coupled-chain analysis carried out in Sec.~\ref{RRsection}.  Note that much of the ensuing discussion applies also to the bosonic $(221)$-state with minor modifications; this bosonic setup will be briefly addressed later in Secs.~\ref{sec:FibPhaseEdge} and~\ref{TQFTsection}.

\subsection{Edge theory}
\label{EdgeTheorySection}

Edge excitations at the boundary between a spin-unpolarized $\nu = 2/3$ droplet and the vacuum can be described with a two-component field $\vec\phi(x) = \left(\phi_\uparrow(x),\phi_\downarrow(x)\right)$, where $x$ is a coordinate along the edge and the subscripts indicate physical electron spin.  In our conventions $\phi_\alpha(x)$ is compact on the interval $[0,2\pi)$; hence physical operators involve either derivatives of $\vec\phi$ or take the form $e^{i{\vec l}\cdot \vec\phi}$ for some integer vector ${\vec l}$.
Commutation relations between these fields follow from an integer-valued $K$-matrix that encodes the charge and statistics for allowed quasiparticles in the theory \cite{WenBook}.
For the case of interest here we have 
\begin{equation}
  [\phi_\alpha(x),\phi_\beta(x')] = i\pi \left[ (K^{-1})_{\alpha\beta} \operatorname{sgn}(x-x') + i\sigma^y_{\alpha\beta} \right]
  \label{Commutation1}
\end{equation}
with
\begin{eqnarray}
  K = \begin{pmatrix} 1 & 2 \\ 2 & 1 \end{pmatrix}.
\end{eqnarray}
The term involving the Pauli matrix $\sigma^y$ corresponds to a Klein factor as discussed below.  Since $\operatorname{det}K <0$ the $\nu = 2/3$ edge supports counterpropagating modes; these can be viewed, roughly, as $\nu = 1$ and $\nu = 1/3$ modes running in opposite directions.  
 
In terms of the `charge vector' $\vec q = (1,1)$ the total electron density for the edge is ${\vec q} \cdot \partial_x\vec\phi/(2\pi)$.  Since we are dealing with an unpolarized state, it is also useful to consider the density for electrons with a definite spin $\alpha = \uparrow,\downarrow$, which is given by 
\begin{equation}
  \rho_\alpha = \frac{\partial_x\phi_\alpha}{2\pi}.
  \label{Density1}
\end{equation}
Equations~\eqref{Commutation1} and \eqref{Density1} allow one to identify 
\begin{equation}
  \psi_\alpha = e^{iK_{\alpha\beta}\phi_\beta}
\end{equation}
as spin-$\alpha$ electron operators.  Indeed, these operators add one unit of electric charge and satisfy appropriate anticommutation relations (note that anticommutation between $\psi_\uparrow$ and $\psi_\downarrow$ requires the Klein factor introduced above).
One can further, with the aid of Eq.~\eqref{Density1}, define a Hamiltonian incorporating explicit density-density interactions via
\begin{equation}
  H = \int_x \frac{1}{4\pi}\sum_{\alpha,\beta = \uparrow,\downarrow} (\partial_x\phi_{\alpha})V_{\alpha\beta}(\partial_x\phi_{\beta}) + \dots \;,
  \label{H_single_edge}
\end{equation}
where $V_{\alpha\beta}$ is a positive-definite matrix describing screened Coulomb interactions and the ellipsis denotes all other allowed quasiparticle processes.

\begin{figure}[t]
	\centering
	\includegraphics[width = \columnwidth]{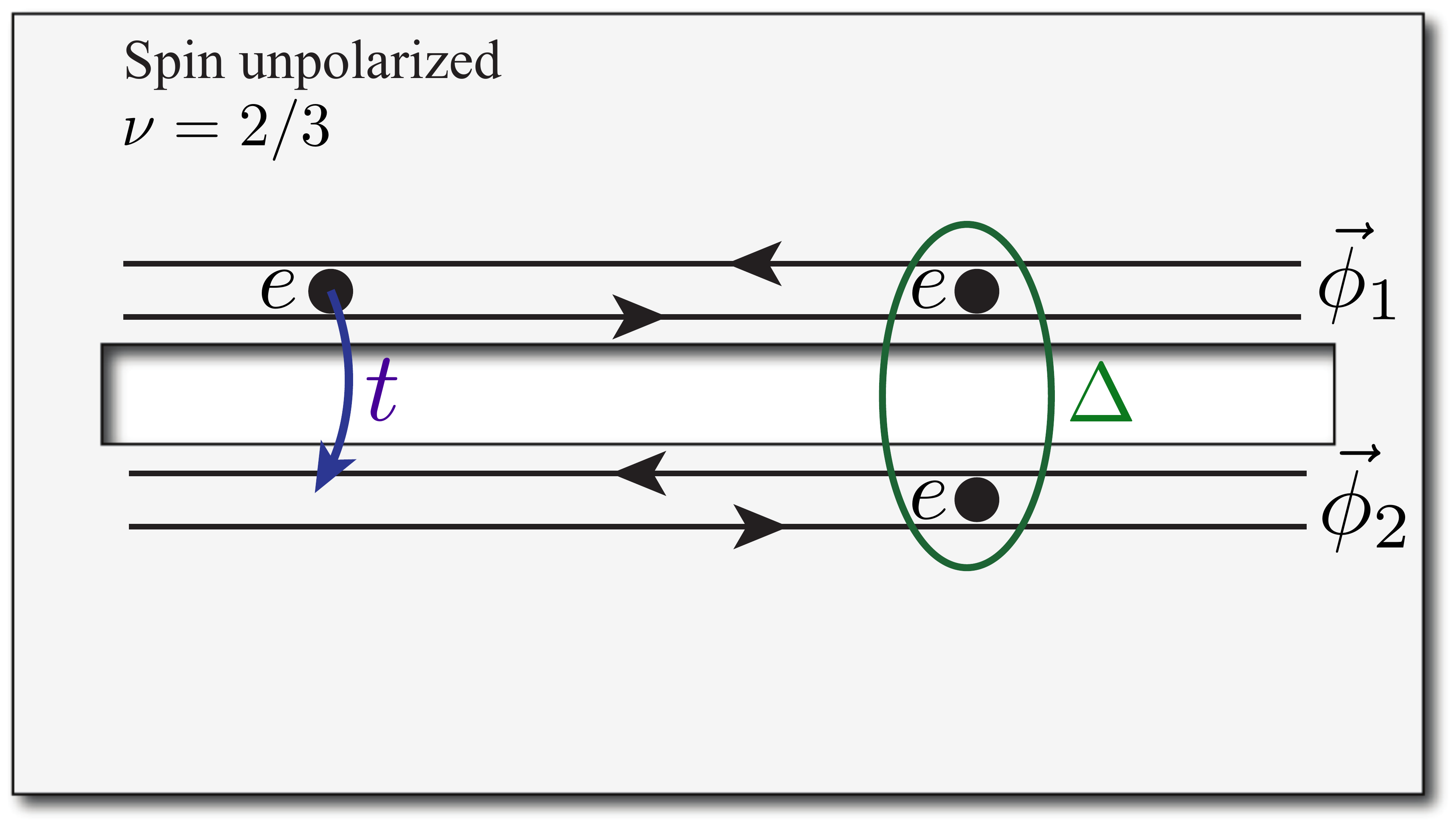}
	\caption{Spin-unpolarized $\nu = 2/3$ setup with a long, narrow trench producing counterpropagating sets of edge states described by fields $\vec\phi_1$ on the top and $\vec\phi_2$ on the bottom.  One way of gapping these modes is through electron backscattering across the interface---which essentially `sews up' the trench.  A second gapping mechanism can arise if an $s$-wave superconductor mediates spin-singlet Cooper pairing of electrons from the top and bottom sides of the trench as illustrated above.  These processes lead to physically distinct gapped states that cannot be smoothly connected, resulting in the formation of $\mathbb{Z}_3$ generalized Majorana zero-modes at domain walls separating the two.}
	\label{setup}
\end{figure}

These preliminary definitions allow us to readily treat the following more interesting setup.  Suppose that one carves out a long, narrow trench from the system as sketched in Fig.~\ref{setup}, thus generating two identical (but oppositely oriented) sets of $\nu = 2/3$ edge states in close proximity to each other.  To describe this `doubled' edge structure we employ fields $\vec\phi_{1} = (\phi_{1\uparrow},\phi_{1\downarrow})$ for the top side of the trench and $\vec\phi_{2} = (\phi_{2\uparrow},\phi_{2\downarrow})$ for the bottom.  The corresponding electron densities for spin $\alpha$ are defined as
\begin{equation}
  \rho_{1\alpha} = \frac{\partial_x\phi_{1\alpha}}{2\pi}, \qquad \rho_{2\alpha} = -\frac{\partial_x\phi_{2\alpha}}{2\pi},
\end{equation}
while the commutation relations read
\begin{align}
  [\phi_{1\alpha}(x),\phi_{1\beta}(x')] &= i\pi[(K^{-1})_{\alpha\beta} \operatorname{sgn}(x-x') + i\sigma^y_{\alpha\beta}] ,
  \nonumber \\ 
  [\phi_{2\alpha}(x),\phi_{2\beta}(x')] &= i\pi[-(K^{-1})_{\alpha\beta} \operatorname{sgn}(x-x') + i\sigma^y_{\alpha\beta}] ,
  \nonumber \\
  [\phi_{1\alpha}(x),\phi_{2\beta}(x')] &= i\pi[-(K^{-1})_{\alpha\beta} + i\sigma^y_{\alpha\beta}] .
  \label{Commutation2}
\end{align}
(The relative minus sign for the density on the bottom side of the trench, along with the commutation relations above, can be understood by viewing $\vec\phi_1$ and $\vec\phi_2$ as essentially the same fields connected at the right end of the trench.)  It follows that the electron operators for the top and bottom sides of the trench are respectively
\begin{equation}
 	\psi_{1\alpha} = e^{iK_{\alpha\beta}\phi_{1\beta}}, \qquad \psi_{2\alpha} = e^{i K_{\alpha\beta}\phi_{2\beta}}.
\end{equation}

Similarly to Eq.~\eqref{H_single_edge}, one can express the Hamiltonian for the edge interface as
\begin{align}
	H &= \int_x \frac{1}{4\pi} \!\sum_{\substack{\alpha,\beta = \uparrow,\downarrow \\ a,b = 1,2}}\! (\partial_x\phi_{a \alpha})V_{a \alpha;b \beta}(\partial_x\phi_{b\beta})
		\;+\; \delta H.
	\label{InterfaceH}
\end{align}
Of crucial importance here are the additional terms present in $\delta H$.  Since the interface carries identical sets of counterpropagating modes, it is always possible for perturbations to gap out the edges entirely.  Here we will invoke two physically distinct gapping mechanisms, similar to our earlier $\nu = 1$ setup: $(i)$ spin conserving electron tunneling across the interface and $(ii)$ spin-singlet Cooper pairing of electrons on opposite sides of the trench, mediated by an $s$-wave superconductor.  These processes are schematically illustrated in Fig.~\ref{setup} and lead to the following perturbations,
\begin{align}\begin{split}
  \delta H &= \int_x \begin{bmatrix} -t(\psi_{1\uparrow}^\dagger \psi_{2\uparrow} + \psi_{1\downarrow}^\dagger \psi_{2\downarrow} + H.c.) 
	  \\ + \Delta(\psi_{1\uparrow}\psi_{2\downarrow} - \psi_{1\downarrow} \psi_{2\uparrow} + H.c.) \end{bmatrix} ,
  \label{deltaHfermions}
\end{split}\end{align}
where $t$ and $\Delta$ are the tunneling and pairing amplitudes.  It is important to emphasize that in this setup tunneling and pairing of \emph{fractional} charges across the trench is not possible---such processes are unphysical since the intervening region separating the top and bottom sides by assumption supports only electronic excitations.  Later, however, we will encounter edges separated by $\nu = 2/3$ quantum Hall fluid, and in such a geometry inter-edge fractional charge tunneling can arise.

Before discussing the fate of the system in the presence of the couplings in $\delta H$ it is useful to introduce a basis change to charge- and spin-sector fields
\begin{align}\begin{split}
	\theta_\rho &= \tfrac{1}{2}(\phi_{1\uparrow}+\phi_{1\downarrow}-\phi_{2\uparrow}-\phi_{2\downarrow}) ,
\\	\phi_\rho &= \tfrac{1}{2}(\phi_{1\uparrow}+\phi_{1\downarrow}+\phi_{2\uparrow}+\phi_{2\downarrow}) ,
\\	\theta_\sigma &= \tfrac{1}{2}(\phi_{1\uparrow}-\phi_{1\downarrow}-\phi_{2\uparrow}+\phi_{2\downarrow}) ,
\\	\phi_\sigma &= \tfrac{1}{2}(\phi_{1\uparrow}-\phi_{1\downarrow}+\phi_{2\uparrow}-\phi_{2\downarrow}) .
	\label{phi_theta}
\end{split}\end{align}
Here $\rho_{+} = \partial_x\theta_\rho/\pi$ and $S_{+} = \partial_x\theta_\sigma/\pi$ respectively denote the total edge electron density and spin density, while $ \rho_- = \partial_x\phi_\rho/\pi$ and $S_- = \partial_x\phi_\sigma/\pi$ are respectively the difference in the electron density and spin density between the top and bottom sides of the trench.  Equations~\eqref{Commutation2} imply that the only nontrivial commutation relations amongst these fields are
\begin{align}\begin{split}
	[\theta_\rho(x),\phi_\rho(x')] &= -\frac{2\pi i}{3}\Theta(x'-x) ,
\\	~[\theta_\sigma(x),\phi_\sigma(x')] &= 2\pi i\Theta(x'-x) ,
\\	~[\phi_\rho(x),\phi_\sigma(x')] &= -2\pi i ,
	\label{Commutation3}
\end{split}\end{align}
where $\Theta$ is the Heaviside step function.
(Contrary to the first two lines, the third is nontrivial only because of Klein factors.)  

In this basis $\delta H$ becomes simply
\begin{align}
  \delta H = \int_x \big[4t \cos\theta_\sigma \sin(3\theta_\rho) - 4\Delta \cos\theta_\sigma \sin(3\phi_\rho)\big].
  \label{deltaH}
\end{align}
The scaling dimensions of the operators above depend on the matrix $V_{a\alpha;b\beta}$ in Eq.~\eqref{InterfaceH} specifying the edge density-density interactions.  In the simplest case $V_{a\alpha;b\beta} = v\delta_{ab}\delta_{\alpha\beta}$ both the tunneling and pairing terms have scaling dimension 2 and hence are marginal (to leading order).  Following Ref.~\onlinecite{KaneFisherPRB}, we have verified that upon tuning $V_{a\alpha;b\beta}$ away from this limit $t$ and $\Delta$ can be made simultaneously relevant.  Hereafter we assume that both terms can drive an instability, either because they are explicitly relevant or possess `order one' bare coupling constants.  

Suppose first that inter-edge tunneling dominates.  In terms of integer-valued operators $\hat{M}$, $\hat{m}$, this coupling pins
\begin{align}\begin{split}
	\theta_\sigma &= \pi \hat{M} ,
\\	\theta_\rho &= \frac{2\pi}{3}\hat{m}+\frac{\pi}{3}\hat{M}-\frac{\pi}{6} \qquad \textrm{(tunneling~gap)},
	\label{TunnelingGap}
\end{split}\end{align}
to minimize the energy, thus fully gapping the charge and spin sectors.  Note that both fields are simultaneously pinnable since $\theta_\sigma$ and $\theta_\rho$ commute with each other.  If the pairing term dominates, however, a gap arises from pinning 
\begin{align}\begin{split}
	\theta_\sigma &= \pi \hat{M} ,
\\	\phi_\rho &= \frac{2\pi}{3}\hat{n}+\frac{\pi}{3}\hat{M}+\frac{\pi}{6} \qquad \textrm{(pairing~gap)},
	\label{PairingGap}
\end{split}\end{align}
where $\hat{n}$ is another integer operator.  Both fields are again simultaneously pinnable, but note that Eqs.~\eqref{TunnelingGap} and \eqref{PairingGap} can \emph{not} be simultaneously fulfilled in the same region of space since $[\theta_\rho(x),\phi_\rho(x')] \neq 0$.
Consequently, the tunneling and pairing terms compete with one another~\footnote{By itself this does not necessarily imply that the phases generated by the tunneling and pairing terms are distinct, but it turns out that this is the case here.}.
The physics is directly analogous to the competing ferromagnetic and superconducting instabilities in a quantum spin Hall edge; there domain walls separating regions gapped by these different means bind Majorana zero-modes~\cite{MajoranaQSHedge}.  Due to the fractionalized nature of the $\nu = 2/3$ host system, in the present context domain walls generate more exotic zero-modes---as in Refs.~\onlinecite{ChernInsulatorParafendleyons, ClarkeParafendleyons,LindnerParafendleyons,ChengParafendleyons,BarkeshliParafendleyons1,BarkeshliParafendleyons2,QuantumWiresParafendleyons,BarkeshliClassification1,BarkeshliClassification2}---that will eventually serve as our building blocks for a $\mathbb{Z}_3$ parafermion CFT.


\subsection{\texorpdfstring{$\mathbb{Z}_3$ zero-modes}{Z3 zero-modes}}
\label{ParafendleyonSection}

\begin{figure}[t]
	\centering
	\includegraphics[width = \columnwidth]{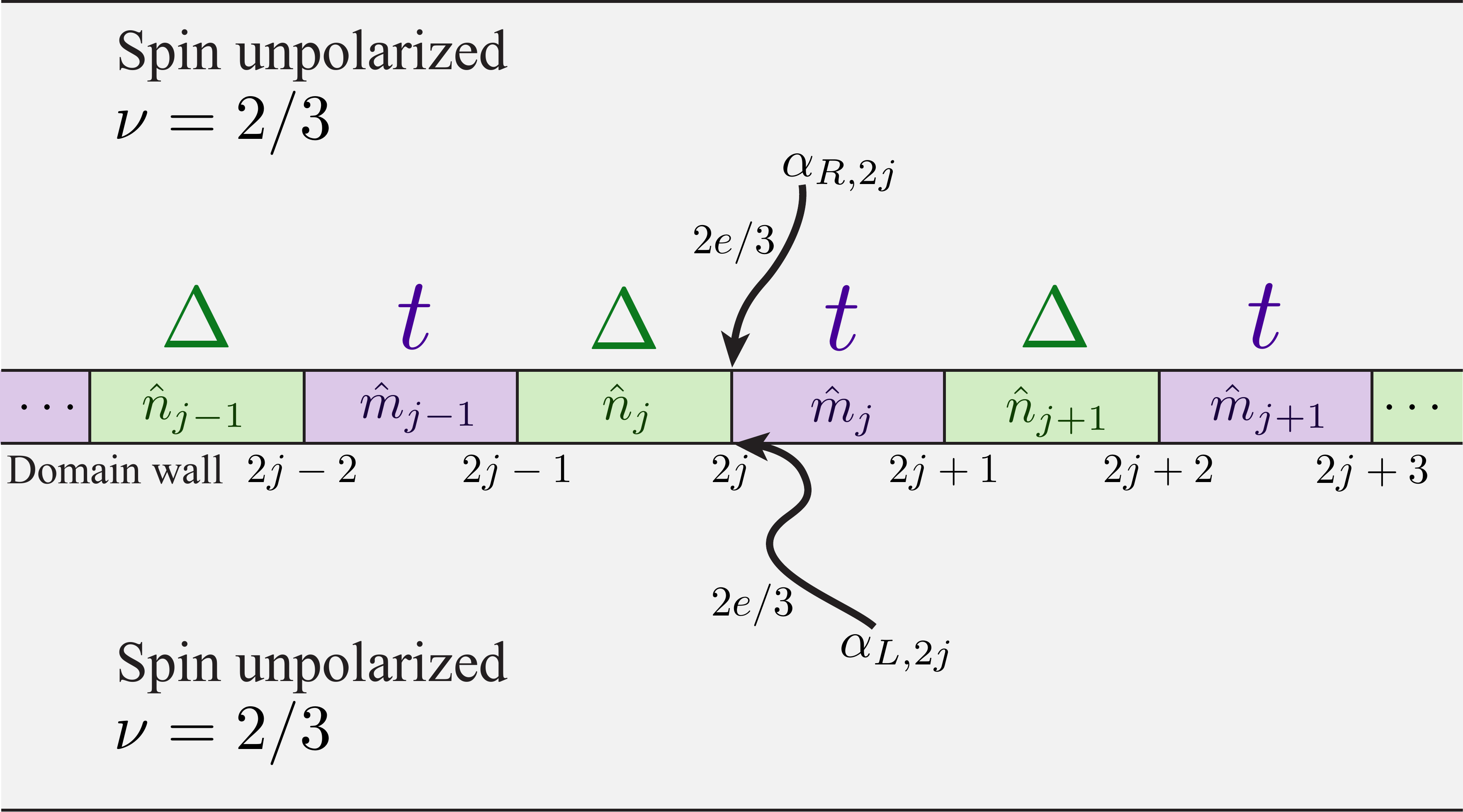}
	\caption{Schematic of a spin-unpolarized $\nu = 2/3$ system hosting a trench in which the edge modes are alternately gapped by electron backscattering $t$ and Cooper pairing $\Delta$.  The integer operators $\hat{m}_i$ and $\hat{n}_i$ in each domain characterize the pinning of the charge-sector fields as specified in Eqs.~\eqref{TunnelingGap} and \eqref{PairingGap}.  Physically, $\hat{m}_i-\hat{m}_{i-1}$ quantifies the total charge (top plus bottom) $Q^+_i$ on the intervening superconducting-gapped region, while $\hat{n}_{i+1}-\hat{n}_{i}$ quantifies the charge difference (top minus bottom) $Q^-_i$ on the intervening tunneling-gapped segment.  The remaining low-energy physics is captured by $\mathbb{Z}_3$ generalizations of Majorana operators $\alpha_{R/L,j}$ bound to each domain wall labeled as above.  These operators cycle the values of $Q^\pm_i$ on the domains by adding charge $2e/3$ (mod $2e$) to the top and bottom trench edges as illustrated in the figure.  Charge $2e/3$ tunneling between neighboring domain walls hybridizes these modes, and can be described by a 1D Hamiltonian [Eq.~\eqref{Hparafendleyons2}] intimately related to the three-state quantum clock model.  The critical point of this Hamiltonian, as in the clock model context, is described by $\mathbb{Z}_3$ parafermion conformal field theory.  }
	\label{Chain_fig}
\end{figure}

As an incremental step towards this goal we would like to now capture these zero-modes by studying an infinite array of long domains alternately gapped by tunneling and pairing as displayed in Fig.~\ref{Chain_fig} \footnote{A discussion of a finite, closed ring of alternating domains can be found in Ref.~\onlinecite{LindnerParafendleyons}.}; note the similarity to the integer quantum Hall setup analyzed in Sec.~\ref{ModulatedCase}.
(For illuminating complementary perspectives on this problem see the references cited at the end of the previous paragraph.)
In each tunneling- and pairing-gapped segment the fields are pinned according to Eqs.~\eqref{TunnelingGap} and \eqref{PairingGap}, respectively.  Since $\theta_\sigma$ is pinned everywhere, in the ground-state sector the integer operator $\hat{M}$ takes on a common value throughout the trench.  (Nonuniformity in $\hat{M}$ requires energetically costly twists in $\theta_\sigma$.)  Conversely, the pinning of $\theta_\rho$ and $\phi_\rho$ is described by independent operators $\hat{m}_j$ and $\hat{n}_j$ in different domains---see Fig.~\ref{Chain_fig} for our labeling conventions.  The commutation relations between the integer operators follow from Eqs.~\eqref{Commutation3}, which yield
\begin{align}
	[\hat{n}_j,\hat{m}_{j'}] = \begin{cases}
		\frac{3}{2\pi}i, & \quad j> j', \\
		0, & \quad j \leq j',
		\end{cases}
  \label{nm_commutator}
\end{align}
while all other commutators vanish.  

The zero-mode operators of interest can be obtained from quasiparticle operators 
$e^{i(\vec{l}_1\cdot\vec\phi_1 + \vec{l}_2\cdot\vec\phi_2)}$ \emph{acting inside of a domain wall}, simply by projecting into the ground-state manifold.  To project nontrivially the dependence on the field $\phi_\sigma$ must drop out since $e^{i\phi_\sigma}$ creates a kink in $\theta_\sigma$ which costs energy.  This condition is satisfied provided
\begin{equation}
  l_{1\uparrow} -l_{1\downarrow}+  l_{2\uparrow} -l_{2\downarrow} = 0.
  \label{lcondition}
\end{equation}
Projection of the remaining fields is achieved by replacing $\theta_\sigma$, $\theta_\rho$, \emph{and} $\phi_\rho$ by their pinned values on the adjacent domains.  The complete set of projected quasiparticle operators obeying Eq.~\eqref{lcondition} can be generated by $e^{i(\vec{l}_1\cdot\vec\phi_1 + \vec{l}_2\cdot\vec\phi_2)}$ with $l_{1\uparrow} = l_{1\downarrow} = 1$, $l_{2\uparrow} = l_{2\downarrow} = 0$ and $l_{1\uparrow} = l_{1\downarrow} = 0$, $l_{2\uparrow} = l_{2\downarrow} = 1$.  Crucially, these correspond to charge-$2e/3$ quasiparticle operators acting on the top and bottom edges of the trench, respectively.  
Suppose that $\mathcal{P}$ is the ground-state projector while $x_j$ denotes a coordinate inside of domain wall $j$.  We then explicitly get
\begin{align}\begin{split}
	\mathcal{P}e^{i[\phi_{1\uparrow}(x_{j})+\phi_{1\downarrow}(x_{j})]}\mathcal{P} &\equiv (-1)^j\alpha_{Rj} ,
\\	\mathcal{P}e^{i[\phi_{2\uparrow}(x_{j})+\phi_{2\downarrow}(x_{j})]}\mathcal{P} &\equiv (-1)^je^{i\pi/3}\alpha_{Lj} ,
	\label{ProjectedOps}
\end{split}\end{align}
where on the right side we have inserted phase factors for later convenience and defined $\mathbb{Z}_3$ generalized Majorana zero-mode operators
\begin{align}\begin{aligned}
	\alpha_{R,2j-1} &= e^{i\frac{2\pi}{3}(\hat{n}_{j}+\hat{M}-1)}e^{i\frac{2\pi}{3}\hat{m}_{j-1}}	&&\textrm{(top edge)} ,
\\	\alpha_{R,2j} &= e^{i\frac{2\pi}{3}(\hat{n}_{j}+\hat{M})}e^{i\frac{2\pi}{3}\hat{m}_{j}}	&&\textrm{(top edge)} ,
\\	\alpha_{L,2j-1} &= e^{i\frac{2\pi}{3}(\hat{n}_j+1)}e^{-i\frac{2\pi}{3}\hat{m}_{j-1}}	&&\textrm{(bottom edge)} ,
\\	\alpha_{L,2j} &= e^{i\frac{2\pi}{3}\hat{n}_j}e^{-i\frac{2\pi}{3}\hat{m}_{j}}	&&\textrm{(bottom edge)} .
	\label{parafendleyons2}
\end{aligned}\end{align}
Above we denote whether a given zero-mode operator adds charge $2e/3$ (mod $2e$) to the top or bottom edge.  The importance of the spatial separation between $\alpha_{Rj}$ and $\alpha_{Lj}$ evident here is hard to overstate and will prove exceedingly valuable in the following section.  Equation~\eqref{nm_commutator} implies that the $\mathbb{Z}_3$ zero-mode operators in our quantum Hall setup satisfy precisely the properties in Eqs.~\eqref{alphaProperties} through \eqref{LRrelation} introduced in the quantum clock model context.  Once again $\alpha_{Rj}$ and $\alpha_{Lj}$ are not independent, but as we will see describing physical processes for coupled trenches in a simple way requires retaining both representations because of their spatial separation.  

The $\mathbb{Z}_3$ zero-modes encode a ground-state degeneracy that admits a simple physical interpretation.  First we note that gauge invariant quantities involve \emph{differences} in the $\hat{m}_j$ or $\hat{n}_j$ operators on different domains.
Consider then the quantity $A(x-x') = e^{i\pi \int_{x}^{x'}\rho_{+}(x'')dx''} = e^{i[\theta_\rho(x')-\theta_\rho(x)]}$, where again $\rho_{+} = \partial_x\theta_\rho/\pi$ denotes the total density.  If $x$ and $x'$ straddle a pairing-gapped domain in which $\hat{n}_j$ is pinned, then Eq.~\eqref{TunnelingGap} yields a ground-state projection 
\begin{equation}
  \mathcal{P}A(x-x')\mathcal{P} = e^{i\frac{2\pi}{3}(\hat{m}_{j}-\hat{m}_{j-1})} = e^{-i\frac{2\pi}{3}}\alpha_{R,2j-1}^\dagger \alpha_{R,2j}.
\end{equation}  
Hence 
\begin{equation}
  Q^{+}_j \equiv \frac{2}{3}(\hat{m}_{j}-\hat{m}_{j-1})
\end{equation}
specifies the total charge (mod $2e$) on the pairing-gapped segment.  A comparison with the more familiar case of Majorana zero-modes along a quantum spin Hall edge is useful here.  In that context the Majoranas encode a two-fold degeneracy between even and odd parity ground states of a superconducting-gapped region of the edge.  Here the physics is richer---a superconducting segment of the $\nu = 2/3$ interface supports ground states with charge $0, 2/3$, or $4/3$ (mod $2e$).
From the density difference $\rho_- = \partial_x\phi_\rho/\pi$ between the top and bottom edges of the trench one can similarly define $B(x-x') = e^{i\pi \int_{x}^{x'}\! \rho_-(x'')dx''} = e^{i[\phi_\rho(x')-\phi_\rho(x)]}$.
With $x$ and $x'$ now straddling an $\hat{m}_j$-pinned tunneling-gapped region, one obtains
\begin{equation}
  \mathcal{P}B(x-x')\mathcal{P} = e^{i\frac{2\pi}{3}(\hat{n}_{j+1}-\hat{n}_{j})} = e^{-i\frac{2\pi}{3}}\alpha_{R,2j}^\dagger \alpha_{R,2j+1}.
\end{equation}  
We thus see that 
\begin{equation}
  Q^-_j \equiv \frac{2}{3}(\hat{n}_{j+1}-\hat{n}_{j})
\end{equation}
represents the charge difference (again mod $2e$) across the trench in a tunneling-gapped region, which can also take on three distinct values.
If desired one can use these definitions to express $\hat{m}_j = \frac{3}{2}\sum_{i\leq j}Q^+_i$ and $\hat{n}_j = \frac{3}{2}\sum_{i<j}Q^-_i$; these forms can then be used to rewrite the $\mathbb{Z}_3$ zero-mode operators of Eq.~\eqref{parafendleyons2} in terms of physical quantities.

To avoid overcounting degeneracy, observe that due to the nontrivial commutator in Eq.~\eqref{nm_commutator} one can specify either the total charge $Q^+_j$ on each superconducting segment \emph{or} the charge difference $Q^-_j$ on each tunneling gapped region---but not both simultaneously.
Consequently, there exists three ground states per pair of domain walls (neglecting possible Hilbert space constraints), yielding a quantum dimension of $\sqrt{3}$ associated with each zero-mode~\footnote{Note that $\hat{M}$ does not produce additional ground-state degeneracy since there is no gauge invariant quantity one can construct from this operator.}.
The action of the zero-mode operators on a given initial state alters $Q^\pm_j$ by integer multiples of $2e/3$, thereby allowing one to cycle through the entire ground-state manifold.  More precisely, the modification of these charges follows from
\begin{equation}
  e^{i\pi(Q^+_j+2/3)}\alpha_{R/L,k} = \alpha_{R/L,k}e^{i\pi Q^+_j}
  \label{Qj_transform}
\end{equation}
for $k = 2j-1$ or $2j$, while
\begin{eqnarray}
  e^{i\pi(Q^-_j\pm2/3)}\alpha_{R/L,k} &=& \alpha_{R/L,k}e^{i\pi Q^-_j}
  \label{deltaQj_transform}
\end{eqnarray}
for $k = 2j$ or $2j+1$.  (At other values of $k$ the zero-modes do not affect $Q^\pm_j$.)  Notice that $\alpha_{R,k}$ and $\alpha_{L,k}$ increment the charge difference $Q^-_j$ in opposing directions because they add quasiparticles to opposite sides of the trench.  

One can now intuitively understand why two nontrivial $R/L$ representations exist for the $\mathbb{Z}_3$ zero-modes whereas the Majorana operators $\gamma_j$ discussed in Sec.~\ref{ModulatedCase} are uniquely defined, up to a sign.  For concreteness let us work in a basis where the ground states are labeled by the set of charges $\{Q^+_j\}$ on the superconducting regions.  The key point is that in the fractional quantum Hall case there are two physically distinct processes that transform the system from one such ground state to another.  Namely, the total charge on a given superconducting segment can be incremented by adding fractional charge either to the upper or lower trench edges.  This distinction is meaningful since fractional charge injected at one edge can not pass to the other because only electrons can tunnel across the trench.  These two processes are implemented precisely by $\alpha_{Rj}$ and $\alpha_{Lj}$, as illustrated in Fig.~\ref{Chain_fig}.  By contrast in the integer quantum Hall case no such distinction exists.  The Majorana operators add one unit of electric charge (mod $2e$) which can readily meander across the trench, so that their representation is essentially unique.  

Finally, we note a curious feature implicit in the zero-modes and ground states: although a $\nu = 2/3$ edge supports charge-$e/3$ excitations, they are evidently frozen out in the low-energy subspace in which we are working.  The doubling of the minimal charge arises because the spin sector is uniformly gapped throughout the trench.  Charge-$e/3$ excitations must therefore come in opposite-spin pairs to circumvent the spin gap.  As a corollary, one cannot define an electron operator in the projected Hilbert space since charge-$e$ excitations are absent for the same reason.
This explains the $\mathbb{Z}_3$ structure arising in the theory---along with the difference from the $\mathbb{Z}_6$ structure found in related studies of $\nu = 1/3$ Laughlin states~\cite{ClarkeParafendleyons,LindnerParafendleyons,ChengParafendleyons,QuantumWiresParafendleyons}.

\subsection{\texorpdfstring{$\mathbb{Z}_3$ parafermion criticality}{Z3 parafermion criticality}}
\label{Z3criticality}

Imagine now that the size of each domain shrinks so that quasiparticle tunneling between neighboring domain walls becomes appreciable.  Such processes lift the ground-state degeneracy described above and can be modeled by an effective Hamiltonian
\begin{equation}
  H_\textrm{eff} = -J_\Delta \sum_j \cos(\pi Q^+_j)- J_t \sum_j \cos(\pi Q^-_j)
\end{equation}
with $J_\Delta,J_t>0$.  The first term reflects a fractional Josephson coupling between adjacent superconducting segments~\cite{1DwiresKitaev,ClarkeParafendleyons,LindnerParafendleyons,ChengParafendleyons}, mediated by charge-$2e/3$ tunneling across the intervening tunneling-gapped region.  This favors pinning $\hat{m}_j$ to uniform values in all superconducting regions, resulting in $Q^+_j = 0$ throughout.  Similarly, the second (competing) term represents a `dual fractional Josephson'~\cite{dual_FJE_Nilsson,dual_FJE_Meng,dual_FJE_Jiang,dual_FJE_Panagiotis} coupling favoring uniform $\hat{n}_j$ in tunneling-gapped regions and hence $Q^-_j = 0$.  In terms of generalized Majorana operators defined in Eq.~\eqref{parafendleyons2} the effective Hamiltonian becomes
\begin{align}\begin{split}
  H_\textrm{eff} &= -J_t \sum_j (e^{i2\pi/3}\alpha_{R,2j+1}^\dagger \alpha_{R,2j} + H.c.)
  \\
  &\quad	- J_\Delta \sum_j (e^{i2\pi/3}\alpha_{R,2j}^\dagger\alpha_{R,2j-1} +H.c.),
  \label{Hparafendleyons2}
\end{split}\end{align} 
which exhibits precisely the same form as the Fradkin-Kadanoff representation of the quantum clock model in Eq.~\eqref{Hparafendleyons1}.  

The connection to the quantum clock model can be further solidified by considering how the various symmetries present in the former are manifested in our $\nu = 2/3$ setup.  Appendix \ref{Symmetries} discusses this important issue and shows that all of these in fact have a transparent physical origin (including the time-reversal operation $\mathcal{T}$ that squares to unity).  To streamline the analysis we have defined the generalized Majorana operators in Eqs.~\eqref{parafendleyons2} such that under each symmetry they transform identically to those defined in the clock model.  

The $\mathbb{Z}_3$ and $\mathbb{Z}_3^\textrm{dual}$ transformations, which send 
\begin{subequations}\begin{align}
	\alpha_{R/Lj} &\rightarrow e^{i 2\pi/3}\alpha_{R/Lj} && (\mathbb{Z}_3) ,
	\label{Z3transformation}
\\	\alpha_{R/Lj} &\rightarrow e^{\pm i 2\pi/3}\alpha_{R/Lj} && (\mathbb{Z}_3^\textrm{dual}) ,
	\label{dualZ3transformation}
\end{align}\end{subequations}
warrant special attention.  Clearly the Hamiltonian in Eq.~\eqref{Hparafendleyons2} preserves both operations.  In our quantum Hall problem these symmetries relate to physical electric charges.  More precisely, they reflect global conservation of the `triality' operators 
\begin{align}
	e^{i\pi Q^+_\textrm{tot}} \equiv e^{i\pi \sum_j Q^+_j},
	\qquad e^{i\pi Q^-_\textrm{tot}} \equiv e^{i\pi \sum_j Q^-_j},
	\label{trialities}
\end{align}
which generalize the notion of parity and take on three distinct values.  The trialities respectively constitute conserved $\mathbb{Z}_3$ and $\mathbb{Z}_3^\textrm{dual}$ quantities that specify (mod 2) the sum and difference of the total electric charge on each side of the trench.  According to Eqs.~\eqref{Z3transformation} and \eqref{dualZ3transformation}, $\alpha_{Rj}$ and $\alpha_{Lj}$ carry the same $\mathbb{Z}_3$ charge but opposite $\mathbb{Z}_3^\textrm{dual}$ charge; this is sensible given that these operators increment the charge on opposite trench edges [see also Eqs.~\eqref{Qj_transform} and \eqref{deltaQj_transform}].

The correspondence with the clock model allows us to directly import results from Sec.~\ref{Z3CFT} to the present setup.  Most importantly we immediately conclude that the limit $J_\Delta = J_t$ realizes a self-dual critical point described by a $\mathbb{Z}_3$ parafermion CFT.  Furthermore, at the critical point the primary fields relate to the lattice operators through Eqs.~\eqref{alphaR} and \eqref{alphaL}, repeated here for clarity:
\begin{subequations}\begin{align}
	\alpha_{Rj} &\sim a \psi_{R} + (-1)^j b \sigma_{R}\epsilon_L + \dots
	 	&&\textrm{(top~edge)} ,	\label{alphaR2}
\\	\alpha_{Lj} &\sim a \psi_{L} + (-1)^j b \sigma_{L}\epsilon_R + \dots
		&&\textrm{(bottom~edge)}.	\label{alphaL2}
\end{align}\end{subequations}
An important piece of physics that is special to our $\nu = 2/3$ setup is worth emphasizing here.
First we note that $\epsilon_{A}$, with $A = R$ or $L$, represents an electrically neutral field that modifies neither the total charge nor the charge difference across the trench.  This can be understood either from the fusion rule $\epsilon \times \epsilon \sim 1 + \epsilon$---which implies that $\epsilon_{A}$ carries the same (trivial) charge as the identity---or by recalling from Sec.~\ref{Z3CFT} that $\epsilon_{R/L}$ remains invariant under both $\mathbb{Z}_3$ and $\mathbb{Z}_3^\textrm{dual}$.  It follows that $\psi_{R/L}$ and $\sigma_{R/L}$ must carry all of the physical charge of the lattice operators $\alpha_{R/Lj}$.  That is, like their lattice counterparts, $\psi_R$ and $\sigma_R$ add charge $2e/3$ to the top edge of the trench, while $\psi_L$ and $\sigma_L$ add charge $2e/3$ to the bottom trench edge.  In this sense the $\psi$ and $\sigma$ fields inherit the spatial separation exhibited by $\alpha_{R/Lj}$.  The next section explores stacks of critical chains, and there this property will severely restrict the perturbations that couple fields from neighboring chains, ultimately enabling us to access a superconducting analogue of the Read-Rezayi state in a rather natural way.


\section{\texorpdfstring{Fibonacci phase: a superconducting analogue of the $\mathbb{Z}_3$ Read-Rezayi state}{Fibonacci phase}}
\label{RRsection}

\begin{figure}[t]
	\centering
	\subfigure[\label{fig:CoupledChaina}]{ \includegraphics[width = 0.99\columnwidth]{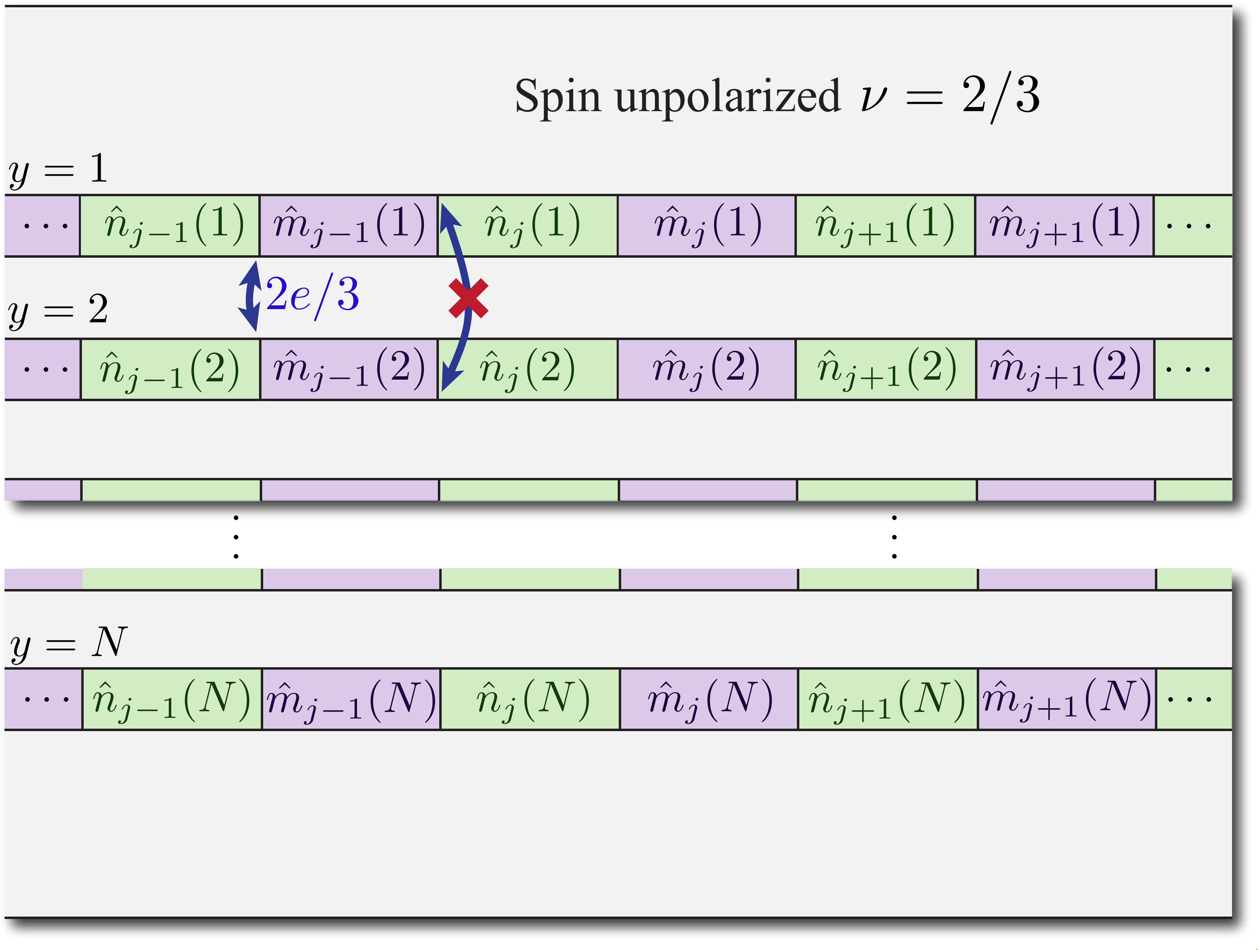} }
	\\
	\subfigure[\label{fig:CoupledChainb}]{ \includegraphics[width = 0.8\columnwidth]{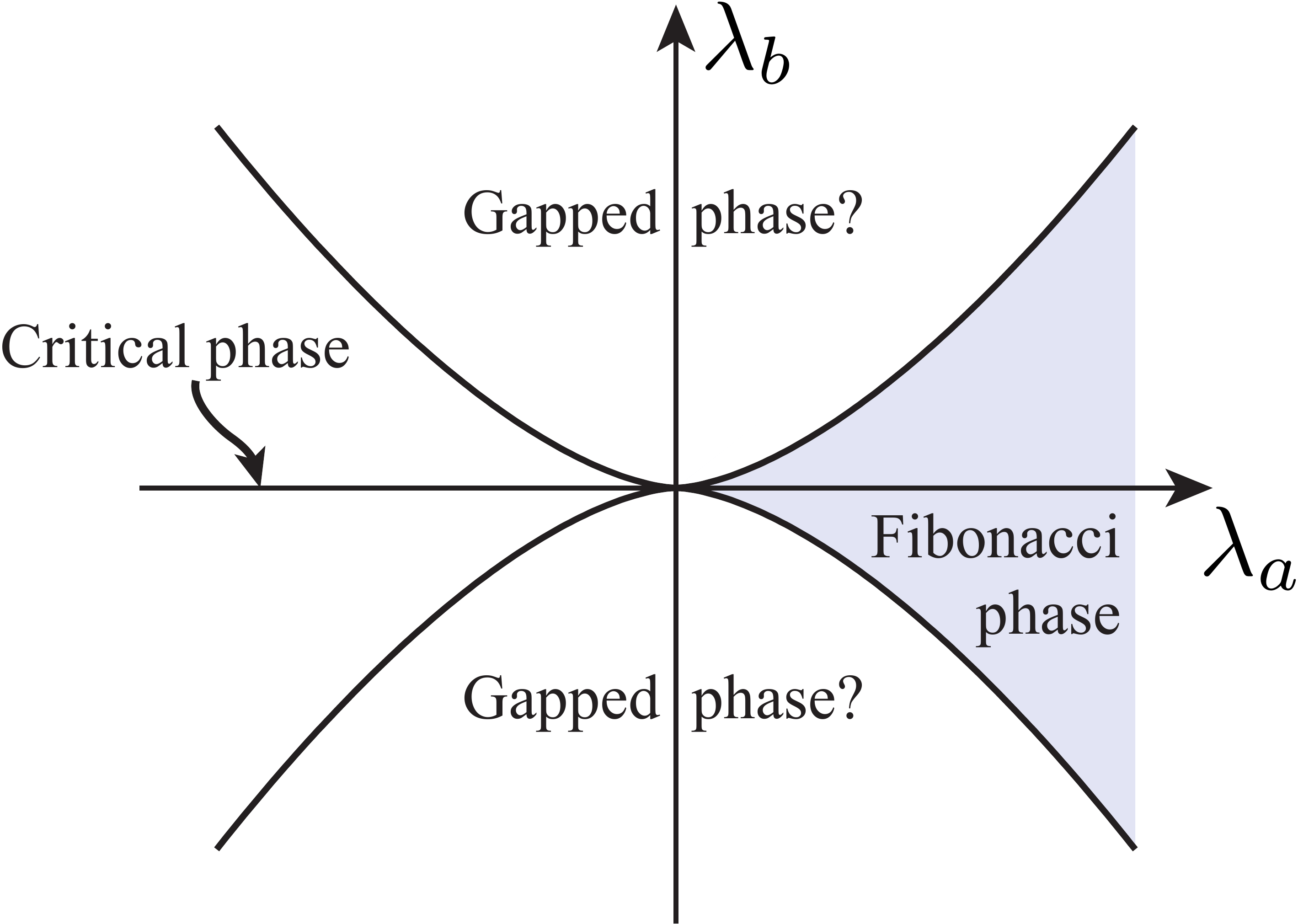} \quad }
	\caption{(a) Multi-chain generalization of Fig.~\ref{Chain_fig} in which a sequence of trenches labeled by $y = 1,\ldots,N$ is embedded in a spin-unpolarized $\nu = 2/3$ quantum Hall system.  Once again the edge modes opposite each trench are alternately gapped by electron backscattering and Cooper pairing, with $\hat{m}_i(y)$ and $\hat{n}_i(y)$ characterizing the pinned charge-sector fields in a given domain [see Eqs.~\eqref{TunnelingGap} and \eqref{PairingGap}].  We assume that the  $\mathbb{Z}_3$ generalized Majorana operators bound to each domain wall hybridize strongly within a trench and weakly between neighboring trenches.  Underlying this hybridization is tunneling of $2e/3$ charges which can only take place through the fractional quantum Hall fluid; examples of allowed and disallowed processes are illustrated above.  (b) Phase diagram for this system of weakly coupled chains starting from the limit where each chain is tuned to a critical point described by $\mathbb{Z}_3$ parafermion conformal field theory.  The couplings $\lambda_{a/b}$ represent interchain perturbations defined in Eq.~\eqref{Hperp}.  }
	\label{Coupled_chain_fig}
\end{figure}

Consider now the geometry of Fig.~\subfigref{fig:CoupledChaina} in which a spin-unpolarized $\nu = 2/3$ quantum Hall system hosts an array of $N$ trenches of the type studied in Sec.~\ref{QH_Z3_criticality}.  
Edge excitations on the top and bottom of each trench 
can similarly be described with fields $\phi_{1\alpha}(x,y)$ and $\phi_{2\alpha}(x,y)$, where $\alpha$ denotes spin, $x$ is a coordinate along the edges, and $y = 1,\ldots,N$ labels the trenches.  In the charge- and spin-sector basis defined in Eqs.~\eqref{phi_theta}  the nontrivial commutation relations now read
\begin{align}\begin{split}
	[\theta_\rho(x,y),\phi_\rho(x',y')] &= \begin{cases}
		-\frac{2\pi i}{3}\Theta(x'-x), & y = y', \\
		-\frac{2\pi i}{3}\Theta(y'-y), & y \neq y',
		\end{cases} \\
	~[\theta_\sigma(x,y),\phi_\sigma(x',y')] &= \begin{cases}
		2\pi i\Theta(x'-x), & y = y', \\
		2\pi i\Theta(y'-y), & y \neq y',
		\end{cases} \\
	~[\phi_\rho(x,y),\phi_\sigma(x',y')] &= -2\pi i.
	\label{Commutation4}
\end{split}\end{align}
For $y = y'$ one simply recovers Eqs.~\eqref{Commutation3}.  The additional commutators for $y \neq y'$ ensure proper anticommutation relations between electron operators acting at different trenches but play no important role in our analysis.

We assume that the set of counterpropagating edge modes opposite each trench are alternately gapped by the Cooper pairing and electron backscattering mechanisms discussed in Sec.~\ref{QH_Z3_criticality}.  At low energies the pinning of the charge- and spin-sector fields in each gapped region is again described by Eqs.~\eqref{TunnelingGap} and \eqref{PairingGap}.  Using the labeling scheme in Fig.~\subfigref{fig:CoupledChaina}, we respectively denote the integer operators characterizing $\theta_\sigma$, $\theta_\rho$, and $\phi_\rho$  in a given domain by $\hat{M}(y)$, $\hat{m}_j(y)$, and $\hat{n}_j(y)$.  [Note that $\hat{M}(y)$ depends only on $y$ since the spin sector is gapped uniformly in each trench].  It follows from Eqs.~\eqref{Commutation4} that $\hat{M}(y)$ commutes with all integer operators while
\begin{equation}
   [\hat{n}_j(y),\hat{m}_{j'}(y')] = \begin{cases}
      	\frac{3}{2\pi}i, & \quad y>y' , \\
      	\frac{3}{2\pi}i, & \quad  y = y' ~\textrm{and}~ j> j' , \\
   	   0, & \quad y<y' .
	  \end{cases}
  \label{nm_commutator2}
\end{equation}
The remaining low-energy degrees of freedom for the system are captured by $\mathbb{Z}_3$ generalized Majorana operators $\alpha_{R/L,j}(y)$ bound to the domain walls; these are defined precisely as in Eq.~\eqref{parafendleyons2} upon appending a trench label $y$ to each operator.  In the spirit of Ref.~\onlinecite{TeoKaneChains} we are interested in the situation where these modes hybridize strongly with their neighbors inside of a given trench, and secondarily with neighbors from adjacent trenches.  Just as for the Majorana case discussed in Sec.~\ref{MajoranaCase}, this weakly coupled chain approach allows us to utilize the formalism developed for a single trench in Sec.~\ref{QH_Z3_criticality} to access nontrivial 2D phases.  

Let the effective Hamiltonian describing this setup be
\begin{equation}
  H = H_\textrm{intra} + H_{\perp}.
\end{equation}
The first term incorporates interactions between $\mathbb{Z}_3$ generalized Majorana operators within each trench and essentially reflects $N$ copies of the Hamiltonian in Eq.~\eqref{Hparafendleyons2}:
\begin{align}
  H_\textrm{intra} &= -\sum_{y = 1}^N \bigg\{ J_t \sum_j [e^{i\frac{2\pi}{3}}\alpha^\dag_{R,2j+1}(y) \alpha^{}_{R,2j}(y) + H.c.]
  \nonumber \\
  &\;\;	+ J_\Delta \sum_j [e^{i\frac{2\pi}{3}}\alpha^\dag_{R,2j}(y) \alpha_{R,2j-1}(y)+H.c.] \bigg\}.
  \label{Hintra}
\end{align}
Here $J_\Delta$ and $J_t$ denote superconducting and `dual' fractional Josephson couplings, respectively, mediated by charge-$2e/3$ tunneling across the domains.  

Interchain couplings are encoded in $H_{\perp}$ and similarly arise from tunneling of fractional charges between adjacent trenches. Consider, for example, the perturbations
\begin{equation}
  \sum_{y=1}^{N-1}\sum_{j,j'}\left[\zeta_{jj'} e^{-i\vec{l}\cdot\vec{\phi}_{2}(x_j,y)}e^{i\vec{l}\cdot\vec{\phi}_{1}(x_{j'},y+1)} + H.c.\right]
  \label{InterchainHopping}
\end{equation}
with $\vec{l} = (1,1)$ and $x_k$ corresponding to a coordinate in domain wall $k$ in a given chain.  These terms transfer charge $2e/3$ between the top edge of domain wall $j'$ on trench $y+1$, and the bottom edge of domain wall $j$ on trench $y$.  Such processes are indeed physical since the intervening quantum Hall fluid supports fractionalized excitations.  As emphasized earlier $2e/3$ tunneling across a trench is, by contrast, \emph{not} permitted since the charge would necessarily pass through trivial regions that support only electrons.  For instance, hopping of charge-$2e/3$ quasiparticles from the bottom edge of trench $y+1$ to the top edge of trench $y$ is disallowed for this reason.  Figure~\subfigref{fig:CoupledChaina} schematically illustrates such physical and unphysical processes.  
Symmetry partially constrains the tunneling coefficient $\zeta_{jj'}$ in Eq.~\eqref{InterchainHopping}.  Specifically, enforcing charge conjugation $\mathcal{C}$ (up to a $\mathbb{Z}_3^\textrm{dual}$ transformation) allows one to take $\zeta_{jj'} = e^{i2\pi/3}\zeta_{jj'}^*$.  We will further assume for simplicity that $\zeta_{jj'}$ depends only on $j-j'$, i.e., that the coupling strength between domain walls on adjacent chains depends only on their separation.  The explicit dependence of $\zeta_{jj'}$ on this separation depends on microscopic details but should of course be appropriately short-ranged.  

The action of Eq.~\eqref{InterchainHopping} in the low-energy manifold can be deduced by projecting onto the $\mathbb{Z}_3$ generalized Majorana operators $\alpha_{R/L,j}(y)$ using a trivial extension of Eqs.~\eqref{ProjectedOps} to the multi-chain case~\footnote{Tunneling of $e/3$ (rather than $2e/3$) charge can also in principle arise.  However, those processes have no effect in the low-energy subspace in which we are working since such tunneling operators project trivially.}.
Using this procedure, one can show that the quasiparticle hoppings in Eq.~\eqref{InterchainHopping} generate the following form of the interchain Hamiltonian:
\begin{align}\begin{split}
	H_{\perp} &= -\sum_{y = 1}^{N-1}\sum_{j,j'}(-1)^{j+j'}t_{j-j'}
\\	&\qquad\qquad
		\times \big[ \alpha_{L,j}^\dag(y) \alpha_{R,j'}^{\vphantom{|}}(y+1) + H.c. \big]
	\label{Hinter}
\end{split}\end{align}
with $t_{j-j'}$ real.  The factor of $(-1)^{j+j'}$ above reflects the alternating sign between even and odd domain walls on the right-hand side of the projection in Eqs.~\eqref{ProjectedOps}.  We have chosen to explicitly display this factor to distinguish from possible sign structure in $t_{j-j'}$, which encodes phases acquired by quasiparticles upon tunneling from domain wall $j$ in one chain to $j'$ in another.
Note also the conspicuous absence of terms that couple $\alpha_{R,j}^\dag(y)$ with $\alpha_{L,j'}(y+1)$---which importantly are unphysical.  As stressed in Sec.~\ref{ParafendleyonSection} $\alpha_{Rj}$ and $\alpha_{Lj}$ respectively add fractionalized quasiparticles to the top and bottom edges of a given trench.  Consequently such terms would implement disallowed processes similar to that illustrated in Fig.~\subfigref{fig:CoupledChaina}.  

Suppose that $J_t = J_\Delta$ so that in the decoupled-chain limit each trench resides at a critical point described by a $\mathbb{Z}_3$ parafermion CFT.  Again, this limit is advantageous since arbitrarily weak inter-trench couplings can dramatically impact the properties of the coupled-chain system.  At low energies it is then legitimate to expand the lattice operators $\alpha_{R/L,j}(y)$ in terms of critical fields using Eqs.~\eqref{alphaR2} and \eqref{alphaL2}.  Inserting this expansion into the interchain Hamiltonian yields
\begin{eqnarray}
  H_{\perp} &\sim& -\sum_{y = 1}^{N-1} \int_x \big{[}\lambda_a \psi_L^\dag(y) \psi_R(y+1) 
  \label{Hperp} \\
  &&\quad + \lambda_b \sigma_L(y)\epsilon_R(y)\sigma_R^\dagger(y+1)\epsilon_L(y+1) + H.c.\big{]}
  \notag
\end{eqnarray}
with real couplings
\begin{align}
  \lambda_a = a^2 \sum_j (-1)^j t_j, \qquad \lambda_b = b^2 \sum_j t_j.
\end{align}
Insight into the phases driven by these interchain perturbations---both of which are relevant at the decoupled-chain fixed point---can be gleaned by examining certain extreme limits.  

Consider first the case with $\lambda_a = 0$, $\lambda_b \neq 0$.  Since $\lambda_b$ hybridizes both the right- and left-moving sectors of a given chain with those of its neighbor, we conjecture 
that this coupling drives a flow to a fully gapped 2D phase with no low-energy modes `left behind'.  It is unclear, however, whether this putative gapped state smoothly connects to that generated by moving each individual trench off of criticality by turning on the thermal perturbation $H_T = \sum_y \int_x \lambda_T \epsilon_R(y)\epsilon_L(y)$, where $\lambda_T\sim J_t-J_\Delta$.  This intriguing question warrants further investigation but will not be pursued in this paper.

Instead we concentrate on the opposite limit $\lambda_a \neq 0$, $\lambda_b = 0$, where an immediately more interesting scenario arises.  Here the parafermion fields hybridize in a nontrivial way---left-movers from chain 1 couple only to right-movers in chain 2, left-movers from chain 2 couple only to right-movers in chain 3, and so on.  `Unpaired' right- and left-moving $\mathbb{Z}_3$ parafermion CFT sectors thus remain at the first and last chains, respectively.  The structure of this perturbation parallels the coupling that produced spinless $p+ip$ superconductivity from critical chains in the integer quantum Hall case studied in Sec.~\ref{MajoranaCase}, and furthermore closely resembles that arising in Teo and Kane's construction of Read-Rezayi quantum Hall states from coupled Luttinger liquids \cite{TeoKaneChains}.  In the present context, provided $\lambda_a$ gaps the bulk (which requires $\lambda_a>0$ as discussed below) the system enters a superconducting analogue of the $\mathbb{Z}_3$ Read-Rezayi phase that possesses edge and bulk quasiparticle content similar to its non-Abelian quantum Hall cousin.
For brevity, we hereafter refer to this state as the `Fibonacci phase'---the reason for this nomenclature will become clear later in this section.

One can deduce rough boundaries separating the phases driven by $\lambda_a$ and $\lambda_b$ from scaling.  To leading order, these couplings flow under renormalization according to
\begin{equation}
  \partial_\ell \lambda_{a/b} = (2-\Delta_{a/b})\lambda_{a/b},
\end{equation}
where $\ell$ is a logarithmic rescaling factor and $\Delta_a = 4/3$, $\Delta_b = 14/15$ represent the scaling dimensions of the respective terms.  The physics will be dominated by whichever of these relevant couplings first flows to strong coupling (i.e., values of order some cutoff $\Lambda$).  Equating the renormalization group scales at which $\lambda_{a/b}$ reach strong coupling yields the following phase boundary:
\begin{equation}
  |\lambda_b^*| \propto |\lambda_a^*|^{8/5}
\end{equation}
with $\lambda_{a/b}^*$ the bare couplings at the transition.  Figure~\subfigref{fig:CoupledChaina} sketches the resulting phase diagram, which we expound on below.  

Naturally we are especially interested in the Fibonacci phase favored by $\lambda_a>0$ and flesh out its properties in the remainder of this section.  We do so in several stages.  First, Sec.~\ref{IntegrableLadder} analyzes the properties of a single `ladder' consisting of left-movers from one trench and right-movers from its neighbor.  As we will see this toy problem is already extremely rich and contains seeds of the physics for the 2D Fibonacci phase.  Section~\ref{GSD_quasiparticles} then bootstraps off of the results there to obtain the Fibonacci phase's ground-state degeneracy and quasiparticle content.  The properties of superconducting vortices in this state are addressed in Sec.~\ref{sc_vortices}, and finally Sec.~\ref{sec:FibPhaseEdge} discusses the edge structure between the Fibonacci phase and the vacuum (as opposed to the interface with $\nu = 2/3$ fluid).

\subsection{Energy spectrum of a single `ladder'}
\label{IntegrableLadder}

Until specified otherwise we study the critical trenches perturbed by Eq.~\eqref{Hperp} assuming $\lambda_b = 0$.
This special case allows us to obtain various numerical and exact analytical results that will be used to uncover universal topological properties of the Fibonacci phase that persist much more generally.  Tractability here originates from the fact that with $\lambda_b = 0$ one can rewrite the coupled-chain Hamiltonian as $H = \sum_y H_\textrm{ladder}^{y,y+1}$, where the `ladder' Hamiltonian involves \emph{only} left-moving fields from trench $y$ and right-movers from trench $y+1$.  (Non-zero $\lambda_b$ clearly spoils this decomposition.)  More explicitly, $H_\textrm{ladder}^{y,y+1}$ can be written as
\begin{align}\begin{split}
	H_\textrm{ladder}^{y,y+1} &= H_\textrm{CFT}^L(y) + H_\textrm{CFT}^R(y+1)
		\\ &\quad - \int_x \left[ \lambda_a \psi_L^\dag(y) \psi_R(y+1) + H.c. \right],
	\label{eq:H_lambda_psi}
\end{split}\end{align}
with $H_\textrm{CFT}$ terms describing the dynamics for the unperturbed left- and right-movers from trenches $y$ and $y+1$, respectively.
Although the ladder Hamiltonians  at different values of $y$ act on completely different sectors, the problem does not quite decouple: there remains an important constraint between their Hilbert spaces which will become crucial in Sec.~\ref{GSD_quasiparticles}.  For the rest of this subsection we explore the structure of $H_\textrm{ladder}^{y,y+1}$ for a single ladder.  The information gleaned here will then allow us to address the full 2D problem.  

\begin{figure}[t]
	\subfigure[\label{fig:lambda_psi_phase}]{
		\includegraphics[width=0.45\columnwidth]{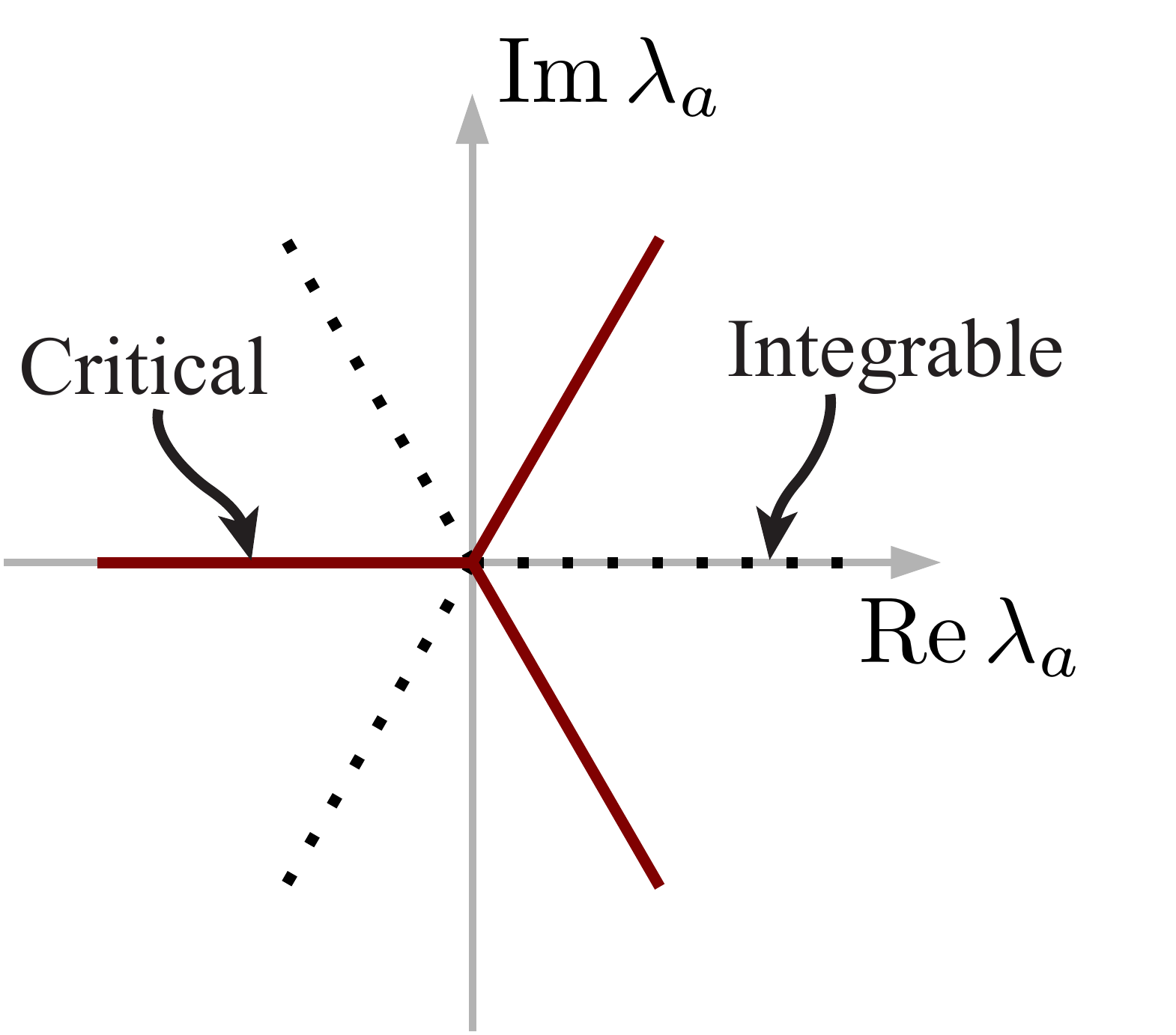}
	}
	\subfigure[\label{fig:lambda_psi_kink}]{
		\includegraphics[width=0.5\columnwidth]{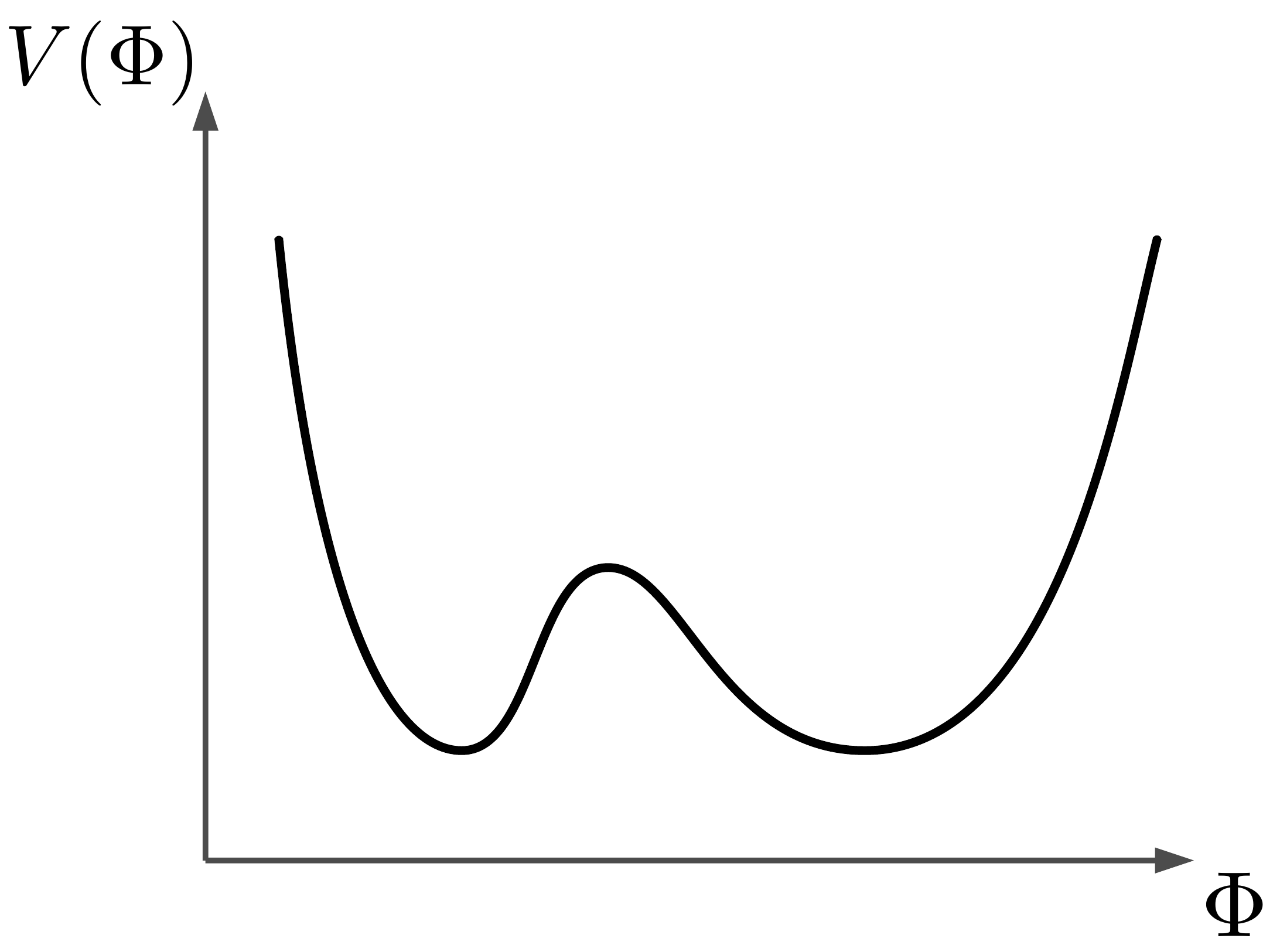}
	}
	\\[3mm]
	\subfigure[\label{fig:lambda_psi_tcsa}]{ \begin{minipage}{\columnwidth}
		\centering
		\begin{tabular}{p{0.31\columnwidth} p{0.31\columnwidth} p{0.31\columnwidth}} $[\bbone\bar\bbone]$ sector & $[\bbone\bar\varepsilon]$ sector & $[\varepsilon\bar\varepsilon]$ sector \end{tabular}\\
		\includegraphics[width=0.97\columnwidth]{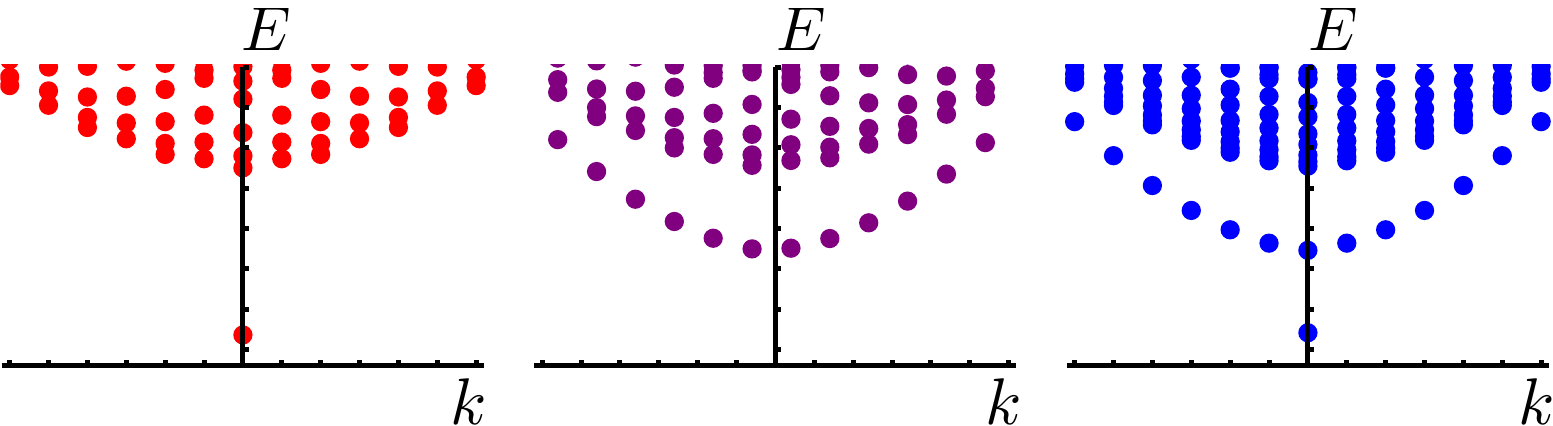}
	\end{minipage} }
	\caption{%
		(a) Phase diagram of the `ladder' Hamiltonian in Eq.~\eqref{eq:H_lambda_psi} for complex $\lambda_a$.  At $\lambda_a = 0$, the ladder resides at a $\mathbb{Z}_3$ parafermion critical point.  Along the three solid lines the ladder remains gapless, but flows instead to the tricritical Ising point.  Everywhere else the system is gapped and exhibits two symmetry-unrelated ground states together with the `Fibonacci kink spectrum' described in the main text.  The dotted lines indicate integrability.
		(b) Effective double-well Ginzburg-Landau potential of the ladder Hamiltonian, which provides an intuitive picture for the ground-state degeneracy and Fibonacci kink spectrum.  The equal-depth wells represent the two ground state sectors.  Excitations in these sectors are non-degenerate, and correspond to massive modes about the asymmetric well minima.  `Kinks' and `antikinks' interpolate between ground states, and turn out to have the same energy as the `oscillator' excitations in one of the ground states.  This is the hallmark of the Fibonacci kink spectrum.  
		(c) Energy versus momentum obtained via the truncated conformal space approach for each superselection sector.  (The $[\varepsilon\bar\bbone]$ spectrum is identical to that of $[\bbone\bar \varepsilon]$ with $k\rightarrow -k$.)  Notice the two ground states, the nearly identical single-particle bands in $[\bbone\bar \varepsilon]$ and $[\varepsilon\bar \varepsilon]$, as well as the multi-particle continuum in all sectors.  
	}
	\label{fig:FibKink}
\end{figure}

Although $\lambda_a$ as defined earlier is real it will be useful to now allow for complex values---not all of which yield distinct spectra.  Because correlators in the critical theory with $\lambda_a=0$ are non-zero only when each of the total  $\mathbb{Z}_3$ charges is trivial, perturbing around the critical point shows that the partition function can only depend on the combinations  $(\lambda_a)^3$, $(\lambda_a^*)^3$ and $|\lambda_a |^2$. 
Thus Hamiltonians related by the mapping $\lambda_a \rightarrow e^{i2\pi /3}\lambda_a$ are equivalent. The physics does, however, differ dramatically for $\lambda_a$ positive and negative \cite{Fateev1990,FateevZamo1991}. For $\lambda_a<0$ the model flows to another critical point, which turns out to fall in the universality class of the tricritical Ising model. In CFT language, this is an example of a flow between minimal models via the $\Phi_{1,3}$ operator \cite{LudwigCardy}; here the flow is from central charge $c=4/5$ to $c=7/10$ theories.  The solid lines in Fig.~\subfigref{fig:lambda_psi_phase} correspond to $\lambda_a$ values for which the ladder remains gapless.  These results imply that the full coupled-chain model with $\lambda_a<0$ and $\lambda_b = 0$ realizes a critical phase as denoted in Fig.~\subfigref{fig:CoupledChaina}.  

For $\lambda_a$ non-negative (and not with phase $\pm \pi/3$) the spectrum of a single ladder is gapped. We focus on this case from now on---especially the limit of $\lambda_a$ real and positive (modulo a phase of $2\pi/3$), where the field theory is integrable \cite{Fateev1990}.  These special values are indicated by dotted lines in Fig.~\subfigref{fig:lambda_psi_phase}.  Integrability provides a valuable tool for understanding the physics as it allows one to obtain exact results for the ladder spectrum.  Namely, the spectrum can be described in terms of quasiparticles with known scattering matrices and degeneracies.  References \onlinecite{Smirnov1991,Fateev1990} determined these via the indirect method of finding the simplest solution of the integrability constraints adhering to known properties of a Hamiltonian equivalent to Eq.~\eqref{eq:H_lambda_psi}.
This analysis is fairly technical, using tools from the representation theory of quantum groups \cite{ReshetikhinSmirnov}. While this language is probably unfamiliar to most condensed-matter physicists, the results are not: they are the rules for fusing anyons! The connection between the quasiparticle spectrum  and scattering matrix of a 1+1D integrable quantum field theory and the fusing and braiding of anyons in a 2+1D topological phase is explained in depth in Ref.~\onlinecite{FendleyFradkin}. For the $\mathbb{Z}_3$ parafermion case of interest here, the implications of integrability are striking but quite simple to understand.

To illustrate the results it is useful to first characterize the Hilbert space for a critical clock chain reviewed in Sec.~\ref{Z3CFT}, and then identify the (related but not identical) Hilbert space for a single ladder.  Consider for the moment the familiar three-state quantum clock model.  As discussed in Sec.~\ref{Z3CFT}, the entire spectrum at the critical point can be organized into sectors labeled by the chiral primary fields. With periodic boundary conditions the allowed left- and right-moving Hilbert spaces correspond to conjugate pairs $\mathcal{H}_\mathcal{F}^L \otimes \mathcal{H}_{\mathcal{F}^\dagger}^R$, where $\mathcal{F}$ signifies one of the six fields $I, \psi, \psi^\dag, \epsilon, \sigma, \sigma^\dag$.  Perturbing the critical clock model with a perturbation $\delta H \propto \int_x (\psi^\dagger_L \psi_R + H.c.)$ analogous to the $\lambda_a$ term in our ladder Hamiltonian mixes these sectors, but not completely.  Two decoupled sectors remain.  This follows from the fusion algebra described in Sec.~\ref{Z3CFT}: the key property here is that fusing with $\psi$ or $\psi^\dagger$ does not mix the first three of the six fields above with the last three. Thus when the critical clock Hamiltonian is perturbed by $\delta H$, the Hilbert space can still be divided into the following `superselection' sectors,
\begin{align}\begin{split}
	[\bbone\bar\bbone] &= \mathcal{H}_I^L \otimes \mathcal{H}_I^R \;\oplus\; \mathcal{H}_\psi^L \otimes \mathcal{H}_{\psi^\dag}^R \;\oplus\; \mathcal{H}_{\psi^\dag}^L \otimes \mathcal{H}_{\psi}^R ,
\\	[\varepsilon \bar \varepsilon] &= \mathcal{H}_\epsilon^L \otimes \mathcal{H}_\epsilon^R \;\oplus\; \mathcal{H}_\sigma^L \otimes \mathcal{H}_{\sigma^\dag}^R \;\oplus\; \mathcal{H}_{\sigma^\dag}^L \otimes \mathcal{H}_{\sigma}^R .
  \label{sectors1}
\end{split}\end{align}

Next we return to the ladder Hamiltonian given in Eq.~\eqref{eq:H_lambda_psi}.
In this case the superselection sectors above still appear, \emph{but now the left- and right-moving Hilbert spaces correspond to different trenches}.  For this reason the constraints between the left- and right-movers are relaxed, resulting in sectors not present in the periodic clock chain.  Specifically, there are two additional superselection sectors given by
\begin{align}\begin{split}
	[\bbone\bar \varepsilon] &= \mathcal{H}_I^L \otimes \mathcal{H}_\epsilon^R \;\oplus\; \mathcal{H}_\psi^L \otimes \mathcal{H}_{\sigma^\dag}^R \;\oplus\; \mathcal{H}_{\psi^\dag}^L \otimes \mathcal{H}_{\sigma}^R ,
\\	[\varepsilon\bar \bbone] &= \mathcal{H}_\epsilon^L \otimes \mathcal{H}_I^R \;\oplus\; \mathcal{H}_\sigma^L \otimes \mathcal{H}_{\psi^\dag}^R \;\oplus\; \mathcal{H}_{\sigma^\dag}^L \otimes \mathcal{H}_{\psi}^R ,
  \label{sectors2}
\end{split}\end{align}
where again $L$ and $R$ refer to different trenches.  
Note that we forbid combinations such as $\mathcal{H}_I^L \otimes \mathcal{H}_\psi^R$, which would require net fractional charge in the $\nu=2/3$ strip between the trenches; for a more detailed discussion see Sec.~\ref{sc_vortices}.  The upshot of this perturbed CFT analysis is that the Hilbert space for a single ladder can be split into the four distinct sectors defined in Eqs.~\eqref{sectors1} and \eqref{sectors2}. 

Exploiting integrability of the Hamiltonian in Eq.~\eqref{eq:H_lambda_psi} at $\lambda_a>0$ provides both an intuitive way of understanding the spectrum, and reveals remarkable degeneracies among the sectors that are far from apparent \emph{a priori}.  One important feature is that the integrable model admits two degenerate ground states \emph{not} related by any local symmetry (actually this property survives for rather general $\lambda_a$---see below).  
We confirmed the presence of two ground states by analyzing the spectrum numerically in two complementary ways.  The first method employed the density matrix renormalization group (DMRG) on an integrable lattice model; this analysis will be detailed elsewhere \cite{LatticeCFTrelation}. The second method utilized the truncated conformal space approach (TCSA), which directly simulates the field theory \cite{YurovZamolodchikov1990,YurovZamolodchikov1991}. Here the eigenstates and operator product rules of the CFT are used to characterize the Hilbert space and the action of the perturbation on these states. By truncating the Hilbert space, one obtains a finite-dimensional matrix that can be diagonalized numerically.
Results of this analysis appear in Fig.~\subfigref{fig:lambda_psi_tcsa}, which displays the energy $E$ versus momentum $k$ for three of the physical superselection sectors (the spectrum of the fourth, $[\varepsilon\bar\bbone]$, follows from that of $[\bbone\bar \varepsilon])$.  These plots clearly reveal a degeneracy between the ground states in $[\bbone\bar\bbone]$ and $[\varepsilon\bar \varepsilon]$ sectors, and a gap to all excited states.
Since there is no symmetry of the fusion algebra between the identity and $\varepsilon$ sectors, however, gapped excitations about the two ground states are not degenerate.  This too is readily apparent from our TCSA numerics in Fig.~\subfigref{fig:lambda_psi_tcsa}.  

To understand the situation more intuitively,  it is useful to imagine a Ginzburg-Landau-type effective potential following Refs.~\onlinecite{Lassig1990,Zamolodchikov:1990xc}, where the same spectrum as the ladder Hamiltonian arises (but starting from a different model). 
Two non-symmetry-related vacua together with the low-lying excitations can be described by a double-well potential, where the two wells have the same depth but exhibit different curvature as in Fig.~\subfigref{fig:lambda_psi_kink}.  From this effective potential, one can understand the four sectors in the ladder spectrum as follows.  Two of the sectors, $[\bbone\bar\bbone]$ and $[\varepsilon\bar \varepsilon]$, correspond to the degenerate minima and massive fluctuations thereabout.  The different curvature of the wells leads to non-degenerate massive modes---similar to our TCSA numerical data where $[\varepsilon \bar\varepsilon]$ exhibits the smaller gap.  In fact, there `one-particle' states occur, whereas the gap in the $[\bbone\bar\bbone]$ sector is about twice as large and appears to consist of a multi-particle continuum.  The remaining two sectors correspond to `kinks' interpolating between the ground states.  A kink is a field configuration where the field takes on one minimum to the left of some point in space and a different minimum on the right; the excitation energy is then localized at the region where the field changes.  There are two possible configurations, related by parity, and we will label these here as kinks and antikinks.  It is natural to expect that these parity conjugates occur in the $[\bbone\bar \varepsilon]$ and $[\varepsilon\bar\bbone]$ sectors.  This is indeed consistent with our numerical work displayed in Fig.~\subfigref{fig:lambda_psi_tcsa}. 


Aside from the two ground states, there exists another remarkable degeneracy between two very different quasiparticle excitations: \emph{the gap in the $[\varepsilon \bar \varepsilon]$ sector is the same as the minimum kink or antikink energy} \cite{Smirnov1991,Fateev1990}.  One can see this either directly from the numerics in Fig.~\subfigref{fig:lambda_psi_tcsa}, or from an analysis exploiting integrability. The latter shows that the kink, antikink, and `oscillator' excitation in the $[\varepsilon\bar \varepsilon]$ sector exhibit identical dispersion as well.  The entire spectrum is then built up from these fundamental excitations.
For instance, the lowest excited states in the $[\bbone\bar\bbone]$ sector form a two-particle continuum originating from kink/antikink pairs (as opposed to another species of single-particle excitations), consistent with the numerically determined spectrum.

Even though there are three flavors of excitations, the number of states in the spectrum with $N$ quasiparticles actually grows more slowly than $3^N$.  (By `quasiparticle' we mean a localized excitation that takes the form of either a kink, antikink, or oscillator mode.)  The reason is that the spatial order in which different excitation flavors occur is constrained.  Viewing the problem in terms of the double-well potential described above, the following rules are evident.
Going (say) left to right, a kink can be followed by an antikink or an oscillator excitation; an oscillator can be followed by an antikink or another oscillator; and an antikink can only be followed by a kink.  Because of these restrictions the number of states grows asymptotically with $N$ as $\varphi^N$, where again $\varphi \equiv (1+\sqrt{5})/2$ is the golden ratio.
We therefore dub the features described here as the `Fibonacci kink' spectrum.

Integrability turns out to provide a sufficient but not necessary condition for these striking degeneracies.  We have verified numerically using the TCSA method that the two symmetry-unrelated ground states and the Fibonacci kink spectrum persist even for $\lambda_a$ lying away from the dashed lines in Fig.~\subfigref{fig:lambda_psi_phase} that mark the integrable points \cite{LatticeCFTrelation}.  For instance, with $\lambda_a = e^{i\pi/5}$ the spectra are nearly indistinguishable from those in Fig.~\subfigref{fig:lambda_psi_tcsa}.  Hence for almost all $\lambda_a$ (the exception occurring where the the system is critical) the ladder Hamiltonian realizes a gapped phase with the properties noted above.  
It is useful to comment that one can, in principle, spoil this structure: terms such as $\sigma_L(y) \sigma_R(y+1) + H.c.$ break the degeneracies---but are nonlocal in our setup and thus do not reflect physical perturbations.

We should emphasize here that the preceding discussion applies only to a single ladder Hamiltonian defined in Eq.~\eqref{eq:H_lambda_psi}.  By itself this 1D model does not support Fibonacci anyons as stable excitations in any meaningful sense.  Nevertheless the tantalizing similarities are by no means accidental.  In fact the remarkable Fibonacci kink spectrum should be viewed as a precursor to both the topological order and Fibonacci anyons that do appear in the full 2D coupled-trench system.  This will be elucidated in the next subsection which uses the results obtained here to deduce the ground-state degeneracy and particle content of the Fibonacci phase.

\subsection{Ground state degeneracy and quasiparticle content}
\label{GSD_quasiparticles}

We now show that in the 2D Fibonacci phase the coupled-chain system exhibits a two-fold ground-state degeneracy on a torus.
Consider $N$ parallel trenches labeled by $y$, coupled to their neighbors via $\lambda_a>0$ (we continue to assume $\lambda_b = 0$).  To form the torus geometry each chain is itself periodic and the first and last chains at $y=1,N$ couple as well.  The system is therefore described by $H = \sum_{y = 1}^N H_\textrm{ladder}^{y,y+1}$ with periodic boundary conditions along the $x$ and $y$ directions; the ladder Hamiltonian is defined in Eq.~\eqref{eq:H_lambda_psi} and was studied for a single $y$ in the last subsection.   

Given that \emph{for a single ladder} Eq.~\eqref{eq:H_lambda_psi} already exhibits a two-fold ground-state degeneracy, one might naively expect a $2^N$-fold degeneracy for the full $N$-trench system.  This conclusion is incorrect, however, as such naive counting ignores Hilbert space constraints between the left- and right-movers within a given trench.  
In particular, combinations $\mathcal{H}_\mathcal{F}^R(y) \otimes \mathcal{H}_{\mathcal{F}'}^L(y)$ with $\mathcal{F}\in \{I,\psi,\psi^\dagger\}$ and $\mathcal{F}'\in \{\epsilon,\sigma,\sigma^\dagger\}$ (or vice versa) are forbidden for any physical boundary conditions on trench $y$ \cite{CardyBoundaryOp:1986}.  Here we have explicitly denoted that $\mathcal{H}^{R/L}$ correspond to the same chain $y$ to avoid possible confusion with the previous subsection (where the right- and left-moving Hilbert spaces correspond to different trenches).  
Thus the allowed CFT superselection sectors in each chain must have either $\mathcal{F},\mathcal{F}'\in \{I,\psi,\psi^\dagger\}$ or $\mathcal{F},\mathcal{F}'\in \{\epsilon,\sigma,\sigma^\dagger\}$; in other words,
\begin{align}
	\textrm{CFT sector}^R(y) \sim \textrm{CFT sector}^L(y) \bmod \psi .
	\label{eq:intrachain_constraint}
\end{align}
Note that this includes sectors such as $\mathcal{H}_I^R(y) \otimes \mathcal{H}_\psi^L(y)$, which are physical since fractional charges can hop between trenches. 

Now recall from Sec.~\ref{IntegrableLadder} that the ground states for a single ladder occur in the sectors $[\bbone\bar\bbone]$ and $[\varepsilon \bar \varepsilon]$ as defined in Eq.~\eqref{sectors1}, where again $\mathcal{H}^{R}$ and $\mathcal{H}^{L}$ correspond to chains $y$ and $y+1$.  In order for the 2D coupled-trench system to reside in a ground state, the superselection sectors between adjacent chains must therefore match, i.e.,
\begin{align}
	\textrm{CFT sector}^L(y) \sim \textrm{CFT sector}^R(y+1)^\dag .
	\label{eq:interchain_constraint}
\end{align}
Combining with Eq.~\eqref{eq:intrachain_constraint} this locks the Hilbert spaces of every chain together, yielding two ground states as claimed.
We label the ground states as $\ket{\bbone}$ and $\ket{\varepsilon}$, which denotes the corresponding sectors in the chains.  


Our aim next is to unambiguously identify the anyon content of our coupled-chain phase.  The Fibonacci kink spectrum identified in the ladder problem in Sec.~\ref{IntegrableLadder} already strongly hints that a Fibonacci anyon is present, though we will derive this explicitly in what follows.  To do so it will be instructive to review a few facts regarding topological states on a cylinder (instead of a torus).
On an \emph{infinite} cylinder, the ground state degeneracy equals the number of anyon types.
For every anyon $\alpha$ there is an associated ground state $\ket{\alpha}$, the set of which forms an orthogonal basis for the ground state Hilbert space.
Physically, these states are defined with a fixed anyon charge at infinity, or equivalently, as eigenstates of Wilson loop/anyon flux operators around the circumference of the cylinder.
(They are also referred to as `minimum entangled states' \cite{YZhangVishwanathMES}.)
Anyon excitations are trapped at the domain wall between ground states that are consistent with the fusion rules.
More precisely, using $y$ as a coordinate in the infinite direction of the cylinder, let the wavefunction for $y > 0$ be $\ket{\alpha^+}$ and for $y < 0$ be $\ket{\alpha^-}$.
At least one anyon must be trapped on the circle $y = 0$, with total topological charge $\beta$ satisfying the fusion relation $\alpha^- \times \beta \sim \alpha^+ + \dots$.

Applying this discussion to our setup we now consider an infinite number of trenches, each forming a ring around the cylinder.
This gives us an infinite number of chains labeled by $y \in \mathbb{Z}$, coupled via Eq.~\eqref{eq:H_lambda_psi}.
By the same logic as for the torus geometry, there are again two ground states $\ket{\bbone}$, $\ket{\varepsilon}$ that arise from different superselection sector on each chain.  Keep in mind that for the time being $\bbone$ and $\varepsilon$ are merely labels derived from the coupled-chain construction; we have not yet made the association with anyons.

Recall in our argument for the two ground states that Eq.~\eqref{eq:intrachain_constraint} is an unyielding requirement which follows from the boundary condition, while Eq.~\eqref{eq:interchain_constraint} follows from energetics.
Hence when studying excited states, we can relax the second condition on specific ladders where localized excitations exist.  Let us examine the three flavors of fundamental ladder excitations---kink, antikink, and `oscillator'---identified in Sec.~\ref{IntegrableLadder}.  
Suppose first that there is a single kink between trenches $y = 0,1$---i.e., that the corresponding ladder resides in the $[\bbone\bar\varepsilon]$ sector defined in Eq.~\eqref{sectors2}.  The chains then lie in the $\bbone$ sector for $y \leq 0$ and the $\varepsilon$ sector for $y \geq 1$.
For an antikink, the sectors are $\varepsilon$ and $\bbone$ for $y \leq 0$ and $y \geq 1$, respectively.
Finally, for an oscillator excitation every chain must be in the $\varepsilon$ sector (that excitation type exists only in the $[\varepsilon \bar\varepsilon]$ ladder sector).

Since the three excitations possess the same mass and dispersion, it is natural to identify all of these as the same nontrivial anyon (which we label as $\bullet$ for the time being).  The discussion above then implies that a $\bullet$ anyon can occur at a domain wall between $|\varepsilon\rangle$ and $|\bbone\rangle$ on the cylinder, or simply between two $|\varepsilon\rangle$ regions---but \emph{not} between two $|\bbone\rangle$ states. 
Accordingly, the allowed fusion channels follow as
$\bbone \times \bullet \sim \varepsilon$ and $\varepsilon \times \bullet \sim \bbone + \varepsilon$, whereas $\bbone \times \bullet \rightarrow \bbone$ is forbidden.
We can rewrite these rules as a tensor $N^a_{\bullet, b}$ with integer entries, where $N^a_{\bullet, b} = 1$ if $b \times \bullet \rightarrow a$ is admissible and zero otherwise.
In the basis of $\bbone$ and $\varepsilon$ ground states, the fusion matrix for the excitation is
\begin{align}
	N^a_{\bullet, b} = \begin{pmatrix} 0 & 1 \\ 1 & 1 \end{pmatrix}^a_{\;b}
\end{align}
with dominant eigenvalue, or quantum dimension, equal to the golden ratio: $d_\varepsilon=\varphi\equiv(1+\sqrt{5})/2$.
Hence in addition to being associated with CFT sectors, we can identify $\bbone$ as the trivial anyon and $\varepsilon = \bullet$ as the Fibonacci anyon.

\begin{figure}[t]
	\subfigure[\label{fig:cyl_bipart}]{
		\includegraphics[width=0.40\columnwidth]{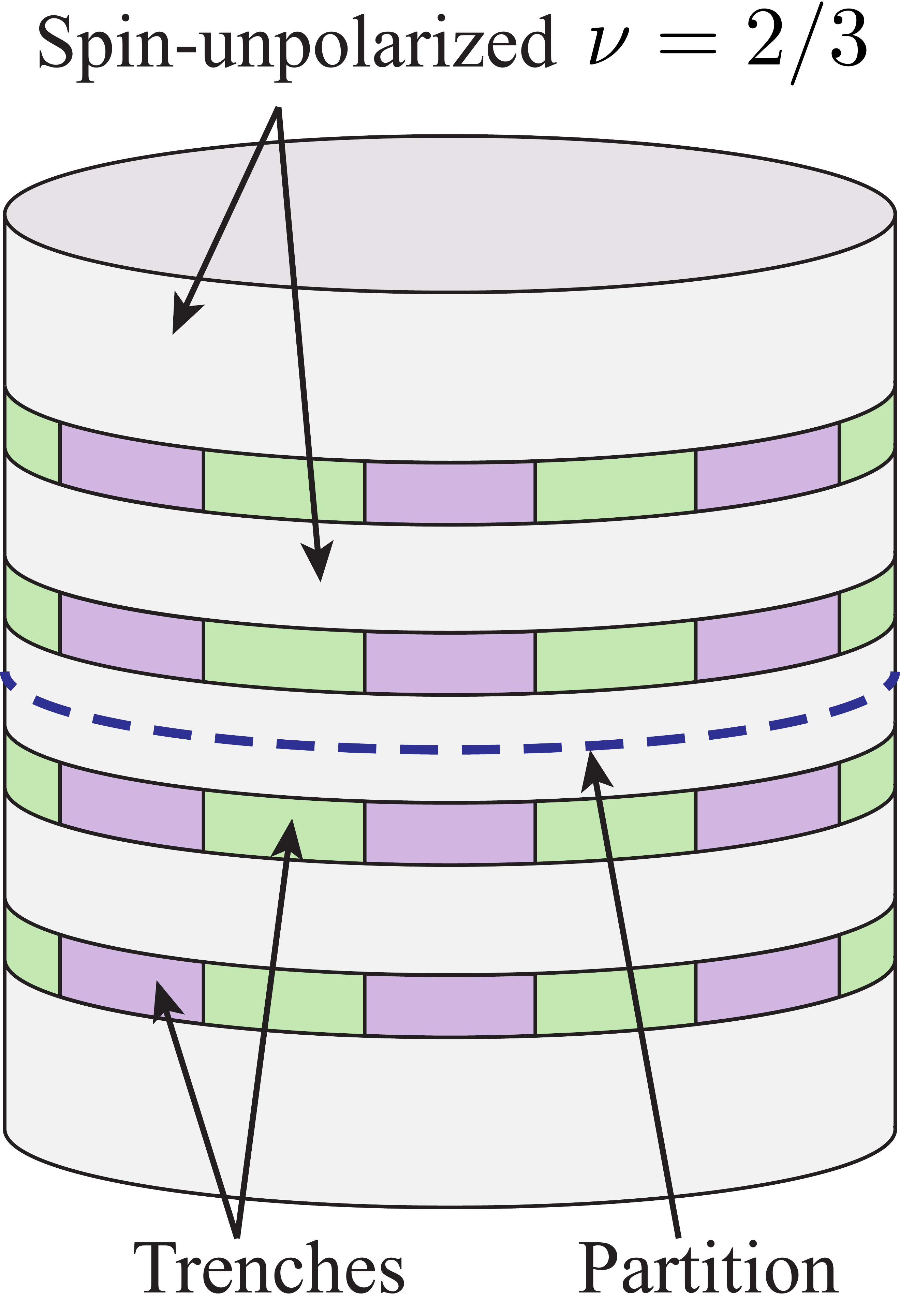}
	}
	\quad
	\subfigure[\label{fig:cyl_entropy}]{
		\includegraphics[width=0.48\columnwidth]{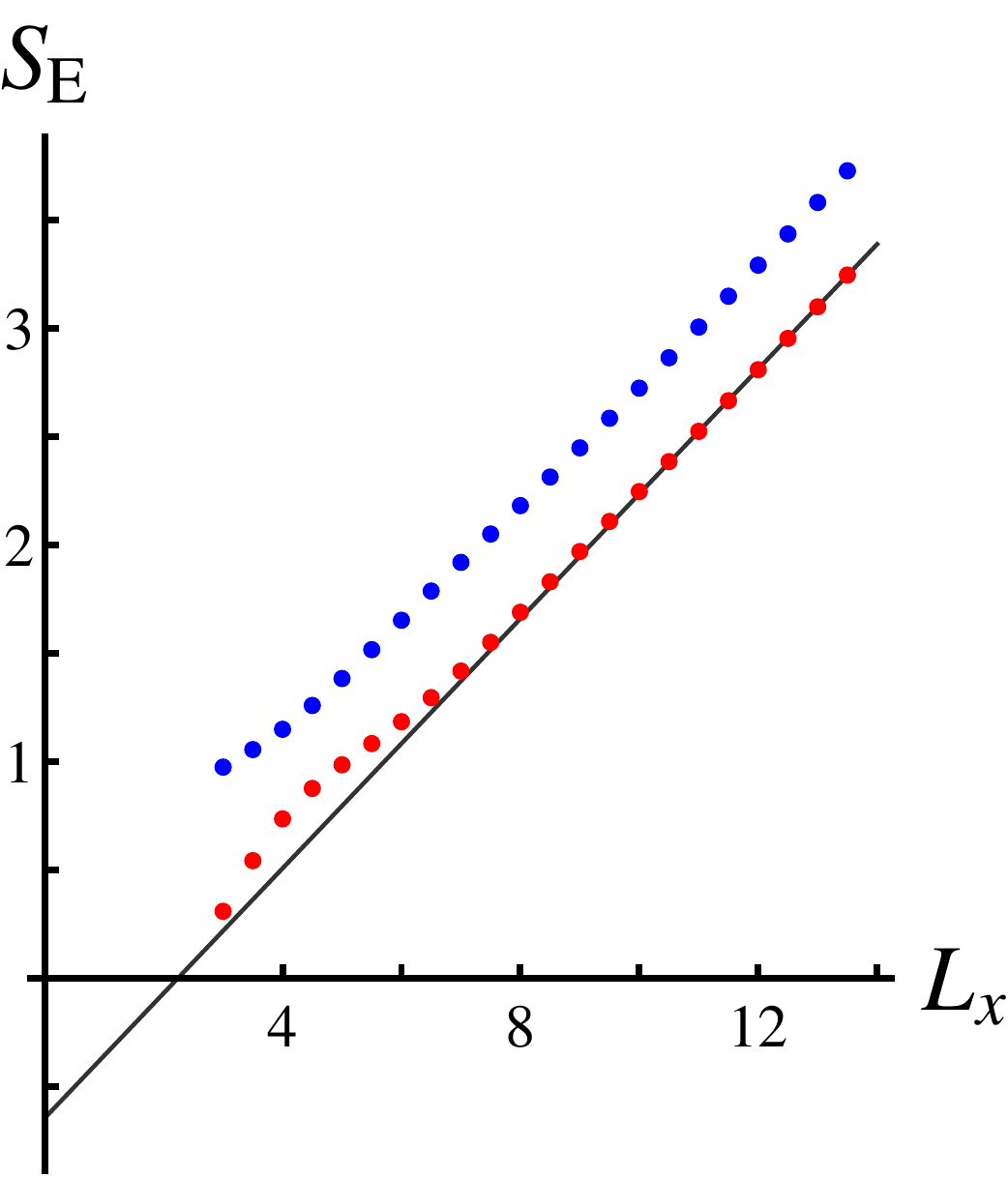}
	}
	\caption{%
		(a) Bipartition of the superstructure that cuts between two chains on a cylinder.
		(b) Entanglement entropy $S_E$ of the $\ket{\bbone}$ (red) and $\ket{\varepsilon}$ (blue) ground states of the 2D Fibonacci phase as a function of the cylinder circumference $L_x$, computed numerically via the truncated conformal space approach.  Fitting $S_E$ for state $\ket{\bbone}$ to the form $sL_x - \gamma$ at large $L_x$, we extract the intercept $-\gamma \approx -0.65$; see solid line the figure.  This yields a total quantum dimension $\mathcal{D}\approx 1.9$ for the Fibonacci phase.  Taking the difference $S_E[\ket{\varepsilon}] - S_E[\ket{\bbone}] = \log d_\varepsilon$, we deduce the quantum dimension $d_\varepsilon \approx 1.62 \approx \varphi$ which confirms that $\varepsilon$ corresponds to the Fibonacci anyon.
	}
\end{figure}

We further corroborate this result through numerical evaluation of the `topological entanglement entropy'.
Suppose that we partition the cylinder between chains $y_c$ and $y_c+1$ as illustrated schematically in Fig.~\subfigref{fig:cyl_bipart}.  The entanglement entropy is given by $S_E = -\operatorname{Tr}_{y > y_c}[ \rho \log \rho ]$, where $\rho = \operatorname{Tr}_{y \leq y_c} \ket{\Psi} \bra{\Psi}$ is the reduced density matrix that comes from a partial trace of the wavefunction $\ket{\Psi}$.
For a ground state of any gapped system, this quantity scales linearly with the cylinder circumference $L_x$: $S_E \sim sL_x - \gamma + \dots$ (up to terms that decay exponentially with $L_x$).  The slope $s$ is identical for all ground states of the same Hamiltonian but depends on non-universal microscopic details.  By contrast, the intercept $\gamma$ defines the `topological entanglement entropy' \cite{KitaevPreskill,LevinWen-2006}---a universal topological invariant of the ground state used in the computation.  This invariant can be further decomposed as $\gamma = \log(\mathcal{D} / d_\Psi)$, where $d_\Psi$ is the quantum dimension of the quasiparticle corresponding to the state $\ket{\Psi}$, and $\mathcal{D}$ is the `total quantum dimension' of the phase \cite{KitaevPreskill,LevinWen-2006,Dong-2008}.  

In the geometry illustrated in Fig.~\subfigref{fig:cyl_bipart}, the only contribution to entanglement comes from the left-movers of chain $y = y_c$ and right-movers of chain $y = y_c + 1$, as all other degrees of freedom decouple at $\lambda_b = 0$.
Hence the entanglement entropy arising from a bipartition of the cylinder is equivalent to that arising from a bipartition of a single ladder into left- and right-movers.
(This setup bears much resemblance to the so-called `spin-1 AKLT' chain \cite{AKLT:1987}.  There each spin fractionalizes into a pair of spin-$\frac12$'s, and in the ground state the `right' spin-$\frac12$ for a given site forms a singlet with the `left' spin-$\frac12$ at the next site over.  An entanglement cut between two adjacent sites thus breaks apart exactly one spin singlet into its left and right spin-$\frac12$'s.)

We used our TCSA simulations of Eq.~\eqref{eq:H_lambda_psi} to evaluate $S_E$ for the two ground states $\ket{\bbone}$ and $\ket{\varepsilon}$; the data appear in Fig.~\subfigref{fig:cyl_entropy}.
By fitting $S_E$ versus $L_x$ for ground state $\ket{\bbone}$ (which corresponds to $d_\bbone = 1$) we extract the total quantum dimension $\mathcal{D} = 1.9 \pm 0.1$. 
One can in principle perform a similar fit for the other ground state $\ket{\varepsilon}$ to extract $d_\varepsilon / \mathcal{D}$.   However, a far more precise value for $d_\varepsilon$ follows from the difference $\delta S_E \equiv S_E[\ket{\varepsilon}]-S_E[\ket{\bbone}]$ of entanglement entropies for the two ground states; the linear term in $L_x$ cancels here leaving $\delta S_E = \log({d_\varepsilon}/{d_\bbone})$.  In this way we obtain quantum dimension $d_\varepsilon = 1.619 \pm 0.002$.  These values are in excellent agreement with those of a Fibonacci anyon model with just one nontrivial particle, for which $\mathcal{D} = \sqrt{d_\bbone^2 + d_\varepsilon^2} = \sqrt{1 + \varphi^2} \approx 1.902$ and $d_\varepsilon = \varphi \approx 1.618$.


The ground state degeneracy on the torus, fusion rules, and topological entanglement entropy computed above are sufficient in this case
to uniquely identify the 2D topological phase that the system enters.
Indeed, there are only two topological phases of fermions with two-fold ground state degeneracy on the torus \cite{Rowell09}.
The nontrivial particle can be either a semion or a Fibonacci anyon.
We can distinguish between these possibilities with either the fusion rules or topological entanglement entropy;
both indicate that our coupled-trench system supports the Fibonacci anyon---which justifies the name `Fibonacci phase' christened here.

Given the particle types and fusion rules, the universal topological properties of this phase can be determined by solving the pentagon and hexagon identities;
they may be summarized as follows (for a concise review, see Ref.~\onlinecite{Trebst08}). 
The Fibonacci phase admits only the two particle types deduced above:
the trivial particle, $\I$, and a Fibonacci anyon, $\varepsilon$.
They have topological spins $\theta_\I=1$, $\theta_\varepsilon=e^{4\pi i/5}$
and satisfy the fusion rule $\varepsilon \times \varepsilon \sim \I + \varepsilon$
	\footnote{The term `Fibonacci anyon' is often used for the $\varepsilon$-particle and for the phase that supports it; the meaning is usually clear from the context.}.
As a result of this fusion rule, the dimension of the low-energy Hilbert space of $(n+1)$ $\varepsilon$-particles with total topological charge $\I$ is the $n$\textsuperscript{th} Fibonacci number, $F_n$, which grows asymptotically as $\varphi^n / \sqrt{5}$; thus its quantum dimension is $d_\varepsilon = \varphi$, as we saw previously.
(This is the same quantity that enters the formulas for the entanglement entropy used above.)  When two Fibonacci anyons are exchanged, the resulting phase acquired is either
	$R^{\varepsilon\varepsilon}_\I = e^{-4\pi i/5}$ or
	$R^{\varepsilon\varepsilon}_\varepsilon = e^{3\pi i/5}$,
	depending on the fusion channel of the two particles denoted in the subscript.
The result of an exchange can thereby be deduced
if we can bring an arbitrary state into a basis in which the two $\varepsilon$-particles
in question have a definite fusion channel. This can be accomplished with the
$F$-symbols, which effect such basis changes. The only nontrivial one is:
\begin{equation}
	F^{\varepsilon\varepsilon\varepsilon}_\varepsilon
	= \begin{pmatrix}
		\varphi^{-1} & \varphi^{-1/2} \\ \varphi^{-1/2} & -\varphi^{-1}
	\end{pmatrix}
\end{equation}
written in the basis $\{\I, \varepsilon\}$ for the central fusion channel.
From these relatively simple rules follows a remarkable fact:
these anyons support universal topological quantum
computation \cite{Freedman02a,Freedman02b}.

While the aforementioned analysis was carried out for $\lambda_b = 0$, the gapped topological phase that we have constructed must be stable up to some finite $\lambda_b$.  Rough phase boundaries for this state were estimated earlier; see Fig.~\subfigref{fig:CoupledChaina}.  However, directly exploring the physics with $\lambda_b \neq 0$, either analytically or numerically, is highly nontrivial since we then lose integrability and can no longer distill the problem into individual `ladders' with a Hilbert space constraint.  Progress could instead be made by employing DMRG simulations to map out the phase diagram more completely, which would certainly be interesting to pursue in follow-up work.

\subsection{Superconducting vortices}
\label{sc_vortices}


Since the Fibonacci phase arises in a superconducting system, it is also important to investigate the properties of $h/2e$ vortices---despite the fact that, unlike Fibonacci anyons, they are confined.  Before turning to this problem it will be useful to briefly recall the corresponding physics in a spinless 2D $p+ip$ superconductor \cite{ReadGreen,StoneRoy,FendleyFisherNayak,GrosfeldStern}.  One way of understanding the nontrivial structure of vortices there is by considering the chiral Majorana edge states of a $p+ip$ superconductor on a cylinder.  Finite-size effects quantize their energy spectrum in a manner that depends on boundary conditions exhibited by the edge Majorana fermions.  With anti-periodic boundary conditions the spectrum is gapped, while in the periodic case an isolated zero-mode appears at each cylinder edge.  Threading integer multiples of $h/2e$ flux through the cylinder axis toggles between these boundary conditions, thereby creating and removing zero-modes.  This reflects the familiar result that $h/2e$ vortices in a planar $p+ip$ superconductor bind Majorana zero-modes and consequently form Ising anyons.

\begin{figure}[t]
	\includegraphics[width=80mm]{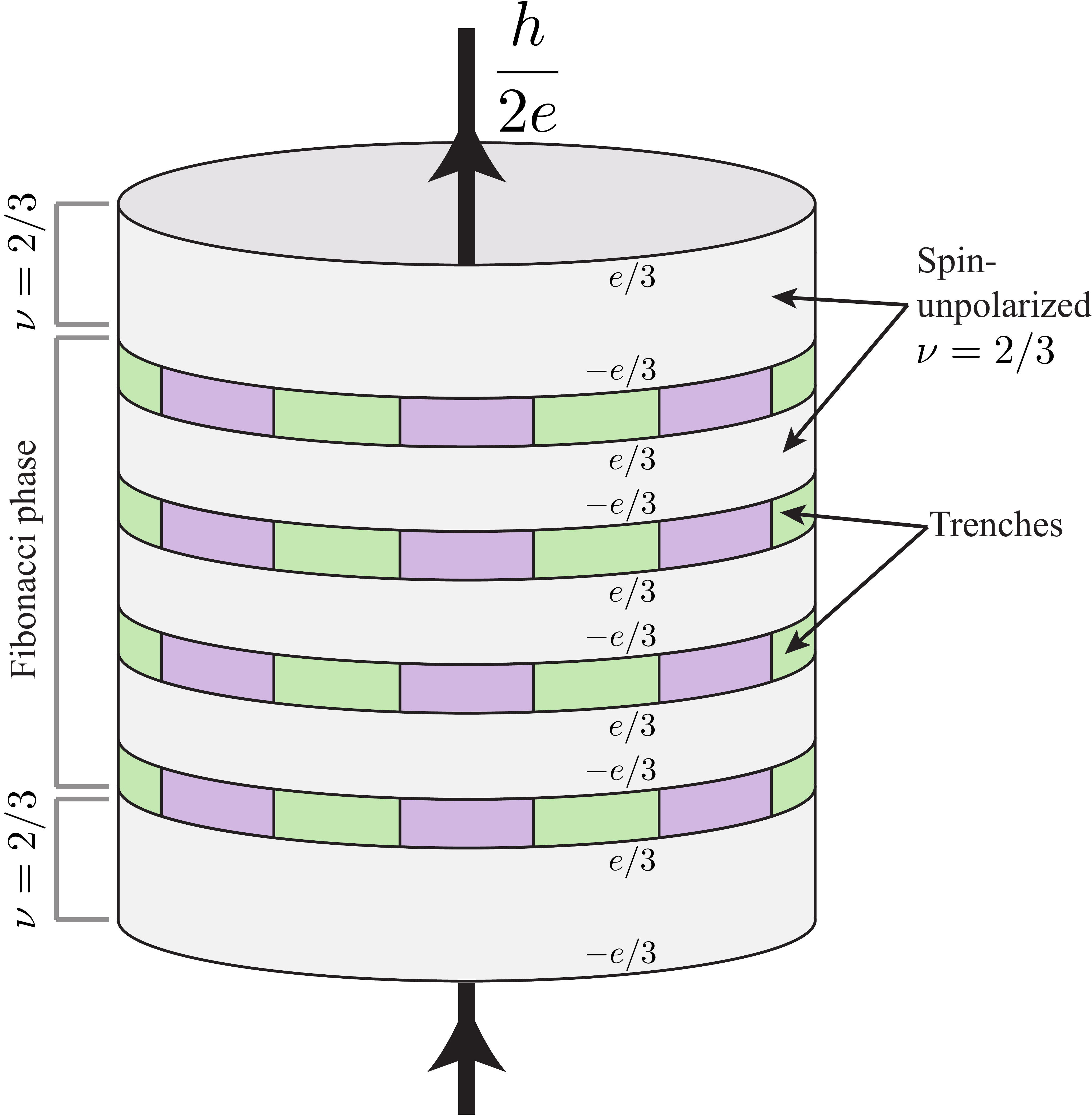}
	\caption{%
		Cylinder geometry used to deduce the properties of $h/2e$ superconducting vortices in the Fibonacci phase.  We initially assume that pure $\nu = 2/3$ quantum Hall states border the Fibonacci phase from above and below.  This results in two well-defined boundaries: the Fibonacci phase/quantum Hall edge, and the quantum Hall/vacuum edge.  Adiabatically inserting $h/2e$ flux through the cylinder (which is topologically equivalent to an $h/2e$ vortex in the bulk of a planar Fibonacci phase) pumps charge $e/3$ across each quantum Hall region as shown above.  Because the charge difference across the trenches then changes, the upper Fibonacci phase/quantum Hall edge binds either a $\psi$ or $\sigma$ excitation that carries charge $2e/3 \bmod 2e$.  The upper quantum Hall/vacuum edge, however, binds charge $e/3$ so that in total the vortex carries only fermion parity.  If one shrinks the pure quantum Hall regions so that the two boundaries hybridize, $\psi$ and $\sigma$ lose their meaning since other sectors mix in.  The final conclusion is that an $h/2e$ vortex traps either a trivial particle or Fibonacci anyon depending on non-universal details, but does not lead to new quasiparticle types.  
	}
	\label{Flux_insertion_fig}
\end{figure}

We will deduce the properties of vortices in the Fibonacci phase by similarly deforming our $\nu = 2/3$ quantum Hall setup into a cylinder as sketched in Fig.~\ref{Flux_insertion_fig}.  In principle the physics can be analyzed by deriving the influence of flux on boundary conditions for the $\mathbb{Z}_3$ parafermionic edge modes supported by this state, though such an approach will not be followed here.  Instead we develop a related adiabatic flux-insertion argument that allows us to obtain the result with minimal formalism.  We proceed by first assuming that the Fibonacci phase is bordered by `wide' $\nu = 2/3$ regions on the upper and lower parts of the cylinder, as Fig.~\ref{Flux_insertion_fig} indicates.  This will allow us to separately address the effect of flux on $(i)$ the gapless $\mathbb{Z}_3$ parafermion modes at the interface between the Fibonacci phase and $\nu = 2/3$ regions, and $(ii)$ the outermost $\nu = 2/3$ edge states that border the vacuum.  One can then couple these sectors to determine the final vortex structure.  Following this logic we will show that in contrast to the $p+ip$ case $h/2e$ flux does \emph{not} introduce new topological anyons beyond the trivial and Fibonacci particles already discussed.  A vortex may, however, provide a local potential that happens to trap a deconfined Fibonacci anyon, though whether or not this transpires is a non-universal question of energetics.  (Note that the same could be said for, say, an impurity, so one should not attach any deep meaning to this statement.)

Let us first a consider a cylinder with no flux, in the limit where each trench is tuned to $\mathbb{Z}_3$ parafermion criticality and interchain coupling is temporarily turned off.  For concreteness we also assert that each $\nu = 2/3$ edge contains no net electric charge mod $2e$.  The sum and difference of the total charge on the two sides of each trench, $Q^\pm_\textrm{tot}$ [see Eq.~\eqref{trialities}], must also then vanish mod $2e$.  This restricts the possible CFT sectors present in the trenches to either $I_R\times I_L$ or $\epsilon_R\times \epsilon_L$; all other physical sectors contain the wrong charge.  
Next we adiabatically increase the flux through the cylinder from 0 to $h/2e$
	\footnote{Superconductivity technically does not allow for continuous ramping of the flux, but this barrier can be easily avoided.  For the purpose of this thought experiment one can imagine temporarily snaking the flux so that it threads the $\nu = 2/3$ regions but avoids passing through the trenches.  Once a value of $h/2e$ is reached, the flux can then be moved entirely within the cylinder.}.
Because of the nontrivial Hall conductivity in the $\nu = 2/3$ fluids, charge $e/3$ pumps from the bottom to top edge of each quantum Hall region in response to the flux insertion, as Fig.~\ref{Flux_insertion_fig} illustrates.  The pumping leaves the total charge $Q^+_\textrm{tot}$ on each trench intact but alters the total charge difference for each trench to $Q^-_\textrm{tot} = -2/3$ mod 2.  The only allowed sectors consistent with this charge arrangement are $\psi_R\times\psi_L^\dagger$ and $\sigma_R\times \sigma_L^\dagger$.  
Finally we turn on the interchain perturbation $\lambda_a$ in Eq.~\eqref{Hperp} to enter the Fibonacci phase.  The CFT sectors in the bulk that are gapped by this coupling will clearly then mix.  However, the gapless right-movers from the top trench and left-movers from the bottom remain unaffected by $\lambda_a$; the former necessarily realizes either $\psi_R$ or $\sigma_R$, while the latter realizes $\psi_L^\dagger$ or $\sigma_L^\dagger$.  

Focusing on the top half of the system, this argument shows that an $h/2e$ superconducting vortex traps an Abelian $\psi$ or non-Abelian $\sigma$ particle at the interface between $\nu = 2/3$ fluid and the Fibonacci phase.
Importantly, we must additionally account for the quantum Hall edge at the top of the cylinder, which also responds to the flux and influences the structure of a vortex in a crucial way as we will see.  Figure~\ref{Flux_insertion_fig} shows that the flux induces charge $+e/3$ at the uppermost cylinder edge.  Together, we see that an $h/2e$ vortex gives rise to edge excitations $\braket{\psi, 1/3}$ or $\braket{\sigma, 1/3}$ when the Fibonacci phase is bordered by wide Abelian quantum Hall fluid.  Here and below $\braket{\mathcal{F}, q}$ indicates that $\nu = 2/3$ liquid/Fibonacci phase interface traps particle type $\mathcal{F}$, while the quantum Hall edge bordering the vacuum binds charge $q\bmod 1$.  
Recalling the $2e/3$ charge associated with $\psi$ and $\sigma$, we conclude that the $h/2e$ vortex carries total charge $e \bmod 2e$---which is not fractional.  
Next we discuss the fate of the $\psi$ and $\sigma$ particles at the Fibonacci phase boundary when we include coupling to the outer quantum Hall edge.


If one assumes that the $\mathbb{Z}_3$ parafermion edge states and outer $\nu = 2/3$ edge modes decouple, then the system can in principle reside in six possible edge sectors: $\braket{ I, 0 }$, $\braket{ \psi, 1/3 }$, $\braket{ \psi^\dag, 2/3 }$, $\braket{ \epsilon, 0 }$, $\braket{ \sigma, 1/3 }$, and $\braket{ \sigma^\dag, 2/3 }$.  (This statement is independent of vorticity, and simply tells one which states have physical charge configurations.)  Suppose now that the pure quantum Hall region at the top of Fig.~\ref{Flux_insertion_fig} shrinks to allow fractional charge tunneling between the parafermion and $\nu = 2/3$ edge modes.  Some of the edge sectors above then mix and hence are no longer distinguishable.  For instance, transferring $e/3$ charge from the vacuum edge to the boundary of the Fibonacci phase can send $\braket{\sigma,1/3}\rightarrow\braket{\epsilon,0}$.  In fact only two inequivalent edge sectors remain---the triplet $\braket{ I, 0 }$, $\braket{ \psi, 1/3 }$, $\braket{ \psi^\dag, 2/3 }$ that is associated with the identity particle, and the remaining set $\braket{ \epsilon, 0 }$, $\braket{ \sigma, 1/3 }$, $\braket{ \sigma^\dag, 2/3 }$ associated with the $\varepsilon$ non-Abelian anyon.

Applying the above discussion to vortices, we infer that $h/2e$ flux does not generically bind a $\psi$ or $\sigma$ in any meaningful way once the parafermion and outer $\nu = 2/3$ edge modes hybridize.  The vortex can trap a trivial or Fibonacci anyon but exhibits no finer $\mathbb{Z}_3$ structure---which is entirely consistent with the fact that it carries only fermion parity.  Which of the two particle types occurs in practice depends on non-universal microscopic details, though both cases are guaranteed to be possible because $\varepsilon$ is deconfined.
(If a vortex binds a trivial particle one can always bring in a Fibonacci anyon from elsewhere and attach it to the vortex to obtain the $\varepsilon$ case, or vice versa.)

In fact a similar state of affairs occurs for any phase that supports a Fibonacci anyon, including the $\mathbb{Z}_3$ Read-Rezayi state.
Because of the fusion rule $\varepsilon \times \varepsilon \sim \I + \varepsilon$, the Fibonacci anyon $\varepsilon$ must carry the same local quantum numbers (such as charge and vorticity) as the trivial particle.
Thus any Abelian anyon $\mathcal{A}$ can fuse with the neutral Fibonacci anyon to form a non-Abelian particle with identical local quantum numbers: $\mathcal{A} \times \varepsilon \sim \mathcal{A}\varepsilon$
	\footnote{In fact, modularity requires that the set of all anyons decomposes into pairs $(\mathcal{A}, \mathcal{A}\varepsilon)$, where the Fibonacci anyon completely factorizes within the fusion rules.  That is, $\mathcal{A}\varepsilon \times \mathcal{B} \sim (\mathcal{A}\times\mathcal{B})\varepsilon$ and $\mathcal{A}\varepsilon \times \mathcal{B}\varepsilon \sim (\mathcal{A}\times\mathcal{B})(\I+\varepsilon)$.
	}.
For example, in the case of the $\mathbb{Z}_3$ Read-Rezayi state at filling $\nu = 13/5$, there are two anyons with electric charge $e/5$: one Abelian and the other non-Abelian with quantum dimension $\varphi$.  The latter quasiparticle may be obtained by fusing the former with a neutral Fibonacci anyon.  Or equivalently, the former may be obtained from the latter by fusing two non-Abelian $e/5$ quasiparticles with a $-e/5$ quasihole.  Which of these $e/5$ excitations has lowest energy is \emph{a priori} non-universal.  Details of such energetics issues are interesting but left to future work.



Finally, we remark that the $\mathbb{Z}_3$ structure at the edge between the Fibonacci phase and $\nu=2/3$ state arises solely from the fractional quantum Hall side.  The corresponding fractionally charged quasiparticles indeed do not exist within the Fibonacci phase, as evidenced by the absence of $\psi$ or $\sigma$ particles in the bulk.  Our coupled-chain construction provides an intuitive way of understanding this: $2e/3$ excitations are naturally confined in the Fibonacci phase since the trenches provide a barrier that prevents fractional charge from tunneling between adjacent quantum Hall regions.  The Fibonacci anyon is neutral, by contrast, and thus suffers no such obstruction.

\subsection{Excitations of the edge between the Fibonacci phase and the vacuum}
\label{sec:FibPhaseEdge}

Bulk properties strongly constrain the edge excitations of a topological phase.
In particular, the edge bordering the vacuum must support as many anyon types as the bulk.
This correspondence is simplest when the bulk is fully chiral.  Edge excitations are then described by a CFT (possibly
deformed by marginal perturbations so that some of the velocities are unequal) that exhibits precisely the same number of primary fields
as the bulk has anyon types. These fields possess fractional scaling dimensions, and
all other fields have scaling dimensions that differ from these by integers.
Therefore, one can view an arbitrary field as creating an anyon (via a primary
operator) together with some additional bosonic excitations.
It is important to note that the edge may have additional symmetry generators beyond just the
Virasoro generators derived from the energy-momentum tensor.
These additional symmetry generators have their scaling dimensions fixed to $1$ (Kac-Moody algebras) or some other integer (e.g., $\mathcal{W}$-algebras) \cite{Zamolodchikov:1985}.

Since the Fibonacci phase has only two particle types, $\I$ and $\varepsilon$, the minimal possible edge theory describing the boundary with the vacuum has two primary fields which we denote as $1$ and $\tilde\epsilon$.  (The tilde is used to distinguish from the field $\epsilon$ that lives at the boundary between the Fibonacci phase and the parent quantum Hall fluid.)  
At first glance, however, our quantum Hall/superconductor heterostructure appears to exhibit a much more complicated edge structure than the quasiparticle content suggests.
The interface between the Fibonacci phase and the spin-unpolarized $\nu=2/3$ state is described by a $\mathbb{Z}_3$ parafermion CFT, and
 the boundary between the $\nu=2/3$ state and the vacuum is described by a CFT for two bosons with $K$-matrix
\begin{align}
 	K &= \begin{pmatrix} 1 & 2 \\ 2 & 1 \end{pmatrix} && \textrm{[fermionic (112) state]}.
	\label{eqn:fermionic-K112}
\end{align}
The former CFT has six primary fields while the latter has three.  One can obtain a direct interface between the Fibonacci phase and vacuum by simply shrinking the outer $\nu = 2/3$ fluid until it disappears altogether; the resulting boundary is then naively characterized by a product of these two edge theories.
However, in the previous subsection we argued that of the 18 primary fields in the product CFT, only a subset of 6 are physical from charge constraints, and these combine to just two primary fields.
Here we explicitly construct a chiral CFT with exactly these two primary fields.
Furthermore we demonstrate that upon edge reconstruction the Fibonacci phase/vacuum interface is described by this CFT combined with unfractionalized fermionic edge modes, in precise correspondence with the bulk quasiparticle types supported by the Fibonacci phase.

It is useful to first examine the simpler case of a $\nu=2/3$ state built out of underlying charge-$e$ bosons.
This allows us to replace the $K$-matrix of Eq.~\eqref{eqn:fermionic-K112} with
\begin{align}
	K = \begin{pmatrix} 2 & 1 \\ 1 & 2 \end{pmatrix}  && \textrm{[bosonic (221) state]}.
	\label{eqn:bosonic-K221}
\end{align}
For brevity we refer to this bosonic quantum Hall phase as the (221) state.
Most of the preceding analysis, including the appearance of a descendant Fibonacci phase, is unchanged by this modification.
However, by working with a bosonic theory we can appeal to modular invariance to connect the bulk quasiparticle structure
to the edge chiral central charge $c_R-c_L$:
\begin{equation}
	\frac{1}{\cal D}\sum_a \theta_a {d_a^2} = e^{\frac{2\pi i}{8}(c_R-c_L)} ,
\end{equation}
where $a$ sums over the two anyon types and $c_{R/L}$ denote the central charges for right/left-movers.
Using results from Sec.~\ref{GSD_quasiparticles}---in particular, ${\cal D}=\sqrt{1+\varphi^2}, d_\I = 1, d_\varepsilon = \varphi$
and $\theta_\I = 1, \theta_\varepsilon=e^{4\pi i/5}$---the chiral central charge follows as $c_R-c_L \equiv 14/5 \bmod 8$.
Thus, the minimal edge theory describing the boundary with the vacuum is purely chiral with $c_R = 14/5$ and $c_L = 0$.
We now show that the bosonic Fibonacci phase/vacuum edge is consistent with these scaling dimensions and central charges.

The key physical observation was made in the previous subsection: fractional charge
and the resulting $\mathbb{Z}_3$ structure are features of the $\nu=2/3$ state, not the Fibonacci phase.
Equivalently, not all of the excitations of the combined $\mathbb{Z}_3$ parafermion CFT and the $(221)$ edge states
are allowed in the Fibonacci phase because we cannot transfer fractional charge from one edge of the system to the other via the bulk.
Fractional charge can pass only between the Fibonacci phase/(221) state and (221) state/vacuum interfaces; together, these two edges form the Fibonacci-to-vacuum edge.
As such the total charge of the Fibonacci phase/vacuum edge must be an integer, which dictates the set of physical operators that appear.

In terms of the $\mathbb{Z}_3$ parafermion operators and the edge fields $\phi_\uparrow, \phi_\downarrow$ of the (221) state, the most relevant operators that transfer fractional charge within an edge are
\begin{subequations} \label{eq:G2Spin1}
\begin{align}
	\psi e^{i\phi_\uparrow}, \psi e^{i\phi_\downarrow}, \psi e^{-i\phi_\uparrow-i\phi_\downarrow},
	\psi^\dag e^{-i\phi_\uparrow}, \psi^\dag e^{-i\phi_\downarrow}, \psi^\dag e^{i\phi_\uparrow+i\phi_\downarrow}.
	\label{eq:G2transfer-charge}
\end{align}
Note that these all have scaling dimension $1$.  
There are six additional dimension-$1$ operators that add integer charge to an edge:
\begin{multline}
	e^{i\phi_\uparrow+2i\phi_\downarrow}, e^{2i\phi_\uparrow+i\phi_\downarrow}, e^{i\phi_\uparrow-i\phi_\downarrow}, \\
	e^{-i\phi_\uparrow-2i\phi_\downarrow}, e^{-2i\phi_\uparrow-i\phi_\downarrow}, e^{-i\phi_\uparrow+i\phi_\downarrow}.
	\label{eq:G2add-charge}
\end{multline}
Finally, the two charge current operators
\begin{align}
	i\sqrt{\frac32}\partial \phi_\uparrow, \quad \frac{i}{\sqrt{2}}\partial \phi_\uparrow + i\sqrt{2}\partial \phi_\downarrow
	\label{eq:G2current}
\end{align}
\end{subequations}
also have scaling dimension 1.
The $14$ operators in Eqs.~\eqref{eq:G2transfer-charge} through \eqref{eq:G2current} satisfy the Kac-Moody algebra for the Lie group $G_2$ at level-$1$:
\begin{align}
	J^a(z) J^b(w) = \frac{\delta_{\bar{a}}^b}{(z-w)^2} + \frac{f^{abc} J^c(w)}{z-w} + \dots \;,
\end{align}
where $f^{abc}$ are the structure constants for the $G_2$ Lie algebra, normalized such that the Killing form $f^{acd}(f^{bcd})^\ast = 8\delta^{ab}$.
The two charge currents form the Cartan subalgebra for $G_2$ while the operators in Eqs.~\eqref{eq:G2transfer-charge}
and \eqref{eq:G2add-charge} correspond to the non-zero roots of $G_2$, as follows:
\begin{align}
	\quad
	\newcommand{\xytag}[2]{\bullet \save[]+<#2,3.3mm>*\txt{$#1$} \restore}
	\raisebox{48mm}{ \xymatrix @!0 @M=0mm @R=7mm @C=8.6mm{
		&&&&
		{} \save[]+<4mm,3mm>*\txt{$\vec{l}\cdot\big(0,\tfrac{\sqrt{3}}{2}\big)$}\restore
	\\	&&&& \xytag{e^{i\phi_\uparrow+2i\phi_\downarrow}}{8mm} &&&&
	\\
	\\	& \xytag{e^{-i\phi_\uparrow+i\phi_\downarrow}}{4mm} && \xytag{\psi e^{i\phi_\downarrow}}{3mm} && \xytag{\psi^\dag e^{i\phi_\uparrow+i\phi_\downarrow}}{5mm} && \xytag{e^{2i\phi_\uparrow+i\phi_\downarrow}}{7mm} &
	\\
	\\	\ar[rrrrrrrr] && \xytag{\psi^\dag e^{-i\phi_\uparrow}}{6mm} && {} && \xytag{\psi e^{i\phi_\uparrow}}{4mm} &&
		{} \save[]+<4mm,3mm>*\txt{$\vec{l}\cdot\big(1,-\tfrac12\big)$}\restore
	\\
	\\	& \xytag{e^{-2i\phi_\uparrow-i\phi_\downarrow}}{3mm} && \xytag{\psi e^{-i\phi_\uparrow-i\phi_\downarrow}}{5.5mm} && \xytag{\psi^\dag e^{-i\phi_\downarrow}}{6mm} && \xytag{e^{i\phi_\uparrow-i\phi_\downarrow}}{6mm} &
	\\
	\\	&&&& \xytag{e^{-i\phi_\uparrow-2i\phi_\downarrow}}{9mm} &&&&
	\\	&&&& \ar[uuuuuuuuuu]
	} }
\end{align}
In the axes the vector $\vec l$ represents the argument of a given (221) state operator written as $e^{i \vec l\cdot \vec\phi}$ [e.g., $\vec l = (2,1)$ for $e^{2i\phi_\uparrow+i\phi_\downarrow}$].  
As an extension to the Virasoro algebra, this Kac-Moody algebra has $c=14/5$ and only two primary fields, the identity $1$ and $\tilde\epsilon = \sigma^\dag e^{i\phi_\uparrow+i\phi_\downarrow}$
	\footnote{%
		It is important to note that $1$ and $\tilde\epsilon$ are the only primary fields when considering the full $G_2$ Kac-Moody algebra.
		The $\tilde\epsilon$ tower $(h_{\tilde\epsilon} = 2/5)$ contains a multiplet of $7$ fields, all with scaling dimension $2/5$, transforming in one of the fundamental representations of $G_2$.
		However, these $7$ fields are related to each other by $G_2$ symmetry transformations or, in other words, are related by operator product expansions with the currents in Eqs.~\eqref{eq:G2Spin1}.
		Physically, the fields differ by bosonic excitations at the edge and therefore correspond to the same bulk anyon.
	}.
All other fields of the CFT can be constructed by combining one of the primaries with the generators in Eqs.~\eqref{eq:G2Spin1}, e.g., $\epsilon$ arises from the operator product expansion between $\tilde\epsilon$ and $\psi e^{-i\phi_\uparrow-i\phi_\downarrow}$.
The identity field has scaling dimension $h_1 = 0$ and transforms trivially under the $G_2$ action, while the nontrivial field $\tilde\epsilon$ has scaling dimension $h_{\tilde\epsilon} = 2/5$ and belongs in the 7-dimensional fundamental representation of $G_2$.
Here we can see that the bulk-edge correspondence is consistent with our identification of the bulk as the Fibonacci phase; for example, the topological spins of $\I$ and $\varepsilon$ are related to the scaling dimensions of the fields $1$ and $\tilde\epsilon$ via $\theta_{\I,\varepsilon} = e^{2\pi ih_{1,\tilde\epsilon}}$.

We now return to the fermionic case, where the $\nu = 2/3$-to-vacuum edge is characterized by the $K$-matrix in Eq.~\eqref{eqn:fermionic-K112}.
The allowed operators that transfer charge in the fermionic Fibonacci phase/vacuum edge are once again given by Eqs.~\eqref{eq:G2Spin1}.
Unlike in the bosonic case, however, these operators are non-chiral because the fermionic $\nu = 2/3$ state supports counterpropagating edge modes at the interface with the vacuum.  Nevertheless they remain spin-$1$ operators as in the bosonic setup.  
Moreover, the fermionic Fibonacci-to-vacuum edge exhibits a phase that bears a simple relation to the bosonic edge, as we now demonstrate.

This phase occurs when the edge reconstructs such that an additional non-chiral pair of unfractionalized modes comes down in energy
and hybridizes with the modes of the $\nu = 2/3$-to-vacuum edge.  In the limit where these modes are
gapless, the $K$-matrix becomes
\begin{eqnarray}
	K^e = \begin{pmatrix} 1 & 2 & 0 & 0 \\ 2 & 1 & 0 & 0 \\ 0 & 0 & 1 & 0 \\ 0 & 0 & 0 & -1 \end{pmatrix} .
	\label{eqn:fermionic-K112-extended}
\end{eqnarray}
The $\nu = 2/3$-to-vacuum edge is then described by the effective field theory
\begin{equation}
	S = \frac{1}{4\pi}\int_{t,x}\!\!
		\big[ {K^e_{IJ}}\,{\partial_t}{\phi_I}\,{\partial_x}{\phi_J} - {V_{IJ}}\,{\partial_x}{\phi_I}\,{\partial_x}{\phi_J} \big] + \dots \;.
\end{equation}
Here the ellipsis represents quasiparticle tunneling processes; indices $I,J$ label the field components such that $\phi_1,\phi_2$ denote the original spin-up and spin-down modes while $\phi_3,\phi_4$ represent the new counterpropagating modes added to the edge; and $V_{IJ}$ is a symmetric matrix that characterizes density-density interactions amongst all four modes.
If $V_{IJ}$ is small for $I=1,2$ and $J=3,4$, then the additional $\phi_{3,4}$ fields generically acquire a gap because one of the tunneling perturbations $\cos({\phi_3} \pm {\phi_4})$ will be relevant \footnote{Both perturbations are allowed since charge is conserved only mod $2e$ in our system.}.  However, when these off-diagonal entries in $V_{IJ}$ are appreciable the edge can enter the new phase that we seek.  

To describe this phase, it is convenient to invoke a basis change to ${\tilde K}^e = W K^e W^T$ and ${\tilde V} = W V W^T$, where
\begin{eqnarray}
	W = \begin{pmatrix} 1 & 0 & 1 & 0 \\ 0 & 1 & -1 & 0 \\ -1 & 1 & -1 & 0 \\ 0 & 0 & 0 & 1 \end{pmatrix} ,
\end{eqnarray}
and
\begin{eqnarray}
	\tilde{K}^e = \begin{pmatrix} 2 & 1 & 0 & 0 \\ 1 & 2 & 0 & 0 \\ 0 & 0 & -1 & 0 \\ 0 & 0 & 0 & -1 \end{pmatrix} .
	\label{Ketilde}
\end{eqnarray}
Suppose for the moment that ${\tilde V}_{IJ}=0$ for $I=1,2$ and $J=3,4$.  By comparing Eqs.~\eqref{eqn:bosonic-K221} and \eqref{Ketilde} one sees that the fermionic edge is then equivalent to the bosonic case examined earlier, supplemented by two Dirac fermion modes running in the opposite direction relative to the chiral modes of the (221) state.  This correspondence allows us to immediately deduce that the fermionic Fibonacci-to-vacuum edge is described by the $G_2$ Kac-Moody theory at level-$1$ together with two
backwards-propagating Dirac fermions (or, equivalently, four backwards-propagating Majorana fermions).
More generally, when ${\tilde V}_{IJ}$ is small but non-zero for $I=1,2$ and $J=3,4$, the $G_2$ theory and the
backwards-propagating fermions hybridize through the marginal
couplings ${\tilde V}_{IJ}$. Once again, we find a correspondence
between the bulk and the edge with the vacuum: both have Fibonacci anyons as well as fermionic excitations \footnote{Although we noted earlier that one cannot define an electron operator in the low-energy subspace spanned by the generalized Majorana operators, electrons can still of course be added at high energies anywhere in the system's bulk.}.


\section{Topological Quantum Field Theory Interpretation}
\label{TQFTsection}

We will now provide an alternative topological quantum field theory (TQFT) interpretation of the Fibonacci phase introduced in the preceding sections.  Although less connected to microscopics, the perspective developed here cuts more directly to the elegant topological properties enjoyed by this state.
Our discussion will draw significantly on the earlier works of Gils \emph{et al}.~\cite{InteractingAnyons1} and especially Ludwig \emph{et al}.~\cite{InteractingAnyons2}.
As already mentioned in the introduction our construction of the Fibonacci phase from superconducting islands embedded in a $\nu =2/3$ quantum Hall state bears some resemblance to these studies.  Starting from parent \emph{non-Abelian} systems Refs.~\onlinecite{InteractingAnyons1,InteractingAnyons2} investigated descendant phases emerging in the interior of the fluid due to interaction amongst a macroscopic collection of non-Abelian anyons.  We followed a similar approach in that the domain walls in our spatially modulated trenches correspond to extrinsic non-Abelian defects \cite{ClarkeParafendleyons,LindnerParafendleyons,ChengParafendleyons,BarkeshliParafendleyons1,BarkeshliParafendleyons2} by virtue of the $\mathbb{Z}_3$ zero-modes that they bind; moreover, we likewise hybridized these defects to access the (descendant) Fibonacci phase within a (parent) $\nu = 2/3$ state.  This common underlying philosophy suggests a deep relationship with Refs.~\onlinecite{InteractingAnyons1,InteractingAnyons2}.

Of course the most glaring difference stems from the Abelian nature of our parent state.  We will show below that one can blur this (certainly important) distinction, however, by developing a non-standard view of the spin-unpolarized $\nu = 2/3$ quantum Hall state---namely, as emerging from some non-Abelian phase upon condensation of a boson that confines the non-Abelian particles.  Such an interpretation might initially seem rather unnatural, but provides an illuminating perspective in situations where one can externally supply the energy necessary to generate these confined non-Abelian excitations in a meaningful way.  This is indeed precisely what we accomplish by forcing superconducting islands into the $\nu = 2/3$ fluid to nucleate the domain walls that trap $\mathbb{Z}_3$ zero-modes.  We will employ such a picture to sharpen the connection with earlier work and, in the process, develop a TQFT view of the Fibonacci phase generated within a $\nu = 2/3$ state.  
In the discussion to follow, we ignore the fermion present in the (112) state, which leads to subtle consequences that we address at the end of this section.
[In fact, our conclusions will apply more directly to the analogous bosonic (221) state.]  

\begin{figure}[t]
	\centering
	\includegraphics[width = \columnwidth]{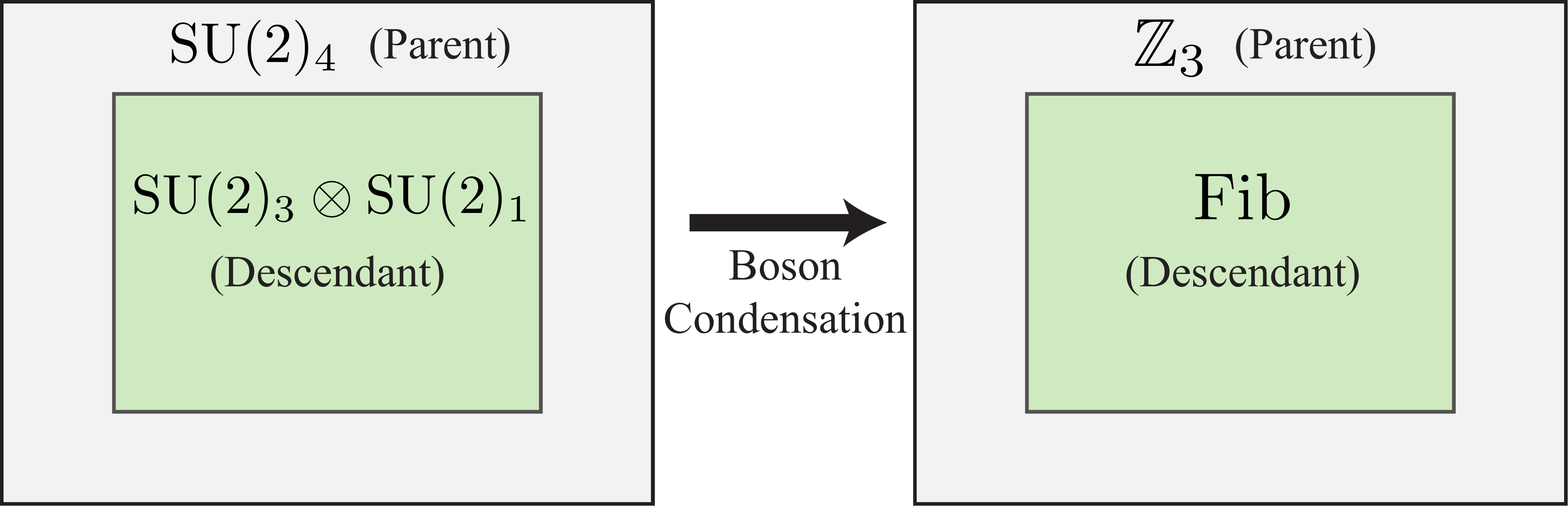}
	\caption{Boson condensation picture leading to a topological quantum field theory (TQFT) interpretation of the Fibonacci phase.  On the left a parent non-Abelian $\SUtwo_4$ phase hosts a descendant $\SUtwo_{3}\otimes\SUtwo_1$ state arising from interacting anyons within the fluid~\cite{InteractingAnyons2}.  Condensing a single boson throughout the system produces the setup on the right in which an Abelian $\mathbb{Z}_3$ parent state gives rise to a descendant phase described by a pure Fibonacci TQFT.  The latter system very closely relates to our spin-unpolarized $\nu = 2/3$ state with superconducting islands that generate the Fibonacci phase inside of the quantum Hall medium, in that the quasiparticle content (modulo the electron) is identical.  An even more precise analogy occurs in the case where the Fibonacci phase resides in a bosonic (221) quantum Hall state; here the TQFT's from the right side of the figure describe the universal topological physics exactly.  }
	\label{TQFT_fig}
\end{figure}

As a first step we summarize the results from Ref.~\onlinecite{InteractingAnyons2} that will be relevant for our discussion.  Consider a parent non-Abelian phase described by an $\SUtwo_4$ TQFT.  Table~\ref{su24} lists the properties of the gapped topological excitations of this phase---including the $\SUtwo$ spin $j$, conformal spin $h$, quantum dimension $d$, and nontrivial fusion rules for each field.  Ludwig \emph{et al}.~found that antiferromagnetically coupling a 2D array of non-Abelian anyons in this parent state produces a gapped descendant phase described by an $\SUtwo_{3}\otimes\SUtwo_1$ TQFT, as sketched in the left half of Fig.~\ref{TQFT_fig}.  See Table~\ref{su231} for the corresponding properties of $\SUtwo_3$ and $\SUtwo_1$.  The interface between these parent and descendant phases supports a gapless $\frac{\SUtwo_{3}\times \SUtwo_1}{\SUtwo_4}$ edge state, which exhibits central charge $c=4/5$ and ten fields corresponding exactly to those of the so-called $\mathcal{M}(6,5)$ minimal model.  Note that this edge theory is distinct from the $\mathbb{Z}_3$ parafermion CFT arising in our setup, which possesses only six fields.  Nevertheless, there are hints of a relation with our work present already here: $\SUtwo_4$ supports non-Abelian anyons with quantum dimension $\sqrt{3}$ (like the non-Abelian defects in our trenches), and the descendant $\SUtwo_{3}\otimes\SUtwo_1$ region supports a Fibonacci anyon (as in our Fibonacci phase).

\begin{table}[htb]
	{\setlength{\topsep}{-\parskip}
	\setlength{\partopsep}{0pt}
	\begin{tabularx}{.85\columnwidth}{|c|*{5}{|X}|}
	\hline
	\multicolumn{6}{|l|}{\hspace{2mm}$\SUtwo_4$ \hspace{8mm} $c=2$}\\
	\hline
	 Field & $\I$ & $X$  & $Y$ & $X'$ & $Z$\\
	\hline
	$j$ & $0$ & $1/2$ & $1$ & $3/2$ & $2$\\
	$h$ & $0$ & $1/8$ & $1/3$ & $5/8$ & $1$\\
	$d$ & $1$ & $\sqrt{3}$ & $2$ & $\sqrt{3}$ & $1$\\
	\hline\hline
	\multicolumn{6}{|l|}{Fusion rules}\\
	\hline
	\multicolumn{6}{|p{.73\columnwidth}|}{
	\begin{tabbing}
	$X\times X\sim \I + Y$ \hspace{13mm}\=$X'\times X'\sim \I + Y$\\
	$X\times Y\sim X + X'$\>$X'\times Y \sim X + X'$\\
	$X\times Z\sim X'$ \>$X'\times Z\sim X$ \\
	$X\times X'\sim Z + Y$ \>$Y\times Z\sim Y$\\
	$Y\times Y\sim \I + Y + Z$ \>$Z\times Z\sim \I$
	\end{tabbing}}\\
	\hline
	\end{tabularx}}
	\caption{Fields of $\SUtwo_4$, along with their corresponding $\SUtwo$ label $j$, conformal spin $h$, quantum dimension $d$, and nontrivial fusion rules. The chiral central charge associated with $\SUtwo_4$ is $c=2$.  The parent state on the left side of Fig.~\ref{TQFT_fig} is described by this TQFT.}
	\label{su24}
\end{table}

\begin{table}[htb]
	{\setlength{\topsep}{-\parskip}
	\setlength{\partopsep}{0pt}
	\begin{tabularx}{.75\columnwidth}{|c|*{4}{|X}|}
	\hline
	\multicolumn{5}{|l|}{\hspace{2mm}$\SUtwo_3$\hspace{8mm}$c=9/5$}\\
	\hline
	 Field  & $\I$ & $\varepsilon'$  & $\varepsilon$ & $\xi$\\
	\hline
	$j$ & $0$ & $1/2$ & $1$ & $3/2$\\
	$h$ & $0$ & $3/20$ & $2/5$ & $3/4$\\
	$d$ & $1$ & $\varphi$ & $\varphi$ & $1$\\
	\hline\hline
	\multicolumn{5}{|l|}{Fusion rules}\\
	\hline
	\multicolumn{5}{|p{.73\columnwidth}|}{
	\begin{tabbing}
	$\varepsilon \times \varepsilon\sim \I + \varepsilon$ \hspace{10mm}\=$\varepsilon'\times \varepsilon'\sim \I + \varepsilon$\\
	$\varepsilon\times \xi\sim\varepsilon'$\>$\varepsilon'\times \xi \sim\varepsilon$\\
	$\varepsilon \times \varepsilon'\sim\xi+\varepsilon'$ \>$\xi\times \xi\sim \I$
	\end{tabbing}}\\
	\hline
	\end{tabularx}
	\\[1.5mm]
	\begin{tabularx}{.75\columnwidth}{|c||X|X|}
	\hline
	\multicolumn{3}{|l|}{\hspace{2mm}$\SUtwo_1$\hspace{8mm}$c=1$}\\
	\hline
	 Field & $\I$ & $\eta$\\
	\hline
	$j$ & $0$ & $1/2$\\
	$h$ & $0$ & $1/4$\\
	$d$ & $1$ &  $1$\\
	\hline\hline
	\multicolumn{3}{|l|}{Fusion rule}\\
	\hline
	\multicolumn{3}{|p{.73\columnwidth}|}{
	\begin{tabbing}
	$\eta\times \eta\sim \I$
	\end{tabbing}}\\
	\hline
	\end{tabularx}}
	\caption{Properties of $\SUtwo_3$ and $\SUtwo_1$ topological quantum field theories, which describe the descendant phase on the left side of Fig.~\ref{TQFT_fig}.  In the table $c$ is the chiral central charge, $j$ is an $\SUtwo$ spin label, $h$ denotes conformal spin, $d$ represents the quantum dimension, and $\varphi$ is the golden ratio.}
	\label{su231}
\end{table}

At this point it is worth speculating on the field content expected from a hypothetical TQFT describing our $\nu = 2/3$ state with domain walls binding $\mathbb{Z}_3$ zero-modes.  First one should have Abelian fields $Y_1$ and $Y_2$ corresponding to charge $2e/3$ and $4e/3$ excitations (which can live either on the gapped regions of the trenches or in the bulk of the quantum Hall fluid). Conservation of charge mod $2e$ suggests the fusion rules $Y_1\times Y_1\sim Y_2$, $Y_2\times Y_2\sim Y_1$, and $Y_1\times Y_2\sim \I$, where $\I$ denotes the neutral identity channel.  One also might expect non-Abelian fields $\widetilde X$ corresponding to domain walls separating pairing- and tunneling-gapped regions of the trenches.  Recalling that the Cooper-paired regions can carry charge $0$, $2e/3$, or $4e/3$ mod $2e$, the merger of two adjacent superconducting islands in a trench should be captured by the fusion rule $\widetilde{X}\times \widetilde{X}\sim \I+Y_1+Y_2$.  From this perspective $\widetilde{X}$ quite clearly possesses a quantum dimension of $d=\sqrt{3}$ (consistent with deductions based on ground-state counting), since $\I$, $Y_1$, and $Y_2$ are Abelian fields with $d=1$.  No other fields are immediately evident.  This picture cannot possibly be complete, however, as there is \emph{no} TQFT with four fields obeying these fusion rules~\cite{ParsaThesis}.

The difficulty with identifying a TQFT using the preceding logic stems from the fact that $\widetilde{X}$ differs fundamentally from the other fields in that it does not represent a point-like excitation.  Rather, $\widetilde X$ occurs only at the end of a `string' formed by a superconducting region within our trenches; since these strings are physically measurable $\widetilde X$ is confined and exhibits only projective non-Abelian statistics.  One could---at least in principle---envision quantum mechanically smearing out the location of the superconductors to elevate $\widetilde{X}$ to the status of a deconfined point-like quantum particle belonging to some genuine non-Abelian TQFT.  Or by turning the problem on its head
one can instead view confined excitations like $\widetilde X$ as remnants of that non-Abelian TQFT after a phase transition.  In the latter viewpoint the mechanism leading to the transition---and the accompanying confinement---is boson condensation, which was described in detail by Bais and Slingerland in the context of topologically ordered phases~\cite{BaisSlingerland}.  

To be precise we will define a boson here as a field possessing integer conformal spin and quantum dimension $d = 1$ \footnote{Bais and Slingerland consider a more general case of `bosons' with $d>1$, though we will not need to consider this more complex situation here.}.
Suppose that a boson $B$ with these properties condenses.  When this happens the condensed boson is identified with the vacuum $\I$, and any fields related to one another by fusion with $B$ are correspondingly identified with each other.  For instance, if $A \times B \sim C$ then fields $A$ and $C$ are equivalent in the condensed theory.  The nature of such fields that are related by the boson $B$ depends on their relative conformal spin.  If their conformal spins differ by an integer, they braid trivially with the new vacuum and represent deconfined excitations.  Otherwise it is no longer possible to define in a gauge-invariant manner the conformal spin for that type of excitation; it braids nontrivially with the new vacuum and therefore must be confined by a physically measurable string.

\begin{table}[htb]
	{\setlength{\topsep}{-\parskip}
	\setlength{\partopsep}{0pt}
	\begin{tabularx}{.85\columnwidth}{|c|*{4}{|X}|}
	\hline
	\multicolumn{5}{|l|}{$\SUtwo_4$ with $Z$ boson condensed}\\
	\hline
	 Field & $\I$ & $\widetilde{X}$  & $Y_1$ & $Y_2$\\
	\hline
	$h$ & $0$ & ill-defined & $1/3$ & $1/3$ \\
	$d$ & $1$ & $\sqrt{3}$ & $1$ & $1$\\
	\hline\hline
	\multicolumn{5}{|l|}{Fusion rules}\\
	\hline
	\multicolumn{5}{|p{.73\columnwidth}|}{
	\begin{tabbing}
	$Y_1\times Y_2\sim \I$ \hspace{10mm}\=
	$\widetilde{X}\times \widetilde{X}\sim \I + Y_1+Y_2$\\
	$Y_1\times Y_1\sim Y_{2}$\>
	$\widetilde{X}\times Y_{1}\sim \widetilde X$\\
	$Y_2\times Y_2\sim Y_1$\>
	$\widetilde{X}\times Y_{2}\sim \widetilde X$
	\end{tabbing}}\\
	\hline
	\end{tabularx}}
	\caption{Field content and fusion rules for $\SUtwo_4$ upon condensing the bosonic $Z$ field listed in Table~\ref{su24}.  As in the other tables $j$ is an $\SUtwo$ spin label, $h$ denotes conformal spin, and $d$ represents the quantum dimension for each particle.  The $\widetilde{X}$ field is confined by the condensation and hence exhibits an ill-defined conformal spin; this field obeys the same fusion rules and projective non-Abelian statistics as the (also confined) domain wall defects in our $\nu = 2/3$ trenches.  Additionally $Y_1$ and $Y_2$ represent Abelian fields that correspond to charge $2e/3$ and $4e/3$ excitations in our quantum Hall setup.  
	If one ignores the confined excitation $\widetilde{X}$, the remainder is a pure Abelian $\mathbb{Z}_3$ theory with only $\I$, $Y_1$, and $Y_2$ particles.}  
	\label{su24broken}
\end{table}

Let us now apply this discussion to the parent $\SUtwo_4$ TQFT described earlier, assuming the $Z$ field condenses (from Table~\ref{su24} we see that this is the only nontrivial boson in the TQFT).  The resulting theory was already discussed extensively by Bais and Slingerland and will be briefly summarized here.  First of all the fusion rules tell us that condensation of $Z$ identifies $X$ and $X'$; anticipating a connection with our $\nu = 2/3$ extrinsic defects, we will label the corresponding excitation by $\widetilde{X}$.  Indeed, $\widetilde{X}$ is confined (because the conformal spins of $X$ and $X'$ differ by a non-integer), possesses a quantum dimension of $\sqrt{3}$, and exhibits the same projective non-Abelian braiding statistics as our quantum Hall domain wall defects~\cite{ClarkeParafendleyons,LindnerParafendleyons,ChengParafendleyons,BarkeshliParafendleyons1}.  
As for the $Y$ field, it can fuse into the vacuum in two different ways when $Z$ condenses (since $Z\rightarrow\I$), and so must split into two Abelian fields with conformal spin $2/3$ mod 1~\cite{BaisSlingerland}.  We will denote these two fields $Y_1$ and $Y_2$ as they exhibit the same characteristics as the charge $2e/3$ and $4e/3$ excitations in our quantum Hall problem.  
The properties of this `broken $\SUtwo_4$' theory~\cite{BaisSlingerland}, including the confined $\widetilde{X}$ excitation, appear in Table~\ref{su24broken}.  From the table it is apparent that this condensed theory reproduces exactly the structure anticipated from our $\nu = 2/3$ setup decorated with superconducting islands that generate $\mathbb{Z}_3$ zero-modes.  Hence the fusion rules and braiding statistics for our parent state can be viewed as inherited (projectively) from $\SUtwo_4$.  Note, however, that `broken $\SUtwo_4$' is not a pure TQFT; focusing only on deconfined excitations, we are left with a simple $\mathbb{Z}_3$ Abelian theory with only $\I$, $Y_1$, and $Y_2$.  

So far we have shown that the parent $\SUtwo_4$ theory discussed by Ludwig \emph{et al}.~recovers the particle content of our parent $\nu = 2/3$ system upon condensing the $Z$ field.  Next we explore the fate of their descendant $\SUtwo_{3}\otimes\SUtwo_1$ phase upon boson condensation.  Let us denote fields from $\SUtwo_{3}\otimes\SUtwo_1$ as $(A,B)$, where $A$ and $B$ respectively belong to $\SUtwo_3$ and $\SUtwo_1$, and explore the consequences of $(\xi,\eta)$ condensing.  (According to Table~\ref{su231} this field is indeed bosonic).  Aside from the identity we need only consider three fields after condensation---$(\varepsilon,\I)$, $(\varepsilon,\eta)$, and $(\xi,\I)$---since all others are related to these by the condensed boson.  The latter two are, however, confined as one can deduce by examining their conformal spin before and after fusing with $(\xi,\eta)$.  The lone deconfined field that remains is $(\varepsilon,\I)$, which is described by a pure Fibonacci theory.  Table~\ref{Fib} summarizes the main features of this TQFT, denoted here by `$\mathrm{Fib}$'.  This theory is analogous to that describing the descendant Fibonacci phase that we obtained by hybridizing arrays of $\mathbb{Z}_3$ zero-modes in our parent $\nu = 2/3$ system.  

\begin{table}[tb]
	{\setlength{\topsep}{-\parskip}
	\setlength{\partopsep}{0pt}
	\begin{tabularx}{.55\columnwidth}{|c||X|X|}
	\hline
	\multicolumn{3}{|l|}{\hspace{2mm}$\mathrm{Fib}$\hspace{8mm}$c=14/5$}\\
	\hline
	 Field & $\I$ & $\varepsilon$\\
	\hline
	$h$ & $0$ & $2/5$\\
	$d$ & $1$ &  $\varphi$\\
	\hline\hline
	\multicolumn{3}{|l|}{Fusion rule}\\
	\hline
	\multicolumn{3}{|p{.53\columnwidth}|}{
	\begin{tabbing}
	$\varepsilon\times \varepsilon\sim \I+\varepsilon$
	\end{tabbing}}\\
	\hline
	\end{tabularx}}
	\caption{The fields of $\mathrm{Fib}$, along with their corresponding conformal spin $h$, quantum dimension $d$, and nontrivial fusion rule.  This TQFT arises from $\SUtwo_{3}\otimes\SUtwo_1$ upon condensing the boson $(\xi,\eta)$ in Table~\ref{su231}, and describes the topologically ordered sector of the Fibonacci phase in our $\nu = 2/3$ setup.  }
	\label{Fib}
\end{table}

While it is not yet apparent, the condensation transitions that we discussed separately in the parent and descendant phases are in fact intimately related.  This connection becomes evident upon examining (from a particular point of view) the structure of the $\mathcal{M}(6,5)$ minimal model describing the boundary between the pure $\SUtwo_4$ and $\SUtwo_{3}\otimes\SUtwo_1$ phases prior to the transitions.  Appendix \ref{M65appendix} shows that \emph{at that boundary} the $Z$ and $(\xi,\eta)$ bosons are identified, which is reasonable since their $\SUtwo$ spin labels, conformal spins, and quantum dimensions all match.  Thus one can move the $Z$ boson smoothly from the parent to the descendant region where it `becomes' $(\xi,\eta)$---or vice versa.  It follows that the transitions in the parent and descendant phases are not independent, but rather can be viewed as arising from the condensation of a single common boson.  

Figure~\ref{TQFT_fig} summarizes the final physical picture that we obtain.
The left-hand side represents the parent $\SUtwo_4$ with descendant $\SUtwo_{3}\otimes\SUtwo_1$ setup analyzed by Ludwig \emph{et al}.~\cite{InteractingAnyons2}, which exhibits quite different physics from what we captured in this paper.  Condensing a single boson throughout that system leads to the parent $\mathbb{Z}_3$ with descendant $\mathrm{Fib}$ configuration illustrated on the right side of the figure.  These parent and descendant states do, by contrast, closely relate to our $\nu = 2/3$ quantum Hall setup with superconducting islands that drive the interior into the Fibonacci phase, in the sense that both systems exhibit the same deconfined bulk excitations in each region.  
There are, however, subtle differences between the system on the right side of Fig.~\ref{TQFT_fig} and our specific quantum Hall architecture that deserve mention.  

First, the Abelian $\mathbb{Z}_3$ TQFT technically does not quite describe the spin-unpolarized $\nu = 2/3$ state: the theory must be augmented to accommodate the electron in this fermionic quantum Hall phase~\cite{ParsaThesis}.  Moreover, the edge structure for the $\mathbb{Z}_3$ TQFT admits a chiral central charge $c = 2$, whereas the $\nu = 2/3$ state has $c = 0$ (because there are counterpropagating modes).  Both of these issues are relatively minor for the purposes of our discussion, however, and in any case can easily be sidestepped by considering a bosonic parent system.  In particular, as alluded to earlier the bosonic (221) state---which provides an equally valid backdrop for the descendant Fibonacci phase---exhibits a chiral central charge of $c = 2$ and is described by a $\mathbb{Z}_3$ TQFT with no modification.  
The $\mathrm{Fib}$ TQFT denoted on the right side of Fig.~\ref{TQFT_fig} also does not exactly describe our Fibonacci phase because this state exhibits a local order parameter (and hence is not strictly described by any TQFT).  This actually poses a far more minor issue than those noted above.  Recall from Sec.~\ref{sc_vortices} that superconducting vortices do not generate additional nontrivial quasiparticles in the Fibonacci phase.  Consequently the order parameter physics `factors out' and essentially decouples from the topological sector.  More formally, one can envision quantum disordering the superconductor by condensing vortices to eradicate the order parameter altogether \emph{without} affecting the quasiparticles supported by the Fibonacci phase that we have constructed
	\footnote{Vortex condensation can actually generate \emph{new} quasiparticles depending on the precise structure of the condensate; for interesting recent examples see Ref.~\onlinecite{SurfaceTopologicalOrder1,SurfaceTopologicalOrder2,SurfaceTopologicalOrder3,SurfaceTopologicalOrder4}.}.

The TQFT perspective on our results espoused in this section has a number of virtues.  For one it clearly illustrates the simplicity underlying the end product of our construction, and also unifies several related works that may at first glance appear somewhat distantly related.  Another benefit is that the condensation picture used along the way naturally captures the confined non-Abelian domain wall defects supported by $\nu = 2/3$ trenches with superconductivity.  More generally, viewing Abelian phases as remnants of non-Abelian TQFT's as we have done here may be useful in various other settings as a way of similarly identifying nontrivial phases accessible from interacting extrinsic defects.


\section{Fibonacci phase from uniform trenches}
\label{UniformTrenches}

In Sec.~\ref{MajoranaCase} we identified two closely linked routes to spinless $p+ip$ superconductivity from an integer quantum Hall system.  The first utilized trenches with spatially uniform Cooper pairing and electron backscattering perturbations present simultaneously; the second considered trenches alternately gapped by pairing and backscattering, yielding chains of hybridized Majorana modes.  In either case the trenches could be tuned to an Ising critical point, at which interchain coupling then naturally generated $p+ip$ superconductivity.  To construct a superconducting $\mathbb{Z}_3$ Read-Rezayi analogue (the Fibonacci phase), Secs.~\ref{QH_Z3_criticality} and \ref{RRsection} adopted the second approach and analyzed chains of $\mathbb{Z}_3$ generalized Majorana modes nucleated in a $\nu = 2/3$ fractional quantum Hall fluid.  This route enabled us to exploit the results of Ref.~\onlinecite{LatticeCFTrelation}, which derived the relation between lattice and CFT operators at the $\mathbb{Z}_3$ parafermion critical point for a single chain, to controllably study the 2D coupled-chain system.  Here we will argue that as in the integer quantum Hall case the same physics can also be obtained from spatially uniform $\nu = 2/3$ trenches.  This is eminently reasonable since on the long length scales relevant at criticality the detailed structure of the trenches should become unimportant.  

The analysis proceeds in two stages.  First we will use results from Lecheminant, Gogolin, and Nersesyan (LGN) \cite{LGN} to argue that a $\nu = 2/3$ trench with uniform pairing and backscattering perturbations also supports a $\mathbb{Z}_3$ parafermion critical point.  The relation between bosonized fields and CFT operators at criticality will then be deduced by coarse-graining the corresponding relationship obtained in Sec.~\ref{QH_Z3_criticality} in the spatially non-uniform case.  At that stage our results from Sec.~\ref{RRsection} carry over straightforwardly, allowing us to immediately deduce the existence of a Fibonacci phase in the uniform-trench setup.  

We start by reviewing the critical properties~\cite{LGN} for a toy Hamiltonian of the form
\begin{align}\begin{split}
	H_\textrm{LGN} &= \int_x\bigg{\{} \frac{v}{2\pi}[(\partial_x\phi)^2 + (\partial_x\theta)^2] 
\\	&\qquad\quad + u_1 \cos(3\theta)+ u_2 \cos(3\phi)\bigg{\}},
	\label{H_LGN}
\end{split}\end{align}
where the fields satisfy \footnote{We use rescaled fields compared to Ref.~\onlinecite{LGN} to highlight the relationship with our $\nu = 2/3$ problem.}
\begin{equation}
  [\theta(x),\phi(x')] = -\frac{2\pi i}{3}\Theta(x'-x).  
  \label{CommutatorLGN}
\end{equation}
The $u_{1,2}$ perturbations in $H_\textrm{LGN}$ are both relevant at the Gaussian fixed point and favor locking $\theta$ and $\phi$ to the three distinct minima of the respective cosines.  Because of the nontrivial commutator above, however, these terms compete and favor physically distinct gapped phases---very much like the tunneling and pairing terms in our quantum Hall trenches.  Using complementary non-perturbative methods, LGN showed that the self-dual limit corresponding to $u_1 = u_2$ realizes the same $\mathbb{Z}_3$ parafermion critical point as the three-state quantum clock model~\cite{LGN}.  
 
To expose the connection to our quantum Hall setup, consider the Hamiltonian introduced in Sec.~\ref{EdgeTheorySection} for a single trench in a $\nu = 2/3$ fluid with backscattering and Cooper pairing induced \emph{uniformly}:
\begin{align}\begin{split}
	H &= \int_x \bigg{\{}\sum_{a = \rho,\sigma}\frac{v_a}{2\pi}[(\partial_x\phi_a)^2 + (\partial_x \theta_a)^2]
\\	&\qquad\quad + 4 \cos\theta_\sigma[t  \sin(3\theta_\rho) - \Delta \sin(3\phi_\rho)]\bigg{\}}.
	\label{H_QH}
\end{split}\end{align}
As before $\phi_{\rho/\sigma}$ and $\theta_{\rho/\sigma}$ represent fields for the charge/spin sectors, while $t$ and $\Delta$ denote the tunneling and pairing strengths.  In writing the first line of $H$ we have assumed a particularly simple form for edge density-density interactions that can be described with velocities $v_{\rho/\sigma}$.  Upon comparison of Eqs.~\eqref{Commutation3} and \eqref{CommutatorLGN} one sees that the charge-sector fields obey the same commutation relation as those in the model studied by LGN.  Furthermore, modulo the spin-sector parts, the $u_{1,2}$ perturbations in Eq.~\eqref{H_LGN} have the same form as the tunneling and pairing terms above.  This hints at common critical behavior for the two models.  

The simplest way to make this relation precise is to include a perturbation that explicitly gaps the spin sector (while leaving the charge sector intact).  One such perturbation arises from correlated spin-flip processes described by $\delta H = \int_x (\Gamma \psi_{1\uparrow}^\dagger \psi_{2\downarrow}^\dagger \psi_{2\uparrow}\psi_{1\downarrow} + H.c.)$, where $\psi_{1\alpha}$ and $\psi_{2\alpha}$ are spin-$\alpha$ electron operators acting on the top and bottom sides of the trench, respectively.  In bosonized language this yields a term of the form
\begin{equation}
  \delta H = u_\sigma\int_x \cos(2\theta_\sigma).
\end{equation}
Suppose that the coupling $u_\sigma$ dominates over $t,\Delta$ and drives an instability in which $\theta_\sigma$ is pinned by the cosine potential above.  At low energies the Hamiltonian $H$ in Eq.~\eqref{H_QH} that describes the remaining charge degrees of freedom then maps onto the LGN Hamiltonian in Eq.~\eqref{H_LGN}.  Consequently, the self-dual critical point at which $|t| = |\Delta|$ is likewise described by $\mathbb{Z}_3$ parafermion CFT.  

For the following reasons we believe that it is likely that the same critical physics arises without explicitly invoking the $u_\sigma$ perturbation.  Recall that both $t$ and $\Delta$ favor pinning the spin-sector field $\theta_\sigma$ in precisely the same fashion, but gap the charge sector in incompatible ways [see Eqs.~\eqref{TunnelingGap} and \eqref{PairingGap}].  Suppose that we start from a phase in which tunneling $t$ gaps both sectors.  Increasing $\Delta$ at fixed $t$ must eventually induce a phase transition in the charge sector.  Provided the spin sector remains gapped throughout it suffices to replace the $\cos\theta_\sigma$ term in Eq.~\eqref{H_QH} by a constant across the transition.  The model then once again reduces to $H_\textrm{LGN}$ and hence exhibits a $\mathbb{Z}_3$ parafermion critical point at $|t| = |\Delta|$.  We stress that although it is difficult to make rigorous statements about this nontrivial, strongly coupled field theory, this outcome is nevertheless intuitively very natural given our results for criticality in spatially modulated trenches.  

Our primary interest lies in `stacking' such critical 1D systems to access new exotic 2D phases.  Physical interchain perturbations can easily be constructed in terms of bosonized fields, as in Sec.~\ref{RRsection}, though at the $\mathbb{Z}_3$ parafermion critical point these fields no longer constitute the `right' low-energy degrees of freedom.  An essential technical step is identifying the correspondence between bosonized and CFT operators at criticality so that one can systematically disentangle high- and low-energy physics.  We will now deduce this relationship for quasiparticle creation operators that are relevant for interchain processes in our $\nu = 2/3$ setup with uniform trenches.  

To do so we first revisit the non-uniform system analyzed in Sec.~\ref{QH_Z3_criticality}.  By combining Eqs.~\eqref{ProjectedOps}, \eqref{alphaR2} and \eqref{alphaL2} we obtain the following expansions valid at the parafermion critical point:
\begin{align}
	e^{i[\phi_{1\uparrow}(x_{j})+\phi_{1\downarrow}(x_{j})]} &\sim a(-1)^j \psi_R + b\sigma_R \epsilon_L + \dots \;,
	\label{ProjectedOps2}
\\	e^{i[\phi_{2\uparrow}(x_{j})+\phi_{2\downarrow}(x_{j})]} &\sim e^{i\pi/3}[a(-1)^j\psi_{L} + b \sigma_L \epsilon_R] + \dots \;.
	\notag
\end{align}
We remind the reader that the operators on the left-hand side create charge-$2e/3$ quasiparticles on the top and bottom trench edges, at position $x_j$ in domain wall $j$ [$\phi_{1/2\alpha}$ relates to the charge- and spin-sector fields through Eqs.~\eqref{phi_theta}].  Moreover, on the right side $a,b$ again denote non-universal constants while the ellipses represent terms with subleading scaling dimensions.  Connection with the uniform trench can now be made upon coarse-graining the expressions above---specifically by averaging over sums and differences of quasiparticle operators at adjacent domain walls in a given unit cell.  (Each unit cell contains two domains as shown in Fig.~\ref{Chain_fig}.)  The oscillating terms clearly cancel for the sum, leaving
\begin{align}\begin{split}
	e^{i[\phi_{1\uparrow}(x)+\phi_{1\downarrow}(x)]} &\sim \sigma_R \epsilon_L + \dots \;,
\\	e^{i[\phi_{2\uparrow}(x)+\phi_{2\downarrow}(x)]} &\sim e^{i\pi/3}\sigma_L \epsilon_R + \dots \;,
	\label{ProjectedOps3}
\end{split}\end{align}
where $x$ now denotes a continuous coordinate.  One can isolate the parafermion fields by instead averaging over differences of quasiparticle operators at neighboring domain walls, which yields
\begin{align}\begin{split}
	\partial_x e^{i[\phi_{1\uparrow}(x)+\phi_{1\downarrow}(x)]} &\sim \psi_R + \dots \;,
\\	\partial_x e^{i[\phi_{2\uparrow}(x)+\phi_{2\downarrow}(x)]} &\sim e^{i\pi/3}\psi_L + \dots \;.
	\label{ProjectedOps4}
\end{split}\end{align}
The extra derivatives on the left-hand side reflect the fact that the parafermions acquire a relative minus sign under parity $P$ compared to the fields on the right side of Eqs.~\eqref{ProjectedOps3}~\cite{LatticeCFTrelation}.  More generally, the coarse-graining procedure used here merely ensures that the quantum numbers carried by the bosonized and CFT operators agree with one another.   

We are now in position to recover the physics discussed in Sec.~\ref{RRsection}, but instead from a system of spatially uniform critical trenches.  Equations~\eqref{ProjectedOps3} and \eqref{ProjectedOps4} allow us to construct interchain quasi-particle hoppings that reproduce the $\lambda_{a,b}$ terms in Eq.~\eqref{Hperp}.  The effective low-energy Hamiltonians in the two closely related setups are then identical---and hence so are the resulting phase diagrams.  In particular, as Fig.~\subfigref{fig:CoupledChaina} illustrates if the interchain coupling $\lambda_a>0$ dominates then the uniform-trench system flows to the Fibonacci phase.  Determining the microscopic parameters (in terms of the underlying electronic system) required to enter this phase remains an interesting open issue, though such a state is in principle physically possible in either setup that we have explored.


\section{Summary and Discussion}
\label{Discussion}

The introduction to this paper provided a broad overview of the main physical results derived here.  Having now completed the rather lengthy analysis, we will begin this discussion with a complementary and slightly more technical summary:

Our setup begins with a spin-unpolarized $\nu=2/3$ Abelian fractional quantum Hall state---also known as the (112) state---as the backbone of our heterostructure.
The (112) state is a strongly correlated phase built from spin-up and spin-down electrons partially occupying their lowest Landau level.
At the boundary with the vacuum its edge structure consists of a charge mode (described by $\phi_\uparrow + \phi_\downarrow$) and a counterpropagating  neutral mode (described by $\phi_\uparrow - \phi_\downarrow$).
We first showed that a long rectangular hole---a `trench'---in this fractional quantum Hall system realizes a $\mathbb{Z}_3$ parafermion critical point when coupled to an ordinary $s$-wave superconductor.
This nontrivial critical theory with central charge $c_L = c_R = 4/5$ is well-known from earlier studies of the three-state quantum clock model, and moreover is important for characterizing edge states of the $\mathbb{Z}_3$ Read-Rezayi phase whose properties we sought to emulate.  
We presented two related constructions.  The first utilizes an alternating pattern of superconducting and non-superconducting regions in the trench, as described in Sec.~\ref{QH_Z3_criticality}, to essentially engineer a non-local representation of the three-state clock model.
The second, explored in Sec.~\ref{UniformTrenches}, employs a `coarse-grained' variation wherein the trench couples uniformly to a superconductor throughout.
Tuning to the $\mathbb{Z}_3$ parafermion critical point follows by adjusting the coupling between domain walls (in the case of modulated trenches), or electron tunneling across the trench (in the uniform-trench setup).
In both scenarios, the neutral excitations are gapped out while the charge modes provide the low-energy degrees of freedom.
One remarkable feature of our mapping is that we can identify the relation between `high-energy' operators and chiral fields describing low-energy physics near criticality, given by Eqs.~\eqref{eq:alphaCFT} for the lattice construction and Eqs.~\eqref{ProjectedOps3} and \eqref{ProjectedOps4} for the continuum version.
This key technical step enabled us to perform calculations parallel to those for coupled Majorana chains described in Sec.~\ref{MajoranaCase}---but at a nontrivial strongly interacting critical point.

To construct a 2D non-Abelian phase reminiscent of the $\mathbb{Z}_3$ Read-Rezayi state we consider an array of these critical trenches in the $\nu = 2/3$ quantum Hall fluid, with neighboring trenches coupled via charge-$2e/3$ quasiparticle hopping (see Fig.~\ref{Coupled_chain_fig} for the lattice setup).
With the correspondence between quasiparticle operators and CFT fields in hand, we find that the second-most-relevant interchain coupling corresponds to a term that couples the right-moving parafermion field $\psi_R(y)$ from trench $y$ with the adjacent left-mover $\psi_L(y+1)$ from trench $y+1$.  This perturbation gaps out each critical trench except for the first right-mover and the final left-mover.
The system then enters a stable 2D chiral topological state, as shown in Sec.~\ref{RRsection}, which we dubbed the `Fibonacci phase'.  Since this phase exhibits a bulk gap its topological properties are stable; therefore it is neither necessary to tune the individual chains exactly to criticality, nor to set the most-relevant interchain coupling precisely to zero.

We uniquely established the universal topological properties of the Fibonacci phase by identifying its two-fold ground state degeneracy on a torus (which implies two anyon species), fusion rules, and quantum dimensions via the topological entanglement entropy.
The quasiparticle structure present here is elegant in its simplicity yet rich in content, consisting of a trivial particle $\I$ and a Fibonacci anyon $\varepsilon$ obeying the simple fusion rule $\varepsilon \times \varepsilon \sim \I + \varepsilon$.
One of the truly remarkable features of this state is that the ability to exchange Fibonacci anyons, and to distinguish the Fibonacci anyon from the vacuum, is sufficient to perform any desired quantum computation in a completely fault-tolerant manner \cite{Freedman02a,Freedman02b}.

The Fibonacci phase supports gapless edge excitations.  When this state borders the parent Abelian quantum Hall fluid from which it descends [as in Fig.~\ref{Summary}(b)], they are described by a chiral $\mathbb{Z}_3$ parafermion CFT with central charge $c = 4/5$---exactly as in the $\mathbb{Z}_3$ Read-Rezayi phase modulo the charge sector.  The edge states arising at the interface with the vacuum can be obtained upon shrinking the outer Abelian quantum Hall liquid, thereby hybridizing the parafermion and quantum Hall edge fields.
If the Fibonacci phase descends from a bosonic analogue of the spin-unpolarized $\nu = 2/3$ state, i.e., the bosonic (221) state, then the boundary with the vacuum exhibits edge modes described by the $G_2$ Kac-Moody algebra at level-$1$.
This fully chiral edge theory has central charge $c=14/5$, contains two primary fields associated with the bulk excitations $\I$ and $\varepsilon$, and occurs also in the pure Fibonacci topological quantum field theory discussed in Sec.~\ref{TQFTsection}.
If instead the Fibonacci phase emerges out of the fermionic (112) state, then the corresponding edge is not fully chiral and does not in general admit a decomposition into independent left- and right-movers.
However, we find that the edge theory may be reconstructed such that it factorizes into two left-moving fermions with central charge $c_L=2$ and a right-moving sector identical to the bosonic case with central charge $c_R = 14/5$.

Because of the superconductivity in our setup the Fibonacci phase admits gapless order parameter phase fluctuations but is otherwise fully gapped away from the edge. 
Nevertheless, its low-energy Hilbert space consists of a tensor product of states for a topologically trivial superconductor and those of a gapped topological phase.  In this sense the superconductivity is peripheral: it provides an essential ingredient in our microscopic construction, but does not influence the Fibonacci phase's universal topological properties.
This stands in stark contrast with the case of a spinless $p+ip$ superconductor.  There an $h/2e$ superconducting vortex binds a Majorana zero-mode and thus exhibits many characteristics of $\sigma$ particles (i.e., Ising anyons), despite being logarithmically confined by order parameter energetics.  If superconductivity is destroyed by the condensation
of double-strength $h/e$ vortices, then the $h/2e$ vortex becomes a bona fide deconfined $\sigma$ particle 
in the resulting insulating phase.  On the other hand, destroying superconductivity by condensing single-strength $h/2e$ vortices produces a trivial phase.
The physics is completely different in the Fibonacci phase where an $h/2e$ vortex braids trivially with an $\varepsilon$ particle.
	(Here we assume that the vortex does not `accidentally' trap a Fibonacci anyon.)
Condensation of $h/2e$ vortices therefore simply leaves the pure
Fibonacci phase with no residual order parameter physics. It is interesting to note that richer physics arises upon condensing $nh/2e$ vortices, which yields the Fibonacci phase tensored with a $\mathbb{Z}_n$ gauge theory; this additional sector is, however, clearly independent of the Fibonacci phase. 

A number of similarities exist between our Fibonacci phase and previously constructed models that harbor Fibonacci anyons.
We have already emphasized several parallels with the $\mathbb{Z}_3$ Read-Rezayi state.  Teo and Kane's coupled-wire construction of this non-Abelian quantum Hall phase is particularly close in spirit to this paper (and indeed motivated many of the technical developments used here).  The $\mathbb{Z}_3$ Read-Rezayi state,
however, certainly represents a distinct state of matter with different universal topological properties.
For instance, there the fields $\psi$ and $\sigma$ (with appropriate bosonic factors) represent deconfined, electrically charged quasiparticles, whereas the Fibonacci anyon $\varepsilon$ provides the only nontrivial quasiparticle in the Fibonacci phase.
Fibonacci anyons also occur in the exactly soluble lattice model of Levin and Wen \cite{Levin05}.  Important differences arise here too: their model is non-chiral, and has the same topological properties as two opposite-chirality copies of the Fibonacci phase constructed in this paper.
(See also the related works of Refs.~\onlinecite{Fidkowski06,Fendley13} for loop gas models that may support such a non-chiral phase.)
Finally, recent unpublished work by Qi \emph{et al}.\ accessed a phase with Fibonacci anyons using $\mathbb{Z}_n$ lattice operators as building blocks, similar to those that arise in our spatially modulated trenches \cite{QiFibonacci}.
It would thus be interesting to explore possible connections with our study.  

We now turn to several other outstanding questions and future directions raised by our results, placing particular emphasis on experimental issues.

Realizing non-Abelian anyons with universal braid statistics in any setting carries great challenges yet correspondingly great rewards if they can be overcome.  Our proposal is no exception.  The price that one must pay to realize Fibonacci anyons as we envision here
is that a fractional quantum Hall system must intimately contact an $s$-wave superconductor.  For several reasons, however, accessing the Fibonacci phase may be less daunting than it appears.  First of all Abelian fractional quantum Hall states appear in many materials---and not just in buried quantum wells such as $\mathrm{GaAs}$.  Among the several possible canvases graphene stands out as particularly promising due to the relative ease with which a proximity effect can be introduced \cite{HeerscheGrapheneSupercurrent,DuSkachkoAndrei08,BouchiatGrapheneSNS09}.  Graphene can also be grown on metallic substrates \cite{Li09}, and if such a substrate undergoes a superconducting transition a strong proximity effect may result.  

Another point worth emphasizing is that weak magnetic fields are not required, which is crucial given that our proposal relies on the fractional quantum Hall effect.  This stems from the fact that superconducting vortices in the Fibonacci phase need not carry topologically nontrivial particles.  Assuming that Fibonacci anyons do not happen to energetically bind to vortex cores--which again they need not---then any field strength up to the (type II) superconductor's upper critical field $H_{c2}$ should suffice.
By contrast, in the case of a spinless $p+ip$ superconductor the density of vortices must remain low because they \emph{necessarily} support Majorana modes.
Appreciable tunneling between these, which will arise if the spacing between vortices becomes too small, therefore destabilizes the Ising phase.  

We also reiterate that preparing precisely the
somewhat elaborate, fine-tuned setups explored here is certainly not necessary for accessing the Fibonacci phase.  
Many of the features we invoked in our analysis---including the multi-trench geometry and all of the fine-tuning that went with it---served purely as a theoretical crutch that enabled us to decisively show that our model supports this state and identify its properties.  The Fibonacci phase is stable to (at least) small perturbations, and the extent of its stability remains a very interesting open question.  It seems quite possible that this stability regime extends across a large swath of
the parameter space for a quantum Hall state coupled to a superconductor.  Hinting that this may be so is the fact that the Fibonacci phase that we have constructed
is actually isotropic and translationally invariant in the long-wavelength limit.  
Hence, it is even possible that a completely `smeared' Abelian quantum Hall/superconductor
heterostructure enters this phase even in the absence
of trenches. Although the methods used in this paper are not applicable to
this case, it may be possible to study such a scenario by applying
exact diagonalization or the density-matrix renormalization group to
small systems of electrons in the lowest Landau level. Numerical studies along these lines are analogous to
previous studies of the fractional quantum Hall effect, but with the added wrinkle
that $\mathrm{U(1)}$ charge conservation symmetry is broken.  This almost entirely untapped area seems ripe for discovery.

As a final remark on experimental realizations, we stress that superconductivity may be altogether inessential---even at the microscopic level.  To see why it is useful to recall that the superconductors in our construction simply provide a mechanism for gapping the edge states opposite a trench that is `incompatible' with the gapping favored by ordinary electronic backscattering.  When balanced these competing terms thus drive the system to a nontrivial critical point that we bootstrapped off of to enter the Fibonacci phase.  In beautiful theoretical studies Refs.~\onlinecite{BarkeshliParafendleyons1,BarkeshliParafendleyons2} showed that similar incompatible gap-generating processes can arise in certain quantum Hall bilayers \emph{without} Cooper pairing; for instance, if one cuts a trench in the bilayer, electrons can backscatter by tunneling from `top to bottom' or `side to side'.  It may thus be possible to realize the Fibonacci phase in a bilayer fractional quantum Hall setup by regulating the inter- and intra-layer tunneling terms along trenches, following Refs.~\onlinecite{BarkeshliParafendleyons1,BarkeshliParafendleyons2}.
Such an avenue would provide another potentially promising route to Fibonacci anyons that is complementary to the superconductor/quantum Hall heterostructures that we focused on here.  

Our construction naturally suggests other interesting generalizations as well.  The $\nu = 2/3$ state is not the only spin-singlet fractional quantum Hall phase---another can occur, e.g., at $\nu = 2/5$.
These may provide equally promising platforms for the Fibonacci phase or relatives thereof.
Moreover, our construction is by no means  limited to fermionic quantum Hall phases.
As we noted earlier the bosonic (221) state, for instance, leads to nearly identical physics (which is actually simpler in some respects).
By following a similar route to that described here, it may be possible
to build on these quantum Hall states to construct other non-Abelian
topological phases, perhaps realizing $\mathbb{Z}_k$ parafermions, $\SUtwo_k$, or yet more exotic phases. 

To conclude we briefly discuss the longer-term prospects of exploiting our model for quantum computation.  
Quantum information can be encoded in a many-$\varepsilon$ state
using either a dense or sparse encoding. There are
two states of three $\varepsilon$ particles with total
charge $\varepsilon$ and also two states of four $\varepsilon$
particles with total charge $\I$, and either pair can be used as a qubit.
The unitary transformations generated
by braiding are dense within the projective unitary group
on the many-anyon Hilbert space and, therefore, within the unitary group
on the computational subspace \cite{Freedman02a,Freedman02b}.
However, this presupposes that we can create pairs of Fibonacci anyons
at will, and braid and detect them. Since they carry neither electric charge nor any flux, this is challenging.
In this respect, the rather featureless $\varepsilon$ particles are analogous to $\psi$ particles
in an Ising anyon phase. This suggests the following approach.
Consider the case of a single Ising or three-state clock model on a ring.
If we make one of the bond couplings equal to $-\infty$, then it breaks the
ring into a line segment and the spins at the two ends are required
to have opposite values. In the Ising case, this means that if one end
is `spin up', the other is `spin down', and vice versa. This forces a $\psi$ into the chain.
However, this particle is not localized and can move freely.
If we now couple many such chains, some of which have $\psi$'s,
then they can also move between chains and annihilate.
However, we can in principle trap a $\psi$ by reducing the gap
at various locations. In the $\mathbb{Z}_3$ clock case, if one end
of a chain is $A$, then the other end is `not-$A$'. (Here, we are calling the
three states $A, B, C$.) This forces an $\varepsilon$ particle into a single chain.
It is plausible that when the chains are coupled through their parafermion operators, these $\varepsilon$ particles will be able to move freely between chains.
They could then similarly be trapped by locally suppressing the gap, as in the Ising case. Showing that this scenario is correct or designing an alternate protocol for manipulating Fibonacci anyons poses an important challenge for future work.

\acknowledgments{%
We are grateful to Parsa Bonderson, Alexey Gorshkov, Victor Gurarie, Roni Ilan, Lesik Motrunich, Hirosi Ooguri, John Preskill, Miles Stoudenmire, and Krysta Svore for illuminating conversations.
This work was supported in part by the NSF under Grant No.\ PHYS-1066293 and the hospitality of the Aspen Center for Physics, where the idea for this work was conceived.
We also acknowledge funding from the NSF through grants DMR-1341822 (D.\ C.\ \& J.\ A.), DMR-MPS1006549 (P.\ F.), DMR-0748925 (K.\ S.), and DMR-1101912 (M.\ F.); the Alfred P.\ Sloan Foundation (J.\ A.); the Sherman Fairchild Foundation (R.\ M.); the DARPA QuEST program (C.\ N. \& K.\ S.); the AFOSR under grant FA9550-10-1-0524 (C.\ N.); the Israel Science Foundation (Y.\ O.); the Paul and Tina Gardner fund for Weizmann-TAMU collaboration (Y.\ O.); the US--Israel Binational Science Foundation (A. S.); the Minerva Foundation (A. S.); Microsoft Research Station Q; 
and the Caltech Institute for Quantum Information and Matter, an NSF Physics Frontiers Center with support of the Gordon and Betty Moore Foundation.
}

\clearpage
\appendix

\section{Symmetries in the quantum Hall setup}
\label{Symmetries}

The quantum clock model reviewed in Sec.~\ref{Z3CFT} exhibits a number of symmetries, preserving $\mathbb{Z}_3$ and $\mathbb{Z}_3^\textrm{dual}$ transformations, translations $T_x$, parity $P$, charge conjugation $\mathcal{C}$, and a time-reversal transformation $\mathcal{T}$.  In this Appendix we illustrate that each of these symmetries exhibits a physical analogue in the quantum Hall architectures discussed in Secs.~\ref{QH_Z3_criticality} and \ref{RRsection}.  To this end consider the geometry of Fig.~\ref{Chain_fig}, in which a single trench hosted by a $\nu = 2/3$ system yields a chain of coupled $\mathbb{Z}_3$ generalized Majorana operators; the Hamiltonian describing the hybridization of these modes is given in Eq.~\eqref{Hparafendleyons2}.  Below we identify the realization of the clock-model symmetries in this specific setup.  The results apply straightforwardly to the multi-trench case as well.  Note that we frequently make reference to the bosonized fields, and the integer operators describing their pinning induced by tunneling $t$ or pairing $\Delta$, defined in Sec.~\ref{QH_Z3_criticality}.  

$(i)$ In the limit where $\Delta = t = 0$ the electron number on each side of the trench is separately conserved.  This is reflected in independent global $\mathrm{U(1)}$ symmetries that send $\theta_\rho \rightarrow \theta_\rho + a_1$ and $\phi_\rho \rightarrow \phi_\rho + a_2$ for arbitrary constants $a_{1,2}$.  Restoring $\Delta$ and $t$ to non-zero values breaks these continuous symmetries down to a pair of discrete $\mathbb{Z}_3$ symmetries, which is immediately apparent from Eq.~\eqref{deltaH}.  The remaining invariance under $\phi_\rho \rightarrow \phi_\rho + 2\pi/3$, which transforms $\hat{n}_j\rightarrow 1+\hat{n}_j$, corresponds to the clock model symmetry $\mathbb{Z}_3$; similarly, the transformation $\theta_\rho \rightarrow \theta_\rho + 2\pi/3$ sends $\hat{m}_j\rightarrow 1+\hat{m}_j$ and corresponds to $\mathbb{Z}_3^\textrm{dual}$.  

$(ii)$ The symmetry $T_x$ corresponds to a simple translation along the trench that shifts $\hat{m}_j\rightarrow\hat{m}_{j+1}$ and $\hat{n}_j\rightarrow\hat{n}_{j+1}$.  

$(iii)$ In the clock model parity $P$ corresponds to a reflection that interchanges the generalized Majorana operators $\alpha_{Rj}$ and $\alpha_{Lj}$.  Since the analogous operators defined in Eqs.~\eqref{parafendleyons2} involve quasiparticles from opposite sides of the trench, here the equivalent of $P$ corresponds to a $\pi$ \emph{rotation} in the plane of the quantum Hall system.  We seek an implementation of this rotation that leaves the total charge and spin densities $\rho_+$, $S_+$ invariant; changes the sign of the density differences $\rho_-$, $S_-$; and preserves the bosonized form of $\delta H$ in Eq.~\eqref{deltaH}.  The following satisfies all of these properties: $\theta_\rho(x) \rightarrow-\theta_\rho(-x)-\pi/3$, $\phi_\rho(x) \rightarrow \phi_\rho(-x) + 4\pi/3$, $\theta_\sigma(x)\rightarrow -\theta_\sigma(-x)$, and $\phi_\sigma(x)\rightarrow \phi_\sigma(-x)$.  (We have included the factor of $4\pi/3$ in the transformation of $\phi_\rho$ so that the generalized Majorana operators in our quantum Hall problem transform as in the clock model under $P$.  This factor transforms all electron operators trivially and thus corresponds to an unimportant global gauge transformation.)  Taking the rotation about the midpoint of a pairing-gapped section, the integer operators transform as $\hat{M}\rightarrow -\hat{M}$, $\hat{m}_j \rightarrow -\hat{m}_{-j-1}$, $\hat{n}_j\rightarrow \hat{n}_{-j}+\hat{M}+2$ under this operation.  

$(iv)$ Charge conjugation $\mathcal{C}$ arises from a particle-hole transformation on the electron operators $\psi_{1\alpha}\rightarrow \psi_{1\alpha}^\dagger$, $\psi_{2\alpha}\rightarrow-\psi_{2\alpha}^\dagger$, which leaves the perturbations in Eq.~\eqref{deltaHfermions} invariant.  In bosonized language this corresponds to $\theta_\rho \rightarrow -\theta_\rho-\pi/3$, $\phi_\rho \rightarrow -\phi_\rho + \pi/3$, $\theta_\sigma \rightarrow -\theta_\sigma$, and $\phi_\sigma \rightarrow -\phi_\sigma$.  The integer operators in turn transform as $\hat{M}\rightarrow - \hat{M}$, $\hat{m}_j\rightarrow-\hat{m}_j$, and $\hat{n}_j\rightarrow -\hat{n}_j$ under $\mathcal{C}$.  Note that it is easy to imagine adding perturbations that violate this symmetry in the original edge Hamiltonian (e.g., spin flips acting on one side of the trench); however, such perturbations project trivially into the ground-state manifold.  Hence one should view $\mathcal{C}$ as an emergent symmetry valid in the low-energy subspace in which we are interested.  

$(v)$ Finally, for the equivalent of the clock-model symmetry $\mathcal{T}$ we need to identify an antiunitary transformation exhibited by our $\nu = 2/3$ setup that squares to unity in the ground-state subspace and swaps the $\alpha_{Rj}$ and $\alpha_{Lj}$ operators.  Physical electronic time-reversal $\mathcal{T}_\textrm{ph}$ composed with a reflection $R_y$ about the length of the trench (which can be a symmetry for electrons in a magnetic field) has precisely these properties---i.e., $\mathcal{T} = \mathcal{T}_\textrm{ph} R_y$.  This operation transforms the electron operators as $\psi_{1\alpha}\rightarrow i\sigma^y_{\alpha\beta}\psi_{2\beta}$, $\psi_{2\alpha}\rightarrow i\sigma^y_{\alpha\beta}\psi_{1\beta}$ and sends the bosonized fields to $\theta_\rho\rightarrow \theta_\rho$, $\phi_\rho \rightarrow -\phi_\rho + \pi/3$, $\theta_\sigma \rightarrow -\theta_\sigma$, and $\phi_\sigma \rightarrow \phi_\sigma + \pi$.  The integer operators correspondingly transform under $\mathcal{T}$ as $\hat{M} \rightarrow - \hat{M}$, $\hat{m}_j \rightarrow \hat{m}_j + \hat{M}$, and $\hat{n}_j \rightarrow -\hat{n}_j$.  Notice that whereas this composite operation squares to $-1$ when acting on the original electron operators, in the projected subspace $(\mathcal{T}_\textrm{ph} R_y)^2 = +1$ as desired.

\section{\texorpdfstring{$\mathcal{M}(6,5)$ edge structure via boson condensation}{M(6,5)}}
\label{M65appendix}

This Appendix deals with the setup shown in the left side of Fig.~\ref{TQFT_fig}, in which a parent state described by an $\SUtwo_4$ TQFT hosts a descendant $\SUtwo_{3}\otimes\SUtwo_1$ phase \cite{InteractingAnyons2}; see Tables~\ref{su24} and \ref{su231} for summaries of the field content in each region.  Our specific goal is to substantiate the claim made in Sec.~\ref{TQFTsection} that the $Z$ and $(\xi,\eta)$ bosons supported in the bulk of the parent and descendant states, respectively, are equivalent at their interface.  [We are again using notation where fields from $\SUtwo_{3}\otimes\SUtwo_1$ are labeled $(A,B)$, with $A$ in $\SUtwo_{3}$ and $B$ in $\SUtwo_1$.]  To meet this objective we will describe how one can recover, via edge boson condensation, the $\mathcal{M}(6,5)$ minimal model describing gapless modes at the interface between the parent and descendant phases.  As we will see this viewpoint makes the identification of the $Z$ and $(\xi,\eta)$ bosons immediately obvious.

First, observe that the gapless modes bordering $\SUtwo_4$ and $\SUtwo_{3}\otimes\SUtwo_1$ topological liquids are naively captured by an \mbox{$\SUtwo_3\otimes \SUtwo_1\otimes \overline{\SUtwo_4}$} CFT, where the overline indicates a reversed chirality.  For concreteness we will assume that the sector with an overline describes left-movers while others correspond to right-movers.  Adopting similar notation as above we describe fields from the product edge theory theory as triplets of fields from the constituent sectors, e.g., $(\varepsilon,\eta,X)$.
(Note that this Appendix will employ the same symbols for primary fields at the interface and bulk anyons to facilitate the connection with Sec.~\ref{TQFTsection}.)
In total, forty such triplets exist---far more than the ten fields found in $\mathcal{M}(6,5)$.  Any non-chiral boson in this edge theory can, however, condense at the interface thereby reducing the number of distinct deconfined fields.  To avoid possible confusion, we stress that in contrast to Sec.~\ref{TQFTsection} we assume throughout this appendix that the \emph{bulk} properties of the parent and descendant phases remain intact.

Ignoring chirality for the moment, we find only three such bosonic combinations (i.e., triplets with integer conformal spin and quantum dimension $d=1$).  They are $(\I,\I,Z)$, $(\xi,\eta,\I)$, and $(\xi,\eta,Z)$.  The right- and left-moving conformal dimensions of these fields are respectively given by $(0,1)$, $(1,0)$, and $(1,1)$.  Consequently, the first two fields form chiral bosons and so cannot condense without an accompanying bulk phase transition in the parent or nucleated liquid---which again we preclude here.  The last field, $(\xi,\eta,Z)$, represents a non-chiral $\mathbb{Z}_2$ boson, and as we now argue when condensed results in the $\mathcal{M}(6,5)$ minimal model on the edge.

To see this, note that one can divide the forty fields of \mbox{$\SUtwo_3\otimes \SUtwo_1\otimes \overline{\SUtwo_4}$} into sets of fields $A_i$ and $B_i$ (with $i = 1,\ldots,20$) related by fusion with the $\mathbb{Z}_2$ boson $(\xi,\eta,Z)$. That is,
\begin{align}\begin{split}
  A_i\times (\xi,\eta,Z) &\sim B_i ,
  \\
  B_i\times (\xi,\eta,Z) &\sim A_i .
\end{split}\end{align}
This reduces the number of fields from forty to twenty---still more than are present in the $\mathcal{M}(6,5)$ minimal model.  There is, however, an additional criterion that one needs to consider.  Namely, only when the conformal spins of $A_i$ and $B_i$ match (mod 1) can a well-defined spin be assigned to the new field $A_i\equiv B_i$ following the condensation of $(\xi,\eta,Z)$; otherwise those fields become confined.  One can readily verify that there are ten pairs of fields $A_i$ and $B_i$ for which the conformal spins agree in the above sense, and these deconfined fields correspond to the ten fields of the $\mathcal{M}(6,5)$ minimal model.  

This picture of $\mathcal{M}(6,5)$ as an \mbox{$\SUtwo_3\otimes \SUtwo_1\otimes \overline{\SUtwo_4}$} edge theory with $(\xi,\eta,Z)$ condensed is very useful.  In particular, since $(\I,\I,Z) \times (\xi,\eta,Z) \sim (\xi,\eta,\I)$, it follows that the $Z$ and $(\xi,\eta)$ bosons native to the parent and descendant phases are indeed identified at their interface, which is what we set out to show.  


\hbadness 10000	
\bibliography{Parafermion_refs}

\begin{thebibliography}{174}%
\makeatletter
\providecommand \@ifxundefined [1]{%
 \@ifx{#1\undefined}
}%
\providecommand \@ifnum [1]{%
 \ifnum #1\expandafter \@firstoftwo
 \else \expandafter \@secondoftwo
 \fi
}%
\providecommand \@ifx [1]{%
 \ifx #1\expandafter \@firstoftwo
 \else \expandafter \@secondoftwo
 \fi
}%
\providecommand \natexlab [1]{#1}%
\providecommand \enquote  [1]{``#1''}%
\providecommand \bibnamefont  [1]{#1}%
\providecommand \bibfnamefont [1]{#1}%
\providecommand \citenamefont [1]{#1}%
\providecommand \href@noop [0]{\@secondoftwo}%
\providecommand \href [0]{\begingroup \@sanitize@url \@href}%
\providecommand \@href[1]{\@@startlink{#1}\@@href}%
\providecommand \@@href[1]{\endgroup#1\@@endlink}%
\providecommand \@sanitize@url [0]{\catcode `\\12\catcode `\$12\catcode
  `\&12\catcode `\#12\catcode `\^12\catcode `\_12\catcode `\%12\relax}%
\providecommand \@@startlink[1]{}%
\providecommand \@@endlink[0]{}%
\providecommand \url  [0]{\begingroup\@sanitize@url \@url }%
\providecommand \@url [1]{\endgroup\@href {#1}{\urlprefix }}%
\providecommand \urlprefix  [0]{URL }%
\providecommand \Eprint [0]{\href }%
\providecommand \doibase [0]{http://dx.doi.org/}%
\providecommand \selectlanguage [0]{\@gobble}%
\providecommand \bibinfo  [0]{\@secondoftwo}%
\providecommand \bibfield  [0]{\@secondoftwo}%
\providecommand \translation [1]{[#1]}%
\providecommand \BibitemOpen [0]{}%
\providecommand \bibitemStop [0]{}%
\providecommand \bibitemNoStop [0]{.\EOS\space}%
\providecommand \EOS [0]{\spacefactor3000\relax}%
\providecommand \BibitemShut  [1]{\csname bibitem#1\endcsname}%
\let\auto@bib@innerbib\@empty
\bibitem [{\citenamefont {Leinaas}\ and\ \citenamefont
  {Myrheim}(1977)}]{Leinaas}%
  \BibitemOpen
  \bibfield  {author} {\bibinfo {author} {\bibfnamefont {J.~M.}\ \bibnamefont
  {Leinaas}}\ and\ \bibinfo {author} {\bibfnamefont {J.}~\bibnamefont
  {Myrheim}},\ }\bibfield  {title} {\enquote {\bibinfo {title} {{On the theory
  of identical particles}},}\ }\href {\doibase 10.1007/BF02727953} {\bibfield
  {journal} {\bibinfo  {journal} {Nuovo Cimento Soc.\ Ital.\ Fis.\ B}\ }\textbf
  {\bibinfo {volume} {37}},\ \bibinfo {pages} {1--23} (\bibinfo {year}
  {1977})}\BibitemShut {NoStop}%
\bibitem [{\citenamefont {Wilczek}(1982)}]{WilczekAnyons}%
  \BibitemOpen
  \bibfield  {author} {\bibinfo {author} {\bibfnamefont {Frank}\ \bibnamefont
  {Wilczek}},\ }\bibfield  {title} {\enquote {\bibinfo {title} {Magnetic flux,
  angular momentum, and statistics},}\ }\href {\doibase%
  10.1103/PhysRevLett.48.1144} {\bibfield  {journal} {\bibinfo  {journal}
  {Phys. Rev. Lett.}\ }\textbf {\bibinfo {volume} {48}},\ \bibinfo {pages}
  {1144--1146} (\bibinfo {year} {1982})}\BibitemShut {NoStop}%
\bibitem [{\citenamefont {Bais}(1980)}]{Bais}%
  \BibitemOpen
  \bibfield  {author} {\bibinfo {author} {\bibfnamefont {F.~A.}\ \bibnamefont
  {Bais}},\ }\bibfield  {title} {\enquote {\bibinfo {title} {Flux
  metamorphosis},}\ }\href {\doibase 10.1016/0550-3213(80)90474-5} {\bibfield
  {journal} {\bibinfo  {journal} {Nucl. Phys. B}\ }\textbf {\bibinfo {volume}
  {170}},\ \bibinfo {pages} {32--43} (\bibinfo {year} {1980})}\BibitemShut
  {NoStop}%
\bibitem [{\citenamefont {Goldin}\ \emph {et~al.}(1981)\citenamefont {Goldin},
  \citenamefont {Menikoff},\ and\ \citenamefont {Sharp}}]{Goldin81}%
  \BibitemOpen
  \bibfield  {author} {\bibinfo {author} {\bibfnamefont {G.~A.}\ \bibnamefont
  {Goldin}}, \bibinfo {author} {\bibfnamefont {R.}~\bibnamefont {Menikoff}},
  and\ \bibinfo {author} {\bibfnamefont {D.~H.}\ \bibnamefont {Sharp}},\
  }\bibfield  {title} {\enquote {\bibinfo {title} {{Representations of a local
  current algebra in nonsimply connected space and the Aharonov--Bohm
  effect}},}\ }\href {\doibase 10.1063/1.525110} {\bibfield  {journal}
  {\bibinfo  {journal} {J. Math. Phys.}\ }\textbf {\bibinfo {volume} {22}},\
  \bibinfo {pages} {1664--1668} (\bibinfo {year} {1981})}\BibitemShut {NoStop}%
\bibitem [{\citenamefont {Goldin}\ \emph {et~al.}(1985)\citenamefont {Goldin},
  \citenamefont {Menikoff},\ and\ \citenamefont {Sharp}}]{Goldin84}%
  \BibitemOpen
  \bibfield  {author} {\bibinfo {author} {\bibfnamefont {Gerald~A.}\
  \bibnamefont {Goldin}}, \bibinfo {author} {\bibfnamefont {Ralph}\
  \bibnamefont {Menikoff}}, and\ \bibinfo {author} {\bibfnamefont {David~H.}\
  \bibnamefont {Sharp}},\ }\bibfield  {title} {\enquote {\bibinfo {title}
  {{Comments on ``General Theory for Quantum Statistics in Two
  Dimensions''}},}\ }\href {\doibase 10.1103/PhysRevLett.54.603} {\bibfield
  {journal} {\bibinfo  {journal} {Phys. Rev. Lett.}\ }\textbf {\bibinfo
  {volume} {54}},\ \bibinfo {pages} {603--603} (\bibinfo {year}
  {1985})}\BibitemShut {NoStop}%
\bibitem [{\citenamefont {Moore}\ and\ \citenamefont
  {Seiberg}(1988)}]{MooreSeiberg1}%
  \BibitemOpen
  \bibfield  {author} {\bibinfo {author} {\bibfnamefont {Gregory}\ \bibnamefont
  {Moore}}\ and\ \bibinfo {author} {\bibfnamefont {Nathan}\ \bibnamefont
  {Seiberg}},\ }\bibfield  {title} {\enquote {\bibinfo {title} {Polynomial
  equations for rational conformal field theories},}\ }\href {\doibase%
  10.1016/0370-2693(88)91796-0} {\bibfield  {journal} {\bibinfo  {journal}
  {Phys. Lett. B}\ }\textbf {\bibinfo {volume} {212}},\ \bibinfo {pages}
  {451--460} (\bibinfo {year} {1988})}\BibitemShut {NoStop}%
\bibitem [{\citenamefont {Moore}\ and\ \citenamefont
  {Seiberg}(1989)}]{MooreSeiberg2}%
  \BibitemOpen
  \bibfield  {author} {\bibinfo {author} {\bibfnamefont {Gregory}\ \bibnamefont
  {Moore}}\ and\ \bibinfo {author} {\bibfnamefont {Nathan}\ \bibnamefont
  {Seiberg}},\ }\bibfield  {title} {\enquote {\bibinfo {title} {Classical and
  quantum conformal field theory},}\ }\href {\doibase 10.1007/BF01238857}
  {\bibfield  {journal} {\bibinfo  {journal} {Commun. Math. Phys.}\ }\textbf
  {\bibinfo {volume} {123}},\ \bibinfo {pages} {177--254} (\bibinfo {year}
  {1989})}\BibitemShut {NoStop}%
\bibitem [{\citenamefont {Witten}(1989)}]{Witten}%
  \BibitemOpen
  \bibfield  {author} {\bibinfo {author} {\bibfnamefont {Edward}\ \bibnamefont
  {Witten}},\ }\bibfield  {title} {\enquote {\bibinfo {title} {{Quantum field
  theory and the Jones polynomial}},}\ }\href {\doibase 10.1007/BF01217730}
  {\bibfield  {journal} {\bibinfo  {journal} {Commun. Math. Phys.}\ }\textbf
  {\bibinfo {volume} {121}},\ \bibinfo {pages} {351--399} (\bibinfo {year}
  {1989})}\BibitemShut {NoStop}%
\bibitem [{\citenamefont {Fredenhagen}\ \emph {et~al.}(1989)\citenamefont
  {Fredenhagen}, \citenamefont {Rehren},\ and\ \citenamefont
  {Schroer}}]{Fredenhagen}%
  \BibitemOpen
  \bibfield  {author} {\bibinfo {author} {\bibfnamefont {K.}~\bibnamefont
  {Fredenhagen}}, \bibinfo {author} {\bibfnamefont {K.~H.}\ \bibnamefont
  {Rehren}}, and\ \bibinfo {author} {\bibfnamefont {B.}~\bibnamefont
  {Schroer}},\ }\bibfield  {title} {\enquote {\bibinfo {title} {{Superselection
  sectors with braid group statistics and exchange algebras}},}\ }\href
  {\doibase 10.1007/BF01217906} {\bibfield  {journal} {\bibinfo  {journal}
  {Commun.\ Math.\ Phys.}\ }\textbf {\bibinfo {volume} {125}},\ \bibinfo
  {pages} {201} (\bibinfo {year} {1989})}\BibitemShut {NoStop}%
\bibitem [{\citenamefont {Fr\"{o}hlich}\ and\ \citenamefont
  {Gabbiani}(1990)}]{Froehlich}%
  \BibitemOpen
  \bibfield  {author} {\bibinfo {author} {\bibfnamefont {J.}~\bibnamefont
  {Fr\"{o}hlich}}\ and\ \bibinfo {author} {\bibfnamefont {F.}~\bibnamefont
  {Gabbiani}},\ }\bibfield  {title} {\enquote {\bibinfo {title} {Braid
  statistics in local quantum theory},}\ }\href {\doibase%
  10.1142/S0129055X90000107} {\bibfield  {journal} {\bibinfo  {journal} {Rev.\
  Math.\ Phys.}\ }\textbf {\bibinfo {volume} {2}},\ \bibinfo {pages} {251}
  (\bibinfo {year} {1990})}\BibitemShut {NoStop}%
\bibitem [{\citenamefont {Imbo}\ \emph {et~al.}(1990)\citenamefont {Imbo},
  \citenamefont {Imbo},\ and\ \citenamefont {Sudarshan}}]{Imbo89}%
  \BibitemOpen
  \bibfield  {author} {\bibinfo {author} {\bibfnamefont {Tom~D.}\ \bibnamefont
  {Imbo}}, \bibinfo {author} {\bibfnamefont {Chandni~Shah}\ \bibnamefont
  {Imbo}}, and\ \bibinfo {author} {\bibfnamefont {E.~C.~G.}\ \bibnamefont
  {Sudarshan}},\ }\bibfield  {title} {\enquote {\bibinfo {title} {{Identical
  particles, exotic statistics and braid groups}},}\ }\href {\doibase%
  10.1016/0370-2693(90)92010-G} {\bibfield  {journal} {\bibinfo  {journal}
  {Phys. Lett. B}\ }\textbf {\bibinfo {volume} {234}},\ \bibinfo {pages}
  {103--107} (\bibinfo {year} {1990})}\BibitemShut {NoStop}%
\bibitem [{\citenamefont {Alford}\ \emph {et~al.}(1991)\citenamefont {Alford},
  \citenamefont {Benson}, \citenamefont {Coleman}, \citenamefont
  {March-Russell},\ and\ \citenamefont {Wilczek}}]{Alford1}%
  \BibitemOpen
  \bibfield  {author} {\bibinfo {author} {\bibfnamefont {Mark~G.}\ \bibnamefont
  {Alford}}, \bibinfo {author} {\bibfnamefont {Katherine}\ \bibnamefont
  {Benson}}, \bibinfo {author} {\bibfnamefont {Sidney~R.}\ \bibnamefont
  {Coleman}}, \bibinfo {author} {\bibfnamefont {John}\ \bibnamefont
  {March-Russell}}, and\ \bibinfo {author} {\bibfnamefont {Frank}\
  \bibnamefont {Wilczek}},\ }\bibfield  {title} {\enquote {\bibinfo {title}
  {{Zero modes of nonabelian vortices}},}\ }\href {\doibase%
  10.1016/0550-3213(91)90331-Q} {\bibfield  {journal} {\bibinfo  {journal}
  {Nucl. Phys. B}\ }\textbf {\bibinfo {volume} {349}},\ \bibinfo {pages}
  {414--438} (\bibinfo {year} {1991})}\BibitemShut {NoStop}%
\bibitem [{\citenamefont {Alford}\ \emph {et~al.}(1992)\citenamefont {Alford},
  \citenamefont {Lee}, \citenamefont {March-Russell},\ and\ \citenamefont
  {Preskill}}]{Alford2}%
  \BibitemOpen
  \bibfield  {author} {\bibinfo {author} {\bibfnamefont {Mark~G.}\ \bibnamefont
  {Alford}}, \bibinfo {author} {\bibfnamefont {Kai-Ming}\ \bibnamefont {Lee}},
  \bibinfo {author} {\bibfnamefont {John}\ \bibnamefont {March-Russell}},
  and\ \bibinfo {author} {\bibfnamefont {John}\ \bibnamefont {Preskill}},\
  }\bibfield  {title} {\enquote {\bibinfo {title} {{Quantum field theory of
  non-Abelian strings and vortices}},}\ }\href {\doibase%
  10.1016/0550-3213(92)90468-Q} {\bibfield  {journal} {\bibinfo  {journal}
  {Nucl. Phys. B}\ }\textbf {\bibinfo {volume} {384}},\ \bibinfo {pages}
  {251--317} (\bibinfo {year} {1992})}\BibitemShut {NoStop}%
\bibitem [{\citenamefont {Kitaev}(2003)}]{kitaev}%
  \BibitemOpen
  \bibfield  {author} {\bibinfo {author} {\bibfnamefont {Alexei~Yu}\
  \bibnamefont {Kitaev}},\ }\bibfield  {title} {\enquote {\bibinfo {title}
  {{Fault-tolerant quantum computation by anyons}},}\ }\href {\doibase%
  10.1016/S0003-4916(02)00018-0} {\bibfield  {journal} {\bibinfo  {journal}
  {Ann.\ Phys.}\ }\textbf {\bibinfo {volume} {303}},\ \bibinfo {pages} {2--30}
  (\bibinfo {year} {2003})}\BibitemShut {NoStop}%
\bibitem [{\citenamefont {Freedman}(1998)}]{Freedman98}%
  \BibitemOpen
  \bibfield  {author} {\bibinfo {author} {\bibfnamefont {Michael~H.}\
  \bibnamefont {Freedman}},\ }\bibfield  {title} {\enquote {\bibinfo {title}
  {{P/NP, and the quantum field computer}},}\ }\href
  {http://www.pnas.org/content/95/1/98.abstract} {\bibfield  {journal}
  {\bibinfo  {journal} {Proc.\ Natl.\ Acad.\ Sci.}\ }\textbf {\bibinfo {volume}
  {95}},\ \bibinfo {pages} {98--101} (\bibinfo {year} {1998})}\BibitemShut
  {NoStop}%
\bibitem [{\citenamefont {Freedman}\ \emph {et~al.}(2003)\citenamefont
  {Freedman}, \citenamefont {Kitaev}, \citenamefont {Larsen},\ and\
  \citenamefont {Wang}}]{Freedman03}%
  \BibitemOpen
  \bibfield  {author} {\bibinfo {author} {\bibfnamefont {Michael~H.}\
  \bibnamefont {Freedman}}, \bibinfo {author} {\bibfnamefont {Alexei}\
  \bibnamefont {Kitaev}}, \bibinfo {author} {\bibfnamefont {Michael~J.}\
  \bibnamefont {Larsen}}, and\ \bibinfo {author} {\bibfnamefont {Zhenghan}\
  \bibnamefont {Wang}},\ }\bibfield  {title} {\enquote {\bibinfo {title}
  {Topological quantum computation},}\ }\href {\doibase%
  10.1090/S0273-0979-02-00964-3} {\bibfield  {journal} {\bibinfo  {journal}
  {Bull.\ Amer.\ Math.\ Soc.}\ }\textbf {\bibinfo {volume} {40}},\ \bibinfo
  {pages} {31--38} (\bibinfo {year} {2003})}\BibitemShut {NoStop}%
\bibitem [{\citenamefont {Bonderson}\ \emph {et~al.}(2008)\citenamefont
  {Bonderson}, \citenamefont {Freedman},\ and\ \citenamefont
  {Nayak}}]{MeasurementOnlyTQC}%
  \BibitemOpen
  \bibfield  {author} {\bibinfo {author} {\bibfnamefont {Parsa}\ \bibnamefont
  {Bonderson}}, \bibinfo {author} {\bibfnamefont {Michael}\ \bibnamefont
  {Freedman}}, and\ \bibinfo {author} {\bibfnamefont {Chetan}\ \bibnamefont
  {Nayak}},\ }\bibfield  {title} {\enquote {\bibinfo {title} {{Measurement-Only
  Topological Quantum Computation}},}\ }\href {\doibase%
  10.1103/PhysRevLett.101.010501} {\bibfield  {journal} {\bibinfo  {journal}
  {Phys.\ Rev.\ Lett.}\ }\textbf {\bibinfo {volume} {101}},\ \bibinfo {pages}
  {010501} (\bibinfo {year} {2008})}\BibitemShut {NoStop}%
\bibitem [{\citenamefont {Nayak}\ \emph {et~al.}(2008)\citenamefont {Nayak},
  \citenamefont {Simon}, \citenamefont {Stern}, \citenamefont {Freedman},\ and\
  \citenamefont {Das~Sarma}}]{TQCreview}%
  \BibitemOpen
  \bibfield  {author} {\bibinfo {author} {\bibfnamefont {Chetan}\ \bibnamefont
  {Nayak}}, \bibinfo {author} {\bibfnamefont {Steven~H.}\ \bibnamefont
  {Simon}}, \bibinfo {author} {\bibfnamefont {Ady}\ \bibnamefont {Stern}},
  \bibinfo {author} {\bibfnamefont {Michael}\ \bibnamefont {Freedman}}, and\
  \bibinfo {author} {\bibfnamefont {Sankar}\ \bibnamefont {Das~Sarma}},\
  }\bibfield  {title} {\enquote {\bibinfo {title} {{Non-Abelian anyons and
  topological quantum computation}},}\ }\href {\doibase%
  10.1103/RevModPhys.80.1083} {\bibfield  {journal} {\bibinfo  {journal} {Rev.
  Mod. Phys.}\ }\textbf {\bibinfo {volume} {80}},\ \bibinfo {pages}
  {1083--1159} (\bibinfo {year} {2008})}\BibitemShut {NoStop}%
\bibitem [{\citenamefont {Stern}(2008)}]{SternReview}%
  \BibitemOpen
  \bibfield  {author} {\bibinfo {author} {\bibfnamefont {Ady}\ \bibnamefont
  {Stern}},\ }\bibfield  {title} {\enquote {\bibinfo {title} {{Anyons and the
  quantum Hall effect---A pedagogical review}},}\ }\href {\doibase%
  10.1016/j.aop.2007.10.008} {\bibfield  {journal} {\bibinfo  {journal} {Annals
  of Physics}\ }\textbf {\bibinfo {volume} {323}},\ \bibinfo {pages} {204 --
  249} (\bibinfo {year} {2008})}\BibitemShut {NoStop}%
\bibitem [{\citenamefont {Tsui}\ \emph {et~al.}(1982)\citenamefont {Tsui},
  \citenamefont {Stormer},\ and\ \citenamefont {Gossard}}]{GaAsFQHE}%
  \BibitemOpen
  \bibfield  {author} {\bibinfo {author} {\bibfnamefont {D.~C.}\ \bibnamefont
  {Tsui}}, \bibinfo {author} {\bibfnamefont {H.~L.}\ \bibnamefont {Stormer}},
  and\ \bibinfo {author} {\bibfnamefont {A.~C.}\ \bibnamefont {Gossard}},\
  }\bibfield  {title} {\enquote {\bibinfo {title} {Two-dimensional
  magnetotransport in the extreme quantum limit},}\ }\href {\doibase%
  10.1103/PhysRevLett.48.1559} {\bibfield  {journal} {\bibinfo  {journal}
  {Phys. Rev. Lett.}\ }\textbf {\bibinfo {volume} {48}},\ \bibinfo {pages}
  {1559--1562} (\bibinfo {year} {1982})}\BibitemShut {NoStop}%
\bibitem [{\citenamefont {Du}\ \emph {et~al.}(2009)\citenamefont {Du},
  \citenamefont {Skachko}, \citenamefont {Duerr}, \citenamefont {Luican},\ and\
  \citenamefont {Andrei}}]{GrapheneFQHE1}%
  \BibitemOpen
  \bibfield  {author} {\bibinfo {author} {\bibfnamefont {Xu}~\bibnamefont
  {Du}}, \bibinfo {author} {\bibfnamefont {Ivan}\ \bibnamefont {Skachko}},
  \bibinfo {author} {\bibfnamefont {Fabian}\ \bibnamefont {Duerr}}, \bibinfo
  {author} {\bibfnamefont {Adina}\ \bibnamefont {Luican}}, and\ \bibinfo
  {author} {\bibfnamefont {Eva~Y.}\ \bibnamefont {Andrei}},\ }\bibfield
  {title} {\enquote {\bibinfo {title} {{Fractional quantum Hall effect and
  insulating phase of Dirac electrons in graphene}},}\ }\href {\doibase%
  10.1038/nature08522} {\bibfield  {journal} {\bibinfo  {journal} {Nature}\
  }\textbf {\bibinfo {volume} {462}},\ \bibinfo {pages} {192--195} (\bibinfo
  {year} {2009})}\BibitemShut {NoStop}%
\bibitem [{\citenamefont {Bolotin}\ \emph {et~al.}(2009)\citenamefont
  {Bolotin}, \citenamefont {Ghahari}, \citenamefont {Shulman}, \citenamefont
  {Stormer},\ and\ \citenamefont {Kim}}]{GrapheneFQHE2}%
  \BibitemOpen
  \bibfield  {author} {\bibinfo {author} {\bibfnamefont {Kirill~I.}\
  \bibnamefont {Bolotin}}, \bibinfo {author} {\bibfnamefont {Fereshte}\
  \bibnamefont {Ghahari}}, \bibinfo {author} {\bibfnamefont {Michael~D.}\
  \bibnamefont {Shulman}}, \bibinfo {author} {\bibfnamefont {Horst~L.}\
  \bibnamefont {Stormer}}, and\ \bibinfo {author} {\bibfnamefont {Philip}\
  \bibnamefont {Kim}},\ }\bibfield  {title} {\enquote {\bibinfo {title}
  {{Observation of the fractional quantum Hall effect in graphene}},}\ }\href
  {\doibase 10.1038/nature08582} {\bibfield  {journal} {\bibinfo  {journal}
  {Nature}\ }\textbf {\bibinfo {volume} {462}},\ \bibinfo {pages} {196--199}
  (\bibinfo {year} {2009})}\BibitemShut {NoStop}%
\bibitem [{\citenamefont {Tsukazaki}\ \emph {et~al.}(2010)\citenamefont
  {Tsukazaki}, \citenamefont {Akasaka}, \citenamefont {Nakahara}, \citenamefont
  {Ohno}, \citenamefont {Ohno}, \citenamefont {Maryenko}, \citenamefont
  {Ohtomo},\ and\ \citenamefont {Kawasaki}}]{OxideFQHE1}%
  \BibitemOpen
  \bibfield  {author} {\bibinfo {author} {\bibfnamefont {A.}~\bibnamefont
  {Tsukazaki}}, \bibinfo {author} {\bibfnamefont {S.}~\bibnamefont {Akasaka}},
  \bibinfo {author} {\bibfnamefont {K.}~\bibnamefont {Nakahara}}, \bibinfo
  {author} {\bibfnamefont {Y.}~\bibnamefont {Ohno}}, \bibinfo {author}
  {\bibfnamefont {H.}~\bibnamefont {Ohno}}, \bibinfo {author} {\bibfnamefont
  {D.}~\bibnamefont {Maryenko}}, \bibinfo {author} {\bibfnamefont
  {A.}~\bibnamefont {Ohtomo}}, and\ \bibinfo {author} {\bibfnamefont
  {M.}~\bibnamefont {Kawasaki}},\ }\bibfield  {title} {\enquote {\bibinfo
  {title} {{Observation of the fractional quantum Hall effect in an oxide}},}\
  }\href {\doibase 10.1038/nmat2874} {\bibfield  {journal} {\bibinfo  {journal}
  {Nat. Mater.}\ }\textbf {\bibinfo {volume} {9}},\ \bibinfo {pages} {889}
  (\bibinfo {year} {2010})}\BibitemShut {NoStop}%
\bibitem [{\citenamefont {Hwang}\ \emph {et~al.}(2012)\citenamefont {Hwang},
  \citenamefont {Iwasa}, \citenamefont {Kawasaki}, \citenamefont {Keimer},
  \citenamefont {Nagaosa},\ and\ \citenamefont {Tokura}}]{OxideFQHE2}%
  \BibitemOpen
  \bibfield  {author} {\bibinfo {author} {\bibfnamefont {H.~Y.}\ \bibnamefont
  {Hwang}}, \bibinfo {author} {\bibfnamefont {Y.}~\bibnamefont {Iwasa}},
  \bibinfo {author} {\bibfnamefont {M.}~\bibnamefont {Kawasaki}}, \bibinfo
  {author} {\bibfnamefont {B.}~\bibnamefont {Keimer}}, \bibinfo {author}
  {\bibfnamefont {N.}~\bibnamefont {Nagaosa}}, and\ \bibinfo {author}
  {\bibfnamefont {Y.}~\bibnamefont {Tokura}},\ }\bibfield  {title} {\enquote
  {\bibinfo {title} {Emergent phenomena at oxide interfaces},}\ }\href
  {\doibase 10.1038/nmat3223} {\bibfield  {journal} {\bibinfo  {journal} {Nat.
  Mater.}\ }\textbf {\bibinfo {volume} {11}},\ \bibinfo {pages} {103} (\bibinfo
  {year} {2012})}\BibitemShut {NoStop}%
\bibitem [{\citenamefont {Piot}\ \emph {et~al.}(2010)\citenamefont {Piot},
  \citenamefont {Kunc}, \citenamefont {Potemski}, \citenamefont {Maude},
  \citenamefont {Betthausen}, \citenamefont {Vogl}, \citenamefont {Weiss},
  \citenamefont {Karczewski},\ and\ \citenamefont {Wojtowicz}}]{CdTeFQHE}%
  \BibitemOpen
  \bibfield  {author} {\bibinfo {author} {\bibfnamefont {B.~A.}\ \bibnamefont
  {Piot}}, \bibinfo {author} {\bibfnamefont {J.}~\bibnamefont {Kunc}}, \bibinfo
  {author} {\bibfnamefont {M.}~\bibnamefont {Potemski}}, \bibinfo {author}
  {\bibfnamefont {D.~K.}\ \bibnamefont {Maude}}, \bibinfo {author}
  {\bibfnamefont {C.}~\bibnamefont {Betthausen}}, \bibinfo {author}
  {\bibfnamefont {A.}~\bibnamefont {Vogl}}, \bibinfo {author} {\bibfnamefont
  {D.}~\bibnamefont {Weiss}}, \bibinfo {author} {\bibfnamefont
  {G.}~\bibnamefont {Karczewski}}, and\ \bibinfo {author} {\bibfnamefont
  {T.}~\bibnamefont {Wojtowicz}},\ }\bibfield  {title} {\enquote {\bibinfo
  {title} {{Fractional quantum Hall effect in $\mathrm{CdTe}$}},}\ }\href
  {\doibase 10.1103/PhysRevB.82.081307} {\bibfield  {journal} {\bibinfo
  {journal} {Phys. Rev. B}\ }\textbf {\bibinfo {volume} {82}},\ \bibinfo
  {pages} {081307} (\bibinfo {year} {2010})}\BibitemShut {NoStop}%
\bibitem [{\citenamefont {Moore}\ and\ \citenamefont
  {Read}(1991)}]{MooreReadNonabelion91}%
  \BibitemOpen
  \bibfield  {author} {\bibinfo {author} {\bibfnamefont {Gregory}\ \bibnamefont
  {Moore}}\ and\ \bibinfo {author} {\bibfnamefont {Nicholas}\ \bibnamefont
  {Read}},\ }\bibfield  {title} {\enquote {\bibinfo {title} {{Nonabelions in
  the fractional quantum Hall effect}},}\ }\href {\doibase%
  10.1016/0550-3213(91)90407-O} {\bibfield  {journal} {\bibinfo  {journal}
  {Nucl. Phys. B}\ }\textbf {\bibinfo {volume} {360}},\ \bibinfo {pages}
  {362--396} (\bibinfo {year} {1991})}\BibitemShut {NoStop}%
\bibitem [{\citenamefont {Milovanovi\'{c}}\ and\ \citenamefont
  {Read}(1996)}]{Milovanovic96}%
  \BibitemOpen
  \bibfield  {author} {\bibinfo {author} {\bibfnamefont {M.}~\bibnamefont
  {Milovanovi\'{c}}}\ and\ \bibinfo {author} {\bibfnamefont {N.}~\bibnamefont
  {Read}},\ }\bibfield  {title} {\enquote {\bibinfo {title} {Edge excitations
  of paired fractional quantum {Hall} states},}\ }\href {\doibase%
  10.1103/PhysRevB.53.13559} {\bibfield  {journal} {\bibinfo  {journal} {Phys.
  Rev. B}\ }\textbf {\bibinfo {volume} {53}},\ \bibinfo {pages} {13559--13582}
  (\bibinfo {year} {1996})}\BibitemShut {NoStop}%
\bibitem [{Note1()}]{Note1}%
  \BibitemOpen
  \bibinfo {note} {The term `Ising anyon' refers to a non-Abelian particle
  whose nontrivial braiding statistics derives from bound Majorana zero-modes.
  Strictly speaking Ising anyons have a particular overall $\protect \mathrm
  {U(1)}$ phase associated with their braiding, though we will use this
  terminology even when this overall phase is ill-defined.}\BibitemShut {Stop}%
\bibitem [{\citenamefont {Nayak}\ and\ \citenamefont
  {Wilczek}(1996)}]{Nayak96c}%
  \BibitemOpen
  \bibfield  {author} {\bibinfo {author} {\bibfnamefont {Chetan}\ \bibnamefont
  {Nayak}}\ and\ \bibinfo {author} {\bibfnamefont {Frank}\ \bibnamefont
  {Wilczek}},\ }\bibfield  {title} {\enquote {\bibinfo {title} {{$2n$-quasihole
  states realize $2^{n-1}$-dimensional spinor braiding statistics in paired
  quantum Hall states}},}\ }\href {\doibase 10.1016/0550-3213(96)00430-0}
  {\bibfield  {journal} {\bibinfo  {journal} {Nucl. Phys. B}\ }\textbf
  {\bibinfo {volume} {479}},\ \bibinfo {pages} {529--553} (\bibinfo {year}
  {1996})}\BibitemShut {NoStop}%
\bibitem [{\citenamefont {Gurarie}\ and\ \citenamefont
  {Nayak}(1997)}]{Gurarie1997}%
  \BibitemOpen
  \bibfield  {author} {\bibinfo {author} {\bibfnamefont {V.}~\bibnamefont
  {Gurarie}}\ and\ \bibinfo {author} {\bibfnamefont {C.}~\bibnamefont
  {Nayak}},\ }\bibfield  {title} {\enquote {\bibinfo {title} {{A plasma analogy
  and Berry matrices for non-Abelian quantum Hall states}},}\ }\href {\doibase%
  10.1016/S0550-3213(97)00612-3} {\bibfield  {journal} {\bibinfo  {journal}
  {Nucl. Phys. B}\ }\textbf {\bibinfo {volume} {506}},\ \bibinfo {pages}
  {685--694} (\bibinfo {year} {1997})}\BibitemShut {NoStop}%
\bibitem [{\citenamefont {Tserkovnyak}\ and\ \citenamefont
  {Simon}(2003)}]{Tserkovnyak03}%
  \BibitemOpen
  \bibfield  {author} {\bibinfo {author} {\bibfnamefont {Yaroslav}\
  \bibnamefont {Tserkovnyak}}\ and\ \bibinfo {author} {\bibfnamefont
  {Steven~H.}\ \bibnamefont {Simon}},\ }\bibfield  {title} {\enquote {\bibinfo
  {title} {{Monte Carlo Evaluation of Non-Abelian Statistics}},}\ }\href
  {\doibase 10.1103/PhysRevLett.90.016802} {\bibfield  {journal} {\bibinfo
  {journal} {Phys. Rev. Lett.}\ }\textbf {\bibinfo {volume} {90}},\ \bibinfo
  {pages} {016802} (\bibinfo {year} {2003})}\BibitemShut {NoStop}%
\bibitem [{\citenamefont {Seidel}(2008)}]{Seidel08}%
  \BibitemOpen
  \bibfield  {author} {\bibinfo {author} {\bibfnamefont {Alexander}\
  \bibnamefont {Seidel}},\ }\bibfield  {title} {\enquote {\bibinfo {title}
  {{Pfaffian Statistics through Adiabatic Transport in the 1D Coherent State
  Representation}},}\ }\href {\doibase 10.1103/PhysRevLett.101.196802}
  {\bibfield  {journal} {\bibinfo  {journal} {Phys. Rev. Lett.}\ }\textbf
  {\bibinfo {volume} {101}},\ \bibinfo {pages} {196802} (\bibinfo {year}
  {2008})}\BibitemShut {NoStop}%
\bibitem [{\citenamefont {Read}(2009)}]{Read2008}%
  \BibitemOpen
  \bibfield  {author} {\bibinfo {author} {\bibfnamefont {N.}~\bibnamefont
  {Read}},\ }\bibfield  {title} {\enquote {\bibinfo {title} {{Non-Abelian
  adiabatic statistics and Hall viscosity in quantum Hall states and
  ${p}_{x}+i{p}_{y}$ paired superfluids}},}\ }\href {\doibase%
  10.1103/PhysRevB.79.045308} {\bibfield  {journal} {\bibinfo  {journal} {Phys.
  Rev. B}\ }\textbf {\bibinfo {volume} {79}},\ \bibinfo {eid} {045308}
  (\bibinfo {year} {2009})}\BibitemShut {NoStop}%
\bibitem [{\citenamefont {Baraban}\ \emph {et~al.}(2009)\citenamefont
  {Baraban}, \citenamefont {Zikos}, \citenamefont {Bonesteel},\ and\
  \citenamefont {Simon}}]{Baraban09}%
  \BibitemOpen
  \bibfield  {author} {\bibinfo {author} {\bibfnamefont {M.}~\bibnamefont
  {Baraban}}, \bibinfo {author} {\bibfnamefont {G.}~\bibnamefont {Zikos}},
  \bibinfo {author} {\bibfnamefont {N.}~\bibnamefont {Bonesteel}}, and\
  \bibinfo {author} {\bibfnamefont {S.~H.}\ \bibnamefont {Simon}},\ }\bibfield
  {title} {\enquote {\bibinfo {title} {{Numerical Analysis of Quasiholes of the
  Moore-Read Wave Function}},}\ }\href {\doibase%
  10.1103/PhysRevLett.103.076801} {\bibfield  {journal} {\bibinfo  {journal}
  {Phys. Rev. Lett.}\ }\textbf {\bibinfo {volume} {103}},\ \bibinfo {pages}
  {076801} (\bibinfo {year} {2009})}\BibitemShut {NoStop}%
\bibitem [{\citenamefont {Prodan}\ and\ \citenamefont
  {Haldane}(2009)}]{Prodan09}%
  \BibitemOpen
  \bibfield  {author} {\bibinfo {author} {\bibfnamefont {Emil}\ \bibnamefont
  {Prodan}}\ and\ \bibinfo {author} {\bibfnamefont {F.~D.~M.}\ \bibnamefont
  {Haldane}},\ }\bibfield  {title} {\enquote {\bibinfo {title} {{Mapping the
  braiding properties of the Moore-Read state}},}\ }\href {\doibase%
  10.1103/PhysRevB.80.115121} {\bibfield  {journal} {\bibinfo  {journal} {Phys.
  Rev. B}\ }\textbf {\bibinfo {volume} {80}},\ \bibinfo {pages} {115121}
  (\bibinfo {year} {2009})}\BibitemShut {NoStop}%
\bibitem [{\citenamefont {Bonderson}\ \emph {et~al.}(2011)\citenamefont
  {Bonderson}, \citenamefont {Gurarie},\ and\ \citenamefont
  {Nayak}}]{Bonderson11b}%
  \BibitemOpen
  \bibfield  {author} {\bibinfo {author} {\bibfnamefont {Parsa}\ \bibnamefont
  {Bonderson}}, \bibinfo {author} {\bibfnamefont {Victor}\ \bibnamefont
  {Gurarie}}, and\ \bibinfo {author} {\bibfnamefont {Chetan}\ \bibnamefont
  {Nayak}},\ }\bibfield  {title} {\enquote {\bibinfo {title} {{Plasma analogy
  and non-Abelian statistics for Ising-type quantum Hall states}},}\ }\href
  {\doibase 10.1103/PhysRevB.83.075303} {\bibfield  {journal} {\bibinfo
  {journal} {Phys. Rev. B}\ }\textbf {\bibinfo {volume} {83}},\ \bibinfo
  {pages} {075303} (\bibinfo {year} {2011})}\BibitemShut {NoStop}%
\bibitem [{\citenamefont {Levin}\ \emph {et~al.}(2007)\citenamefont {Levin},
  \citenamefont {Halperin},\ and\ \citenamefont {Rosenow}}]{antiPfaffian1}%
  \BibitemOpen
  \bibfield  {author} {\bibinfo {author} {\bibfnamefont {Michael}\ \bibnamefont
  {Levin}}, \bibinfo {author} {\bibfnamefont {Bertrand~I.}\ \bibnamefont
  {Halperin}}, and\ \bibinfo {author} {\bibfnamefont {Bernd}\ \bibnamefont
  {Rosenow}},\ }\bibfield  {title} {\enquote {\bibinfo {title} {Particle-hole
  symmetry and the pfaffian state},}\ }\href {\doibase%
  10.1103/PhysRevLett.99.236806} {\bibfield  {journal} {\bibinfo  {journal}
  {Phys. Rev. Lett.}\ }\textbf {\bibinfo {volume} {99}},\ \bibinfo {pages}
  {236806} (\bibinfo {year} {2007})}\BibitemShut {NoStop}%
\bibitem [{\citenamefont {Lee}\ \emph {et~al.}(2007)\citenamefont {Lee},
  \citenamefont {Ryu}, \citenamefont {Nayak},\ and\ \citenamefont
  {Fisher}}]{antiPfaffian2}%
  \BibitemOpen
  \bibfield  {author} {\bibinfo {author} {\bibfnamefont {Sung-Sik}\
  \bibnamefont {Lee}}, \bibinfo {author} {\bibfnamefont {Shinsei}\ \bibnamefont
  {Ryu}}, \bibinfo {author} {\bibfnamefont {Chetan}\ \bibnamefont {Nayak}},
  and\ \bibinfo {author} {\bibfnamefont {Matthew P.~A.}\ \bibnamefont
  {Fisher}},\ }\bibfield  {title} {\enquote {\bibinfo {title} {{Particle-Hole
  Symmetry and the $\nu=\frac{5}{2}$ Quantum Hall State}},}\ }\href {\doibase%
  10.1103/PhysRevLett.99.236807} {\bibfield  {journal} {\bibinfo  {journal}
  {Phys. Rev. Lett.}\ }\textbf {\bibinfo {volume} {99}},\ \bibinfo {pages}
  {236807} (\bibinfo {year} {2007})}\BibitemShut {NoStop}%
\bibitem [{\citenamefont {Willett}\ \emph {et~al.}(1987)\citenamefont
  {Willett}, \citenamefont {Eisenstein}, \citenamefont {St\"ormer},
  \citenamefont {Tsui}, \citenamefont {Gossard},\ and\ \citenamefont
  {English}}]{FiveHalvesDiscovery}%
  \BibitemOpen
  \bibfield  {author} {\bibinfo {author} {\bibfnamefont {R.}~\bibnamefont
  {Willett}}, \bibinfo {author} {\bibfnamefont {J.~P.}\ \bibnamefont
  {Eisenstein}}, \bibinfo {author} {\bibfnamefont {H.~L.}\ \bibnamefont
  {St\"ormer}}, \bibinfo {author} {\bibfnamefont {D.~C.}\ \bibnamefont {Tsui}},
  \bibinfo {author} {\bibfnamefont {A.~C.}\ \bibnamefont {Gossard}}, and\
  \bibinfo {author} {\bibfnamefont {J.~H.}\ \bibnamefont {English}},\
  }\bibfield  {title} {\enquote {\bibinfo {title} {Observation of an
  even-denominator quantum number in the fractional quantum hall effect},}\
  }\href {\doibase 10.1103/PhysRevLett.59.1776} {\bibfield  {journal} {\bibinfo
   {journal} {Phys. Rev. Lett.}\ }\textbf {\bibinfo {volume} {59}},\ \bibinfo
  {pages} {1776--1779} (\bibinfo {year} {1987})}\BibitemShut {NoStop}%
\bibitem [{\citenamefont {Willett}\ \emph {et~al.}(2009)\citenamefont
  {Willett}, \citenamefont {Pfeiffer},\ and\ \citenamefont {West}}]{Willett1}%
  \BibitemOpen
  \bibfield  {author} {\bibinfo {author} {\bibfnamefont {R.~L.}\ \bibnamefont
  {Willett}}, \bibinfo {author} {\bibfnamefont {L.~N.}\ \bibnamefont
  {Pfeiffer}}, and\ \bibinfo {author} {\bibfnamefont {K.~W.}\ \bibnamefont
  {West}},\ }\bibfield  {title} {\enquote {\bibinfo {title} {{Measurement of
  filling factor 5/2 quasiparticle interference with observation of charge e/4
  and e/2 period oscillations}},}\ }\href {\doibase 10.1073/pnas.0812599106}
  {\bibfield  {journal} {\bibinfo  {journal} {Proc. Nat. Acad. Sci.}\ }\textbf
  {\bibinfo {volume} {106}},\ \bibinfo {pages} {8853--8858} (\bibinfo {year}
  {2009})}\BibitemShut {NoStop}%
\bibitem [{\citenamefont {Willett}\ \emph {et~al.}(2010)\citenamefont
  {Willett}, \citenamefont {Pfeiffer},\ and\ \citenamefont {West}}]{Willett2}%
  \BibitemOpen
  \bibfield  {author} {\bibinfo {author} {\bibfnamefont {R.~L.}\ \bibnamefont
  {Willett}}, \bibinfo {author} {\bibfnamefont {L.~N.}\ \bibnamefont
  {Pfeiffer}}, and\ \bibinfo {author} {\bibfnamefont {K.~W.}\ \bibnamefont
  {West}},\ }\bibfield  {title} {\enquote {\bibinfo {title} {Alternation and
  interchange of e/4 and e/2 period interference oscillations consistent with
  filling factor 5/2 non-abelian quasiparticles},}\ }\href {\doibase%
  10.1103/PhysRevB.82.205301} {\bibfield  {journal} {\bibinfo  {journal} {Phys.
  Rev. B}\ }\textbf {\bibinfo {volume} {82}},\ \bibinfo {pages} {205301}
  (\bibinfo {year} {2010})}\BibitemShut {NoStop}%
\bibitem [{\citenamefont {Radu}\ \emph {et~al.}(2008)\citenamefont {Radu},
  \citenamefont {Miller}, \citenamefont {Marcus}, \citenamefont {Kastner},
  \citenamefont {Pfeiffer},\ and\ \citenamefont {West}}]{FiveHalvesTunneling}%
  \BibitemOpen
  \bibfield  {author} {\bibinfo {author} {\bibfnamefont {Iuliana~P.}\
  \bibnamefont {Radu}}, \bibinfo {author} {\bibfnamefont {J.~B.}\ \bibnamefont
  {Miller}}, \bibinfo {author} {\bibfnamefont {C.~M.}\ \bibnamefont {Marcus}},
  \bibinfo {author} {\bibfnamefont {M.~A.}\ \bibnamefont {Kastner}}, \bibinfo
  {author} {\bibfnamefont {L.~N.}\ \bibnamefont {Pfeiffer}}, and\ \bibinfo
  {author} {\bibfnamefont {K.~W.}\ \bibnamefont {West}},\ }\bibfield  {title}
  {\enquote {\bibinfo {title} {{Quasi-Particle Properties from Tunneling in the
  $\nu = 5/2$ Fractional Quantum Hall State}},}\ }\href {\doibase%
  10.1126/science.1157560} {\bibfield  {journal} {\bibinfo  {journal}
  {Science}\ }\textbf {\bibinfo {volume} {320}},\ \bibinfo {pages} {899--902}
  (\bibinfo {year} {2008})}\BibitemShut {NoStop}%
\bibitem [{\citenamefont {Dolev}\ \emph {et~al.}(2008)\citenamefont {Dolev},
  \citenamefont {Heiblum}, \citenamefont {Umansky}, \citenamefont {Stern},\
  and\ \citenamefont {Mahalu}}]{FiveHalvesCharge}%
  \BibitemOpen
  \bibfield  {author} {\bibinfo {author} {\bibfnamefont {M.}~\bibnamefont
  {Dolev}}, \bibinfo {author} {\bibfnamefont {M.}~\bibnamefont {Heiblum}},
  \bibinfo {author} {\bibfnamefont {V.}~\bibnamefont {Umansky}}, \bibinfo
  {author} {\bibfnamefont {Ady}\ \bibnamefont {Stern}}, and\ \bibinfo
  {author} {\bibfnamefont {D.}~\bibnamefont {Mahalu}},\ }\bibfield  {title}
  {\enquote {\bibinfo {title} {{Observation of a quarter of an electron charge
  at the $\nu = 5/2$ quantum Hall state}},}\ }\href {\doibase%
  10.1038/nature06855} {\bibfield  {journal} {\bibinfo  {journal} {Nature}\
  }\textbf {\bibinfo {volume} {452}},\ \bibinfo {pages} {829--834} (\bibinfo
  {year} {2008})}\BibitemShut {NoStop}%
\bibitem [{\citenamefont {Bid}\ \emph {et~al.}(2010)\citenamefont {Bid},
  \citenamefont {Ofek}, \citenamefont {Inoue}, \citenamefont {Heiblum},
  \citenamefont {Kane}, \citenamefont {Umansky},\ and\ \citenamefont
  {Mahalu}}]{FiveHalvesNeutralModes}%
  \BibitemOpen
  \bibfield  {author} {\bibinfo {author} {\bibfnamefont {Aveek}\ \bibnamefont
  {Bid}}, \bibinfo {author} {\bibfnamefont {N.}~\bibnamefont {Ofek}}, \bibinfo
  {author} {\bibfnamefont {H.}~\bibnamefont {Inoue}}, \bibinfo {author}
  {\bibfnamefont {M.}~\bibnamefont {Heiblum}}, \bibinfo {author} {\bibfnamefont
  {C.~L.}\ \bibnamefont {Kane}}, \bibinfo {author} {\bibfnamefont
  {V.}~\bibnamefont {Umansky}}, and\ \bibinfo {author} {\bibfnamefont
  {D.}~\bibnamefont {Mahalu}},\ }\bibfield  {title} {\enquote {\bibinfo {title}
  {{Observation of neutral modes in the fractional quantum Hall regime}},}\
  }\href {\doibase 10.1038/nature09277} {\bibfield  {journal} {\bibinfo
  {journal} {Nature}\ }\textbf {\bibinfo {volume} {466}},\ \bibinfo {pages}
  {585--590} (\bibinfo {year} {2010})}\BibitemShut {NoStop}%
\bibitem [{\citenamefont {An}\ \emph {et~al.}(2011)\citenamefont {An},
  \citenamefont {Jiang}, \citenamefont {Choi}, \citenamefont {Kang},
  \citenamefont {Simon}, \citenamefont {Pfeiffer}, \citenamefont {West},\ and\
  \citenamefont {Baldwin}}]{FiveHalvesKang}%
  \BibitemOpen
  \bibfield  {author} {\bibinfo {author} {\bibfnamefont {Sanghun}\ \bibnamefont
  {An}}, \bibinfo {author} {\bibfnamefont {P.}~\bibnamefont {Jiang}}, \bibinfo
  {author} {\bibfnamefont {H.}~\bibnamefont {Choi}}, \bibinfo {author}
  {\bibfnamefont {W.}~\bibnamefont {Kang}}, \bibinfo {author} {\bibfnamefont
  {S.~H.}\ \bibnamefont {Simon}}, \bibinfo {author} {\bibfnamefont {L.~N.}\
  \bibnamefont {Pfeiffer}}, \bibinfo {author} {\bibfnamefont {K.~W.}\
  \bibnamefont {West}}, and\ \bibinfo {author} {\bibfnamefont {K.~W.}\
  \bibnamefont {Baldwin}},\ }\href@noop {} {\enquote {\bibinfo {title}
  {{Braiding of Abelian and Non-Abelian Anyons in the Fractional Quantum Hall
  Effect}},}\ } (\bibinfo {year} {2011}),\ \bibinfo {note} {unpublished},\
  \Eprint {http://arxiv.org/abs/1112.3400} {arXiv:1112.3400
  [cond-mat.mes-hall]} \BibitemShut {NoStop}%
\bibitem [{\citenamefont {Tiemann}\ \emph {et~al.}(2012)\citenamefont
  {Tiemann}, \citenamefont {Gamez}, \citenamefont {Kumada},\ and\ \citenamefont
  {Muraki}}]{FiveHalvesSpinPolarization}%
  \BibitemOpen
  \bibfield  {author} {\bibinfo {author} {\bibfnamefont {Lars}\ \bibnamefont
  {Tiemann}}, \bibinfo {author} {\bibfnamefont {Gerardo}\ \bibnamefont
  {Gamez}}, \bibinfo {author} {\bibfnamefont {Norio}\ \bibnamefont {Kumada}},
  and\ \bibinfo {author} {\bibfnamefont {Koji}\ \bibnamefont {Muraki}},\
  }\bibfield  {title} {\enquote {\bibinfo {title} {{Unraveling the spin
  polarization of the $\nu = 5/2$ fractional quantum Hall state}},}\ }\href
  {\doibase 10.1126/science.1216697} {\bibfield  {journal} {\bibinfo  {journal}
  {Science}\ }\textbf {\bibinfo {volume} {335}},\ \bibinfo {pages} {828--831}
  (\bibinfo {year} {2012})}\BibitemShut {NoStop}%
\bibitem [{\citenamefont {Willett}\ \emph {et~al.}(2013)\citenamefont
  {Willett}, \citenamefont {Nayak}, \citenamefont {Shtengel}, \citenamefont
  {Pfeiffer},\ and\ \citenamefont {West}}]{Willett3}%
  \BibitemOpen
  \bibfield  {author} {\bibinfo {author} {\bibfnamefont {R.~L.}\ \bibnamefont
  {Willett}}, \bibinfo {author} {\bibfnamefont {C.}~\bibnamefont {Nayak}},
  \bibinfo {author} {\bibfnamefont {K.}~\bibnamefont {Shtengel}}, \bibinfo
  {author} {\bibfnamefont {L.~N.}\ \bibnamefont {Pfeiffer}}, and\ \bibinfo
  {author} {\bibfnamefont {K.~W.}\ \bibnamefont {West}},\ }\href@noop {}
  {\enquote {\bibinfo {title} {{Magnetic field-tuned Aharonov--Bohm
  oscillations and evidence for non-Abelian anyons at $\nu=5/2$}},}\ }
  (\bibinfo {year} {2013}),\ \bibinfo {note} {unpublished},\ \Eprint
  {http://arxiv.org/abs/1301.2639} {arXiv:1301.2639 [cond-mat.mes-hall]}
  \BibitemShut {NoStop}%
\bibitem [{\citenamefont {Read}\ and\ \citenamefont
  {Rezayi}(1999)}]{ReadRezayi}%
  \BibitemOpen
  \bibfield  {author} {\bibinfo {author} {\bibfnamefont {N.}~\bibnamefont
  {Read}}\ and\ \bibinfo {author} {\bibfnamefont {E.}~\bibnamefont {Rezayi}},\
  }\bibfield  {title} {\enquote {\bibinfo {title} {Beyond paired quantum hall
  states: Parafermions and incompressible states in the first excited landau
  level},}\ }\href {\doibase 10.1103/PhysRevB.59.8084} {\bibfield  {journal}
  {\bibinfo  {journal} {Phys. Rev. B}\ }\textbf {\bibinfo {volume} {59}},\
  \bibinfo {pages} {8084--8092} (\bibinfo {year} {1999})}\BibitemShut {NoStop}%
\bibitem [{\citenamefont {Fendley}\ \emph {et~al.}(2009)\citenamefont
  {Fendley}, \citenamefont {Fisher},\ and\ \citenamefont
  {Nayak}}]{FendleyFisherNayak2}%
  \BibitemOpen
  \bibfield  {author} {\bibinfo {author} {\bibfnamefont {Paul}\ \bibnamefont
  {Fendley}}, \bibinfo {author} {\bibfnamefont {Matthew P.~A.}\ \bibnamefont
  {Fisher}}, and\ \bibinfo {author} {\bibfnamefont {Chetan}\ \bibnamefont
  {Nayak}},\ }\bibfield  {title} {\enquote {\bibinfo {title} {{Boundary
  Conformal Field Theory and Tunneling of Edge Quasiparticles in non-Abelian
  Topological States}},}\ }\href {\doibase 10.1016/j.aop.2009.03.005}
  {\bibfield  {journal} {\bibinfo  {journal} {Annals Phys.}\ }\textbf {\bibinfo
  {volume} {324}},\ \bibinfo {pages} {1547} (\bibinfo {year}
  {2009})}\BibitemShut {NoStop}%
\bibitem [{\citenamefont {Bishara}\ \emph {et~al.}(2008)\citenamefont
  {Bishara}, \citenamefont {Fiete},\ and\ \citenamefont
  {Nayak}}]{AntiReadRezayi}%
  \BibitemOpen
  \bibfield  {author} {\bibinfo {author} {\bibfnamefont {Waheb}\ \bibnamefont
  {Bishara}}, \bibinfo {author} {\bibfnamefont {Gregory~A.}\ \bibnamefont
  {Fiete}}, and\ \bibinfo {author} {\bibfnamefont {Chetan}\ \bibnamefont
  {Nayak}},\ }\bibfield  {title} {\enquote {\bibinfo {title} {{Quantum Hall
  states at $\nu=\frac{2}{k+2}$: Analysis of the particle-hole conjugates of
  the general level-$k$ Read-Rezayi states}},}\ }\href {\doibase%
  10.1103/PhysRevB.77.241306} {\bibfield  {journal} {\bibinfo  {journal} {Phys.
  Rev. B}\ }\textbf {\bibinfo {volume} {77}},\ \bibinfo {pages} {241306}
  (\bibinfo {year} {2008})}\BibitemShut {NoStop}%
\bibitem [{\citenamefont {Pan}\ \emph {et~al.}(1999)\citenamefont {Pan},
  \citenamefont {Xia}, \citenamefont {Shvarts}, \citenamefont {Adams},
  \citenamefont {Stormer}, \citenamefont {Tsui}, \citenamefont {Pfeiffer},
  \citenamefont {Baldwin},\ and\ \citenamefont {West}}]{Pan}%
  \BibitemOpen
  \bibfield  {author} {\bibinfo {author} {\bibfnamefont {W.}~\bibnamefont
  {Pan}}, \bibinfo {author} {\bibfnamefont {J.-S.}\ \bibnamefont {Xia}},
  \bibinfo {author} {\bibfnamefont {V.}~\bibnamefont {Shvarts}}, \bibinfo
  {author} {\bibfnamefont {D.~E.}\ \bibnamefont {Adams}}, \bibinfo {author}
  {\bibfnamefont {H.~L.}\ \bibnamefont {Stormer}}, \bibinfo {author}
  {\bibfnamefont {D.~C.}\ \bibnamefont {Tsui}}, \bibinfo {author}
  {\bibfnamefont {L.~N.}\ \bibnamefont {Pfeiffer}}, \bibinfo {author}
  {\bibfnamefont {K.~W.}\ \bibnamefont {Baldwin}}, and\ \bibinfo {author}
  {\bibfnamefont {K.~W.}\ \bibnamefont {West}},\ }\bibfield  {title} {\enquote
  {\bibinfo {title} {{Exact Quantization of the Even-Denominator Fractional
  Quantum Hall State at $\nu=5/2$ Landau Level Filling Factor}},}\ }\href
  {\doibase 10.1103/PhysRevLett.83.3530} {\bibfield  {journal} {\bibinfo
  {journal} {Phys. Rev. Lett.}\ }\textbf {\bibinfo {volume} {83}},\ \bibinfo
  {pages} {3530--3533} (\bibinfo {year} {1999})}\BibitemShut {NoStop}%
\bibitem [{\citenamefont {Xia}\ \emph {et~al.}(2004)\citenamefont {Xia},
  \citenamefont {Pan}, \citenamefont {Vicente}, \citenamefont {Adams},
  \citenamefont {Sullivan}, \citenamefont {Stormer}, \citenamefont {Tsui},
  \citenamefont {Pfeiffer}, \citenamefont {Baldwin},\ and\ \citenamefont
  {West}}]{Xia}%
  \BibitemOpen
  \bibfield  {author} {\bibinfo {author} {\bibfnamefont {J.~S.}\ \bibnamefont
  {Xia}}, \bibinfo {author} {\bibfnamefont {W.}~\bibnamefont {Pan}}, \bibinfo
  {author} {\bibfnamefont {C.~L.}\ \bibnamefont {Vicente}}, \bibinfo {author}
  {\bibfnamefont {E.~D.}\ \bibnamefont {Adams}}, \bibinfo {author}
  {\bibfnamefont {N.~S.}\ \bibnamefont {Sullivan}}, \bibinfo {author}
  {\bibfnamefont {H.~L.}\ \bibnamefont {Stormer}}, \bibinfo {author}
  {\bibfnamefont {D.~C.}\ \bibnamefont {Tsui}}, \bibinfo {author}
  {\bibfnamefont {L.~N.}\ \bibnamefont {Pfeiffer}}, \bibinfo {author}
  {\bibfnamefont {K.~W.}\ \bibnamefont {Baldwin}}, and\ \bibinfo {author}
  {\bibfnamefont {K.~W.}\ \bibnamefont {West}},\ }\bibfield  {title} {\enquote
  {\bibinfo {title} {Electron correlation in the second landau level: A
  competition between many nearly degenerate quantum phases},}\ }\href
  {\doibase 10.1103/PhysRevLett.93.176809} {\bibfield  {journal} {\bibinfo
  {journal} {Phys. Rev. Lett.}\ }\textbf {\bibinfo {volume} {93}},\ \bibinfo
  {pages} {176809} (\bibinfo {year} {2004})}\BibitemShut {NoStop}%
\bibitem [{\citenamefont {Pan}\ \emph {et~al.}(2008)\citenamefont {Pan},
  \citenamefont {Xia}, \citenamefont {Stormer}, \citenamefont {Tsui},
  \citenamefont {Vicente}, \citenamefont {Adams}, \citenamefont {Sullivan},
  \citenamefont {Pfeiffer}, \citenamefont {Baldwin},\ and\ \citenamefont
  {West}}]{SecondLL}%
  \BibitemOpen
  \bibfield  {author} {\bibinfo {author} {\bibfnamefont {W.}~\bibnamefont
  {Pan}}, \bibinfo {author} {\bibfnamefont {J.~S.}\ \bibnamefont {Xia}},
  \bibinfo {author} {\bibfnamefont {H.~L.}\ \bibnamefont {Stormer}}, \bibinfo
  {author} {\bibfnamefont {D.~C.}\ \bibnamefont {Tsui}}, \bibinfo {author}
  {\bibfnamefont {C.}~\bibnamefont {Vicente}}, \bibinfo {author} {\bibfnamefont
  {E.~D.}\ \bibnamefont {Adams}}, \bibinfo {author} {\bibfnamefont {N.~S.}\
  \bibnamefont {Sullivan}}, \bibinfo {author} {\bibfnamefont {L.~N.}\
  \bibnamefont {Pfeiffer}}, \bibinfo {author} {\bibfnamefont {K.~W.}\
  \bibnamefont {Baldwin}}, and\ \bibinfo {author} {\bibfnamefont {K.~W.}\
  \bibnamefont {West}},\ }\bibfield  {title} {\enquote {\bibinfo {title}
  {{Experimental studies of the fractional quantum Hall effect in the first
  excited Landau level}},}\ }\href {\doibase 10.1103/PhysRevB.77.075307}
  {\bibfield  {journal} {\bibinfo  {journal} {Phys. Rev. B}\ }\textbf {\bibinfo
  {volume} {77}},\ \bibinfo {pages} {075307} (\bibinfo {year}
  {2008})}\BibitemShut {NoStop}%
\bibitem [{\citenamefont {Read}\ and\ \citenamefont {Green}(2000)}]{ReadGreen}%
  \BibitemOpen
  \bibfield  {author} {\bibinfo {author} {\bibfnamefont {N.}~\bibnamefont
  {Read}}\ and\ \bibinfo {author} {\bibfnamefont {Dmitry}\ \bibnamefont
  {Green}},\ }\bibfield  {title} {\enquote {\bibinfo {title} {{Paired states of
  fermions in two dimensions with breaking of parity and time-reversal
  symmetries and the fractional quantum Hall effect}},}\ }\href {\doibase%
  10.1103/PhysRevB.61.10267} {\bibfield  {journal} {\bibinfo  {journal} {Phys.
  Rev. B}\ }\textbf {\bibinfo {volume} {61}},\ \bibinfo {pages} {10267--10297}
  (\bibinfo {year} {2000})}\BibitemShut {NoStop}%
\bibitem [{Note2()}]{Note2}%
  \BibitemOpen
  \bibinfo {note} {Throughout, when referring to spinless $p$-wave
  superconductivity we implicitly mean the topologically nontrivial weak
  pairing phase.}\BibitemShut {Stop}%
\bibitem [{\citenamefont {Ivanov}(2001)}]{Ivanov}%
  \BibitemOpen
  \bibfield  {author} {\bibinfo {author} {\bibfnamefont {Dmitri~A.}\
  \bibnamefont {Ivanov}},\ }\bibfield  {title} {\enquote {\bibinfo {title}
  {{Non-Abelian Statistics of Half-Quantum Vortices in $p$-Wave
  Superconductors}},}\ }\href {\doibase 10.1103/PhysRevLett.86.268} {\bibfield
  {journal} {\bibinfo  {journal} {Phys. Rev. Lett.}\ }\textbf {\bibinfo
  {volume} {86}},\ \bibinfo {pages} {268--271} (\bibinfo {year}
  {2001})}\BibitemShut {NoStop}%
\bibitem [{\citenamefont {Freedman}\ \emph {et~al.}(2011)\citenamefont
  {Freedman}, \citenamefont {Hastings}, \citenamefont {Nayak}, \citenamefont
  {Qi}, \citenamefont {Walker},\ and\ \citenamefont
  {Wang}}]{ProjectiveStatistics}%
  \BibitemOpen
  \bibfield  {author} {\bibinfo {author} {\bibfnamefont {Michael}\ \bibnamefont
  {Freedman}}, \bibinfo {author} {\bibfnamefont {Matthew~B.}\ \bibnamefont
  {Hastings}}, \bibinfo {author} {\bibfnamefont {Chetan}\ \bibnamefont
  {Nayak}}, \bibinfo {author} {\bibfnamefont {Xiao-Liang}\ \bibnamefont {Qi}},
  \bibinfo {author} {\bibfnamefont {Kevin}\ \bibnamefont {Walker}}, and\
  \bibinfo {author} {\bibfnamefont {Zhenghan}\ \bibnamefont {Wang}},\
  }\bibfield  {title} {\enquote {\bibinfo {title} {Projective ribbon
  permutation statistics: A remnant of non-abelian braiding in higher
  dimensions},}\ }\href {\doibase 10.1103/PhysRevB.83.115132} {\bibfield
  {journal} {\bibinfo  {journal} {Phys. Rev. B}\ }\textbf {\bibinfo {volume}
  {83}},\ \bibinfo {pages} {115132} (\bibinfo {year} {2011})}\BibitemShut
  {NoStop}%
\bibitem [{\citenamefont {Barkeshli}\ \emph
  {et~al.}(2013{\natexlab{a}})\citenamefont {Barkeshli}, \citenamefont {Jian},\
  and\ \citenamefont {Qi}}]{BarkeshliParafendleyons1}%
  \BibitemOpen
  \bibfield  {author} {\bibinfo {author} {\bibfnamefont {Maissam}\ \bibnamefont
  {Barkeshli}}, \bibinfo {author} {\bibfnamefont {Chao-Ming}\ \bibnamefont
  {Jian}}, and\ \bibinfo {author} {\bibfnamefont {Xiao-Liang}\ \bibnamefont
  {Qi}},\ }\bibfield  {title} {\enquote {\bibinfo {title} {{Twist defects and
  projective non-Abelian braiding statistics}},}\ }\href {\doibase%
  10.1103/PhysRevB.87.045130} {\bibfield  {journal} {\bibinfo  {journal} {Phys.
  Rev. B}\ }\textbf {\bibinfo {volume} {87}},\ \bibinfo {pages} {045130}
  (\bibinfo {year} {2013}{\natexlab{a}})}\BibitemShut {NoStop}%
\bibitem [{\citenamefont {Kitaev}(2001)}]{1DwiresKitaev}%
  \BibitemOpen
  \bibfield  {author} {\bibinfo {author} {\bibfnamefont {Alexei~Yu}\
  \bibnamefont {Kitaev}},\ }\bibfield  {title} {\enquote {\bibinfo {title}
  {{Unpaired Majorana fermions in quantum wires}},}\ }\href {\doibase%
  10.1070/1063-7869/44/10S/S29} {\bibfield  {journal} {\bibinfo  {journal}
  {Sov. Phys.--Uspeki}\ }\textbf {\bibinfo {volume} {44}},\ \bibinfo {pages}
  {131} (\bibinfo {year} {2001})}\BibitemShut {NoStop}%
\bibitem [{\citenamefont {Alicea}\ \emph {et~al.}(2011)\citenamefont {Alicea},
  \citenamefont {Oreg}, \citenamefont {Refael}, \citenamefont {{von Oppen}},\
  and\ \citenamefont {Fisher}}]{AliceaBraiding}%
  \BibitemOpen
  \bibfield  {author} {\bibinfo {author} {\bibfnamefont {Jason}\ \bibnamefont
  {Alicea}}, \bibinfo {author} {\bibfnamefont {Yuval}\ \bibnamefont {Oreg}},
  \bibinfo {author} {\bibfnamefont {Gil}\ \bibnamefont {Refael}}, \bibinfo
  {author} {\bibfnamefont {Felix}\ \bibnamefont {{von Oppen}}}, and\ \bibinfo
  {author} {\bibfnamefont {Matthew P.~A.}\ \bibnamefont {Fisher}},\ }\bibfield
  {title} {\enquote {\bibinfo {title} {{Non-Abelian statistics and topological
  quantum information processing in 1D wire networks}},}\ }\href {\doibase%
  10.1038/nphys1915} {\bibfield  {journal} {\bibinfo  {journal} {Nat. Phys.}\
  }\textbf {\bibinfo {volume} {7}},\ \bibinfo {pages} {412--417} (\bibinfo
  {year} {2011})}\BibitemShut {NoStop}%
\bibitem [{\citenamefont {Halperin}\ \emph {et~al.}(2012)\citenamefont
  {Halperin}, \citenamefont {Oreg}, \citenamefont {Stern}, \citenamefont
  {Refael}, \citenamefont {Alicea},\ and\ \citenamefont {von
  Oppen}}]{HalperinBraiding}%
  \BibitemOpen
  \bibfield  {author} {\bibinfo {author} {\bibfnamefont {Bertrand~I.}\
  \bibnamefont {Halperin}}, \bibinfo {author} {\bibfnamefont {Yuval}\
  \bibnamefont {Oreg}}, \bibinfo {author} {\bibfnamefont {Ady}\ \bibnamefont
  {Stern}}, \bibinfo {author} {\bibfnamefont {Gil}\ \bibnamefont {Refael}},
  \bibinfo {author} {\bibfnamefont {Jason}\ \bibnamefont {Alicea}}, and\
  \bibinfo {author} {\bibfnamefont {Felix}\ \bibnamefont {von Oppen}},\
  }\bibfield  {title} {\enquote {\bibinfo {title} {{Adiabatic manipulations of
  Majorana fermions in a three-dimensional network of quantum wires}},}\ }\href
  {\doibase 10.1103/PhysRevB.85.144501} {\bibfield  {journal} {\bibinfo
  {journal} {Phys. Rev. B}\ }\textbf {\bibinfo {volume} {85}},\ \bibinfo
  {pages} {144501} (\bibinfo {year} {2012})}\BibitemShut {NoStop}%
\bibitem [{\citenamefont {Clarke}\ \emph {et~al.}(2011)\citenamefont {Clarke},
  \citenamefont {Sau},\ and\ \citenamefont {Tewari}}]{ClarkeBraiding}%
  \BibitemOpen
  \bibfield  {author} {\bibinfo {author} {\bibfnamefont {David~J.}\
  \bibnamefont {Clarke}}, \bibinfo {author} {\bibfnamefont {Jay~D.}\
  \bibnamefont {Sau}}, and\ \bibinfo {author} {\bibfnamefont {Sumanta}\
  \bibnamefont {Tewari}},\ }\bibfield  {title} {\enquote {\bibinfo {title}
  {{Majorana fermion exchange in quasi-one-dimensional networks}},}\ }\href
  {\doibase 10.1103/PhysRevB.84.035120} {\bibfield  {journal} {\bibinfo
  {journal} {Phys. Rev. B}\ }\textbf {\bibinfo {volume} {84}},\ \bibinfo
  {pages} {035120} (\bibinfo {year} {2011})}\BibitemShut {NoStop}%
\bibitem [{\citenamefont {Bonderson}(2013)}]{BondersonBraiding}%
  \BibitemOpen
  \bibfield  {author} {\bibinfo {author} {\bibfnamefont {Parsa}\ \bibnamefont
  {Bonderson}},\ }\bibfield  {title} {\enquote {\bibinfo {title}
  {Measurement-only topological quantum computation via tunable
  interactions},}\ }\href {\doibase 10.1103/PhysRevB.87.035113} {\bibfield
  {journal} {\bibinfo  {journal} {Phys. Rev. B}\ }\textbf {\bibinfo {volume}
  {87}},\ \bibinfo {pages} {035113} (\bibinfo {year} {2013})}\BibitemShut
  {NoStop}%
\bibitem [{\citenamefont {Fu}\ and\ \citenamefont {Kane}(2008)}]{FuKane}%
  \BibitemOpen
  \bibfield  {author} {\bibinfo {author} {\bibfnamefont {Liang}\ \bibnamefont
  {Fu}}\ and\ \bibinfo {author} {\bibfnamefont {C.~L.}\ \bibnamefont {Kane}},\
  }\bibfield  {title} {\enquote {\bibinfo {title} {{Superconducting Proximity
  Effect and Majorana Fermions at the Surface of a Topological Insulator}},}\
  }\href {\doibase 10.1103/PhysRevLett.100.096407} {\bibfield  {journal}
  {\bibinfo  {journal} {Phys. Rev. Lett.}\ }\textbf {\bibinfo {volume} {100}},\
  \bibinfo {pages} {096407} (\bibinfo {year} {2008})}\BibitemShut {NoStop}%
\bibitem [{\citenamefont {Fu}\ and\ \citenamefont
  {Kane}(2009)}]{MajoranaQSHedge}%
  \BibitemOpen
  \bibfield  {author} {\bibinfo {author} {\bibfnamefont {Liang}\ \bibnamefont
  {Fu}}\ and\ \bibinfo {author} {\bibfnamefont {C.~L.}\ \bibnamefont {Kane}},\
  }\bibfield  {title} {\enquote {\bibinfo {title} {{Josephson current and noise
  at a superconductor/quantum-spin-Hall-insulator/superconductor junction}},}\
  }\href {\doibase 10.1103/PhysRevB.79.161408} {\bibfield  {journal} {\bibinfo
  {journal} {Phys.\ Rev.\ B}\ }\textbf {\bibinfo {volume} {79}},\ \bibinfo
  {pages} {161408(R)} (\bibinfo {year} {2009})}\BibitemShut {NoStop}%
\bibitem [{\citenamefont {Sau}\ \emph {et~al.}(2010)\citenamefont {Sau},
  \citenamefont {Lutchyn}, \citenamefont {Tewari},\ and\ \citenamefont {{Das
  Sarma}}}]{Sau}%
  \BibitemOpen
  \bibfield  {author} {\bibinfo {author} {\bibfnamefont {Jay~D.}\ \bibnamefont
  {Sau}}, \bibinfo {author} {\bibfnamefont {Roman~M.}\ \bibnamefont {Lutchyn}},
  \bibinfo {author} {\bibfnamefont {Sumanta}\ \bibnamefont {Tewari}}, and\
  \bibinfo {author} {\bibfnamefont {S.}~\bibnamefont {{Das Sarma}}},\
  }\bibfield  {title} {\enquote {\bibinfo {title} {{Generic New Platform for
  Topological Quantum Computation Using Semiconductor Heterostructures}},}\
  }\href {\doibase 10.1103/PhysRevLett.104.040502} {\bibfield  {journal}
  {\bibinfo  {journal} {Phys.\ Rev.\ Lett.}\ }\textbf {\bibinfo {volume}
  {104}},\ \bibinfo {pages} {040502} (\bibinfo {year} {2010})}\BibitemShut
  {NoStop}%
\bibitem [{\citenamefont {Alicea}(2010)}]{Alicea}%
  \BibitemOpen
  \bibfield  {author} {\bibinfo {author} {\bibfnamefont {Jason}\ \bibnamefont
  {Alicea}},\ }\bibfield  {title} {\enquote {\bibinfo {title} {{Majorana
  fermions in a tunable semiconductor device}},}\ }\href {\doibase%
  10.1103/PhysRevB.81.125318} {\bibfield  {journal} {\bibinfo  {journal} {Phys.
  Rev. B}\ }\textbf {\bibinfo {volume} {81}},\ \bibinfo {pages} {125318}
  (\bibinfo {year} {2010})}\BibitemShut {NoStop}%
\bibitem [{\citenamefont {Lutchyn}\ \emph {et~al.}(2010)\citenamefont
  {Lutchyn}, \citenamefont {Sau},\ and\ \citenamefont
  {Das~Sarma}}]{1DwiresLutchyn}%
  \BibitemOpen
  \bibfield  {author} {\bibinfo {author} {\bibfnamefont {Roman~M.}\
  \bibnamefont {Lutchyn}}, \bibinfo {author} {\bibfnamefont {Jay~D.}\
  \bibnamefont {Sau}}, and\ \bibinfo {author} {\bibfnamefont
  {S.}~\bibnamefont {Das~Sarma}},\ }\bibfield  {title} {\enquote {\bibinfo
  {title} {{Majorana Fermions and a Topological Phase Transition in
  Semiconductor-Superconductor Heterostructures}},}\ }\href {\doibase%
  10.1103/PhysRevLett.105.077001} {\bibfield  {journal} {\bibinfo  {journal}
  {Phys.\ Rev.\ Lett.}\ }\textbf {\bibinfo {volume} {105}},\ \bibinfo {pages}
  {077001} (\bibinfo {year} {2010})}\BibitemShut {NoStop}%
\bibitem [{\citenamefont {Oreg}\ \emph {et~al.}(2010)\citenamefont {Oreg},
  \citenamefont {Refael},\ and\ \citenamefont {{von Oppen}}}]{1DwiresOreg}%
  \BibitemOpen
  \bibfield  {author} {\bibinfo {author} {\bibfnamefont {Yuval}\ \bibnamefont
  {Oreg}}, \bibinfo {author} {\bibfnamefont {Gil}\ \bibnamefont {Refael}},
  and\ \bibinfo {author} {\bibfnamefont {Felix}\ \bibnamefont {{von Oppen}}},\
  }\bibfield  {title} {\enquote {\bibinfo {title} {{Helical Liquids and
  Majorana Bound States in Quantum Wires}},}\ }\href {\doibase%
  10.1103/PhysRevLett.105.177002} {\bibfield  {journal} {\bibinfo  {journal}
  {Phys.\ Rev.\ Lett.}\ }\textbf {\bibinfo {volume} {105}},\ \bibinfo {pages}
  {177002} (\bibinfo {year} {2010})}\BibitemShut {NoStop}%
\bibitem [{\citenamefont {Cook}\ and\ \citenamefont {Franz}(2011)}]{CookFranz}%
  \BibitemOpen
  \bibfield  {author} {\bibinfo {author} {\bibfnamefont {A.}~\bibnamefont
  {Cook}}\ and\ \bibinfo {author} {\bibfnamefont {M.}~\bibnamefont {Franz}},\
  }\bibfield  {title} {\enquote {\bibinfo {title} {Majorana fermions in a
  topological-insulator nanowire proximity-coupled to an $s$-wave
  superconductor},}\ }\href {\doibase 10.1103/PhysRevB.84.201105} {\bibfield
  {journal} {\bibinfo  {journal} {Phys. Rev. B}\ }\textbf {\bibinfo {volume}
  {84}},\ \bibinfo {pages} {201105} (\bibinfo {year} {2011})}\BibitemShut
  {NoStop}%
\bibitem [{\citenamefont {Beenakker}(2013)}]{BeenakkerReview}%
  \BibitemOpen
  \bibfield  {author} {\bibinfo {author} {\bibfnamefont {C.~W.~J.}\
  \bibnamefont {Beenakker}},\ }\bibfield  {title} {\enquote {\bibinfo {title}
  {{Search for Majorana fermions in superconductors}},}\ }\href {\doibase%
  10.1146/annurev-conmatphys-030212-184337} {\bibfield  {journal} {\bibinfo
  {journal} {Annu. Rev. Con. Mat. Phys.}\ }\textbf {\bibinfo {volume} {4}},\
  \bibinfo {pages} {113--136} (\bibinfo {year} {2013})}\BibitemShut {NoStop}%
\bibitem [{\citenamefont {Alicea}(2012)}]{AliceaReview}%
  \BibitemOpen
  \bibfield  {author} {\bibinfo {author} {\bibfnamefont {Jason}\ \bibnamefont
  {Alicea}},\ }\bibfield  {title} {\enquote {\bibinfo {title} {{New directions
  in the pursuit of Majorana fermions in solid state systems}},}\ }\href
  {\doibase 10.1088/0034-4885/75/7/076501} {\bibfield  {journal} {\bibinfo
  {journal} {Rep. Prog. Phys.}\ }\textbf {\bibinfo {volume} {75}},\ \bibinfo
  {pages} {076501} (\bibinfo {year} {2012})}\BibitemShut {NoStop}%
\bibitem [{\citenamefont {Mourik}\ \emph {et~al.}(2012)\citenamefont {Mourik},
  \citenamefont {Zuo}, \citenamefont {Frolov}, \citenamefont {Plissard},
  \citenamefont {Bakkers},\ and\ \citenamefont {Kouwenhoven}}]{mourik12}%
  \BibitemOpen
  \bibfield  {author} {\bibinfo {author} {\bibfnamefont {V.}~\bibnamefont
  {Mourik}}, \bibinfo {author} {\bibfnamefont {K.}~\bibnamefont {Zuo}},
  \bibinfo {author} {\bibfnamefont {S.~M.}\ \bibnamefont {Frolov}}, \bibinfo
  {author} {\bibfnamefont {S.~R.}\ \bibnamefont {Plissard}}, \bibinfo {author}
  {\bibfnamefont {E.~P. A.~M.}\ \bibnamefont {Bakkers}}, and\ \bibinfo
  {author} {\bibfnamefont {L.~P.}\ \bibnamefont {Kouwenhoven}},\ }\bibfield
  {title} {\enquote {\bibinfo {title} {{Signatures of Majorana Fermions in
  Hybrid Superconductor-Semiconductor Nanowire Devices}},}\ }\href {\doibase%
  10.1126/science.1222360} {\bibfield  {journal} {\bibinfo  {journal}
  {Science}\ }\textbf {\bibinfo {volume} {336}},\ \bibinfo {pages} {1003--1007}
  (\bibinfo {year} {2012})}\BibitemShut {NoStop}%
\bibitem [{\citenamefont {Das}\ \emph {et~al.}(2012)\citenamefont {Das},
  \citenamefont {Ronen}, \citenamefont {Most}, \citenamefont {Oreg},
  \citenamefont {Heiblum},\ and\ \citenamefont {Shtrikman}}]{das12}%
  \BibitemOpen
  \bibfield  {author} {\bibinfo {author} {\bibfnamefont {A.}~\bibnamefont
  {Das}}, \bibinfo {author} {\bibfnamefont {Y.}~\bibnamefont {Ronen}}, \bibinfo
  {author} {\bibfnamefont {Y.}~\bibnamefont {Most}}, \bibinfo {author}
  {\bibfnamefont {Y.}~\bibnamefont {Oreg}}, \bibinfo {author} {\bibfnamefont
  {M.}~\bibnamefont {Heiblum}}, and\ \bibinfo {author} {\bibfnamefont
  {H.}~\bibnamefont {Shtrikman}},\ }\bibfield  {title} {\enquote {\bibinfo
  {title} {{Zero-bias peaks and splitting in an $\mathrm{Al}$-$\mathrm{InAs}$
  nanowire topological superconductor as a signature of Majorana fermions}},}\
  }\href {\doibase 10.1038/nphys2479} {\bibfield  {journal} {\bibinfo
  {journal} {Nat. Phys.}\ }\textbf {\bibinfo {volume} {8}},\ \bibinfo {pages}
  {887--895} (\bibinfo {year} {2012})}\BibitemShut {NoStop}%
\bibitem [{\citenamefont {Rokhinson}\ \emph {et~al.}(2012)\citenamefont
  {Rokhinson}, \citenamefont {Liu},\ and\ \citenamefont {Furdyna}}]{Rokhinson}%
  \BibitemOpen
  \bibfield  {author} {\bibinfo {author} {\bibfnamefont {Leonid~P.}\
  \bibnamefont {Rokhinson}}, \bibinfo {author} {\bibfnamefont {Xinyu}\
  \bibnamefont {Liu}}, and\ \bibinfo {author} {\bibfnamefont {Jacek~K.}\
  \bibnamefont {Furdyna}},\ }\bibfield  {title} {\enquote {\bibinfo {title}
  {{The fractional a.c. {J}osephson effect in a semiconductor-superconductor
  nanowire as a signature of {M}ajorana particles}},}\ }\href {\doibase%
  10.1038/nphys2429} {\bibfield  {journal} {\bibinfo  {journal} {Nat. Phys.}\
  }\textbf {\bibinfo {volume} {8}},\ \bibinfo {pages} {795--799} (\bibinfo
  {year} {2012})}\BibitemShut {NoStop}%
\bibitem [{\citenamefont {Deng}\ \emph {et~al.}(2012)\citenamefont {Deng},
  \citenamefont {Yu}, \citenamefont {Huang}, \citenamefont {Larsson},
  \citenamefont {Caroff},\ and\ \citenamefont {Xu}}]{deng12}%
  \BibitemOpen
  \bibfield  {author} {\bibinfo {author} {\bibfnamefont {M.~T.}\ \bibnamefont
  {Deng}}, \bibinfo {author} {\bibfnamefont {C.~L.}\ \bibnamefont {Yu}},
  \bibinfo {author} {\bibfnamefont {G.~Y.}\ \bibnamefont {Huang}}, \bibinfo
  {author} {\bibfnamefont {M.}~\bibnamefont {Larsson}}, \bibinfo {author}
  {\bibfnamefont {P.}~\bibnamefont {Caroff}}, and\ \bibinfo {author}
  {\bibfnamefont {H.~Q.}\ \bibnamefont {Xu}},\ }\bibfield  {title} {\enquote
  {\bibinfo {title} {{Anomalous Zero-Bias Conductance Peak in a
  $\mathrm{Nb}$-$\mathrm{InSb}$ Nanowire-$\mathrm{Nb}$ Hybrid Device}},}\
  }\href {\doibase 10.1021/nl303758w} {\bibfield  {journal} {\bibinfo
  {journal} {Nano Lett.}\ }\textbf {\bibinfo {volume} {12}},\ \bibinfo {pages}
  {6414} (\bibinfo {year} {2012})}\BibitemShut {NoStop}%
\bibitem [{\citenamefont {Finck}\ \emph {et~al.}(2013)\citenamefont {Finck},
  \citenamefont {Van~Harlingen}, \citenamefont {Mohseni}, \citenamefont
  {Jung},\ and\ \citenamefont {Li}}]{finck12}%
  \BibitemOpen
  \bibfield  {author} {\bibinfo {author} {\bibfnamefont {A.~D.~K.}\
  \bibnamefont {Finck}}, \bibinfo {author} {\bibfnamefont {D.~J.}\ \bibnamefont
  {Van~Harlingen}}, \bibinfo {author} {\bibfnamefont {P.~K.}\ \bibnamefont
  {Mohseni}}, \bibinfo {author} {\bibfnamefont {K.}~\bibnamefont {Jung}},
  and\ \bibinfo {author} {\bibfnamefont {X.}~\bibnamefont {Li}},\ }\bibfield
  {title} {\enquote {\bibinfo {title} {{Anomalous Modulation of a Zero-Bias
  Peak in a Hybrid Nanowire-Superconductor Device}},}\ }\href {\doibase%
  10.1103/PhysRevLett.110.126406} {\bibfield  {journal} {\bibinfo  {journal}
  {Phys. Rev. Lett.}\ }\textbf {\bibinfo {volume} {110}},\ \bibinfo {pages}
  {126406} (\bibinfo {year} {2013})}\BibitemShut {NoStop}%
\bibitem [{\citenamefont {Churchill}\ \emph {et~al.}(2013)\citenamefont
  {Churchill}, \citenamefont {Fatemi}, \citenamefont {Grove-Rasmussen},
  \citenamefont {Deng}, \citenamefont {Caroff}, \citenamefont {Xu},\ and\
  \citenamefont {Marcus}}]{Churchill}%
  \BibitemOpen
  \bibfield  {author} {\bibinfo {author} {\bibfnamefont {H.~O.~H.}\
  \bibnamefont {Churchill}}, \bibinfo {author} {\bibfnamefont {V.}~\bibnamefont
  {Fatemi}}, \bibinfo {author} {\bibfnamefont {K.}~\bibnamefont
  {Grove-Rasmussen}}, \bibinfo {author} {\bibfnamefont {M.~T.}\ \bibnamefont
  {Deng}}, \bibinfo {author} {\bibfnamefont {P.}~\bibnamefont {Caroff}},
  \bibinfo {author} {\bibfnamefont {H.~Q.}\ \bibnamefont {Xu}}, and\ \bibinfo
  {author} {\bibfnamefont {C.~M.}\ \bibnamefont {Marcus}},\ }\bibfield  {title}
  {\enquote {\bibinfo {title} {{Superconductor-nanowire devices from tunneling
  to the multichannel regime: Zero-bias oscillations and magnetoconductance
  crossover}},}\ }\href {\doibase 10.1103/PhysRevB.87.241401} {\bibfield
  {journal} {\bibinfo  {journal} {Phys. Rev. B}\ }\textbf {\bibinfo {volume}
  {87}},\ \bibinfo {pages} {241401} (\bibinfo {year} {2013})}\BibitemShut
  {NoStop}%
\bibitem [{\citenamefont {Clarke}\ \emph {et~al.}(2013)\citenamefont {Clarke},
  \citenamefont {Alicea},\ and\ \citenamefont
  {Shtengel}}]{ClarkeParafendleyons}%
  \BibitemOpen
  \bibfield  {author} {\bibinfo {author} {\bibfnamefont {David~J.}\
  \bibnamefont {Clarke}}, \bibinfo {author} {\bibfnamefont {Jason}\
  \bibnamefont {Alicea}}, and\ \bibinfo {author} {\bibfnamefont {Kirill}\
  \bibnamefont {Shtengel}},\ }\bibfield  {title} {\enquote {\bibinfo {title}
  {{Exotic non-Abelian anyons from conventional fractional quantum Hall
  states}},}\ }\href {\doibase 10.1038/ncomms2340} {\bibfield  {journal}
  {\bibinfo  {journal} {Nature Commun.}\ }\textbf {\bibinfo {volume} {4}},\
  \bibinfo {pages} {1348} (\bibinfo {year} {2013})}\BibitemShut {NoStop}%
\bibitem [{\citenamefont {Lindner}\ \emph {et~al.}(2012)\citenamefont
  {Lindner}, \citenamefont {Berg}, \citenamefont {Refael},\ and\ \citenamefont
  {Stern}}]{LindnerParafendleyons}%
  \BibitemOpen
  \bibfield  {author} {\bibinfo {author} {\bibfnamefont {Netanel~H.}\
  \bibnamefont {Lindner}}, \bibinfo {author} {\bibfnamefont {Erez}\
  \bibnamefont {Berg}}, \bibinfo {author} {\bibfnamefont {Gil}\ \bibnamefont
  {Refael}}, and\ \bibinfo {author} {\bibfnamefont {Ady}\ \bibnamefont
  {Stern}},\ }\bibfield  {title} {\enquote {\bibinfo {title} {{Fractionalizing
  Majorana Fermions: Non-Abelian Statistics on the Edges of Abelian Quantum
  Hall States}},}\ }\href {\doibase 10.1103/PhysRevX.2.041002} {\bibfield
  {journal} {\bibinfo  {journal} {Phys. Rev. X}\ }\textbf {\bibinfo {volume}
  {2}},\ \bibinfo {pages} {041002} (\bibinfo {year} {2012})}\BibitemShut
  {NoStop}%
\bibitem [{\citenamefont {Cheng}(2012)}]{ChengParafendleyons}%
  \BibitemOpen
  \bibfield  {author} {\bibinfo {author} {\bibfnamefont {Meng}\ \bibnamefont
  {Cheng}},\ }\bibfield  {title} {\enquote {\bibinfo {title} {Superconducting
  proximity effect on the edge of fractional topological insulators},}\ }\href
  {\doibase 10.1103/PhysRevB.86.195126} {\bibfield  {journal} {\bibinfo
  {journal} {Phys. Rev. B}\ }\textbf {\bibinfo {volume} {86}},\ \bibinfo
  {pages} {195126} (\bibinfo {year} {2012})}\BibitemShut {NoStop}%
\bibitem [{\citenamefont {Vaezi}(2013)}]{VaeziParafendleyons}%
  \BibitemOpen
  \bibfield  {author} {\bibinfo {author} {\bibfnamefont {Abolhassan}\
  \bibnamefont {Vaezi}},\ }\bibfield  {title} {\enquote {\bibinfo {title}
  {{Fractional topological superconductor with fractionalized Majorana
  fermions}},}\ }\href {\doibase 10.1103/PhysRevB.87.035132} {\bibfield
  {journal} {\bibinfo  {journal} {Phys. Rev. B}\ }\textbf {\bibinfo {volume}
  {87}},\ \bibinfo {pages} {035132} (\bibinfo {year} {2013})}\BibitemShut
  {NoStop}%
\bibitem [{\citenamefont {Barkeshli}\ and\ \citenamefont
  {Qi}(2013)}]{BarkeshliParafendleyons2}%
  \BibitemOpen
  \bibfield  {author} {\bibinfo {author} {\bibfnamefont {Maissam}\ \bibnamefont
  {Barkeshli}}\ and\ \bibinfo {author} {\bibfnamefont {Xiao-Liang}\
  \bibnamefont {Qi}},\ }\href@noop {} {\enquote {\bibinfo {title} {{Synthetic
  Topological Qubits in Conventional Bilayer Quantum Hall Systems}},}\ }
  (\bibinfo {year} {2013}),\ \bibinfo {note} {unpublished},\ \Eprint
  {http://arxiv.org/abs/1302.2673} {arXiv:1302.2673 [cond-mat.mes-hall]}
  \BibitemShut {NoStop}%
\bibitem [{\citenamefont {Barkeshli}\ and\ \citenamefont
  {Qi}(2012)}]{ChernInsulatorParafendleyons}%
  \BibitemOpen
  \bibfield  {author} {\bibinfo {author} {\bibfnamefont {Maissam}\ \bibnamefont
  {Barkeshli}}\ and\ \bibinfo {author} {\bibfnamefont {Xiao-Liang}\
  \bibnamefont {Qi}},\ }\bibfield  {title} {\enquote {\bibinfo {title}
  {Topological nematic states and non-abelian lattice dislocations},}\ }\href
  {\doibase 10.1103/PhysRevX.2.031013} {\bibfield  {journal} {\bibinfo
  {journal} {Phys. Rev. X}\ }\textbf {\bibinfo {volume} {2}},\ \bibinfo {pages}
  {031013} (\bibinfo {year} {2012})}\BibitemShut {NoStop}%
\bibitem [{\citenamefont {Oreg}\ \emph {et~al.}(2013)\citenamefont {Oreg},
  \citenamefont {Sela},\ and\ \citenamefont
  {Stern}}]{QuantumWiresParafendleyons}%
  \BibitemOpen
  \bibfield  {author} {\bibinfo {author} {\bibfnamefont {Yuval}\ \bibnamefont
  {Oreg}}, \bibinfo {author} {\bibfnamefont {Eran}\ \bibnamefont {Sela}},
  and\ \bibinfo {author} {\bibfnamefont {Ady}\ \bibnamefont {Stern}},\
  }\href@noop {} {\enquote {\bibinfo {title} {{Fractional Helical Liquids and
  Non-Abelian Anyons in Quantum Wires}},}\ } (\bibinfo {year} {2013}),\
  \bibinfo {note} {unpublished},\ \Eprint {http://arxiv.org/abs/1301.7335}
  {arXiv:1301.7335 [cond-mat.str-el]} \BibitemShut {NoStop}%
\bibitem [{\citenamefont {Fendley}(2012)}]{Fendley}%
  \BibitemOpen
  \bibfield  {author} {\bibinfo {author} {\bibfnamefont {Paul}\ \bibnamefont
  {Fendley}},\ }\bibfield  {title} {\enquote {\bibinfo {title} {{Parafermionic
  edge zero modes in $\mathbb{Z}_n$-invariant spin chains}},}\ }\href {\doibase%
  10.1088/1742-5468/2012/11/P11020} {\bibfield  {journal} {\bibinfo  {journal}
  {J. Stat. Mech.}\ }\textbf {\bibinfo {volume} {2012}},\ \bibinfo {pages}
  {11020} (\bibinfo {year} {2012})}\BibitemShut {NoStop}%
\bibitem [{Note3()}]{Note3}%
  \BibitemOpen
  \bibinfo {note} {Some references refer to these generalizations as
  parafermion zero-modes. We intentionally avoid this nomenclature here to
  avoid confusion with the rather different (though related) parafermions that
  appear in conformal field theory, particularly since both contexts frequently
  arise in this paper.}\BibitemShut {Stop}%
\bibitem [{\citenamefont {Hastings}\ \emph {et~al.}(2013)\citenamefont
  {Hastings}, \citenamefont {Nayak},\ and\ \citenamefont {Wang}}]{Hastings}%
  \BibitemOpen
  \bibfield  {author} {\bibinfo {author} {\bibfnamefont {Matthew~B.}\
  \bibnamefont {Hastings}}, \bibinfo {author} {\bibfnamefont {Chetan}\
  \bibnamefont {Nayak}}, and\ \bibinfo {author} {\bibfnamefont {Zhenghan}\
  \bibnamefont {Wang}},\ }\bibfield  {title} {\enquote {\bibinfo {title}
  {{Metaplectic anyons, Majorana zero modes, and their computational power}},}\
  }\href {\doibase 10.1103/PhysRevB.87.165421} {\bibfield  {journal} {\bibinfo
  {journal} {Phys. Rev. B}\ }\textbf {\bibinfo {volume} {87}},\ \bibinfo
  {pages} {165421} (\bibinfo {year} {2013})}\BibitemShut {NoStop}%
\bibitem [{\citenamefont {Gils}\ \emph {et~al.}(2009)\citenamefont {Gils},
  \citenamefont {Ardonne}, \citenamefont {Trebst}, \citenamefont {Ludwig},
  \citenamefont {Troyer},\ and\ \citenamefont {Wang}}]{InteractingAnyons1}%
  \BibitemOpen
  \bibfield  {author} {\bibinfo {author} {\bibfnamefont {Charlotte}\
  \bibnamefont {Gils}}, \bibinfo {author} {\bibfnamefont {Eddy}\ \bibnamefont
  {Ardonne}}, \bibinfo {author} {\bibfnamefont {Simon}\ \bibnamefont {Trebst}},
  \bibinfo {author} {\bibfnamefont {Andreas W.~W.}\ \bibnamefont {Ludwig}},
  \bibinfo {author} {\bibfnamefont {Matthias}\ \bibnamefont {Troyer}}, and\
  \bibinfo {author} {\bibfnamefont {Zhenghan}\ \bibnamefont {Wang}},\
  }\bibfield  {title} {\enquote {\bibinfo {title} {{Collective States of
  Interacting Anyons, Edge States, and the Nucleation of Topological
  Liquids}},}\ }\href {\doibase 10.1103/PhysRevLett.103.070401} {\bibfield
  {journal} {\bibinfo  {journal} {Phys. Rev. Lett.}\ }\textbf {\bibinfo
  {volume} {103}},\ \bibinfo {pages} {070401} (\bibinfo {year}
  {2009})}\BibitemShut {NoStop}%
\bibitem [{\citenamefont {Ludwig}\ \emph {et~al.}(2011)\citenamefont {Ludwig},
  \citenamefont {Poilblanc}, \citenamefont {Trebst},\ and\ \citenamefont
  {Troyer}}]{InteractingAnyons2}%
  \BibitemOpen
  \bibfield  {author} {\bibinfo {author} {\bibfnamefont {Andreas W.~W.}\
  \bibnamefont {Ludwig}}, \bibinfo {author} {\bibfnamefont {Didier}\
  \bibnamefont {Poilblanc}}, \bibinfo {author} {\bibfnamefont {Simon}\
  \bibnamefont {Trebst}}, and\ \bibinfo {author} {\bibfnamefont {Matthias}\
  \bibnamefont {Troyer}},\ }\bibfield  {title} {\enquote {\bibinfo {title}
  {{Two-dimensional quantum liquids from interacting non-Abelian anyons}},}\
  }\href {\doibase 10.1088/1367-2630/13/4/045014} {\bibfield  {journal}
  {\bibinfo  {journal} {New Journal of Physics}\ }\textbf {\bibinfo {volume}
  {13}},\ \bibinfo {pages} {045014} (\bibinfo {year} {2011})}\BibitemShut
  {NoStop}%
\bibitem [{\citenamefont {Qi}\ \emph {et~al.}(2010)\citenamefont {Qi},
  \citenamefont {Hughes},\ and\ \citenamefont {Zhang}}]{QAHproposal1}%
  \BibitemOpen
  \bibfield  {author} {\bibinfo {author} {\bibfnamefont {Xiao-Liang}\
  \bibnamefont {Qi}}, \bibinfo {author} {\bibfnamefont {Taylor~L.}\
  \bibnamefont {Hughes}}, and\ \bibinfo {author} {\bibfnamefont {Shou-Cheng}\
  \bibnamefont {Zhang}},\ }\bibfield  {title} {\enquote {\bibinfo {title}
  {{Chiral topological superconductor from the quantum Hall state}},}\ }\href
  {\doibase 10.1103/PhysRevB.82.184516} {\bibfield  {journal} {\bibinfo
  {journal} {Phys. Rev. B}\ }\textbf {\bibinfo {volume} {82}},\ \bibinfo
  {pages} {184516} (\bibinfo {year} {2010})}\BibitemShut {NoStop}%
\bibitem [{\citenamefont {Ii}\ \emph {et~al.}(2011)\citenamefont {Ii},
  \citenamefont {Yada}, \citenamefont {Sato},\ and\ \citenamefont
  {Tanaka}}]{QAHproposal2}%
  \BibitemOpen
  \bibfield  {author} {\bibinfo {author} {\bibfnamefont {Akihiro}\ \bibnamefont
  {Ii}}, \bibinfo {author} {\bibfnamefont {Keiji}\ \bibnamefont {Yada}},
  \bibinfo {author} {\bibfnamefont {Masatoshi}\ \bibnamefont {Sato}}, and\
  \bibinfo {author} {\bibfnamefont {Yukio}\ \bibnamefont {Tanaka}},\ }\bibfield
   {title} {\enquote {\bibinfo {title} {Theory of edge states in a quantum
  anomalous hall insulator/spin-singlet $s$-wave superconductor hybrid
  system},}\ }\href {\doibase 10.1103/PhysRevB.83.224524} {\bibfield  {journal}
  {\bibinfo  {journal} {Phys. Rev. B}\ }\textbf {\bibinfo {volume} {83}},\
  \bibinfo {pages} {224524} (\bibinfo {year} {2011})}\BibitemShut {NoStop}%
\bibitem [{\citenamefont {Kukushkin}\ \emph {et~al.}(1999)\citenamefont
  {Kukushkin}, \citenamefont {von Klitzing},\ and\ \citenamefont
  {Eberl}}]{SpinUnpolarizedStates}%
  \BibitemOpen
  \bibfield  {author} {\bibinfo {author} {\bibfnamefont {I.~V.}\ \bibnamefont
  {Kukushkin}}, \bibinfo {author} {\bibfnamefont {K.}~\bibnamefont {von
  Klitzing}}, and\ \bibinfo {author} {\bibfnamefont {K.}~\bibnamefont
  {Eberl}},\ }\bibfield  {title} {\enquote {\bibinfo {title} {{Spin
  Polarization of Composite Fermions: Measurements of the Fermi Energy}},}\
  }\href {\doibase 10.1103/PhysRevLett.82.3665} {\bibfield  {journal} {\bibinfo
   {journal} {Phys. Rev. Lett.}\ }\textbf {\bibinfo {volume} {82}},\ \bibinfo
  {pages} {3665--3668} (\bibinfo {year} {1999})}\BibitemShut {NoStop}%
\bibitem [{\citenamefont {Burrello}\ \emph {et~al.}(2013)\citenamefont
  {Burrello}, \citenamefont {van Heck},\ and\ \citenamefont
  {Cobanera}}]{ParafendleyonLattice}%
  \BibitemOpen
  \bibfield  {author} {\bibinfo {author} {\bibfnamefont {M.}~\bibnamefont
  {Burrello}}, \bibinfo {author} {\bibfnamefont {B.}~\bibnamefont {van Heck}},
  and\ \bibinfo {author} {\bibfnamefont {E.}~\bibnamefont {Cobanera}},\
  }\bibfield  {title} {\enquote {\bibinfo {title} {{Topological phases in
  two-dimensional arrays of parafermionic zero modes}},}\ }\href {\doibase%
  10.1103/PhysRevB.87.195422} {\bibfield  {journal} {\bibinfo  {journal} {Phys.
  Rev. B}\ }\textbf {\bibinfo {volume} {87}},\ \bibinfo {pages} {195422}
  (\bibinfo {year} {2013})}\BibitemShut {NoStop}%
\bibitem [{\citenamefont {Teo}\ and\ \citenamefont
  {Kane}(2011)}]{TeoKaneChains}%
  \BibitemOpen
  \bibfield  {author} {\bibinfo {author} {\bibfnamefont {Jeffrey C.~Y.}\
  \bibnamefont {Teo}}\ and\ \bibinfo {author} {\bibfnamefont {C.~L.}\
  \bibnamefont {Kane}},\ }\href@noop {} {\enquote {\bibinfo {title} {{From
  Luttinger liquid to non-Abelian quantum Hall states}},}\ } (\bibinfo {year}
  {2011}),\ \bibinfo {note} {unpublished},\ \Eprint
  {http://arxiv.org/abs/1111.2617} {arXiv:1111.2617 [cond-mat.mes-hall]}
  \BibitemShut {NoStop}%
\bibitem [{\citenamefont {Baraban}\ \emph {et~al.}(2010)\citenamefont
  {Baraban}, \citenamefont {Bonesteel},\ and\ \citenamefont
  {Simon}}]{Baraban10}%
  \BibitemOpen
  \bibfield  {author} {\bibinfo {author} {\bibfnamefont {M.}~\bibnamefont
  {Baraban}}, \bibinfo {author} {\bibfnamefont {N.~E.}\ \bibnamefont
  {Bonesteel}}, and\ \bibinfo {author} {\bibfnamefont {S.~H.}\ \bibnamefont
  {Simon}},\ }\bibfield  {title} {\enquote {\bibinfo {title} {{Resources
  required for topological quantum factoring}},}\ }\href {\doibase%
  10.1103/PhysRevA.81.062317} {\bibfield  {journal} {\bibinfo  {journal} {Phys.
  Rev. A}\ }\textbf {\bibinfo {volume} {81}},\ \bibinfo {pages} {062317}
  (\bibinfo {year} {2010})}\BibitemShut {NoStop}%
\bibitem [{\citenamefont {Bravyi}(2006)}]{Bravyi06}%
  \BibitemOpen
  \bibfield  {author} {\bibinfo {author} {\bibfnamefont {Sergey}\ \bibnamefont
  {Bravyi}},\ }\bibfield  {title} {\enquote {\bibinfo {title} {{Universal
  quantum computation with the $\nu=5/2$ fractional quantum Hall state}},}\
  }\href {\doibase 10.1103/PhysRevA.73.042313} {\bibfield  {journal} {\bibinfo
  {journal} {Phys. Rev. A}\ }\textbf {\bibinfo {volume} {73}},\ \bibinfo
  {pages} {042313} (\bibinfo {year} {2006})}\BibitemShut {NoStop}%
\bibitem [{Note4()}]{Note4}%
  \BibitemOpen
  \bibinfo {note} {This is dependent on the specific protocol, and the precise
  numbers will vary \cite {Svore}.}\BibitemShut {Stop}%
\bibitem [{\citenamefont {Asahi}\ and\ \citenamefont
  {Nagaosa}(2012)}]{CoupledChainsNagaosa}%
  \BibitemOpen
  \bibfield  {author} {\bibinfo {author} {\bibfnamefont {Daichi}\ \bibnamefont
  {Asahi}}\ and\ \bibinfo {author} {\bibfnamefont {Naoto}\ \bibnamefont
  {Nagaosa}},\ }\bibfield  {title} {\enquote {\bibinfo {title} {{Topological
  indices, defects, and Majorana fermions in chiral superconductors}},}\ }\href
  {\doibase 10.1103/PhysRevB.86.100504} {\bibfield  {journal} {\bibinfo
  {journal} {Phys. Rev. B}\ }\textbf {\bibinfo {volume} {86}},\ \bibinfo
  {pages} {100504} (\bibinfo {year} {2012})}\BibitemShut {NoStop}%
\bibitem [{Note5()}]{Note5}%
  \BibitemOpen
  \bibinfo {note} {If the quantum Hall edge states are completely
  spin-polarized then the superconductor should have a triplet component in
  order to achieve the desired proximity effect.}\BibitemShut {Stop}%
\bibitem [{\citenamefont {Ringel}\ \emph {et~al.}(2012)\citenamefont {Ringel},
  \citenamefont {Kraus},\ and\ \citenamefont
  {Stern}}]{RingelKrausStern:WTI:12}%
  \BibitemOpen
  \bibfield  {author} {\bibinfo {author} {\bibfnamefont {Zohar}\ \bibnamefont
  {Ringel}}, \bibinfo {author} {\bibfnamefont {Yaacov~E.}\ \bibnamefont
  {Kraus}}, and\ \bibinfo {author} {\bibfnamefont {Ady}\ \bibnamefont
  {Stern}},\ }\bibfield  {title} {\enquote {\bibinfo {title} {Strong side of
  weak topological insulators},}\ }\href {\doibase 10.1103/PhysRevB.86.045102}
  {\bibfield  {journal} {\bibinfo  {journal} {Phys. Rev. B}\ }\textbf {\bibinfo
  {volume} {86}},\ \bibinfo {pages} {045102} (\bibinfo {year}
  {2012})}\BibitemShut {NoStop}%
\bibitem [{\citenamefont {Mong}\ \emph {et~al.}(2012)\citenamefont {Mong},
  \citenamefont {Bardarson},\ and\ \citenamefont
  {Moore}}]{MongBardarsonMooreWTI12}%
  \BibitemOpen
  \bibfield  {author} {\bibinfo {author} {\bibfnamefont {Roger S.~K.}\
  \bibnamefont {Mong}}, \bibinfo {author} {\bibfnamefont {Jens~H.}\
  \bibnamefont {Bardarson}}, and\ \bibinfo {author} {\bibfnamefont {Joel~E.}\
  \bibnamefont {Moore}},\ }\bibfield  {title} {\enquote {\bibinfo {title}
  {{Quantum Transport and Two-Parameter Scaling at the Surface of a Weak
  Topological Insulator}},}\ }\href {\doibase 10.1103/PhysRevLett.108.076804}
  {\bibfield  {journal} {\bibinfo  {journal} {Phys. Rev. Lett.}\ }\textbf
  {\bibinfo {volume} {108}},\ \bibinfo {pages} {076804} (\bibinfo {year}
  {2012})}\BibitemShut {NoStop}%
\bibitem [{\citenamefont {Fu}\ and\ \citenamefont
  {Kane}(2012)}]{FuKaneSymplectic}%
  \BibitemOpen
  \bibfield  {author} {\bibinfo {author} {\bibfnamefont {Liang}\ \bibnamefont
  {Fu}}\ and\ \bibinfo {author} {\bibfnamefont {C.~L.}\ \bibnamefont {Kane}},\
  }\bibfield  {title} {\enquote {\bibinfo {title} {Topology, delocalization via
  average symmetry and the symplectic anderson transition},}\ }\href {\doibase%
  10.1103/PhysRevLett.109.246605} {\bibfield  {journal} {\bibinfo  {journal}
  {Phys. Rev. Lett.}\ }\textbf {\bibinfo {volume} {109}},\ \bibinfo {pages}
  {246605} (\bibinfo {year} {2012})}\BibitemShut {NoStop}%
\bibitem [{\citenamefont {Fulga}\ \emph {et~al.}(2012)\citenamefont {Fulga},
  \citenamefont {van Heck}, \citenamefont {Edge},\ and\ \citenamefont
  {Akhmerov}}]{FulgaAkhmerovEdge}%
  \BibitemOpen
  \bibfield  {author} {\bibinfo {author} {\bibfnamefont {I.~C.}\ \bibnamefont
  {Fulga}}, \bibinfo {author} {\bibfnamefont {B.}~\bibnamefont {van Heck}},
  \bibinfo {author} {\bibfnamefont {J.~M.}\ \bibnamefont {Edge}}, and\
  \bibinfo {author} {\bibfnamefont {A.~R.}\ \bibnamefont {Akhmerov}},\
  }\href@noop {} {\enquote {\bibinfo {title} {Statistical topological
  insulators},}\ } (\bibinfo {year} {2012}),\ \bibinfo {note} {unpublished},\
  \Eprint {http://arxiv.org/abs/1212.6191} {arXiv:1212.6191
  [cond-mat.mes-hall]} \BibitemShut {NoStop}%
\bibitem [{Note6()}]{Note6}%
  \BibitemOpen
  \bibinfo {note} {This state is also sometimes referred to as a `weak 2D
  topological superconductor', not to be confused with the weak pairing phase
  of a spinless 2D $p+ip$ superconductor.}\BibitemShut {Stop}%
\bibitem [{Note7()}]{Note7}%
  \BibitemOpen
  \bibinfo {note} {We specifically enforced an anti-unitary symmetry that sends
  $f_{R/L} \rightarrow i f_{L/R}$, which preserves Eqs.~\protect \textup {\hbox
  {\mathsurround \z@ \protect \normalfont (\ignorespaces \ref {H_KE}\unskip
  \@@italiccorr )}} and \protect \textup {\hbox {\mathsurround \z@ \protect
  \normalfont (\ignorespaces \ref {deltaH_Majorana}\unskip \@@italiccorr )}}.
  Consequently, the prefactors in front of the $f_R$ and $f_L$ hopping terms in
  Eq.~\protect \textup {\hbox {\mathsurround \z@ \protect \normalfont
  (\ignorespaces \ref {DiscreteHoppings}\unskip \@@italiccorr )}} must be
  complex conjugates, as written.}\BibitemShut {Stop}%
\bibitem [{Note8()}]{Note8}%
  \BibitemOpen
  \bibinfo {note} {The additional $\lambda _\perp '$ term reflects the reduced
  translation symmetry in the present setup. Consequently, the low-energy
  expansion for $f_{R/L}(y)$ involves both $\gamma _R(y)$ \protect \emph {and}
  $\gamma _L(y)$ in contrast to the uniform-trench system, so that more terms
  arise under projection.}\BibitemShut {Stop}%
\bibitem [{\citenamefont {Temperley}\ and\ \citenamefont
  {Lieb}(1971)}]{Temperley:1971}%
  \BibitemOpen
  \bibfield  {author} {\bibinfo {author} {\bibfnamefont {H.~N.~V.}\
  \bibnamefont {Temperley}}\ and\ \bibinfo {author} {\bibfnamefont {E.~H.}\
  \bibnamefont {Lieb}},\ }\bibfield  {title} {\enquote {\bibinfo {title}
  {{Relations between the `percolation' and `colouring' problem and other
  graph-theoretical problems associated with regular planar lattices: some
  exact results for the `percolation' problem}},}\ }\href
  {http://www.jstor.org/stable/77727} {\bibfield  {journal} {\bibinfo
  {journal} {Proc. Roy. Soc. Lond. A}\ }\textbf {\bibinfo {volume} {322}},\
  \bibinfo {pages} {251--280} (\bibinfo {year} {1971})}\BibitemShut {NoStop}%
\bibitem [{\citenamefont {Zamolodchikov}\ and\ \citenamefont
  {Fateev}(1985)}]{ZamolodchikovParafermion}%
  \BibitemOpen
  \bibfield  {author} {\bibinfo {author} {\bibfnamefont {A.~B.}\ \bibnamefont
  {Zamolodchikov}}\ and\ \bibinfo {author} {\bibfnamefont {V.}~\bibnamefont
  {Fateev}},\ }\bibfield  {title} {\enquote {\bibinfo {title} {{Nonlocal
  (parafermion) currents in two-dimensional conformal quantum field theory and
  self-dual critical points in $\mathbb{Z}_N$-symmetric statistical
  systems}},}\ }\href
  {http://www.jetp.ac.ru/cgi-bin/e/index/e/62/2/p215?a=list} {\bibfield
  {journal} {\bibinfo  {journal} {JETP}\ }\textbf {\bibinfo {volume} {62}},\
  \bibinfo {pages} {215--225} (\bibinfo {year} {1985})},\ \bibinfo {note} {[Zh.
  Eksp. Teor. Fiz., {\bf 89}, 380--399]}\BibitemShut {NoStop}%
\bibitem [{\citenamefont {Fradkin}\ and\ \citenamefont
  {Kadanoff}(1980)}]{FradkinKadanoff}%
  \BibitemOpen
  \bibfield  {author} {\bibinfo {author} {\bibfnamefont {E.}~\bibnamefont
  {Fradkin}}\ and\ \bibinfo {author} {\bibfnamefont {L.~P.}\ \bibnamefont
  {Kadanoff}},\ }\bibfield  {title} {\enquote {\bibinfo {title} {Disorder
  variables and parafermions in two-dimensional statistical mechanics.}}\
  }\href {\doibase 10.1016/0550-3213(80)90472-1} {\bibfield  {journal}
  {\bibinfo  {journal} {Nucl. Phys. B}\ }\textbf {\bibinfo {volume} {170}},\
  \bibinfo {pages} {1} (\bibinfo {year} {1980})}\BibitemShut {NoStop}%
\bibitem [{\citenamefont {Fateev}\ and\ \citenamefont
  {Zamolodchikov}(1987)}]{FateevZamo:W3:87}%
  \BibitemOpen
  \bibfield  {author} {\bibinfo {author} {\bibfnamefont {V.~A.}\ \bibnamefont
  {Fateev}}\ and\ \bibinfo {author} {\bibfnamefont {A.~B.}\ \bibnamefont
  {Zamolodchikov}},\ }\bibfield  {title} {\enquote {\bibinfo {title}
  {{Conformal quantum field theory models in two dimensions having
  $\mathbb{Z}_3$ symmetry}},}\ }\href {\doibase 10.1016/0550-3213(87)90166-0}
  {\bibfield  {journal} {\bibinfo  {journal} {Nucl. Phys. B}\ }\textbf
  {\bibinfo {volume} {280}},\ \bibinfo {pages} {644--660} (\bibinfo {year}
  {1987})}\BibitemShut {NoStop}%
\bibitem [{\citenamefont {Fateev}(1991)}]{Fateev1990}%
  \BibitemOpen
  \bibfield  {author} {\bibinfo {author} {\bibfnamefont {V.~A.}\ \bibnamefont
  {Fateev}},\ }\bibfield  {title} {\enquote {\bibinfo {title} {{Integrable
  deformations in $\mathbb{Z}_N$ symmetrical models of conformal quantum field
  theory}},}\ }\href {\doibase 10.1142/S0217751X91001052} {\bibfield  {journal}
  {\bibinfo  {journal} {Int. J. Mod. Phys.}\ }\textbf {\bibinfo {volume}
  {A6}},\ \bibinfo {pages} {2109--2132} (\bibinfo {year} {1991})}\BibitemShut
  {NoStop}%
\bibitem [{\citenamefont {Fateev}\ and\ \citenamefont
  {Zamolodchikov}(1991)}]{FateevZamo1991}%
  \BibitemOpen
  \bibfield  {author} {\bibinfo {author} {\bibfnamefont {V.~A.}\ \bibnamefont
  {Fateev}}\ and\ \bibinfo {author} {\bibfnamefont {A.~B.}\ \bibnamefont
  {Zamolodchikov}},\ }\bibfield  {title} {\enquote {\bibinfo {title}
  {{Integrable perturbations of $\mathbb{Z}_N$ parafermion models and $O(3)$
  sigma model}},}\ }\href {\doibase 10.1016/0370-2693(91)91283-2} {\bibfield
  {journal} {\bibinfo  {journal} {Phys. Lett.}\ }\textbf {\bibinfo {volume}
  {B271}},\ \bibinfo {pages} {91--100} (\bibinfo {year} {1991})}\BibitemShut
  {NoStop}%
\bibitem [{\citenamefont {Mong}\ \emph {et~al.}(2013)\citenamefont {Mong},
  \citenamefont {Clarke}, \citenamefont {Alicea}, \citenamefont {Lindner},\
  and\ \citenamefont {Fendley}}]{LatticeCFTrelation}%
  \BibitemOpen
  \bibfield  {author} {\bibinfo {author} {\bibfnamefont {Roger S.~K.}\
  \bibnamefont {Mong}}, \bibinfo {author} {\bibfnamefont {David~J.}\
  \bibnamefont {Clarke}}, \bibinfo {author} {\bibfnamefont {Jason}\
  \bibnamefont {Alicea}}, \bibinfo {author} {\bibfnamefont {Netanel~H.}\
  \bibnamefont {Lindner}}, and\ \bibinfo {author} {\bibfnamefont {Paul}\
  \bibnamefont {Fendley}},\ }\href@noop {} {} (\bibinfo {year} {2013}),\
  \bibinfo {note} {in preparation}\BibitemShut {NoStop}%
\bibitem [{\citenamefont {Kane}\ \emph {et~al.}(1994)\citenamefont {Kane},
  \citenamefont {Fisher},\ and\ \citenamefont
  {Polchinski}}]{KaneFisherPolchinski}%
  \BibitemOpen
  \bibfield  {author} {\bibinfo {author} {\bibfnamefont {C.~L.}\ \bibnamefont
  {Kane}}, \bibinfo {author} {\bibfnamefont {Matthew P.~A.}\ \bibnamefont
  {Fisher}}, and\ \bibinfo {author} {\bibfnamefont {J.}~\bibnamefont
  {Polchinski}},\ }\bibfield  {title} {\enquote {\bibinfo {title} {{Randomness
  at the edge: Theory of quantum Hall transport at filling $\nu=2/3$}},}\
  }\href {\doibase 10.1103/PhysRevLett.72.4129} {\bibfield  {journal} {\bibinfo
   {journal} {Phys. Rev. Lett.}\ }\textbf {\bibinfo {volume} {72}},\ \bibinfo
  {pages} {4129--4132} (\bibinfo {year} {1994})}\BibitemShut {NoStop}%
\bibitem [{\citenamefont {Wen}(2004)}]{WenBook}%
  \BibitemOpen
  \bibfield  {author} {\bibinfo {author} {\bibfnamefont {Xiao-Gang}\
  \bibnamefont {Wen}},\ }\href@noop {} {\emph {\bibinfo {title} {{Quantum Field
  Theory of Many-Body Systems}}}},\ Oxford Graduate Texts\ (\bibinfo
  {publisher} {Oxford University Press},\ \bibinfo {address} {Oxford},\
  \bibinfo {year} {2004})\BibitemShut {NoStop}%
\bibitem [{\citenamefont {Kane}\ and\ \citenamefont
  {Fisher}(1995)}]{KaneFisherPRB}%
  \BibitemOpen
  \bibfield  {author} {\bibinfo {author} {\bibfnamefont {C.~L.}\ \bibnamefont
  {Kane}}\ and\ \bibinfo {author} {\bibfnamefont {Matthew P.~A.}\ \bibnamefont
  {Fisher}},\ }\bibfield  {title} {\enquote {\bibinfo {title} {{Impurity
  scattering and transport of fractional quantum Hall edge states}},}\ }\href
  {\doibase 10.1103/PhysRevB.51.13449} {\bibfield  {journal} {\bibinfo
  {journal} {Phys. Rev. B}\ }\textbf {\bibinfo {volume} {51}},\ \bibinfo
  {pages} {13449--13466} (\bibinfo {year} {1995})}\BibitemShut {NoStop}%
\bibitem [{Note9()}]{Note9}%
  \BibitemOpen
  \bibinfo {note} {By itself this does not necessarily imply that the phases
  generated by the tunneling and pairing terms are distinct, but it turns out
  that this is the case here.}\BibitemShut {Stop}%
\bibitem [{\citenamefont {Barkeshli}\ \emph
  {et~al.}(2013{\natexlab{b}})\citenamefont {Barkeshli}, \citenamefont {Jian},\
  and\ \citenamefont {Qi}}]{BarkeshliClassification1}%
  \BibitemOpen
  \bibfield  {author} {\bibinfo {author} {\bibfnamefont {Maissam}\ \bibnamefont
  {Barkeshli}}, \bibinfo {author} {\bibfnamefont {Chao-Ming}\ \bibnamefont
  {Jian}}, and\ \bibinfo {author} {\bibfnamefont {Xiao-Liang}\ \bibnamefont
  {Qi}},\ }\href@noop {} {\enquote {\bibinfo {title} {{Classification of
  Topological Defects in Abelian Topological States}},}\ } (\bibinfo {year}
  {2013}{\natexlab{b}}),\ \bibinfo {note} {unpublished},\ \Eprint
  {http://arxiv.org/abs/1304.7579} {arXiv:1304.7579 [cond-mat.str-el]}
  \BibitemShut {NoStop}%
\bibitem [{\citenamefont {Barkeshli}\ \emph
  {et~al.}(2013{\natexlab{c}})\citenamefont {Barkeshli}, \citenamefont {Jian},\
  and\ \citenamefont {Qi}}]{BarkeshliClassification2}%
  \BibitemOpen
  \bibfield  {author} {\bibinfo {author} {\bibfnamefont {Maissam}\ \bibnamefont
  {Barkeshli}}, \bibinfo {author} {\bibfnamefont {Chao-Ming}\ \bibnamefont
  {Jian}}, and\ \bibinfo {author} {\bibfnamefont {Xiao-Liang}\ \bibnamefont
  {Qi}},\ }\href@noop {} {\enquote {\bibinfo {title} {{Theory of defects in
  Abelian topological states}},}\ } (\bibinfo {year} {2013}{\natexlab{c}}),\
  \bibinfo {note} {unpublished},\ \Eprint {http://arxiv.org/abs/1305.7203}
  {arXiv:1305.7203 [cond-mat.str-el]} \BibitemShut {NoStop}%
\bibitem [{Note10()}]{Note10}%
  \BibitemOpen
  \bibinfo {note} {A discussion of a finite, closed ring of alternating domains
  can be found in Ref.~\protect \rev@citealp
  {LindnerParafendleyons}.}\BibitemShut {Stop}%
\bibitem [{Note11()}]{Note11}%
  \BibitemOpen
  \bibinfo {note} {Note that $\protect \mathaccentV {hat}05E{M}$ does not
  produce additional ground-state degeneracy since there is no gauge invariant
  quantity one can construct from this operator.}\BibitemShut {Stop}%
\bibitem [{\citenamefont {Nilsson}\ \emph {et~al.}(2008)\citenamefont
  {Nilsson}, \citenamefont {Akhmerov},\ and\ \citenamefont
  {Beenakker}}]{dual_FJE_Nilsson}%
  \BibitemOpen
  \bibfield  {author} {\bibinfo {author} {\bibfnamefont {Johan}\ \bibnamefont
  {Nilsson}}, \bibinfo {author} {\bibfnamefont {A.~R.}\ \bibnamefont
  {Akhmerov}}, and\ \bibinfo {author} {\bibfnamefont {C.~W.~J.}\ \bibnamefont
  {Beenakker}},\ }\bibfield  {title} {\enquote {\bibinfo {title} {{Splitting of
  a Cooper Pair by a Pair of Majorana Bound States}},}\ }\href {\doibase%
  10.1103/PhysRevLett.101.120403} {\bibfield  {journal} {\bibinfo  {journal}
  {Phys. Rev. Lett.}\ }\textbf {\bibinfo {volume} {101}},\ \bibinfo {pages}
  {120403} (\bibinfo {year} {2008})}\BibitemShut {NoStop}%
\bibitem [{\citenamefont {Meng}\ \emph {et~al.}(2012)\citenamefont {Meng},
  \citenamefont {Shivamoggi}, \citenamefont {Hughes}, \citenamefont {Gilbert},\
  and\ \citenamefont {Vishveshwara}}]{dual_FJE_Meng}%
  \BibitemOpen
  \bibfield  {author} {\bibinfo {author} {\bibfnamefont {Qinglei}\ \bibnamefont
  {Meng}}, \bibinfo {author} {\bibfnamefont {Vasudha}\ \bibnamefont
  {Shivamoggi}}, \bibinfo {author} {\bibfnamefont {Taylor~L.}\ \bibnamefont
  {Hughes}}, \bibinfo {author} {\bibfnamefont {Matthew~J.}\ \bibnamefont
  {Gilbert}}, and\ \bibinfo {author} {\bibfnamefont {Smitha}\ \bibnamefont
  {Vishveshwara}},\ }\bibfield  {title} {\enquote {\bibinfo {title}
  {{Fractional spin Josephson effect and electrically controlled magnetization
  in quantum spin Hall edges}},}\ }\href {\doibase 10.1103/PhysRevB.86.165110}
  {\bibfield  {journal} {\bibinfo  {journal} {Phys. Rev. B}\ }\textbf {\bibinfo
  {volume} {86}},\ \bibinfo {pages} {165110} (\bibinfo {year}
  {2012})}\BibitemShut {NoStop}%
\bibitem [{\citenamefont {Jiang}\ \emph {et~al.}(2013)\citenamefont {Jiang},
  \citenamefont {Pekker}, \citenamefont {Alicea}, \citenamefont {Refael},
  \citenamefont {Oreg}, \citenamefont {Brataas},\ and\ \citenamefont {von
  Oppen}}]{dual_FJE_Jiang}%
  \BibitemOpen
  \bibfield  {author} {\bibinfo {author} {\bibfnamefont {Liang}\ \bibnamefont
  {Jiang}}, \bibinfo {author} {\bibfnamefont {David}\ \bibnamefont {Pekker}},
  \bibinfo {author} {\bibfnamefont {Jason}\ \bibnamefont {Alicea}}, \bibinfo
  {author} {\bibfnamefont {Gil}\ \bibnamefont {Refael}}, \bibinfo {author}
  {\bibfnamefont {Yuval}\ \bibnamefont {Oreg}}, \bibinfo {author}
  {\bibfnamefont {Arne}\ \bibnamefont {Brataas}}, and\ \bibinfo {author}
  {\bibfnamefont {Felix}\ \bibnamefont {von Oppen}},\ }\bibfield  {title}
  {\enquote {\bibinfo {title} {{Magneto-Josephson effects in junctions with
  Majorana bound states}},}\ }\href {\doibase 10.1103/PhysRevB.87.075438}
  {\bibfield  {journal} {\bibinfo  {journal} {Phys. Rev. B}\ }\textbf {\bibinfo
  {volume} {87}},\ \bibinfo {pages} {075438} (\bibinfo {year}
  {2013})}\BibitemShut {NoStop}%
\bibitem [{\citenamefont {Kotetes}\ \emph {et~al.}(2013)\citenamefont
  {Kotetes}, \citenamefont {Sch{\"o}n},\ and\ \citenamefont
  {Shnirman}}]{dual_FJE_Panagiotis}%
  \BibitemOpen
  \bibfield  {author} {\bibinfo {author} {\bibfnamefont {Panagiotis}\
  \bibnamefont {Kotetes}}, \bibinfo {author} {\bibfnamefont {Gerd}\
  \bibnamefont {Sch{\"o}n}}, and\ \bibinfo {author} {\bibfnamefont
  {Alexander}\ \bibnamefont {Shnirman}},\ }\bibfield  {title} {\enquote
  {\bibinfo {title} {{Engineering and manipulating topological qubits in 1D
  quantum wires}},}\ }\href {\doibase 10.3938/jkps.62.1558} {\bibfield
  {journal} {\bibinfo  {journal} {Journal of the Korean Physical Society}\
  }\textbf {\bibinfo {volume} {62}},\ \bibinfo {pages} {1558--1563} (\bibinfo
  {year} {2013})}\BibitemShut {NoStop}%
\bibitem [{Note12()}]{Note12}%
  \BibitemOpen
  \bibinfo {note} {Tunneling of $e/3$ (rather than $2e/3$) charge can also in
  principle arise. However, those processes have no effect in the low-energy
  subspace in which we are working since such tunneling operators project
  trivially.}\BibitemShut {Stop}%
\bibitem [{\citenamefont {Ludwig}\ and\ \citenamefont
  {Cardy}(1987)}]{LudwigCardy}%
  \BibitemOpen
  \bibfield  {author} {\bibinfo {author} {\bibfnamefont {A.~W.~W.}\
  \bibnamefont {Ludwig}}\ and\ \bibinfo {author} {\bibfnamefont {John~L.}\
  \bibnamefont {Cardy}},\ }\bibfield  {title} {\enquote {\bibinfo {title}
  {{Perturbative Evaluation of the Conformal Anomaly at New Critical Points
  with Applications to Random Systems}},}\ }\href {\doibase%
  10.1016/0550-3213(87)90362-2} {\bibfield  {journal} {\bibinfo  {journal}
  {Nucl. Phys.}\ }\textbf {\bibinfo {volume} {B285}},\ \bibinfo {pages}
  {687--718} (\bibinfo {year} {1987})}\BibitemShut {NoStop}%
\bibitem [{\citenamefont {Smirnov}(1991)}]{Smirnov1991}%
  \BibitemOpen
  \bibfield  {author} {\bibinfo {author} {\bibfnamefont {F.~A.}\ \bibnamefont
  {Smirnov}},\ }\bibfield  {title} {\enquote {\bibinfo {title} {{Exact
  S-matrices for $\phi_{1,2}$-perturbated minimal models of conformal field
  theory}},}\ }\href {\doibase 10.1142/S0217751X91000745} {\bibfield  {journal}
  {\bibinfo  {journal} {Int. J. Mod. Phys.}\ }\textbf {\bibinfo {volume}
  {A6}},\ \bibinfo {pages} {1407--1428} (\bibinfo {year} {1991})}\BibitemShut
  {NoStop}%
\bibitem [{\citenamefont {Reshetikhin}\ and\ \citenamefont
  {Smirnov}(1990)}]{ReshetikhinSmirnov}%
  \BibitemOpen
  \bibfield  {author} {\bibinfo {author} {\bibfnamefont {N.}~\bibnamefont
  {Reshetikhin}}\ and\ \bibinfo {author} {\bibfnamefont {F.}~\bibnamefont
  {Smirnov}},\ }\bibfield  {title} {\enquote {\bibinfo {title} {{Hidden quantum
  group symmetry and integrable perturbations of conformal field theories}},}\
  }\href {\doibase 10.1007/BF02097683} {\bibfield  {journal} {\bibinfo
  {journal} {Commun. Math. Phys.}\ }\textbf {\bibinfo {volume} {131}},\
  \bibinfo {pages} {157--178} (\bibinfo {year} {1990})}\BibitemShut {NoStop}%
\bibitem [{\citenamefont {Fendley}\ and\ \citenamefont
  {Fradkin}(2005)}]{FendleyFradkin}%
  \BibitemOpen
  \bibfield  {author} {\bibinfo {author} {\bibfnamefont {Paul}\ \bibnamefont
  {Fendley}}\ and\ \bibinfo {author} {\bibfnamefont {Eduardo}\ \bibnamefont
  {Fradkin}},\ }\bibfield  {title} {\enquote {\bibinfo {title} {{Realizing
  non-Abelian statistics}},}\ }\href {\doibase 10.1103/PhysRevB.72.024412}
  {\bibfield  {journal} {\bibinfo  {journal} {Phys. Rev.}\ }\textbf {\bibinfo
  {volume} {B72}},\ \bibinfo {pages} {024412} (\bibinfo {year} {2005})},\
  \Eprint {http://arxiv.org/abs/cond-mat/0502071} {arXiv:cond-mat/0502071
  [cond-mat]} \BibitemShut {NoStop}%
\bibitem [{\citenamefont {Yurov}\ and\ \citenamefont
  {Zamolodchikov}(1990)}]{YurovZamolodchikov1990}%
  \BibitemOpen
  \bibfield  {author} {\bibinfo {author} {\bibfnamefont {V.~P.}\ \bibnamefont
  {Yurov}}\ and\ \bibinfo {author} {\bibfnamefont {Al.~B.}\ \bibnamefont
  {Zamolodchikov}},\ }\bibfield  {title} {\enquote {\bibinfo {title}
  {{Truncated conformal space approach to scaling Lee-Yang model}},}\ }\href
  {\doibase 10.1142/S0217751X9000218X} {\bibfield  {journal} {\bibinfo
  {journal} {International Journal of Modern Physics A}\ }\textbf {\bibinfo
  {volume} {05}},\ \bibinfo {pages} {3221--3245} (\bibinfo {year}
  {1990})}\BibitemShut {NoStop}%
\bibitem [{\citenamefont {Yurov}\ and\ \citenamefont
  {Zamolodchikov}(1991)}]{YurovZamolodchikov1991}%
  \BibitemOpen
  \bibfield  {author} {\bibinfo {author} {\bibfnamefont {V.~P.}\ \bibnamefont
  {Yurov}}\ and\ \bibinfo {author} {\bibfnamefont {Al.~B.}\ \bibnamefont
  {Zamolodchikov}},\ }\bibfield  {title} {\enquote {\bibinfo {title}
  {{Truncated-fermionic-space approach to the critical 2D Ising model with
  magnetic field}},}\ }\href {\doibase 10.1142/S0217751X91002161} {\bibfield
  {journal} {\bibinfo  {journal} {International Journal of Modern Physics A}\
  }\textbf {\bibinfo {volume} {06}},\ \bibinfo {pages} {4557--4578} (\bibinfo
  {year} {1991})}\BibitemShut {NoStop}%
\bibitem [{\citenamefont {Lassig}\ \emph {et~al.}(1991)\citenamefont {Lassig},
  \citenamefont {Mussardo},\ and\ \citenamefont {Cardy}}]{Lassig1990}%
  \BibitemOpen
  \bibfield  {author} {\bibinfo {author} {\bibfnamefont {Michael}\ \bibnamefont
  {Lassig}}, \bibinfo {author} {\bibfnamefont {Giuseppe}\ \bibnamefont
  {Mussardo}}, and\ \bibinfo {author} {\bibfnamefont {John~L.}\ \bibnamefont
  {Cardy}},\ }\bibfield  {title} {\enquote {\bibinfo {title} {{The scaling
  region of the tricritical Ising model in two-dimensions}},}\ }\href {\doibase%
  10.1016/0550-3213(91)90206-D} {\bibfield  {journal} {\bibinfo  {journal}
  {Nucl. Phys.}\ }\textbf {\bibinfo {volume} {B348}},\ \bibinfo {pages}
  {591--618} (\bibinfo {year} {1991})}\BibitemShut {NoStop}%
\bibitem [{\citenamefont {Zamolodchikov}(1990)}]{Zamolodchikov:1990xc}%
  \BibitemOpen
  \bibfield  {author} {\bibinfo {author} {\bibfnamefont {A.~B.}\ \bibnamefont
  {Zamolodchikov}},\ }\bibfield  {title} {\enquote {\bibinfo {title} {{S matrix
  of the subleading magnetic perturbation of the tricritical Ising model}},}\
  }\href@noop {} {\bibfield  {journal} {\bibinfo  {journal} {Princeton
  Preprint}\ } (\bibinfo {year} {1990})}\BibitemShut {NoStop}%
\bibitem [{\citenamefont {Cardy}(1986)}]{CardyBoundaryOp:1986}%
  \BibitemOpen
  \bibfield  {author} {\bibinfo {author} {\bibfnamefont {John~L.}\ \bibnamefont
  {Cardy}},\ }\bibfield  {title} {\enquote {\bibinfo {title} {{Effect of
  boundary conditions on the operator content of two-dimensional conformally
  invariant theories}},}\ }\href {\doibase 10.1016/0550-3213(86)90596-1}
  {\bibfield  {journal} {\bibinfo  {journal} {Nucl. Phys. B}\ }\textbf
  {\bibinfo {volume} {275}},\ \bibinfo {pages} {200--218} (\bibinfo {year}
  {1986})}\BibitemShut {NoStop}%
\bibitem [{\citenamefont {Zhang}\ \emph {et~al.}(2012)\citenamefont {Zhang},
  \citenamefont {Grover}, \citenamefont {Turner}, \citenamefont {Oshikawa},\
  and\ \citenamefont {Vishwanath}}]{YZhangVishwanathMES}%
  \BibitemOpen
  \bibfield  {author} {\bibinfo {author} {\bibfnamefont {Yi}~\bibnamefont
  {Zhang}}, \bibinfo {author} {\bibfnamefont {Tarun}\ \bibnamefont {Grover}},
  \bibinfo {author} {\bibfnamefont {Ari}\ \bibnamefont {Turner}}, \bibinfo
  {author} {\bibfnamefont {Masaki}\ \bibnamefont {Oshikawa}}, and\ \bibinfo
  {author} {\bibfnamefont {Ashvin}\ \bibnamefont {Vishwanath}},\ }\bibfield
  {title} {\enquote {\bibinfo {title} {Quasiparticle statistics and braiding
  from ground-state entanglement},}\ }\href {\doibase%
  10.1103/PhysRevB.85.235151} {\bibfield  {journal} {\bibinfo  {journal} {Phys.
  Rev. B}\ }\textbf {\bibinfo {volume} {85}},\ \bibinfo {pages} {235151}
  (\bibinfo {year} {2012})}\BibitemShut {NoStop}%
\bibitem [{\citenamefont {Kitaev}\ and\ \citenamefont
  {Preskill}(2006)}]{KitaevPreskill}%
  \BibitemOpen
  \bibfield  {author} {\bibinfo {author} {\bibfnamefont {Alexei}\ \bibnamefont
  {Kitaev}}\ and\ \bibinfo {author} {\bibfnamefont {John}\ \bibnamefont
  {Preskill}},\ }\bibfield  {title} {\enquote {\bibinfo {title} {{Topological
  Entanglement Entropy}},}\ }\href {\doibase 10.1103/PhysRevLett.96.110404}
  {\bibfield  {journal} {\bibinfo  {journal} {Phys. Rev. Lett.}\ }\textbf
  {\bibinfo {volume} {96}},\ \bibinfo {pages} {110404} (\bibinfo {year}
  {2006})}\BibitemShut {NoStop}%
\bibitem [{\citenamefont {Levin}\ and\ \citenamefont
  {Wen}(2006)}]{LevinWen-2006}%
  \BibitemOpen
  \bibfield  {author} {\bibinfo {author} {\bibfnamefont {Michael}\ \bibnamefont
  {Levin}}\ and\ \bibinfo {author} {\bibfnamefont {Xiao-Gang}\ \bibnamefont
  {Wen}},\ }\bibfield  {title} {\enquote {\bibinfo {title} {{Detecting
  Topological Order in a Ground State Wave Function}},}\ }\href {\doibase%
  10.1103/PhysRevLett.96.110405} {\bibfield  {journal} {\bibinfo  {journal}
  {Phys. Rev. Lett.}\ }\textbf {\bibinfo {volume} {96}},\ \bibinfo {pages}
  {110405} (\bibinfo {year} {2006})}\BibitemShut {NoStop}%
\bibitem [{\citenamefont {Dong}\ \emph {et~al.}(2008)\citenamefont {Dong},
  \citenamefont {Fradkin}, \citenamefont {Leigh},\ and\ \citenamefont
  {Nowling}}]{Dong-2008}%
  \BibitemOpen
  \bibfield  {author} {\bibinfo {author} {\bibfnamefont {Shiying}\ \bibnamefont
  {Dong}}, \bibinfo {author} {\bibfnamefont {Eduardo}\ \bibnamefont {Fradkin}},
  \bibinfo {author} {\bibfnamefont {Robert~G.}\ \bibnamefont {Leigh}}, and\
  \bibinfo {author} {\bibfnamefont {Sean}\ \bibnamefont {Nowling}},\ }\bibfield
   {title} {\enquote {\bibinfo {title} {{Topological entanglement entropy in
  Chern-Simons theories and quantum Hall fluids}},}\ }\href {\doibase%
  10.1088/1126-6708/2008/05/016} {\bibfield  {journal} {\bibinfo  {journal}
  {Journal of High Energy Physics}\ }\textbf {\bibinfo {volume} {2008}},\
  \bibinfo {pages} {016} (\bibinfo {year} {2008})}\BibitemShut {NoStop}%
\bibitem [{\citenamefont {Affleck}\ \emph {et~al.}(1987)\citenamefont
  {Affleck}, \citenamefont {Kennedy}, \citenamefont {Lieb},\ and\ \citenamefont
  {Tasaki}}]{AKLT:1987}%
  \BibitemOpen
  \bibfield  {author} {\bibinfo {author} {\bibfnamefont {Ian}\ \bibnamefont
  {Affleck}}, \bibinfo {author} {\bibfnamefont {Tom}\ \bibnamefont {Kennedy}},
  \bibinfo {author} {\bibfnamefont {Elliott~H.}\ \bibnamefont {Lieb}}, and\
  \bibinfo {author} {\bibfnamefont {Hal}\ \bibnamefont {Tasaki}},\ }\bibfield
  {title} {\enquote {\bibinfo {title} {Rigorous results on valence-bond ground
  states in antiferromagnets},}\ }\href {\doibase 10.1103/PhysRevLett.59.799}
  {\bibfield  {journal} {\bibinfo  {journal} {Phys. Rev. Lett.}\ }\textbf
  {\bibinfo {volume} {59}},\ \bibinfo {pages} {799--802} (\bibinfo {year}
  {1987})}\BibitemShut {NoStop}%
\bibitem [{\citenamefont {{Rowell}}\ \emph {et~al.}(2009)\citenamefont
  {{Rowell}}, \citenamefont {{Stong}},\ and\ \citenamefont
  {{Wang}}}]{Rowell09}%
  \BibitemOpen
  \bibfield  {author} {\bibinfo {author} {\bibfnamefont {E.}~\bibnamefont
  {{Rowell}}}, \bibinfo {author} {\bibfnamefont {R.}~\bibnamefont {{Stong}}},
  and\ \bibinfo {author} {\bibfnamefont {Z.}~\bibnamefont {{Wang}}},\
  }\bibfield  {title} {\enquote {\bibinfo {title} {{On classification of
  modular tensor categories}},}\ }\href {\doibase 10.1007/s00220-009-0908-z}
  {\bibfield  {journal} {\bibinfo  {journal} {Commun. Math. Phys.}\ }\textbf
  {\bibinfo {volume} {292}},\ \bibinfo {pages} {343--389} (\bibinfo {year}
  {2009})}\BibitemShut {NoStop}%
\bibitem [{\citenamefont {{Trebst}}\ \emph {et~al.}(2008)\citenamefont
  {{Trebst}}, \citenamefont {{Troyer}}, \citenamefont {{Wang}},\ and\
  \citenamefont {{Ludwig}}}]{Trebst08}%
  \BibitemOpen
  \bibfield  {author} {\bibinfo {author} {\bibfnamefont {S.}~\bibnamefont
  {{Trebst}}}, \bibinfo {author} {\bibfnamefont {M.}~\bibnamefont {{Troyer}}},
  \bibinfo {author} {\bibfnamefont {Z.}~\bibnamefont {{Wang}}}, and\ \bibinfo
  {author} {\bibfnamefont {A.~W.~W.}\ \bibnamefont {{Ludwig}}},\ }\bibfield
  {title} {\enquote {\bibinfo {title} {{A short introduction to Fibonacci anyon
  models}},}\ }\href {\doibase 10.1143/PTPS.176.384} {\bibfield  {journal}
  {\bibinfo  {journal} {Prog. Theor. Phys. Supp.}\ }\textbf {\bibinfo {volume}
  {176}},\ \bibinfo {pages} {384} (\bibinfo {year} {2008})}\BibitemShut
  {NoStop}%
\bibitem [{Note13()}]{Note13}%
  \BibitemOpen
  \bibinfo {note} {The term `Fibonacci anyon' is often used for the
  $\varepsilon $-particle and for the phase that supports it; the meaning is
  usually clear from the context.}\BibitemShut {Stop}%
\bibitem [{\citenamefont {Freedman}\ \emph
  {et~al.}(2002{\natexlab{a}})\citenamefont {Freedman}, \citenamefont
  {Larsen},\ and\ \citenamefont {Wang}}]{Freedman02a}%
  \BibitemOpen
  \bibfield  {author} {\bibinfo {author} {\bibfnamefont {Michael~H.}\
  \bibnamefont {Freedman}}, \bibinfo {author} {\bibfnamefont {Michael~J.}\
  \bibnamefont {Larsen}}, and\ \bibinfo {author} {\bibfnamefont {Zhenghan}\
  \bibnamefont {Wang}},\ }\bibfield  {title} {\enquote {\bibinfo {title} {{A
  Modular Functor Which is Universal for Quantum Computation}},}\ }\href
  {\doibase 10.1007/s002200200645} {\bibfield  {journal} {\bibinfo  {journal}
  {Commun. Math. Phys.}\ }\textbf {\bibinfo {volume} {227}},\ \bibinfo {pages}
  {605--622} (\bibinfo {year} {2002}{\natexlab{a}})}\BibitemShut {NoStop}%
\bibitem [{\citenamefont {Freedman}\ \emph
  {et~al.}(2002{\natexlab{b}})\citenamefont {Freedman}, \citenamefont
  {Larsen},\ and\ \citenamefont {Wang}}]{Freedman02b}%
  \BibitemOpen
  \bibfield  {author} {\bibinfo {author} {\bibfnamefont {Michael~H.}\
  \bibnamefont {Freedman}}, \bibinfo {author} {\bibfnamefont {Michael~J.}\
  \bibnamefont {Larsen}}, and\ \bibinfo {author} {\bibfnamefont {Zhenghan}\
  \bibnamefont {Wang}},\ }\bibfield  {title} {\enquote {\bibinfo {title} {{The
  Two-Eigenvalue Problem and Density of Jones Representation of Braid
  Groups}},}\ }\href {\doibase 10.1007/s002200200636} {\bibfield  {journal}
  {\bibinfo  {journal} {Commun. Math. Phys.}\ }\textbf {\bibinfo {volume}
  {228}},\ \bibinfo {pages} {177--199} (\bibinfo {year}
  {2002}{\natexlab{b}})}\BibitemShut {NoStop}%
\bibitem [{\citenamefont {Stone}\ and\ \citenamefont {Roy}(2004)}]{StoneRoy}%
  \BibitemOpen
  \bibfield  {author} {\bibinfo {author} {\bibfnamefont {Michael}\ \bibnamefont
  {Stone}}\ and\ \bibinfo {author} {\bibfnamefont {Rahul}\ \bibnamefont
  {Roy}},\ }\bibfield  {title} {\enquote {\bibinfo {title} {Edge modes, edge
  currents, and gauge invariance in ${p}_{x}{+ip}_{y}$ superfluids and
  superconductors},}\ }\href {\doibase 10.1103/PhysRevB.69.184511} {\bibfield
  {journal} {\bibinfo  {journal} {Phys. Rev. B}\ }\textbf {\bibinfo {volume}
  {69}},\ \bibinfo {pages} {184511} (\bibinfo {year} {2004})}\BibitemShut
  {NoStop}%
\bibitem [{\citenamefont {Fendley}\ \emph {et~al.}(2007)\citenamefont
  {Fendley}, \citenamefont {Fisher},\ and\ \citenamefont
  {Nayak}}]{FendleyFisherNayak}%
  \BibitemOpen
  \bibfield  {author} {\bibinfo {author} {\bibfnamefont {Paul}\ \bibnamefont
  {Fendley}}, \bibinfo {author} {\bibfnamefont {Matthew P.~A.}\ \bibnamefont
  {Fisher}}, and\ \bibinfo {author} {\bibfnamefont {Chetan}\ \bibnamefont
  {Nayak}},\ }\bibfield  {title} {\enquote {\bibinfo {title} {{Edge states and
  tunneling of non-Abelian quasiparticles in the $\nu=5/2$ quantum Hall state
  and $p+ip$ superconductors}},}\ }\href {\doibase 10.1103/PhysRevB.75.045317}
  {\bibfield  {journal} {\bibinfo  {journal} {Phys. Rev. B}\ }\textbf {\bibinfo
  {volume} {75}},\ \bibinfo {pages} {045317} (\bibinfo {year}
  {2007})}\BibitemShut {NoStop}%
\bibitem [{\citenamefont {Grosfeld}\ and\ \citenamefont
  {Stern}(2011)}]{GrosfeldStern}%
  \BibitemOpen
  \bibfield  {author} {\bibinfo {author} {\bibfnamefont {Eytan}\ \bibnamefont
  {Grosfeld}}\ and\ \bibinfo {author} {\bibfnamefont {Ady}\ \bibnamefont
  {Stern}},\ }\bibfield  {title} {\enquote {\bibinfo {title} {{Observing
  Majorana bound states of Josephson vortices in topological
  superconductors}},}\ }\href {\doibase 10.1073/pnas.1101469108} {\bibfield
  {journal} {\bibinfo  {journal} {Proc. Natl. Acad. Sci.}\ }\textbf {\bibinfo
  {volume} {108}},\ \bibinfo {pages} {11810--11814} (\bibinfo {year}
  {2011})}\BibitemShut {NoStop}%
\bibitem [{Note14()}]{Note14}%
  \BibitemOpen
  \bibinfo {note} {Superconductivity technically does not allow for continuous
  ramping of the flux, but this barrier can be easily avoided. For the purpose
  of this thought experiment one can imagine temporarily snaking the flux so
  that it threads the $\nu = 2/3$ regions but avoids passing through the
  trenches. Once a value of $h/2e$ is reached, the flux can then be moved
  entirely within the cylinder.}\BibitemShut {Stop}%
\bibitem [{Note15()}]{Note15}%
  \BibitemOpen
  \bibinfo {note} {In fact, modularity requires that the set of all anyons
  decomposes into pairs $(\protect \mathcal {A}, \protect \mathcal
  {A}\varepsilon )$, where the Fibonacci anyon completely factorizes within the
  fusion rules. That is, $\protect \mathcal {A}\varepsilon \times \protect
  \mathcal {B} \sim (\protect \mathcal {A}\times \protect \mathcal
  {B})\varepsilon $ and $\protect \mathcal {A}\varepsilon \times \protect
  \mathcal {B}\varepsilon \sim (\protect \mathcal {A}\times \protect \mathcal
  {B})( {\protect \mathds {1}} +\varepsilon )$.}\BibitemShut {Stop}%
\bibitem [{\citenamefont {Zamolodchikov}(1985)}]{Zamolodchikov:1985}%
  \BibitemOpen
  \bibfield  {author} {\bibinfo {author} {\bibfnamefont {A.~B.}\ \bibnamefont
  {Zamolodchikov}},\ }\bibfield  {title} {\enquote {\bibinfo {title} {{Infinite
  Additional Symmetries in Two-Dimensional Conformal Quantum Field Theory}},}\
  }\href {\doibase 10.1007/BF01036128} {\bibfield  {journal} {\bibinfo
  {journal} {Theor. Math. Phys.}\ }\textbf {\bibinfo {volume} {65}},\ \bibinfo
  {pages} {1205--1213} (\bibinfo {year} {1985})}\BibitemShut {NoStop}%
\bibitem [{Note16()}]{Note16}%
  \BibitemOpen
  \bibinfo {note} {It is important to note that $1$ and $\protect \mathaccentV
  {tilde}07E\epsilon $ are the only primary fields when considering the full
  $G_2$ Kac-Moody algebra. The $\protect \mathaccentV {tilde}07E\epsilon $
  tower $(h_{\protect \mathaccentV {tilde}07E\epsilon } = 2/5)$ contains a
  multiplet of $7$ fields, all with scaling dimension $2/5$, transforming in
  one of the fundamental representations of $G_2$. However, these $7$ fields
  are related to each other by $G_2$ symmetry transformations or, in other
  words, are related by operator product expansions with the currents in
  Eqs.~\protect \textup {\hbox {\mathsurround \z@ \protect \normalfont
  (\ignorespaces \ref {eq:G2Spin1}\unskip \@@italiccorr )}}. Physically, the
  fields differ by bosonic excitations at the edge and therefore correspond to
  the same bulk anyon.}\BibitemShut {Stop}%
\bibitem [{Note17()}]{Note17}%
  \BibitemOpen
  \bibinfo {note} {Both perturbations are allowed since charge is conserved
  only mod $2e$ in our system.}\BibitemShut {Stop}%
\bibitem [{Note18()}]{Note18}%
  \BibitemOpen
  \bibinfo {note} {Although we noted earlier that one cannot define an electron
  operator in the low-energy subspace spanned by the generalized Majorana
  operators, electrons can still of course be added at high energies anywhere
  in the system's bulk.}\BibitemShut {Stop}%
\bibitem [{\citenamefont {Bonderson}(2007)}]{ParsaThesis}%
  \BibitemOpen
  \bibfield  {author} {\bibinfo {author} {\bibfnamefont {Parsa}\ \bibnamefont
  {Bonderson}},\ }\emph {\bibinfo {title} {{Non-Abelian Anyons and
  Interferometry}}},\ \href {http://thesis.library.caltech.edu/2447/} {Ph.D.
  thesis},\ \bibinfo  {school} {California Institute of Technology} (\bibinfo
  {year} {2007})\BibitemShut {NoStop}%
\bibitem [{\citenamefont {Bais}\ and\ \citenamefont
  {Slingerland}(2009)}]{BaisSlingerland}%
  \BibitemOpen
  \bibfield  {author} {\bibinfo {author} {\bibfnamefont {F.~A.}\ \bibnamefont
  {Bais}}\ and\ \bibinfo {author} {\bibfnamefont {J.~K.}\ \bibnamefont
  {Slingerland}},\ }\bibfield  {title} {\enquote {\bibinfo {title}
  {Condensate-induced transitions between topologically ordered phases},}\
  }\href {\doibase 10.1103/PhysRevB.79.045316} {\bibfield  {journal} {\bibinfo
  {journal} {Phys. Rev. B}\ }\textbf {\bibinfo {volume} {79}},\ \bibinfo
  {pages} {045316} (\bibinfo {year} {2009})}\BibitemShut {NoStop}%
\bibitem [{Note19()}]{Note19}%
  \BibitemOpen
  \bibinfo {note} {Bais and Slingerland consider a more general case of
  `bosons' with $d>1$, though we will not need to consider this more complex
  situation here.}\BibitemShut {Stop}%
\bibitem [{Note20()}]{Note20}%
  \BibitemOpen
  \bibinfo {note} {Vortex condensation can actually generate \protect \emph
  {new} quasiparticles depending on the precise structure of the condensate;
  for interesting recent examples see Ref.~\protect \rev@citealp
  {SurfaceTopologicalOrder1,SurfaceTopologicalOrder2,SurfaceTopologicalOrder3,%
SurfaceTopologicalOrder4}.}\BibitemShut {Stop}%
\bibitem [{\citenamefont {Lecheminant}\ \emph {et~al.}(2002)\citenamefont
  {Lecheminant}, \citenamefont {Gogolin},\ and\ \citenamefont
  {Nersesyan}}]{LGN}%
  \BibitemOpen
  \bibfield  {author} {\bibinfo {author} {\bibfnamefont {P.}~\bibnamefont
  {Lecheminant}}, \bibinfo {author} {\bibfnamefont {A.~O.}\ \bibnamefont
  {Gogolin}}, and\ \bibinfo {author} {\bibfnamefont {A.~A.}\ \bibnamefont
  {Nersesyan}},\ }\bibfield  {title} {\enquote {\bibinfo {title} {{Criticality
  in self-dual sine-Gordon models}},}\ }\href {\doibase%
  10.1016/S0550-3213(02)00474-1} {\bibfield  {journal} {\bibinfo  {journal}
  {Nucl. Phys. B}\ }\textbf {\bibinfo {volume} {639}},\ \bibinfo {pages}
  {502--523} (\bibinfo {year} {2002})}\BibitemShut {NoStop}%
\bibitem [{Note21()}]{Note21}%
  \BibitemOpen
  \bibinfo {note} {We use rescaled fields compared to Ref.~\protect
  \rev@citealp {LGN} to highlight the relationship with our $\nu = 2/3$
  problem.}\BibitemShut {Stop}%
\bibitem [{\citenamefont {Levin}\ and\ \citenamefont {Wen}(2005)}]{Levin05}%
  \BibitemOpen
  \bibfield  {author} {\bibinfo {author} {\bibfnamefont {Michael~A.}\
  \bibnamefont {Levin}}\ and\ \bibinfo {author} {\bibfnamefont {Xiao-Gang}\
  \bibnamefont {Wen}},\ }\bibfield  {title} {\enquote {\bibinfo {title}
  {String-net condensation:\quad{}a physical mechanism for topological
  phases},}\ }\href {\doibase 10.1103/PhysRevB.71.045110} {\bibfield  {journal}
  {\bibinfo  {journal} {Phys. Rev. B}\ }\textbf {\bibinfo {volume} {71}},\
  \bibinfo {pages} {045110} (\bibinfo {year} {2005})}\BibitemShut {NoStop}%
\bibitem [{\citenamefont {Fidkowski}\ \emph {et~al.}(2009)\citenamefont
  {Fidkowski}, \citenamefont {Freedman}, \citenamefont {Nayak}, \citenamefont
  {Walker},\ and\ \citenamefont {Wang}}]{Fidkowski06}%
  \BibitemOpen
  \bibfield  {author} {\bibinfo {author} {\bibfnamefont {Lukasz}\ \bibnamefont
  {Fidkowski}}, \bibinfo {author} {\bibfnamefont {Michael}\ \bibnamefont
  {Freedman}}, \bibinfo {author} {\bibfnamefont {Chetan}\ \bibnamefont
  {Nayak}}, \bibinfo {author} {\bibfnamefont {Kevin}\ \bibnamefont {Walker}},
  and\ \bibinfo {author} {\bibfnamefont {Zhenghan}\ \bibnamefont {Wang}},\
  }\bibfield  {title} {\enquote {\bibinfo {title} {{From String Nets to
  Nonabelions}},}\ }\href {\doibase 10.1007/s00220-009-0757-9} {\bibfield
  {journal} {\bibinfo  {journal} {Commun. Math. Phys.}\ }\textbf {\bibinfo
  {volume} {287}},\ \bibinfo {pages} {805--827} (\bibinfo {year}
  {2009})}\BibitemShut {NoStop}%
\bibitem [{\citenamefont {Fendley}\ \emph {et~al.}(2013)\citenamefont
  {Fendley}, \citenamefont {Isakov},\ and\ \citenamefont {Troyer}}]{Fendley13}%
  \BibitemOpen
  \bibfield  {author} {\bibinfo {author} {\bibfnamefont {Paul}\ \bibnamefont
  {Fendley}}, \bibinfo {author} {\bibfnamefont {Sergei~V.}\ \bibnamefont
  {Isakov}}, and\ \bibinfo {author} {\bibfnamefont {Matthias}\ \bibnamefont
  {Troyer}},\ }\bibfield  {title} {\enquote {\bibinfo {title} {Fibonacci
  topological order from quantum nets},}\ }\href {\doibase%
  10.1103/PhysRevLett.110.260408} {\bibfield  {journal} {\bibinfo  {journal}
  {Phys. Rev. Lett.}\ }\textbf {\bibinfo {volume} {110}},\ \bibinfo {pages}
  {260408} (\bibinfo {year} {2013})}\BibitemShut {NoStop}%
\bibitem [{\citenamefont {Qi}\ \emph {et~al.}(2013)\citenamefont {Qi},
  \citenamefont {Jiang}, \citenamefont {Barkeshli},\ and\ \citenamefont
  {Thomale}}]{QiFibonacci}%
  \BibitemOpen
  \bibfield  {author} {\bibinfo {author} {\bibfnamefont {Xiao-Liang}\
  \bibnamefont {Qi}}, \bibinfo {author} {\bibfnamefont {Hongchen}\ \bibnamefont
  {Jiang}}, \bibinfo {author} {\bibfnamefont {Maissam}\ \bibnamefont
  {Barkeshli}}, and\ \bibinfo {author} {\bibfnamefont {Ronny}\ \bibnamefont
  {Thomale}},\ }\href@noop {} {} (\bibinfo {year} {2013}),\ \bibinfo {note} {in
  preparation}\BibitemShut {NoStop}%
\bibitem [{\citenamefont {Heersche}\ \emph {et~al.}(2007)\citenamefont
  {Heersche}, \citenamefont {Jarillo-Herrero}, \citenamefont {Oostinga},
  \citenamefont {Vandersypen},\ and\ \citenamefont
  {Morpurgo}}]{HeerscheGrapheneSupercurrent}%
  \BibitemOpen
  \bibfield  {author} {\bibinfo {author} {\bibfnamefont {Hubert~B.}\
  \bibnamefont {Heersche}}, \bibinfo {author} {\bibfnamefont {Pablo}\
  \bibnamefont {Jarillo-Herrero}}, \bibinfo {author} {\bibfnamefont
  {Jeroen~B.}\ \bibnamefont {Oostinga}}, \bibinfo {author} {\bibfnamefont
  {Lieven M.~K.}\ \bibnamefont {Vandersypen}}, and\ \bibinfo {author}
  {\bibfnamefont {Alberto~F.}\ \bibnamefont {Morpurgo}},\ }\bibfield  {title}
  {\enquote {\bibinfo {title} {{Bipolar supercurrent in graphene}},}\ }\href
  {\doibase 10.1038/nature05555} {\bibfield  {journal} {\bibinfo  {journal}
  {Nature}\ }\textbf {\bibinfo {volume} {446}},\ \bibinfo {pages} {56--59}
  (\bibinfo {year} {2007})}\BibitemShut {NoStop}%
\bibitem [{\citenamefont {Du}\ \emph {et~al.}(2008)\citenamefont {Du},
  \citenamefont {Skachko},\ and\ \citenamefont {Andrei}}]{DuSkachkoAndrei08}%
  \BibitemOpen
  \bibfield  {author} {\bibinfo {author} {\bibfnamefont {Xu}~\bibnamefont
  {Du}}, \bibinfo {author} {\bibfnamefont {Ivan}\ \bibnamefont {Skachko}},
  and\ \bibinfo {author} {\bibfnamefont {Eva~Y.}\ \bibnamefont {Andrei}},\
  }\bibfield  {title} {\enquote {\bibinfo {title} {{Josephson current and
  multiple Andreev reflections in graphene SNS junctions}},}\ }\href {\doibase%
  10.1103/PhysRevB.77.184507} {\bibfield  {journal} {\bibinfo  {journal} {Phys.
  Rev. B}\ }\textbf {\bibinfo {volume} {77}},\ \bibinfo {pages} {184507}
  (\bibinfo {year} {2008})}\BibitemShut {NoStop}%
\bibitem [{\citenamefont {Ojeda-Aristizabal}\ \emph {et~al.}(2009)\citenamefont
  {Ojeda-Aristizabal}, \citenamefont {Ferrier}, \citenamefont {Gu\'eron},\ and\
  \citenamefont {Bouchiat}}]{BouchiatGrapheneSNS09}%
  \BibitemOpen
  \bibfield  {author} {\bibinfo {author} {\bibfnamefont {C.}~\bibnamefont
  {Ojeda-Aristizabal}}, \bibinfo {author} {\bibfnamefont {M.}~\bibnamefont
  {Ferrier}}, \bibinfo {author} {\bibfnamefont {S.}~\bibnamefont {Gu\'eron}},
  and\ \bibinfo {author} {\bibfnamefont {H.}~\bibnamefont {Bouchiat}},\
  }\bibfield  {title} {\enquote {\bibinfo {title} {Tuning the proximity effect
  in a superconductor-graphene-superconductor junction},}\ }\href {\doibase%
  10.1103/PhysRevB.79.165436} {\bibfield  {journal} {\bibinfo  {journal} {Phys.
  Rev. B}\ }\textbf {\bibinfo {volume} {79}},\ \bibinfo {pages} {165436}
  (\bibinfo {year} {2009})}\BibitemShut {NoStop}%
\bibitem [{\citenamefont {{Li}}\ \emph {et~al.}(2009)\citenamefont {{Li}},
  \citenamefont {{Cai}}, \citenamefont {{An}}, \citenamefont {{Kim}},
  \citenamefont {{Nah}}, \citenamefont {{Yang}}, \citenamefont {{Piner}},
  \citenamefont {{Velamakanni}}, \citenamefont {{Jung}}, \citenamefont
  {{Tutuc}}, \citenamefont {{Banerjee}}, \citenamefont {{Colombo}},\ and\
  \citenamefont {{Ruoff}}}]{Li09}%
  \BibitemOpen
  \bibfield  {author} {\bibinfo {author} {\bibfnamefont {X.}~\bibnamefont
  {{Li}}}, \bibinfo {author} {\bibfnamefont {W.}~\bibnamefont {{Cai}}},
  \bibinfo {author} {\bibfnamefont {J.}~\bibnamefont {{An}}}, \bibinfo {author}
  {\bibfnamefont {S.}~\bibnamefont {{Kim}}}, \bibinfo {author} {\bibfnamefont
  {J.}~\bibnamefont {{Nah}}}, \bibinfo {author} {\bibfnamefont
  {D.}~\bibnamefont {{Yang}}}, \bibinfo {author} {\bibfnamefont
  {R.}~\bibnamefont {{Piner}}}, \bibinfo {author} {\bibfnamefont
  {A.}~\bibnamefont {{Velamakanni}}}, \bibinfo {author} {\bibfnamefont
  {I.}~\bibnamefont {{Jung}}}, \bibinfo {author} {\bibfnamefont
  {E.}~\bibnamefont {{Tutuc}}}, \bibinfo {author} {\bibfnamefont {S.~K.}\
  \bibnamefont {{Banerjee}}}, \bibinfo {author} {\bibfnamefont
  {L.}~\bibnamefont {{Colombo}}}, and\ \bibinfo {author} {\bibfnamefont
  {R.~S.}\ \bibnamefont {{Ruoff}}},\ }\bibfield  {title} {\enquote {\bibinfo
  {title} {{Large-Area Synthesis of High-Quality and Uniform Graphene Films on
  Copper Foils}},}\ }\href {\doibase 10.1126/science.1171245} {\bibfield
  {journal} {\bibinfo  {journal} {Science}\ }\textbf {\bibinfo {volume}
  {324}},\ \bibinfo {pages} {1312--1314} (\bibinfo {year} {2009})},\ \Eprint
  {http://arxiv.org/abs/0905.1712} {arXiv:0905.1712 [cond-mat.mtrl-sci]}
  \BibitemShut {NoStop}%
\bibitem [{\citenamefont {Svore}()}]{Svore}%
  \BibitemOpen
  \bibfield  {author} {\bibinfo {author} {\bibfnamefont {K.}~\bibnamefont
  {Svore}},\ }\href@noop {} {}\bibinfo {note} {Private
  communication}\BibitemShut {NoStop}%
\bibitem [{\citenamefont {Bonderson}\ \emph {et~al.}(2013)\citenamefont
  {Bonderson}, \citenamefont {Nayak},\ and\ \citenamefont
  {Qi}}]{SurfaceTopologicalOrder1}%
  \BibitemOpen
  \bibfield  {author} {\bibinfo {author} {\bibfnamefont {Parsa}\ \bibnamefont
  {Bonderson}}, \bibinfo {author} {\bibfnamefont {Chetan}\ \bibnamefont
  {Nayak}}, and\ \bibinfo {author} {\bibfnamefont {Xiao-Liang}\ \bibnamefont
  {Qi}},\ }\href@noop {} {\enquote {\bibinfo {title} {{A Time-Reversal
  Invariant Topological Phase at the Surface of a 3D Topological Insulator}},}\
  } (\bibinfo {year} {2013}),\ \bibinfo {note} {unpublished},\ \Eprint
  {http://arxiv.org/abs/1306.3230} {arXiv:1306.3230 [cond-mat.str-el]}
  \BibitemShut {NoStop}%
\bibitem [{\citenamefont {Wang}\ \emph {et~al.}(2013)\citenamefont {Wang},
  \citenamefont {Potter},\ and\ \citenamefont
  {Senthil}}]{SurfaceTopologicalOrder2}%
  \BibitemOpen
  \bibfield  {author} {\bibinfo {author} {\bibfnamefont {Chong}\ \bibnamefont
  {Wang}}, \bibinfo {author} {\bibfnamefont {Andrew~C.}\ \bibnamefont
  {Potter}}, and\ \bibinfo {author} {\bibfnamefont {T.}~\bibnamefont
  {Senthil}},\ }\href@noop {} {\enquote {\bibinfo {title} {{Classification of
  interacting electronic topological insulators in three dimensions}},}\ }
  (\bibinfo {year} {2013}),\ \bibinfo {note} {unpublished},\ \Eprint
  {http://arxiv.org/abs/1306.3238} {arXiv:1306.3238 [cond-mat.str-el]}
  \BibitemShut {NoStop}%
\bibitem [{\citenamefont {Chen}\ \emph {et~al.}(2013)\citenamefont {Chen},
  \citenamefont {Fidkowski},\ and\ \citenamefont
  {Vishwanath}}]{SurfaceTopologicalOrder3}%
  \BibitemOpen
  \bibfield  {author} {\bibinfo {author} {\bibfnamefont {Xie}\ \bibnamefont
  {Chen}}, \bibinfo {author} {\bibfnamefont {Lukasz}\ \bibnamefont
  {Fidkowski}}, and\ \bibinfo {author} {\bibfnamefont {Ashvin}\ \bibnamefont
  {Vishwanath}},\ }\href@noop {} {\enquote {\bibinfo {title} {{Symmetry
  Enforced Non-Abelian Topological Order at the Surface of a Topological
  Insulator}},}\ } (\bibinfo {year} {2013}),\ \bibinfo {note} {unpublished},\
  \Eprint {http://arxiv.org/abs/1306.3250} {arXiv:1306.3250 [cond-mat.str-el]}
  \BibitemShut {NoStop}%
\bibitem [{\citenamefont {Metlitski}\ \emph {et~al.}(2013)\citenamefont
  {Metlitski}, \citenamefont {Kane},\ and\ \citenamefont
  {Fisher}}]{SurfaceTopologicalOrder4}%
  \BibitemOpen
  \bibfield  {author} {\bibinfo {author} {\bibfnamefont {Max~A.}\ \bibnamefont
  {Metlitski}}, \bibinfo {author} {\bibfnamefont {C.~L.}\ \bibnamefont {Kane}},
  and\ \bibinfo {author} {\bibfnamefont {Matthew P.~A.}\ \bibnamefont
  {Fisher}},\ }\href@noop {} {\enquote {\bibinfo {title} {A symmetry-respecting
  topologically-ordered surface phase of 3d electron topological insulators},}\
  } (\bibinfo {year} {2013}),\ \bibinfo {note} {unpublished},\ \Eprint
  {http://arxiv.org/abs/1306.3286} {arXiv:1306.3286 [cond-mat.str-el]}
  \BibitemShut {NoStop}%
\end{thebibliography}%

\end{document}